# LECTURES ON STATISTICAL MECHANICS

*Allan N. Kaufman[1]*

*Physics Department, University of California*
*Berkeley, California*

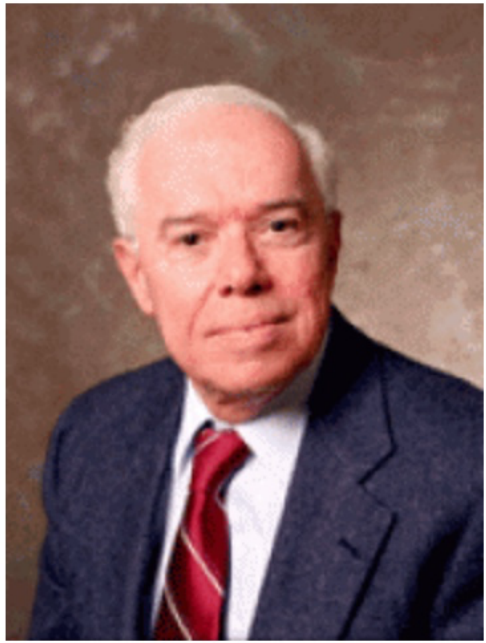

*Allan N. Kaufman, 1927-2022*

*Lecture Notes for Physics 212A and 212B, University of California, Berkeley*

*Transcribed and edited by Bruce I. Cohen[2] and Alain J. Brizard[3]*

[1]Physics Department, University of California Berkeley, CA, USA (deceased)
[2]Physics Division, Lawrence Livermore National Laboratory, CA, USA (semi-retired)
[3]Physics Department, St. Michael's College, Burlington, VT, USA

# Abstract

Presented here is a transcription of the lecture notes from Professor Allan N. Kaufman's graduate statistical mechanics course Physics 212A and 212B at the University of California Berkeley from the 1972-1973 academic year. 212A addressed equilibrium statistical mechanics with topics: fundamentals (micro-canonical and sub-canonical ensembles, adiabatic law and action conservation, fluctuations, pressure and virial theorem), classical fluids and other systems (equation of state, deviations from ideality, virial coefficients and van der Waals potential, canonical ensemble and partition function, quasi-static evolution, grand-canonical ensemble and partition function, chemical potential, simple model of a phase transition, quantum virial expansion, numerical simulation of equations of state and phase transition), chemical equilibrium (systems with multiple species and chemical reactions, law of mass action, Saha equation, chemical equilibrium including ionization and excited states), and long-range interactions (examples of interactions: Coulomb, dipole, gravitational; Debye-Hückel theory and shielding). 212B addressed non-equilibrium statistical mechanics with topics: fundamentals (definitions: realizations, moments, characteristic function, and discrete variables), Brownian motion (Langevin equation, fluctuation-dissipation theorem, spatial diffusion, Boltzmann's H theorem), Liouville and Klimontovich equations, Landau equation (derivation, elaboration and H theorem, irreversibility), Markov processes and Fokker-Planck equation (derivations of the Fokker-Planck equation and a master equation), linear response and transport theory (linear Boltzmann equation, linear response theory of Kubo and Mori, relation of entropy production to electrical conductivity, transport relations and coefficients, normal mode solutions of the transport equations, sketch of a generalized Langevin equation method for transport theory), and an introduction to non-equilibrium quantum statistical mechanics.

# Table of Contents

2. Nonequilibrium statistical mechanics
    a. Fundamentals
        i. Definitions of a realization, moments, characteristic function, and discrete variables
        ii. Derivation of the central limit theorem
        iii. Random processes, spectral density, and correlation function
    b. Brownian motion
        i. Brownian motion – Langevin equation
        ii. Fluctuation-dissipation theorem
        iii. Spatial diffusion and diffusivity
        iv. Boltzmann's H theorem
    c. Liouville and Klimontovich equations
        i. Liouville equation
        ii. Klimontovich phase space and distribution function
    d. Landau equation
        i. Derivation of the Landau equation
        ii. Elaboration of the Landau equation and derivation of an H-theorem
        iii. Irreversibility
    e. Markov processes and the Fokker-Planck equation
        i. Expansion of Chapman-Kolmogorov equation to derive the Fokker-Planck equation
        ii. Discontinuous Markov process and derivation of a master equation
    f. Linear response theory, linear Boltzmann equation, and transport theory
        i. Evolution of velocity angle probability distribution due to scattering
        ii. Linear response fundamentals
        iii. Linear Boltzmann equation
        iv. Collision models and conductivity
        v. Linear response theory and Kubo formulae
        vi. Relation of entropy production to electrical conductivity
        vii. Transport relations and coefficients
        viii. Normal mode solutions of the transport equations
        ix. Generalized Langevin method for transport relations -- sketch
    g. Non-equilibrium quantum statistical mechanics

**Editor's Addendum: Appendix -- Thermodynamic Potentials, Maxwell Relations and Identities**

**References**

# Foreword

Allan Kaufman (1927-2022) grew up in the Hyde Park neighborhood of Chicago not far from the University of Chicago. Allan attended the University of Chicago for both his undergraduate and doctoral degrees in physics. Allan's doctoral thesis advisor was Murph Goldberger who was relatively new to the faculty at Chicago and just five years older than Allan. Allan did a theoretical thesis on a strong-coupling theory of meson-nucleon scattering. Allan published an autobiographical article entitled "A half-century in plasma physics" in A. N. Kaufman, Journal of Physics: Conference Series **169** (2009) 012002.

Allan worked at Lawrence Livermore Laboratory from June 1953 through 1963. While at Livermore Laboratory he taught the one-year graduate course in electricity and magnetism in 1959-1963 at UC Berkeley. In 1963 he first taught the first semester of the graduate course in Theoretical Plasma Physics 242 at Berkeley. He taught the plasma theory course at UCLA in the 1964-1965 school year while on leave from Livermore before joining the faculty at UC Berkeley in the 1965 school year. Allan frequently taught the graduate plasma theory course and the graduate statistical mechanics course Physics 212A and B until his retirement from teaching in 1998.

These lecture notes were from Kaufman's graduate statistical mechanics course in the 1972-1973 academic year. The notes follow the chronological order of the lectures. The equations and derivations are as Kaufman presented, and the text is a reconstruction of Kaufman's discussion and commentary. Equation numbers were added to facilitate the exposition of the derivations. Although the material is fifty years old, the mathematical rigor and elegance of Kaufman's treatment of the subject matter should still be useful to students interested in learning the fundamentals of statistical mechanics. A few of the equations that are important results and conclusions in the analysis are labeled as "Theorems" to draw attention to them; these are not necessarily formal theorems in the mathematical sense but are consistent with terminology in physics textbooks. Editor's Notes, Editor's Addendum, and Reviewer's Comments have been inserted with the goal of providing additional useful material, updates, and references. In this regard we are very much indebted to the three reviewers of these lecture notes who were very energetic and whose suggestions have added considerable value, which deserves attribution and recognition. In particular, we thank Dominique Escande and Martin Lemoine for their many valuable comments in reviewing the manuscript. These lecture notes are intended as a resource.

The focus of Kaufman's research at Berkeley was plasma physics. Although these lecture notes address the general subject of statistical mechanics, there is a definite emphasis on plasma physics in the examples and applications. Statistical mechanics is foundational for plasma physics. Examples of specific material in these lecture notes addressing plasma physics topics are as follows: Hamiltonian theory for a kinetic plasma with Coulomb forces, rigorous derivation of the pressure and virial expansion, partition function and statistics for an unmagnetized plasma in thermal equilibrium with electromagnetic waves, the Bohr-Van Leeuwen theorem in an equilibrium plasma, derivation of the Poisson-Boltzmann equation for a Coulomb model of a plasma in thermal equilibrium, analysis of Debye shielding and quasi-neutrality conditions, derivation of the Maxwell-Boltzmann equilibrium distribution, Hamiltonian theory of a non-equilibrium plasma with electromagnetic fields, Langevin equation model of Brownian motion in a plasma with Coulomb forces, the fluctuation-dissipation theorem, Boltzmann's H theorem, derivations of Liouville, Klimontovich, Vlasov, Landau, Boltzmann, and Fokker-Planck equations in a plasma, linear response theory and derivation of transport equations and coefficients, collisions and conductivities (electrical and thermal) in a plasma.

Bruce Cohen joined Kaufman's research group during the 1971-1972 academic year and received his Ph. D. in 1975.  These lecture notes were word-processed in 2023 after Allan's death in December of 2022.  Allan encouraged Cohen to word-process his notes on plasma theory and statistical mechanics so that they could be shared.   In 2019 Kaufman and Cohen published Cohen's transcription of Kaufman's lecture notes from the graduate plasma physics course at Berkeley, Physics 242: A. N. Kaufman and B. I. Cohen, *Theoretical Plasma Physics*, J. Plasma Phys. **85**, 205850601 (2019), doi:10.1017/S0022377819000667

Alain Brizard worked at Berkeley as a post-doctoral researcher from 1989-1992 with Kaufman, from 1992-1994 with Ken Fowler, and summers 1995-2000 with Kaufman and Jonathan Wurtele.  Alain was a research collaborator with Kaufman for three decades.  Brizard published many papers with Kaufman and the book **Ray Tracing and Beyond**, with E.R. Tracy, A.S. Richardson, and Kaufman , Cambridge University Press (2014).  Brizard reviewed A. N. Kaufman and B. I. Cohen, *Theoretical Plasma Physics*, J. Plasma Phys. **85**, 205850601 (2019) and suggested valuable improvements before its publication.

Professor Kaufman's work on these lecture notes was performed while he was employed as a Professor of Physics at the University of California Berkeley.  Professor Kaufman's separate research activity was funded in part by the United States Department of Energy.  Bruce Cohen's work on these lecture notes was *pro bono*. Cohen's separate research activity has been funded at the Lawrence Livermore National Laboratory by the United States Department of Energy.  Alain Brizard is a Professor of Physics at St. Michael's College in Vermont, and his separate research activity in the present and past has been supported by the United States National Science Foundation and the Department of Energy. Brizard's work on these lecture notes was not funded under his research grants.

Bruce I. Cohen
Alain J. Brizard

# 1. Equilibrium statistical mechanics

[*Editor's Note: In the first lecture of Physics 212A Kaufman discussed the syllabus and schedule for the lectures. Kaufman used CGS units with some customizations throughout his notes, e.g., Boltzmann's constant is set to unity. There was no textbook for the course. Some of the references for his lectures included L. D. Landau and E. M. LIfshitz, **Statistical Physics** (Landau and Lifshitz, 1969); R. C. Tolman, **The Principles of Statistical Mechanics** (Tolman, 1938) ; R. Kubo, **Statistical Mechanics** (Kubo, 1965); J. O. Hirschfelder, C. F. Curtiss, and R. B. Bird, **Molecular Theory of Gases and Liquids** (Hirschfelder, Curtiss, and Bird, 1954); F. Reif, **Fundamentals of Statistical and Thermal Physics** (Reif, 1965); H. B. Callen, **Thermodynamics and an Introduction to Thermostatics** (Callen, 1960); and R. Becker, **Theory of Heat** (Becker, 1967).*]

[*Reviewer Dominique Escande's Comment: Section 2.4 of (Sator, Pavloff, and Couëdel, 2023) provides a series of useful references in order to go further into the foundations of statistical mechanics.*]

## 1.a Fundamentals

Statistical mechanics provides a mathematical framework for bridging the gap between microscopic laws to macroscopic descriptions. Statistical mechanics is confronted with a set of dichotomies: equilibrium vs. non-equilibrium; a range of degrees of freedom from few (~10) to many (~$10^3$), to very many (~$10^{24}$), to a denumerable infinity, to an uncountable infinity; classical vs. quantum; relativistic vs. non-relativistic; closed vs. open systems; inert vs. chemically reactive; and many levels of description, e.g., exact, kinetic $f(\mathbf{x},\mathbf{v},t)$, fluid $n(\mathbf{x},t)$, $\mathbf{v}(\mathbf{x},t)$, $T(\mathbf{x},t)$. Statistical mechanics is home to the fundamental laws of thermodynamics: 0) A=B=C transitivity; 1) conservation of energy; 2) the change of entropy is non-negative $\Delta S \geq 0$; 3) entropy $S \to 0$ as temperature $T \to 0$. Statistical mechanics distinguishes between extensive and intensive properties of matter, *i.e.,* properties that are either volume dependent or independent, respectively.

### 1.a.i Postulate of equal probabilities

Definition: A *macroscopic state* is described by a set of partial information. A *microscopic state* can be described by a set of either classical information or quantum information that is a complete set of detailed information at the finest level including boundary and initial conditions.

Example: Consider N coupled harmonic oscillators with Hamiltonian *H* given by

$$H = \sum_{i=1}^{N} \tfrac{1}{2}(p_i^2 + \omega_i^2 q_i^2) + \lambda \sum_{ijk} c_{ijk} q_i q_j q_k \qquad (1.a.1)$$

with parameters: $N, \lambda, \{c_{ijk}\}, \{\omega_i\}$. In a finite-sized box, the energy eigenstates for an uncoupled system of harmonic oscillators are discretized and representative of quantum systems. Because $H$ in this example has no explicit time dependence, $H$ is a constant of the motion; and the macro-state can be characterized by its energy without knowledge of the initial conditions. There is only partial information available in this example.

How does one relate microscopic information to a macroscopic description?

Postulate: (Fundamental Postulate – R. C. Tolman) All micro-states consistent with the given partial information (macro-state) are equally probable.

Definition: $\Gamma(E_0) \equiv$ number of micro-states with $E < E_0$

At this point we drop the classical picture for a little while for pedagogic reasons, chiefly because counting and summing microstates over a discretized phase space resolves certain mathematical measure complications encountered in classical systems.

Definition: In the discretized quantum picture the probability of one given micro-state is

$$\text{probability} \equiv w_n = \begin{cases} \frac{1}{\Gamma(E_0)}, & E_n < E_0 \\ 0, & E_n > E_0 \end{cases} \tag{1.a.2}$$

In the limit of a very large number $\Gamma(E_0)$ we employ the correspondence principle, and in the volume $\Omega$ of allowed phase space one obtains

$$\Gamma(\Omega) = \lim_{h \to 0} \frac{\int_\Omega dq\,dp}{h^n} \tag{1.a.3}$$

For large $\Gamma$ there are many accessible microstates, and the probability of a given microstate with energy $E_n$ is a relatively smooth function of energy because the granularity of the energy is such fine scale.

### 1.a.ii Example: N uncoupled oscillators

Example: *N* uncoupled simple harmonic oscillators. Let *N*=3 and use a canonical transformation to action-angle variables:

$$\left.\begin{array}{c} q_i \equiv \sqrt{\frac{2J_i}{\omega_i}} \sin \theta_i \\ p_i \equiv \sqrt{2J_i \omega_i} \cos \theta_i \end{array}\right\} \to H = \sum_{i=1}^{3} \omega_i J_i \tag{1.a.4}$$

Further simplify by requiring by requiring $\omega_1 = \omega_2 = \omega_3 = \omega_0$. We note that $\{\theta_1, \theta_2, \theta_3\}$ are ignorable in $H$; hence, the actions $J_i$ are constants of the motion; and the energy is given by $E = \omega_0(J_1 + J_2 + J_3)$. The volume occupied by the system of 3 oscillators in phase space is a triangular solid in *J*-space, with vertices: $\{J_0, 0, 0\}$, $\{0, J_0, 0\}$, and $\{0, 0, J_0\}$ where $J_0 = E_0/\omega_0$ and a rectangular solid in $\theta$-space spanning $[0, 2\pi]$ in each of the three $\theta$ coordinates. The product volume in the $\{J, \theta\}$ phase space divided by $h^3$ yields the number of states:

$$\Gamma(E_0) = \frac{volume}{h^3} = \frac{(2\pi)^3}{h^3} \frac{1}{2}\left(\frac{E_0}{\omega_0}\right)^2 \frac{1}{3}\frac{E_0}{\omega_0} = \frac{1}{3!}\left(\frac{E_0}{\hbar\omega_0}\right)^3 \tag{1.a.5}$$

In (1.a.5) we are assuming that typically the number of states is large, i.e., $E_0 \gg \hbar\omega_0 N$. For the general case of $N$ oscillators,

$$\Gamma(E_0, 3) \to \Gamma(E_0, N) = \frac{1}{N!}\left(\frac{E_0}{\hbar\omega_0}\right)^N \tag{1.a.6}$$

If we allow each of $N$ oscillators to have any one of $M$ possible energy states, where

$$M \equiv \frac{E - \frac{1}{2}\hbar\omega_0}{\hbar\omega_0}$$

then Kubo (Kubo, 1965, p. 38) shows that the number of distinguishable states is given by

$$\Gamma(E, \omega_0) = \sum_M \frac{(N+M-1)!}{(N-1)!M!} \tag{1.a.7}$$

Example: To illustrate (1.a.7) consider N = 3 oscillators and M=4 energy levels, for which there are 15 states {(4,0,0), (3,1,0), (3,0,1), (2,2,0), (2,0,2), (2,1,1), (1,3,0), (1,0,3), (1,2,1), (1,1,2), (0,4,0), (0,0,4), (0,3,1), (0,1,3), (0,2,2)} in agreement with (N + M -1)!/[(N-1)! M!] = 6!/(2! 4!) = 15.

Definition: The specific oscillator energy is $\mathcal{E} \equiv \frac{E}{N}$

Returning to Eq.(1.a.6) and using the definition of the specific oscillator energy one obtains

$$\Gamma(E_0, N) = \frac{1}{N!}\left(\frac{E_0}{\hbar\omega_0}\right)^N = \frac{N^N}{N!}\left(\frac{\mathcal{E}_0}{\hbar\omega_0}\right)^N \approx \frac{e^N}{\sqrt{2\pi N}}\left(\frac{\mathcal{E}_0}{\hbar\omega_0}\right)^N \tag{1.a.8}$$

where in Eq.(1.a.8) we have made use of Stirling's approximation $N! \approx \sqrt{2\pi N}\,(N/e)^N$ for large $N$. We identify $\frac{\mathcal{E}_0}{\hbar\omega_0}$ as a basic quantum number. We note that if the basic quantum number is $O(10)$ and $N \sim 10^{10-20}$, $\Gamma$ is rather large.

Definition: Take the natural logarithm of the number of states and introduce the concept of entropy. If all states in $\Gamma$ are equally probable then define the entropy as

$$S(E_0, N) \equiv \ln\Gamma \to N\ln\left(\frac{\mathcal{E}_0}{\hbar\omega_0}\right) + N - \frac{1}{2}\ln(2\pi N) \approx N\ln\left(\frac{\mathcal{E}_0}{\hbar\omega_0}\right) + N \tag{1.a.9}$$

for large $N$. Thus, $S \sim O(N)$.

<u>Definition</u>: Introduce the specific entropy $\mathcal{S} \equiv S/N$. Hence,

$$\mathcal{S} = \ln\left(\frac{\varepsilon_0}{\hbar\omega_0}\right) + 1 \tag{1.a.10}$$

which has no *N* dependence ("normal dependence") and is a number of order unity.

[*Editor's Note*: *Prof. Kaufman remarked at this point that Problem 2-33 in (Kubo, 1965) addressing the correspondence principle was interesting and not at all obvious*.]

<u>Example</u>: Consider an ensemble of *N* atoms or molecules with a harmonic oscillator Hamiltonian. We will derive the specific entropy for an ideal gas.

The model Hamiltonian for an ideal gas of atoms or molecules is given by

$$H = \sum_{i=1}^{3N} \left[\frac{p_i^2}{2m} + \Phi\right] \tag{1.a.11}$$

where each atom or molecule has 3 degrees of freedom in its motion, and we consider a cube with volume $V = L^3$ and we assume the potential energy $\Phi = 0$. The magnitudes of the momentum components are constrained by the total energy for each oscillator: $p_1^2 + p_2^2 + p_3^2 < \left(\sqrt{2mE}\right)^2$. The phase-space volume is the product of the cubic volume *V* and the spherical volume $\frac{4\pi}{3}\left(\sqrt{2mE}\right)^3$, and the number of states for 3 degrees of freedom per oscillator scales as

$$\Gamma_{f=3} \sim \frac{\frac{4\pi}{3}\left(\sqrt{2mE}\right)^3 L^3}{h^3} \tag{1.a.12}$$

In (1.a.12) we note the quantum discretization. To derive the number of states for *N* atoms or molecules we begin with the volume of a 3*N* dimensional sphere is given by

$$V_{3N}(R) = \pi^{\frac{3N}{2}} R^{3N} / \Gamma\left(\frac{3N}{2} + 1\right)$$

where $\Gamma(z)$ denotes the gamma function. Note that since $\Gamma(n+1) = n!$ for any non-negative integer, the shorthand notation $\Gamma\left(\frac{3N}{2}+1\right) = \left(\frac{3N}{2}\right)!$ is used. Now we introduce the dimensionless "phase-space" radius

$$R = \frac{pL}{h} = \sqrt{2mE}L/h$$

where the volume in configuration space is $V = L^3$. We divide $V_{3N}(R)$ by $N!$ to eliminate permuations of indistinguishable states to obtain

$$\Gamma(E; V, N) = \frac{V_{3N}(R)}{N!} = \frac{(2\pi mE)^{3N/2} V^N}{h^{3N}\left(\frac{3N}{2}\right)!N!} \tag{1.a.13}$$

where *E* is the total energy for *N* particles with 3 degrees of freedom each. At this point we introduce a few definitions to facilitate reducing Eq.(1.a.13) to a more recognizable form.

Definition: The specific energy is $\mathcal{E} = E/N$. The particle density is then $n \equiv N/V$.

With these definitions, use of Stirling's approximation to remove the factorials, $\Gamma(z+1) \cong \sqrt{2\pi z}\left(\frac{z}{e}\right)^z$, for $z \gg 1$, and with $N \gg 1$ we obtain the following result from (1.a.13)

$$\Gamma \approx \left(\frac{4\pi}{3}\frac{m\mathcal{E}}{\bar{p}^2}\right)^{\frac{3N}{2}}\left(\frac{1}{n\Lambda^3}\right)^N \frac{e^{\frac{5N}{2}}}{\sqrt{6\pi N}} = \left(\frac{1}{n\Lambda^3}\right)^N \frac{e^{\frac{5N}{2}}}{\sqrt{6\pi N}} \tag{1.a.14}$$

where the average momentum per particle is $\bar{p} \equiv \sqrt{\frac{4\pi}{3}m\mathcal{E}}$ and the thermal deBroglie wavelength is $\Lambda \equiv \frac{h}{\bar{p}}$. $n\Lambda^3$ is the number of particles in a deBroglie cube, which must be a small number to justify a classical description. From Eq.(1.a.14) we calculate the entropy and recover the specific entropy of an ideal gas.

$$S = \ln \Gamma \rightarrow \mathcal{S} = \frac{S}{N} = \frac{\ln \Gamma}{N} = \frac{5}{2} - \ln(n\Lambda^3) - \frac{\ln N}{N} - \frac{\ln \pi}{N} - \frac{1}{2}\frac{\ln 6}{N} \approx \frac{5}{2} - \ln(n\Lambda^3) \tag{1.a.15}$$

### 1.a.iii Micro-canonical ensemble

Next we introduce the concepts of sub-canonical and micro-canonical ensembles.

Definitions: An ensemble of states for energies $E_N < E$ is a sub-canonical ensemble, and we denote the number of states by $\Gamma_\sigma$. The ensemble of states for energies $E - \delta E < E_n < E + \delta E$ is defined as a micro-canonical ensemble, and its number of states is denoted $\Gamma_\mu$.

Physical sub-canonical ensembles have monotonically increasing $\Gamma_\sigma$ as functions of increasing energy *E*. We can evaluate $\Gamma_\mu$ as follows using Eq.(1.a.15):

$$\Gamma_\mu(E, \delta E) = \Gamma_\sigma(E) - \Gamma_\sigma(E - \delta E) = \Gamma_\sigma(E)\left[1 - \frac{\Gamma_\sigma(E - \delta E)}{\Gamma_\sigma(E)}\right] = \Gamma_\sigma(E)\left[1 - \frac{e^{S(E-\delta E)}}{e^{S(E)}}\right]$$

$$\approx \Gamma_\sigma(E)\left[1 - \frac{e^{S(E) - \delta E \frac{dS}{dE} + \frac{1}{2}(\delta E)^2 \frac{d^2S}{dE^2} + \cdots}}{e^{S(E)}}\right] \approx \Gamma_\sigma(E)\left[1 - e^{-\beta\delta E + \frac{1}{2}(\delta E)^2 \frac{d\beta}{dE}}\right] \tag{1.a.16}$$

where $\beta \equiv \frac{dS}{dE} \equiv \frac{1}{T}$. Note that the specific energy $\frac{E}{N} \sim O(T)$, and hence the last term in the exponential on the right side of Eq.(1.a.16) is small compared to the $\beta\delta E$ term given the constraint $T \ll \delta E \ll E$, so that $\Gamma_\mu(E, \delta E) \sim \Gamma_\sigma(E)(1 - e^{-\beta\delta E})$. Furthermore, $e^{-\beta\delta E}$ is exponentially small; and hence, $\Gamma_\mu \sim \Gamma_\sigma$. The interpretation of this is that the number of states is a sharply increasing function of energy such that the volume of the hyper-sphere is dominated by the volume of the bounding annular shell, i.e., for $V \sim R^N \to \frac{\delta V}{V} \sim N \frac{\delta R}{R}$, $N >> 1$ and $\frac{\delta R}{R} \ll 1$, but $N \frac{\delta R}{R} = O(1)$. For the conditions $\varepsilon \ll \delta E \ll E \to \frac{1}{N} \ll \frac{\delta E}{E} \ll 1$, the system remains on the hyper-surface that can be parametrized in terms of the actions and fills it. The angle space is filled as well.

Now consider the classical entropy after Taylor-series expanding,

$$\Gamma_\mu(E, \delta E) = \Gamma_\sigma(E) - \Gamma_\sigma(E - \delta E) \approx \delta E \frac{d\Gamma_\sigma}{dE} + O(\delta E^2) \tag{1.a.17}$$

after Taylor-series expanding. As compared to (1.a.16), $\Gamma_\mu \propto \delta E$ as $\delta E \to 0$ rather than $\Gamma_\sigma$. The entropy is the logarithm of $\Gamma_\mu$

$$S_\mu = \ln \Gamma_\mu = \ln \delta E + \ln \frac{d\Gamma_\sigma}{dE} \tag{1.a.18}$$

The first term on the right side of Eq.(1.a.18) is a fixed additive term and small compared to the second term which is the natural logarithm of the density of states and is very large.

The classical micro-canonical entropy is to good approximation

$$S_{\mu,class} = \ln \frac{d\Gamma_\sigma}{dE} \tag{1.a.19}$$

<u>Example</u>: Calculate $S_{\mu,class}$ for the harmonic oscillator model of the ideal gas and compare it to the quantum entropy expression. We anticipate that if the two expressions are different, it is only due to constants. The classical micro-canonical entropy is given in Eq.(1.a.18), while the quantum entropy is given by

$$S_{qm} \equiv \ln \Gamma \to \Gamma = e^{S_{qm}} \to \frac{d\Gamma}{dE} = e^{S_{qm}} \frac{dS_{qm}}{dE} \equiv \beta e^{S_{qm}} \tag{1.a.20}$$

Using $\frac{d\Gamma}{dE} = \beta e^{S_{qm}}$ in the last term in Eq.(1.a.18)

$$S_{\mu,class} = \ln \delta E + \ln \beta e^{S_{qm}} = \ln \delta E + \ln \beta + S_{qm} = O(1) + O(1) + O(N) \approx S_{qm} \tag{1.a.21}$$

We note that we should introduce $h^N$ in the denominator in the expressions for $\Gamma$ to give the correct dimensionless units for the phase-space normalized volume. However, this results in no change in the final formulae due to taking the logarithm of a product expands into the sum of logarithms; and the $S_{qm} \approx O(N)$ term remains dominant.

Next consider the phase space of a system with many degrees of freedom whose trajectory in phase space is constrained by a Hamiltonian. Define a sub-domain in this phase space as a shell with thickness $\delta\ell$ and volume defined by $dpdq = dA\delta\ell$, and the thickness of the shell is parametrized by a variation in the total energy:

$$H(p,q) = E - \delta E, \quad \frac{\delta E}{\delta \ell} = |\nabla H|, \quad \delta\ell = \frac{\delta E}{|\nabla H(p,q)|} \quad (1.a.22)$$

which varies as a function of *p* and *q* in phase space. If the probability of the system occupying a given sub-domain in phase space is proportional to the volume of the sub-domain, then

$$\text{Probability} \propto dA\,\delta\ell = dA\frac{\delta E}{|\nabla H(p,q)|} \quad (1.a.23)$$

Theorem: (Boltzmann's Ergodic Hypothesis) The orbit of the system of microstates will completely fill the volume of the accessible phase space given the initial data constraining the degrees of freedom. Over time any sub-domain will be occupied for a time proportional to the sub-domain's volume, i.e., all accessible micro-states are equally probable over a long period of time.

As stated here the Ergodic Hypothesis has the difficulty that the orbit of the system is a one-dimensional manifold embedded within the energy surface and has a different measure than that of the energy surface. Thus, the orbit of the system cannot fill the energy surface in a strict sense. Hence, there is a need for refining the Ergodic Hypothesis as follows.

Theorem: (Quasi-Ergodic Hypothesis due to G. D. Birkhoff (Birkhoff, 1931) Every finite region on the energy surface is accessible.

Theorem: (Ergodic) A system spends equal times in equal volumes (except for a set of measure zero pathological initial conditions).

Corollary: For any integrable function of the phase-space coordinates $f(p,q)$, the time average of $f(p,q)$ is equal to its space average almost everywhere. This is a very important consequence of the ergodic theorem.

### 1.a.iv Non-equilibrium macro-states

We next take up the examination of non-equilibrium macro-states. Consider the simple example of a domain composed of two adjacent contiguous sub-domains I and II occupied by an ideal gas with numbers of particles and energies {$N_I, E_I$} and {$N_{II}, E_{II}$}. We further assume that the ideal gas is described by the same harmonic oscillator Hamiltonian introduced in Eq.(1.a.11). After an invisible membrane is removed we will allow transfer of energy between the two subsystems, but no losses to the exterior world. Thus,

$$E_I + E_{II} = const \equiv E \quad (1.a.24)$$

The accessible number of micro-states for the combined system before the membrane is removed is given by

$$\Gamma(E_I, E_{II}) = \Gamma_I(E_I)\Gamma_{II}(E_{II}) \tag{1.a.25}$$

from which follows that the entropy is given by

$$S(E_I, E_{II}) = \ln \Gamma(E_I, E_{II}) = \ln \Gamma_I(E_I) + \ln \Gamma_I(E_{II}) = S_I(E_I) + S_{II}(E_{II}) \tag{1.a.26}$$

After the membrane is removed, a constraint on the number of states is removed; and the final number of states can exceed the initial number of states:

$$\Gamma_{init} < \Gamma_{final}(E) = \int_0^E \Gamma(E_I, E_{II} = E - E_I) dE_I \tag{1.a.27}$$

Think of the integral in Eq.(1.a.27) as the sum over the number of possible energy states. Clearly the initial and final entropies satisfy $S_{init} < S_{final}$. The probability that the subsystem I has energy $E_I$ subject to the constraint that the total system energy is $E$, is given by the relative fraction of states in system I having energy $E_I$ and system II having energy $E_{II}$, which given by the product of the respective numbers of microstates, divided by the total number of states having energy $E = E_I + E_{II}$.

$$\rho(E_I|E) = \frac{\Gamma(E_I, E_{II})}{\Gamma(E)} = \frac{e^{S(E_I, E_{II}) = S_I(E_I) + S_{II}(E_{II})}}{e^{S(E)}} \propto e^{S_I(E_I)} e^{S_{II}(E_{II} = E - E_I)} \tag{1.a.28}$$

The exponential $e^{S_I(E_I)}$ in Eq.(1.a.28) is a monotonically increasing function of $E_I$, while the exponential $e^{S_{II}(E_{II} = E - E_I)}$ is a monotonically decreasing function of $E_I$. Hence, the probability $\rho$ has a sharp peak at some value $E_I = E_{I*}$ satisfying

$$\frac{\partial \rho}{\partial E_I} \propto \frac{\partial S_I}{\partial E_I} + \frac{\partial S_{II}}{\partial E_{II}}(-1) = 0 \tag{1.a.29}$$

Using our earlier introduced definitions of $\beta$ and $T$, $\beta \equiv \frac{dS}{dE} \equiv \frac{1}{T}$, Eq.(1.a.29) yields the following relation

$$\beta_I(E_I = E_{I*}) = \beta_{II}(E_{II} = E - E_{I*}) \tag{1.a.30}$$

at $E_I = E_{I*}$ where $\rho$ achieves a sharp maximum, i.e., its equilibrium; and $T_I = T_{II}$.

We have not yet specified what systems *I* and *II* are comprised of. For example, recall the expression for the specific entropy of the system comprised of harmonic oscillators in (1.a.10)

$$S = \ln\left(\frac{\varepsilon}{\hbar \omega_0}\right) + 1$$

and the expression for the specific entropy of an ideal gas given in (1.a.15)

$$S = \frac{5}{2} - \ln(n\Lambda^3)$$

where the thermal deBroglie wavelength is $\Lambda \equiv \frac{h}{\bar{p}}$ and $\bar{p} \equiv \sqrt{\frac{4\pi}{3}m\mathcal{E}}$ were introduced earlier preceding Eq.(1.a.14). For systems comprised of an ideal gas each of the three degrees of freedom per atom has $\frac{1}{2}T$ energy, then $\mathcal{E} = \frac{3}{2}T$ in the specific entropy. Moreover,

$$\beta_I = \frac{1}{T_{I*}} = \frac{N_I}{\frac{2}{3}E_{I*}} = \beta_{II} = \frac{1}{T_{II}} = \frac{N_{II}}{\frac{2}{3}(E-E_{I*})} \tag{1.a.31}$$

which determines $E_{I*}$. For a system comprised of one-dimensional harmonic oscillators, the specific potential energy and kinetic energy each have $\frac{1}{2}T$ energy; and thus the specific energy for each oscillator is $\mathcal{E} = T$.

We next consider fluctuations $\delta E_I$ in $E_I$ away from its equilibrium value $E_{I*}$. We examine the formal Taylor-series expansions of $S_I$ and $S_{II}$ with respect to deviations $\delta E_I$ from $E_{I*}$:

$$S_I(E_I) = S(E_{I*}) + \delta E_I \left(\frac{dS_I}{dE_I}\right)_{E_{I*}} + \frac{1}{2}(\delta E_I)^2 \left(\frac{d\beta_I}{dE_I}\right)_{E_{I*}} + \cdots \tag{1.a.32}$$

$$\begin{aligned}S_{II}(E_{II}) &= S(E - E_{I*}) + \delta E_{II} \left(\frac{dS_{II}}{dE_{II}}\right)_{E_{II}} + \frac{1}{2}(\delta E_{II})^2 \left(\frac{d\beta_{II}}{dE_{II}}\right)_{E_{II}} + \cdots \\ &= S(E_{II} = E - E_{I*}) - \delta E_I \left(\frac{dS_{II}}{dE_{II}}\right)_{E_{II}} + \frac{1}{2}(\delta E_{II})^2 \left(\frac{d\beta_{II}}{dE_{II}}\right)_{E-E_{I*}} + \cdots\end{aligned} \tag{1.a.33}$$

Use of Eq.(1.a.28) for the probability and Eqs.(1.a.30), (1.a.32) and (1.a.33) yields

$$\begin{aligned}\rho(\delta E_I) &= e^{\frac{1}{2}(\delta E_I)^2 \left(\frac{d\beta_I}{dE_I} + \frac{d\beta_{II}}{dE_{II}}\right)_{E_{I*}}} \sim e^{-\frac{1}{2}(\beta\delta E_I)^2 O\left(\frac{1}{N_I} + \frac{1}{N_{II}}\right)} \\ &\sim e^{-\frac{\beta^2}{2}(\delta E_I)^2 \left(\frac{1}{C_I N_I} + \frac{1}{C_{II} N_{II}}\right)} \equiv e^{-\frac{(\delta E_I)^2}{2\sigma^2}}\end{aligned} \tag{1.a.34}$$

The probability has a sharp peak around its most probable (equilibrium) value at $E_{I*}$.

Definition: $C$ appears in (1.a.34) and is the derivative of the specific energy with respect to the temperature and depends explicitly on the degrees of freedom,

$$C \equiv \frac{d\mathcal{E}}{dT} = \begin{cases} \frac{3}{2} & \text{ideal gas} \\ 1 & \text{harmonic oscillator} \end{cases}$$

One can read off the standard deviation $\sigma$ around the peak of the probability distribution $\rho(\delta E_I)$ from the right side of Eq.(1.a.34)

$$\sigma = T\left(\frac{1}{C_I N_I} + \frac{1}{C_{II} N_{II}}\right)^{-1/2} \sim T\sqrt{N} \tag{1.a.35}$$

We note that for $N_I \sim N_{II}$ (1.a.35) determines that $\sigma \sim T\sqrt{N}$ and $\frac{\sigma}{E_I} \sim \frac{1}{\sqrt{N}} \ll 1$. In the limit that $N_I \ll N_{II}$, e.g., a heat bath, then $\sigma = \sqrt{C_I N_I} T$ and $\frac{\sigma}{E_I} \sim \frac{1}{\sqrt{N_I}} \ll 1$.

Let's compare $S(E)$ with $S(E_I^*, E_{II}^*)$:

$$\Gamma(E) = \int_0^E dE_I \Gamma(E_I, E_{II}) = \int_0^E dE_I \Gamma(E_I^*, E_{II}^*) e^{-\frac{(\delta E_I)^2}{2\sigma^2}} \approx \Gamma(E_I^*, E_{II}^*)\sqrt{2\pi\sigma^2} \tag{1.a.36}$$

aside from units. Using the relation $S = \ln \Gamma$, Eq.(1.a.36) leads to

$$S(E) = S(E_I^*, E_{II}^*) + \frac{1}{2}\ln(2\pi\sigma^2) \approx S(E_I^*, E_{II}^*) \tag{1.a.37}$$

because $S(E), S(E_I^*, E_{II}^*) \sim O(N) \gg \frac{1}{2}\ln(2\pi\sigma^2) \sim O(\ln N)$. The conclusion is that the entropy is somewhat invariant relative to the system constraints involved.

Example: Consider $N_{II} \gg N_I$ and $E_{II} \gg E_I$, a heat bath if you will. Then

$$\Gamma(E_I, E) = \Gamma_I(E_I)\Gamma_{II}(E - E_I) = \Gamma_I(E_I)e^{S_{II}(E) - E_I \frac{dS_{II}}{dE_{II}} + \frac{1}{2}E_I^2 \frac{d^2 S_{II}}{dE_{II}^2} + \cdots} \tag{1.a.38}$$

We note that $\frac{d^2 S_{II}}{dE_{II}^2} = \frac{d\beta}{dE_{II}} \sim -\frac{N_I^2}{N_{II}}$ in (1.a.38), and we further impose that $N_I^2 \ll N_{II}$ so that this term is small. Hence, Eq.(1.a.38) becomes

$$\Gamma(E_I, E) \cong \Gamma_{II}(E)e^{-\beta_{II}(E_I - T_{II}S_I(E_I))} \equiv \Gamma_{II}(E)e^{-\beta_{II}F_I(E_I, T_{II})} \tag{1.a.39}$$

where we have introduced the definition of the free energy $F_I(E_I, T_{II}) \equiv E_I - T_{II}S_I(E_I)$, so-called because this is the energy available to do work. We recall that for an ideal gas $T_I = \frac{2}{3}\frac{E_I}{N_I}$

The probability is proportional to $\Gamma(E_I, E)$, i.e.,

$$\rho(E_I) \propto e^{-\beta_{II}F_I(E_I, T_{II})} \tag{1.a.40}$$

The peak of the probability distribution determines the most probable value of $E_I$ which corresponds to the minimum of the free energy $F_I$. Minimizing the free energy of the micro-state is equivalent to maximizing the total system entropy.

Example: The free energy of an ideal gas is given by

$$F_I(E_I, T_{II}) \equiv E_I - T_{II} S_I(E_I) = N_I \frac{3}{2} T_I - T_{II} N_I \left[\frac{5}{2} - \ln(n_I \Lambda_I^3(T_I))\right] \quad (1.a.41)$$

Exercise: Show that the most probable temperature is $T_I^* = T_{II}$.

Next we turn to the calculation of the probability of a quantum micro-state. Recall the expression given in Eq.(1.a.28). The probability of a micro-state $n$ in sub-system I in contact with sub-system II is constructed as follows. First, we observe based on (1.a.28)

$$w_n^I \propto \Gamma_{II}(E_{II} = E - E_n^I) = e^{S_{II}(E_{II})(E - E_n^I)} = e^{S_{II}(E_{II})E} e^{-\beta_{II} E_n^I + \cdots} \quad (1.a.42)$$

and we note that the $e^{S_{II}(E_{II})E}$ is just a probability constant that will cancel out after division. Dividing the right side of Eq.(1.a.42) by the sum of $\Gamma_{II}$ over all $n$ yields the probability.

$$w_n^I = \frac{e^{-\beta_{II} E_n^I}}{Z}; \quad Z \equiv Z_I(\beta_{II}) \equiv \sum_n e^{-\beta_{II} E_n^I} \quad (1.a.43)$$

where $Z$ constitutes the Gibbs canonical ensemble, i.e., the statistical ensemble of possible states in equilibrium with a heat bath at fixed temperature.

### 1.a.v Adiabatic law and action conservation

Consider the slow evolution of a system, i.e., an adiabatic change. We refer to Kubo's book for the ideas here.

Example: Assume a slowly varying Hamiltonian for a harmonic oscillator system with $N=1$ modeled by

$$H(p, q; t) = \frac{1}{2}p^2 + \frac{1}{2}\omega_0^2(t)q^2 \quad (1.a.44)$$

and we assume $\frac{d\omega_0}{dt} \ll \omega_0^2$. Energy is not conserved here because the Hamiltonian is time dependent due to $\omega_0(t)$. The elliptical orbit of the system in the $(p,q)$ phase space evolves, but the area of the ellipse is conserved, i.e., there is an adiabatic invariant. The time derivative of the Hamiltonian can be calculated from

$$\frac{dH}{dt} = \frac{\partial H}{\partial t} = \omega_0 \dot{\omega}_0 q^2 \quad (1.a.45)$$

From Eq.(1.a.45) we calculate the time integrated change $\Delta H$ from (1.a.45):

$$\Delta H = \int dt \dot{H} = \int dt \frac{\dot{\omega}_0}{\omega_0} \omega_0^2 q^2 \qquad (1.a.46)$$

For purposes of calculating $\Delta H$ over time durations long compared to the oscillation period, we can assume that $\frac{\dot{\omega}_0}{\omega_0}$ is approximately constant over the oscillation period; and we can average $\omega_0^2 q^2$ over the oscillation period. Noting that $\frac{1}{2}\langle \omega_0^2 q^2 \rangle = \frac{1}{2} <H>$, we conclude:

$$\Delta H \approx \int dt \frac{\dot{\omega}_0}{\omega_0} <H> \quad \text{and} \quad \frac{<\dot{H}>}{<H>} \approx \frac{\dot{\omega}_0}{\omega_0} \qquad (1.a.47)$$

<u>Definition</u>: Introduce the action $\qquad J \equiv \frac{1}{2\pi} \int p\, dq = \frac{<H>}{\omega_0} \qquad (1.a.48)$

<u>Exercise</u>: Calculate the time derivative of $J$ in (1.a.48) and use (1.a.47) to deduce

$$\frac{\dot{J}}{J} = \frac{<\dot{H}>}{<H>} - \frac{\dot{\omega}_0}{\omega_0} \approx 0 \qquad (1.a.49)$$

Hence, the action is "conserved."

<u>Example</u>: $N>1$ and generalize the time dependence of the Hamiltonian: $H(p, q; \lambda(t))$ where $\lambda \to \lambda + \Delta \lambda$ in a time interval $\Delta t$ that is large, i.e., $\lambda$ is assumed to change at a very slow rate compared to $\omega_0$ :

$$H = H(p, q; \lambda(t)), \quad \frac{d \ln \lambda}{dt} << \frac{\dot{\omega}_0}{\omega_0} \to$$

$$\Delta H \equiv \int_0^{\Delta t} \dot{H} dt = \int_0^{\Delta t} \frac{\partial H}{\partial t} dt = \int_0^{\Delta t} \frac{\partial H}{\partial \lambda} \frac{d\lambda}{dt} dt \approx \dot{\lambda} \int_0^{\Delta t} \frac{\partial H}{\partial \lambda} dt \equiv \dot{\lambda}\Delta t \langle \frac{\partial H}{\partial \lambda} \rangle_t = \dot{\lambda}\Delta t \langle \frac{\partial H}{\partial \lambda} \rangle_{\Gamma_\mu(E)} \quad (1.a.50)$$

where the time average has been replaced by an average over the energy surface in phase-space $\Gamma_\mu(E)$.

Hence, $\Delta H = \Delta \lambda \langle \frac{\partial H}{\partial \lambda} \rangle_E$ or $\frac{\Delta E}{\Delta \lambda} = \langle \frac{\partial H}{\partial \lambda} \rangle_E$ where the energy $E$ characterizes the micro-canonical ensemble.

## 1.a.vi Sub-canonical ensemble

The number of states for a sub-canonical ensemble (all states with energies less than a particular energy $E$) is given by

$$\Gamma(E,\lambda) \equiv \int dpdq\, \theta(E - H(p,q;\lambda)) \quad \text{where } \theta \equiv \begin{cases} 1 & H \leq E \\ 0 & H > E \end{cases} \quad (1.a.51)$$

Consider a small change in $\Gamma$ due to a small change in the parameter $\lambda$ and accompanying a change in $E$:

$$\Delta\Gamma = \frac{\partial \Gamma}{\partial E}\Big|_\lambda \Delta E + \frac{\partial \Gamma}{\partial \lambda}\Big|_E \Delta\lambda = \Delta\lambda \left\{ \frac{\partial \Gamma}{\partial E}\Big|_\lambda \langle\frac{\partial H}{\partial \lambda}\rangle_E + \frac{\partial \Gamma}{\partial \lambda}\Big|_E \right\}$$

$$= \Delta\lambda \left\{ \int dpdq\, \delta(E - H) \langle\frac{\partial H}{\partial \lambda}\rangle_E + \int dpdq\, \delta(E - H)\left(-\langle\frac{\partial H}{\partial \lambda}\rangle_E\right) \right\} = 0 \quad (1.a.52)$$

where we have made use of $\frac{d\theta(x)}{dx} = \delta(x)$ and $\delta(x)$ is the Dirac $\delta$- function. Thus, the two terms cancel on the right side of Eq.(1.a.52); and $\Delta\Gamma = 0$ under a slow change in the parameter $\lambda$. In practice, $\frac{d\ln \lambda}{dt}$ is required to be much smaller than the rate of change of anything else in the system.

<u>Corollary</u>: Given $\frac{\Delta E}{\Delta \lambda} = \langle\frac{\partial H}{\partial \lambda}\rangle_E$ for adiabatic changes, then $\Delta\Gamma = 0$, and in consequence $\Delta S = 0$. (Adiabatic Law --- entropy is conserved)

<u>Example</u>: For an ideal gas $\Gamma(E,V) \sim V^N E^{3N/2}$ and an adiabatic change in $V$, then $E \sim V^{-2/3}$ in order that $\Delta\Gamma = 0$. Hence, the pressure is $P \sim E/V \sim 1/V^{5/3}$, i.e., $PV^{5/3} = $ const, the usual adiabatic law for an ideal gas.

**1.a.vii Pressure and the virial theorem**

We next introduce the concept of a generalized force $\{\Lambda_i\}$ to go with parameters $\lambda = \{\lambda_i\}$. Consider an vector array of parameter values $\lambda$ and the Hamiltonian $H(p,q;\lambda)$.

<u>Example</u>: Particles in their own electric field and in an externally applied electric field with electric potential $\phi_0$ have the Hamiltonian:

$$H = \sum_i \frac{p_i^2}{2m_i} + \sum_{i<j} \frac{e_i e_j}{r_{ij}} + \sum_i e_i \phi_0(\mathbf{r}_i) \quad (1.a.53)$$

where $e_i$ are the particle charges, $m_i$ are the particle charges, $p_i$ are the momenta, and $r_{ij}$ are the distances between the $I$ and $j$ particles. We note that

$$\sum_i e_i \phi_0(\mathbf{r}_i) = \int d^3x\, \rho(\mathbf{x}, \{\mathbf{r}_i\}) \phi_0(\mathbf{x}) \quad (1.a.54)$$

and define the charge density as

$$\rho(\mathbf{x}, \mathbf{r}_i) = \sum_i e_i \delta(\mathbf{x} - \mathbf{r}_i) \quad (1.a.55)$$

We can choose $\lambda$ to be whatever attribute of the Hamiltonian is of interest, e.g., $\lambda = \{e_i\}$ or $\{\mathbf{r}_i\}$ or other.

Definition: The generalized force is

$$\mathbf{\Lambda}(p, q; \lambda) \equiv \frac{\partial H(p,q;\lambda)}{\partial \lambda} \quad \text{and} \quad \mathbf{\Lambda}(E; \lambda) \equiv \langle \mathbf{\Lambda}(p, q; \lambda) \rangle_{E, \lambda} \tag{1.a.56}$$

Example: The functional derivative $\frac{\partial H}{\partial \phi_0(x)} = \rho(\mathbf{x}, \{\mathbf{r}_i\})$ [ref. (Schiff, 1968) or (Goldstein, 1950)].

In (1.a.56) $\mathbf{\Lambda}(E; \lambda)$ is the thermodynamic generalized force, and the averaging brackets indicate an average over the accessible phase space for a given energy $E$. Then using the Adiabatic Law:

$$\mathbf{\Lambda}(E; \lambda) = \frac{\partial H(p,q;\lambda)}{\partial \lambda} = \frac{\partial E(S,\lambda)}{\partial \lambda}\bigg|_S \tag{1.a.57}$$

Example: The macroscopic charge density averaged over the phase space constrained by constant energy $E$ and fixed entropy is

$$\langle \rho \rangle(x) = \frac{\partial E(S, \phi_0(\mathbf{x}))}{\partial \phi_0(\mathbf{x})}\bigg|_S \tag{1.a.58}$$

We next introduce the concept of pressure. Let $\lambda = V$ where $V$ is the volume. Then using (1.a.57) the pressure $P$ is

$$P(p, q; V) \equiv -\Lambda = -\frac{\partial H(p,q;V)}{\partial V} \tag{1.a.59}$$

Exercise: Take the model Hamiltonian for an ideal gas or a charged particle plasma and show the consistency of (1.a.59) with the elementary definition $P \equiv F/\text{area}$ where $F$ is the macroscopic force. We note from (1.a.57) and (1.a.59) that $P = -\frac{\partial E(S,V)}{\partial V}\bigg|_S = -\frac{\partial E(S,V)}{\text{area } \partial \ell}\bigg|_S$ and one can identify the force from $-\frac{\partial E(S,V)}{\partial \ell}\bigg|_S$ noting that $dV = \boldsymbol{area} \cdot \boldsymbol{d\ell}$. It is also helpful to note $T \equiv \frac{\partial E(S,V)}{\partial S}\bigg|_V$ and generally $dE(S, \lambda) = TdS + \mathbf{\Lambda} \cdot d\lambda$ At constant entropy, $dE(\lambda)|_S = \mathbf{\Lambda} \cdot d\lambda = -PdV = -P\boldsymbol{area} \cdot \boldsymbol{d\ell} = -\mathbf{F} \cdot \boldsymbol{d\ell} = -dE$. Hence, the pressure at constant entropy is the force divided by the area. We will return to consideration of the pressure subsequently.

[*Editor's Note: Kaufman made the cryptic remark that this exercise is not trivial and alluded to the Ergodic Theorem (Sec. 1.a.iii) without further explanation.*]

Next we introduce the concept of heat. Again consider a physical system composed of two subsystems I and II. The composite Hamiltonian is $H = H_I + H_{II} + H_{int}$ where $H_{int}$ is the interaction Hamiltonian. The energy gained or lost by subsystem I is then

$$\Delta E_I = \int dt \, \dot{H}_I = \int dt \, \{H_I, H\} = \int dt \, \{H_I, H_{int}\} \equiv Q_I \tag{1.a.60}$$

The Poisson bracket is $\{A(p,q), B(p,q)\} = -\sum_i \left(\frac{\partial A}{\partial p_i}\frac{\partial B}{\partial q_i} - \frac{\partial A}{\partial q_i}\frac{\partial B}{\partial p_i}\right)$, and $Q_I$ is the heat transfer. The total time derivative of any quantity can be shown to be

$$\dot{A} = \frac{\partial A}{\partial t} + \frac{\partial A}{\partial p}\dot{p} + \frac{\partial A}{\partial q}\dot{q} = \frac{\partial A}{\partial t} + \{A, H\}$$

<u>Theorem</u>: If thermal equilibrium is maintained during heat input and if $\delta V = \delta \lambda = 0$ then

$$Q = \delta E = T\delta S \;\;\rightarrow\;\; \delta S = Q/T \tag{1.a.61}$$

and for a slow variation of $\lambda$ in the neighborhood of thermal equilibrium:

$$\Delta E = Q + W = Q + R \tag{1.a.62}$$

where $W$ or $R$ equals the work done on the sub-system and $Q$ is the heat or thermal input energy. If $R=-PdV$ for small $dV$, then from $dE(S,V)=TdS-PdV$ we realize that $Q=TdS$ whether or not work is being done on or by the sub-system. If thermal equilibrium is not maintained, then internal processes will drive the system toward equilibrium with $\Delta S > 0$ and $\Delta S \geq Q/T$ where $\Delta S$ is the sum of internal and external heat input. We realize that (1.a.62) is quite general, and $\Delta S = (\Delta E - R)/T$ is generally true.

<u>Theorem</u>: The change in time of $TdS \geq dE + PdV$, and there is equality if the system is in thermal equilibrium.

We return to consideration of the pressure. Consider a surface enveloping a volume and a differential surface area element $d^2\boldsymbol{\sigma}$ with the vector oriented outward and normal to the surface. The sum of forces on a "wall" at the surface of the volume is

$$\sum_i \mathbf{f}_{i,w} = P d^2\boldsymbol{\sigma} \tag{1.a.63}$$

and in consequence of Newton's third law the force of the wall back on the volume is $\sum_i \mathbf{f}_{w,i} = -Pd^2\boldsymbol{\sigma}$.

From Eq.(1.a.63), Newton's law, and the divergence theorem

$$\sum_{i=1}^N \mathbf{f}_{w,i} \cdot \mathbf{r}_i = -\oint P\mathbf{r} \cdot d^2\boldsymbol{\sigma} = -\int_V d^3r \, \nabla \cdot (P\mathbf{r}) \tag{1.a.64}$$

Here the force of the wall back on the system is balanced by the pressure of the particles back on the wall. If we assume the system is in equilibrium then we can also assume that the pressure is uniform and pull $P$ outside the integral in (1.a.64). Hence,

$$\sum_{i=1}^N \mathbf{f}_{w,i} \cdot \mathbf{r}_i = -\int_V d^3r \, \nabla \cdot (P\mathbf{r}) = -P\int_V d^3r \, \nabla \cdot (\mathbf{r}) = -3PV \tag{1.a.65}$$

The last relation in (1.a.65) is the so-called "virial" of the wall. Now consider Newton's third law including forces on the particles on one another and of the wall on the particles:

$$\sum_i \mathbf{r}_i \cdot m_i \dot{\mathbf{v}}_i = \sum_{i \neq j} \sum_j \mathbf{f}_{j,i} \cdot \mathbf{r}_i + \sum_i \mathbf{f}_{w,i} \cdot \mathbf{r}_i \qquad (1.a.66)$$

Using (1.a.65) to replace the last term in (1.a.66) we obtain the following.

$$P = -\frac{1}{3V}\left\{\sum_i \mathbf{r}_i \cdot m_i \dot{\mathbf{v}}_i - \sum_{i \neq j} \sum_j \mathbf{f}_{j,i} \cdot \mathbf{r}_i\right\} \qquad (1.a.67)$$

which can be further manipulated and simplified using

$$\sum_i \mathbf{r}_i \cdot m_i \dot{\mathbf{v}}_i = \frac{d}{dt}\sum_i \mathbf{r}_i \cdot m_i \mathbf{v}_i - \sum_i m_i \mathrm{v}_i^2 \equiv \frac{d}{dt}A - 2K \qquad (1.a.68)$$

where $K$ is the total kinetic energy and $A \equiv \sum_i \mathbf{r}_i \cdot m_i \mathbf{v}_i$ has units of action, and

$$\sum_{i \neq j}\sum_j \mathbf{f}_{j,i} \cdot \mathbf{r}_i = -\sum_{i \neq j}\sum_j \mathbf{f}_{i,j} \cdot \mathbf{r}_i = -\sum_{i \neq j}\sum_j \mathbf{f}_{j,i} \cdot \mathbf{r}_j = \frac{1}{2}\sum_{i \neq j}\sum_j \mathbf{f}_{j,i} \cdot (\mathbf{r}_i - \mathbf{r}_j) \qquad (1.a.69)$$

to obtain

$$P = \frac{1}{3V}\left\{2K - \frac{d}{dt}A + \frac{1}{2}\sum_{i \neq j}\sum_j \mathbf{f}_{j,i} \cdot (\mathbf{r}_i - \mathbf{r}_j)\right\} \qquad (1.a.70)$$

<u>Corollary</u>: The phase-space average $\langle \frac{dA}{dt} \rangle = 0$.

Proof: $A=(q,p)$ then $\langle \frac{dA}{dt} \rangle \equiv \int d\Gamma \rho(q,p) \frac{d}{dt} A(q,p)$ where the integral is over the phase-space volume and $\rho$ is the phase-space probability density; and we can generalize to $A(q,p;t)$. We use

$$\frac{dA}{dt} = \frac{\partial A}{\partial t} + \dot{p}\frac{\partial A}{\partial p} + \dot{q}\frac{\partial A}{\partial q} = \frac{\partial A}{\partial t} + \{A, H\} \qquad (1.a.71)$$

and note

$$\langle \frac{dA}{dt} \rangle \equiv \frac{d}{dt}\int d\Gamma \rho(q,p) A(q,p) - \int \frac{d}{dt}\bigl(d\Gamma \rho(q,p)\bigr) A(q,p) \qquad (1.a.72)$$

as the volume element in phase space may have time dependence. However, we note that $\frac{d\rho}{dt} = 0$ in consequence of Liouville's theorem, and $\frac{d}{dt}\Gamma = 0$ due to conservation of probability volume (which is not independent of Liouville's theorem). Hence, $\langle \frac{dA}{dt} \rangle = \frac{d}{dt}\langle A \rangle$. Finally, at equilibrium with no explicit time dependence, $\frac{\partial A}{\partial t} = 0$ and then $\frac{d}{dt}\langle A \rangle = 0$. We can now calculate the phase-space average of Eq.(1.a.70) at equilibrium which becomes

$$<P> = \frac{1}{3V}\left\{2<K> + \frac{1}{2}<\sum_{i\neq j}\sum_j \mathbf{f}_{j,i}\cdot(\mathbf{r}_i - \mathbf{r}_j)>\right\} \tag{1.a.73}$$

Example: $<K> = \frac{3}{2}NT$ which is valid for an ideal or non-ideal gas (with interactions), and (1.a.73) becomes

$$<P> = nT + \frac{1}{6V}<\sum_{i\neq j}\sum_j \mathbf{f}_{j,i}\cdot(\mathbf{r}_i - \mathbf{r}_j)> \tag{1.a.74}$$

where $n = N/V$.

## 1.b Classical fluids and other systems

### 1.b.i Equation of state and deviations from ideality

Postulate: Consider a general force law of particle $j$ on particle $i$ represented by

$$\mathbf{f}_{ij} = -\hat{\mathbf{r}}_{ij}\frac{\partial}{\partial r_i}\phi(\mathbf{r}_{ij}), \quad \mathbf{r}_{ij} \equiv \mathbf{r}_i - \mathbf{r}_j \tag{1.b.1}$$

where $\left(\frac{2}{3}\right)\left\langle\frac{K}{V}\right\rangle = \left(\frac{N}{V}\right)T \equiv nT$. We justify (1.b.1) based on Newton's third law and symmetry. Then the equilibrium pressure deduced from Eq.(1.a.74) is

$$<P> = nT - \frac{1}{6V}<\sum_{i\neq j}\sum_j \mathbf{r}_{ij}\cdot\hat{\mathbf{r}}_{ij}\frac{d\phi}{dr_{ij}}> = nT - \frac{1}{6V}\sum_{i\neq j}\sum_j <\mathbf{r}_{ij}\cdot\hat{\mathbf{r}}_{ij}\frac{d\phi}{dr_{ij}}> \tag{1.b.2}$$

commuting the sum over $N(N-1)\sim N^2$ pairs of interacting particles with the averaging bracket. Hence,

$$P = nT - \frac{N^2}{6V}\langle r_{12}\phi'(r_{12})\rangle \tag{1.b.3}$$

The $nT$ term is the kinetic pressure and the second term in (1.b.3) is the interaction pressure. The average in the interaction pressure is

$$\langle r_{12}\phi'(r_{12})\rangle \equiv \int d^3 r_{12}\rho(\mathbf{r}_{ij}\to r_{12})\, r_{12}\phi'(r_{12}) \tag{1.b.4}$$

due to isotropy in the probability density and because the interaction force depends only on the scalar separation distance. The probability density can be represented as

$$\rho(r_{12}) = \frac{g(r_{12})}{V} \tag{1.b.5}$$

where $g(r_{12})$ is the pair correlation function such that $g(\infty) \to 1$ and $\int g\, dvol = V$. Then Eq.(1.b.3) becomes

$$P = nT - \frac{N^2}{6V}\langle r_{12}\phi'(r_{12})\rangle = nT - \frac{n^2}{6}\int 4\pi r_{12}^2 dr_{12} g(r_{12})r_{12}\phi'(r_{12}) \tag{1.b.6}$$

If we formulate the energy of a particle (atom or molecule) from first principles by summing the kinetic energy and the potential energy due to interactions over the volume, one obtains

$$\mathcal{E} = \tfrac{3}{2}T + \tfrac{n}{2}\int d^3 r_{12} g(r_{12})\phi(r_{12}) \tag{1.b.7}$$

We can make some qualitative remarks regarding the dependencies of the pair correlation function $g$ and the interaction potential $\phi$ on $r_{12}$ so that the integral in (1.b.7) remains well behaved, and the results for $P$ and $\mathcal{E}$ are physical. Given the constraints $g(\infty) \to 1$ and $\int g\, dvol = V$, $\phi(r_{12})$ must fall off faster than $1/r_{12}^3$ as $r_{12} \to \infty$. For $r_{12} \to 0$, $g(r_{12})\phi(r_{12})$ cannot diverge as fast as $1/r_{12}^3$. As a result, excluded are the Coulomb and gravitational potentials $\sim 1/r$ and the dipole-dipole interaction potential $\sim 1/r^3$.

### 1.b.ii Virial coefficients and van der Waals potential

Consider a dilute gas with only pair interactions and $g(r_{12}) \sim e^{-\beta\phi(r_{12})}$. Particle 1 interacts with particle 2, and the rest of system acts as a heat bath. At high densities when triplet or higher order interactions become important, there are corrections to this correlation function. As $r_{12} \to \infty$, $\phi \to 0$, and $g(\infty) \to 1$. We can substitute this into Eq.(1.b.6) for $P$ and (1.b.7) for $\mathcal{E}$:

$$P(n,T) = nT - \frac{2\pi}{3}n^2 \int_0^\infty s^2 ds\, e^{-\beta\phi(s)} s\phi'(s) + O(n^3)$$

$$= nT + \tfrac{1}{2}n^2 T \int_0^\infty d^3\mathbf{s}\left(1 - e^{-\beta\phi(s)}\right) + O(n^3) \tag{1.b.8}$$

where $s \equiv r_{12}$ and $T = 1/\beta$, and we have integrated by parts. We can represent the result in the standard form

$$\frac{P(n,T)}{T} = n + n^2 b_2(T) + n^3 b_3(T) + \cdots \tag{1.b.9}$$

where $b_\ell(T)$ are "virial" coefficients. In this "classical" example, the second virial coefficient is

$$b_2(T) = \tfrac{1}{2}\int d^3\mathbf{s}\left(1 - e^{-\beta\phi(s)}\right) \tag{1.b.10}$$

The second virial coefficient gives information on the interaction potential of the two particles.

<u>Example</u>: Van der Waals force + hard sphere – Consider the schematic for the electric potential shown in Fig.1.b.1

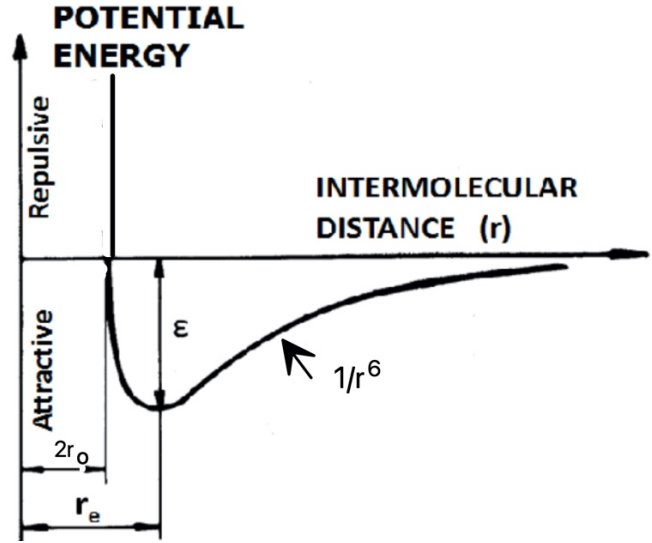

Fig. 1.b.1 Model van der Waals + hard sphere potential

The repulsive force for $r < 2r_0$ is represented as a hard sphere where $2r_0$ is the minimum distance between two hard-sphere centers with $r_0$ the hard-sphere radius. For this model (1.b.10) yields

$$b_2(T) = \frac{1}{2}\frac{4\pi}{3}(2r_0)^3 + \frac{1}{2}\beta \int_{2r_0}^{\infty} d^3s \; \phi(s) = 4V_0 - \frac{\alpha}{T} \tag{1.b.11}$$

where $V_0 \equiv \frac{4\pi}{3}r_0^3 > 0$ due to repulsion, $-2\alpha = \int_{2r_0}^{\infty} d^3s \; \phi(s)$ which is attractive and assumed small, and $1 - e^{-\beta\phi(s)} \approx \beta\phi(s)$ inside the integral. To be consistent with the expansion in (1.b.8) and (1.b.9) we require that $nV_0 \ll 1$.

Exercise:
1. Show that $\mathcal{E} = \frac{3}{2}T - n\alpha$ and there is no contribution from the $V_0$ constant term.
2. Show from $Td\mathcal{S} = d\mathcal{E} + Pd\mathcal{V}$ where $\mathcal{V} \equiv \frac{1}{n}$ that
$$\mathcal{S} = \frac{5}{2} - \ln n\Lambda^3 - 4\left(\frac{V_0}{\mathcal{V}}\right), \text{ where } \Lambda = \frac{h}{\sqrt{2\pi mT}}$$
3. Convert to standard van der Waals form:
$$\left(P + \frac{\alpha}{\mathcal{V}^2}\right)(\mathcal{V} - 4V_0) = T$$

### 1.b.iii Canonical ensemble and the partition function

Any subsystem, micro or macro, in contact with a heat bath at *T* has the attributes as described in (1.a.43) and parametrized by number *N*, volume *V*, and temperature *T*. The ensemble of such states is a canonical ensemble. The probability $w_n$ and partition function *Z* are

$$w_n = \frac{e^{-\beta E_n}}{Z}; \quad Z \equiv \sum_n e^{-\beta E_n}$$

Given the set of probabilities $\{w_n\}$ let us find $S\{w_n\}$.

Example: Let $n=1,2,3$ and $E=E_1, E_2, E_3$, and make $M$ ($M \to \infty$) measurements. As a matter of definition what we mean by $w_i$ is that $n=i$ occurs $w_i M$ times. The number of states for a given number of measurements $M$ is

$$\Gamma_M = \frac{M!}{\prod_n (Mw_n)!} = \frac{M!}{(w_1 M)!(w_2 M)!(w_3 M)!} \tag{1.b.12}$$

and the corresponding entropy is

$$S_M \equiv \ln \Gamma_M = M \ln M - M - \sum_n [Mw_n \ln(Mw_n) - Mw_n] = -M \sum_n w_n \ln(w_n) \tag{1.b.13}$$

where we have used $\sum_n w_n = 1$. From (1.b.13) we define the entropy associated with making a single measurement on the ensemble of three states in equilibrium with a heat bath:

$$S \equiv \lim_{M \to \infty} \left(\frac{S_M}{M}\right) = -\sum_n w_n \ln(w_n) \tag{1.b.14}$$

Example: Suppose all $\Gamma$ states are accessible with equal probability, such that

$$w_n = \frac{1}{\Gamma} \text{ and } S = -\sum_{n=1}^{\Gamma} \frac{1}{\Gamma} \ln \frac{1}{\Gamma} = \ln \Gamma \tag{1.b.15}$$

More generally the entropy for the canonical ensemble using (1.a.43) is

$$S_{can} = -\sum_n \frac{1}{Z} e^{-\beta E_n} \ln \frac{1}{Z} e^{-\beta E_n} = \ln Z + \beta \sum_n w_n E_n = \ln Z + \beta <E> \tag{1.b.16}$$

For a canonical macro ensemble we can convert the sum in the partition function to an integral:

$$Z(\beta) = \int d\Gamma e^{-\beta E} = \int dE \frac{d\Gamma}{dE} e^{-\beta E} \tag{1.b.17}$$

Recall that $S_\mu \sim S_\sigma = \ln \Gamma$ from which follows $\Gamma = e^{S_\sigma}$ and $\frac{d\Gamma}{dE} = e^{S_\sigma} \frac{dS_\sigma}{dE} = \beta e^{S_\sigma}$; then we obtain

$$Z(\beta) = \beta \int dE e^{(-\beta E + S_\sigma)} = \beta \int dE e^{-\beta(E - TS_\sigma(E))} = \beta \int dE e^{-\beta F(E,T)} \tag{1.b.18}$$

where $F$ is the free energy and $T$ is the temperature of the heat bath independent of the energy of the system. That energy for which $F$ is a minimum will maximize the partition function. We require the most probable energy $E^*(T)$ is that which determines $\frac{\partial F(E,T)}{\partial E} = 0$. Then we expand the free energy around $E^*$:

$$F(E,T) = F(E^*, T) + \frac{1}{2} \delta E^2 \frac{\partial^2 F}{\partial E^2} + \cdots \tag{1.b.19}$$

From (1.b.18) and (1.b.19):

$$Z(\beta) \approx \beta e^{-F(E^*,T)} \int_{-\infty}^{\infty} d\delta E \, e^{-\beta \frac{1}{2} \delta E^2 \frac{\partial^2 F}{\partial E^2}} = \beta e^{-F(E^*,T)} \sqrt{2\pi} \sigma_E \equiv \beta e^{-F(E^*,T)} \sqrt{2\pi T/F''} \quad (1.b.20)$$

where $\sigma_E = \sqrt{T/F''}$, and from Eqs.(1.b.16) and (1.b.20)

$$S_{can} = \ln Z + \beta <E> = \ln \beta - \beta F(E^*,T) + \frac{1}{2}\ln(2\pi T/F'') + \beta <E> \approx$$
$$-\beta F(E^*,T) + \beta <E> = -\beta(E^* - TS_\sigma(E^*)) + \beta <E> \quad (1.b.21)$$

where $\ln \beta = O(1)$, $\beta F(E^*,T) = O(N)$, and $\frac{1}{2}\ln(2\pi T/F'') = O(1)$. Note that $F \sim O(E) \sim O(N)$, $F'' \sim O(1/E) \sim O(1/N)$ and $\ln(N) \sim O(1)$ to justify $\frac{1}{2}\ln(2\pi T/F'') = O(1)$.

One rearranges terms in (1.b.21) to obtain

$$S_{can} = S_\sigma(E^*) + \beta(<E> - E^*) \quad (1.b.22)$$

For the canonical ensemble one can calculate <E> as a function of $\beta$

$$<E> = \sum_n w_n E_n = \frac{1}{Z}\sum_n e^{-\beta E_n} E_n = -\frac{1}{Z}\frac{\partial Z}{\partial \beta} = -\frac{\partial \ln Z}{\partial \beta} \quad (1.b.23)$$

We note that $F(E^*,T) = F(T)$ because fluctuations about $E^*$ are small. To $O(N)$ from (1.b.21)

$$\ln Z = -\beta F(E^*,T) = -\beta F(T) \text{ and } F(T) = -T \ln Z(\beta) \quad (1.b.24)$$

For $\ln Z(\beta) > 1$ then $F(T) < 0$. From (1.b.23) and (1.b.24) we deduce

$$<E> = -\frac{\partial \ln Z}{\partial \beta} = \frac{\partial \beta F}{\partial \beta} \text{ and } <E>(T) = E^*(T) \text{ to } O(N) \quad (1.b.25)$$

It also follows that

$$S_{can}(T) = \ln Z - \beta \frac{\partial \ln Z}{\partial \beta} = -\frac{\partial F(T)}{\partial T}, \text{ and hence } dF = -SdT \quad (1.b.26)$$

### 1.b.iv Quasi-static evolution

We next identify a parameter $\lambda$ in the system in contact with a heat bath and consider slow changes of the parameter:

$$w_n(\lambda) = \frac{1}{Z}\sum_n e^{-\beta E_n(\lambda)} \text{ and } Z(\beta, \lambda) = \sum_n e^{-\beta E_n(\lambda)} \quad (1.b.27)$$

For a slow and small change of $\lambda$ the total entropy does not change:

$$\Delta(S_{system} + S_{heat\,bath}) = 0 \text{ and } \Delta(S_{system}) = -\Delta(S_{heat\,bath}) = -\left.\frac{dS}{dE}\right|_{bath} \Delta E_{bath} \quad (1.b.28)$$

Using $\left.\frac{dS}{dE}\right|_{bath} = \frac{1}{T_{bath}}$ and $\Delta E_{bath} = -Q$, where $Q$ is the energy/heat input into the system, then

$$\Delta S_{system} = \frac{Q}{T} \quad (1.b.29)$$

The system energy $E(S, \lambda)$ accrues a small change due to $\Delta\lambda$

$$\Delta E = T\Delta S + \Lambda\Delta\lambda = Q + \Lambda\Delta\lambda \quad (1.b.30)$$

<u>Definition</u>: Let $R \equiv \Delta E - Q = \Lambda\Delta\lambda$ where $\Lambda = \left.\frac{\partial E}{\partial \lambda}\right|_S$

From the definition of the free energy

$$F \equiv E - TS \quad \rightarrow \quad \Delta F = \Delta E - T\Delta S - S\Delta T = T\Delta S + \Lambda\Delta\lambda - T\Delta S - S\Delta T = \Lambda\Delta\lambda = R \quad (1.b.31)$$

because $\Delta T = 0$ due to the contact with the heat bath. We can now make some general remarks regarding systems that are either thermally isolated or in contact with a heat bath (Table 1). We note that if there are no changes in the system parameters, a system in contact with a heat bath experiences no change in $S$ and $E$; and the heat bath is superfluous.

Table 1. Adiabatically evolving systems

|  | Thermal Isolation | Contact with Heat Bath (T=const) |
|---|---|---|
| Slow change in $\lambda$ | $\Delta E = \Lambda\Delta\lambda = R$ | $\Delta F = \Lambda\Delta\lambda = R$ |
| Most probable microstate | $S_{max}$ | $F_{min}$ |
| Approach to equilibrium | $\Delta S > 0$ | $\Delta F < 0$ |

### 1.b.v Mode counting, classical vs quantum systems

<u>Example</u>: $N$ identifiable microsystems, e.g., $N$ weakly interacting harmonic oscillators

$$Z_N(\beta) = \sum_{n=\{n_i\}} e^{-\beta E_n} \quad \text{and} \quad E_{\{n_i\}} = \sum_{n_i=1}^{N} \varepsilon_{n_i}^{(i)} \quad (1.b.32)$$

where $n_i$ can be thought of as the quantum numbers for the energy levels. The partition function becomes

$$Z = \sum_{n_1} \sum_{n_2} \cdots \sum_{n_N} e^{-\beta\left(\varepsilon_{n_1}^{(1)} + \varepsilon_{n_2}^{(2)} + \cdots + \varepsilon_{n_N}^{(N)}\right)} = \prod_{i=1}^{N} \sum_{n_i} e^{-\beta \varepsilon_{n_i}^{(i)}} = \prod_{i=1}^{N} Z_i \qquad (1.b.33)$$

Hence, $\ln Z = \sum_i \ln Z_i$. In the special case where all $N$ subsystems have the same properties, $Z = (Z_1)^N$ and $\ln Z = N \ln Z_1$.

Example: A classical gas or fluid consisting of $N$ indistinguishable particles

$$Z_N(\beta, V) \equiv \int_V \frac{d^{3N}p \, d^{3N}q}{h^{3N} N!} e^{-\beta H(p,q)} \qquad (1.b.34)$$

$$H(p, q) \to \sum_{i=1}^{N} \frac{p_i^2}{2m} + \Phi(\{r_i\}) \qquad (1.b.35)$$

Using (1.b.35) in (1.b.34) one obtains

$$Z_N(\beta, V) = \frac{1}{N!} \left[ V \int \frac{d^3 p}{h^3} e^{-\beta \frac{p^2}{2m}} \right]^N \left[ \int \frac{d^3 r^{(N)}}{V^N} e^{-\beta \Phi(\{r\})} \right]$$

$$\equiv \frac{1}{N!} \left( \frac{V}{\Lambda^3(\beta)} \right)^N Q_N(\beta) \qquad (1.b.36)$$

where $Q_N(\beta) \equiv \left[ \int \frac{dq^N}{V^N} e^{-\beta \Phi(\{r\})} \right]$ is the configurational partition function independent of volume.

For an ideal gas $\Phi(\{r\}) \to 0$ and $Q_N(\beta) = 1$; hence, $Z_N(\beta, V) = \frac{1}{N!} \left( \frac{V}{\Lambda^3(\beta)} \right)^N$.

Exercise: Show the specific free energy is given by $f = \frac{F}{N} = T(\ln n\Lambda^3 - 1)$, $\mathcal{E} = \frac{3}{2} T$, and $\mathcal{S} = \frac{5}{2} - \ln n\Lambda^3$ (Recall Eqs. (1.a.41) and (1.b.11), and the exercise following (1.b.11) in the limit $\Phi \to 0$.)

Example: Single harmonic oscillator ($\ell$) with quantized energy levels

$$\text{Energy levels.} \quad E_n^\ell = \hbar \omega_\ell n + E_0, \quad E_0 = \frac{1}{2} \hbar \omega_\ell \qquad (1.b.37)$$

$$Z_\ell(\beta) = \sum_{n=0}^{\infty} e^{-\beta \hbar \omega_\ell n} = 1 + e^{-x} + e^{-2x} + \cdots = \frac{1}{1 - e^{-x}} = \frac{1}{1 - e^{-\beta \hbar \omega_\ell}} \qquad (1.b.38)$$

We could compare the energy spectrum for the quantum harmonic oscillator in (1.b.37) to a few continuous medium systems:

1. Vibrating string: $\lambda = \frac{2L}{\ell}$, $\ell = 1, 2, 3, \ldots$
2. Drumhead

3. Water waves, e.g., one-dimensional gravity waves in a narrow channel, three-dimensional ocean waves (surface gravity waves, internal waves, …)
4. Electromagnetic waves, e.g., free space ($\omega = kc$), wave-guide modes, cavity modes
5. Plasma waves, e.g., electromagnetic waves ($\omega_k = \sqrt{k^2c^2 + \omega_p^2}$), longitudinal waves, …
6. Fluid sound waves: $\omega_k = kc_s, c_s = \sqrt{\frac{\gamma P}{\rho}}$
7. Waves in a solid: longitudinal sound wave $\omega_k = kc_\ell$, transverse shear wave $\omega_k = kc_t$

In order to calculate the partition function and the statistical properties of any of these systems, one must properly count the distinct modes. Here are two illustrative examples.

(a) One-dimensional standing waves with nodes at x=0 and x=L :
$U(x,t) = A\sin(kx)\sin(\omega t)$, $\omega > 0$, $\frac{\lambda}{2} = \frac{L}{\ell}, k = \frac{2\pi}{\lambda}$ and $\sum_{\ell=1}^{\infty} \to \int_0^\infty d\ell = \frac{L}{\pi}\int_0^\infty dk$

(b) One-dimensional traveling waves with periodic boundary conditions at x=0 and x=L:
$U(x,t) = A\sin(kx - \omega t)$, $\omega > 0$, $\lambda = \frac{L}{\ell}, k = \frac{2\pi}{\lambda}$ and $\sum_{\ell=-\infty}^{\infty} \to \int_{-\infty}^{\infty} d\ell = \frac{L}{2\pi}\int_{-\infty}^{\infty} dk$. If the traveling wave spectrum is symmetric with respect to positive and negative k, then $\frac{L}{2\pi}\int_{-\infty}^{\infty} dk \to \frac{L}{\pi}\int_0^\infty dk$.

In three spatial dimensions $\sum_{modes} \to \frac{1}{(2\pi)^3}\int d^3k$

Example: Classical non-interacting oscillators with $H_\ell(J_\ell) = J_\ell \omega_\ell$

$$Z_\ell(\beta) = \int \frac{dpdq}{h} e^{-\beta H(p,q)} = \frac{2\pi}{h}\int_0^\infty dJ_\ell e^{-\beta J_\ell \omega_\ell} = \frac{2\pi}{h\beta\omega_\ell} = \frac{T}{\hbar\omega_\ell} \qquad (1.b.39)$$

We note that the result Eq.(1.b.38) for the partition function for the quantized harmonic oscillator recovers the classical limit in (1.b.39) in the limit $\hbar\omega_\ell \ll T$:

$$Z_\ell(\beta)_{quantum} = \lim_{\hbar\omega_\ell \ll T} \frac{1}{1-e^{-\beta\hbar\omega_\ell}} \to \frac{T}{\hbar\omega_\ell} \qquad (1.b.40)$$

From Eq.(1.b.40) we can calculate the average energy per mode:

$$<E_\ell>(T) = -\frac{\partial \ln Z}{\partial \beta} = \frac{\hbar\omega_\ell}{e^{\beta\hbar\omega_\ell}-1} \qquad (1.b.41)$$

which yields $<E_\ell>(T) \to T$ in the classical limit and recovers the Rayleigh-Jeans classical result. For black-body radiation, each of the infinite number of modes has energy $T$ in the classical limit which leads to an infinite total energy when summing over all of the modes, i.e., the ultra-violet catastrophe!

## 1.b.vi Electromagnetic modes and interaction of particles

Consider electromagnetic waves in an unmagnetized plasma. The dispersion relation for transverse waves in an unmagnetized plasma is

$$\omega_k^2 = k^2 c^2 + \omega_p^2 \qquad (1.b.42)$$

and the total average wave energy summing over modes is

$$W = \sum_\ell \frac{\hbar \omega_\ell}{e^{\beta \hbar \omega_\ell} - 1} = 2V \int \frac{d^3 k}{(2\pi)^3} \frac{\hbar \omega_\ell}{e^{\beta \hbar \omega_\ell} - 1} \qquad (1.b.43)$$

where the factor 2 in front of the integral in Eq.(1.b.43) takes into account the sum over right and left circularly polarized waves. The energy density derived from Eq.(1.b.43) is

$$\frac{W}{V} = 2 \int_0^\infty \frac{4\pi k^2 dk}{(2\pi)^3} \frac{\hbar \omega_\ell}{e^{\beta \hbar \omega_\ell} - 1} = \text{fn}(T, \omega_p, \hbar c) \qquad (1.b.44)$$

<u>Definition</u>: The Wien wavelength and its inverse $k_w$ are defined by $\bar{\lambda} = \frac{1}{k_w} = \frac{\hbar c}{T}$.

In the limit that the plasma density vanishes $\omega_p \to 0$ then the right side of Eq.(1.b.44) becomes

$$\frac{W}{V} = \frac{\pi^2}{15} \frac{T}{\bar{\lambda}^3} = 4 \frac{\sigma T^4}{c}, \quad \sigma = \frac{\pi^2}{60} \frac{c}{(\hbar c)^3} \qquad (1.b.45)$$

Eq.(1.b.45) is the Stefan-Boltzmann law.

We note in (Jackson, 1975) it is shown that the wave energy density is related to the spatially averaged magnetic field energy density $\frac{<B^2>}{8\pi}$ by the relation:

$$\frac{W}{V} = \frac{<B^2>}{4\pi} \frac{1}{\epsilon} \qquad (1.b.46)$$

where $\epsilon$ is the longitudinal plasma dielectric function; $\epsilon = \frac{k^2 c^2}{\omega^2} = 1 - \frac{\omega_p^2}{\omega^2}$ in a cold plasma. For a wave packet the energy flux density is the product of the wave energy density and the group velocity $v_g = \frac{d\omega}{dk} = \frac{kc^2}{\omega}$.

Consider electromagnetic traveling waves in system with a finite volume and periodic boundary conditions with magnetic field represented by

$$\mathbf{B}(\mathbf{x}, t) = \sqrt{2} \sum_{\mathbf{k},\hat{\mathbf{e}}} B_{\mathbf{k},\hat{\mathbf{e}}} \hat{\mathbf{e}} \sin(\mathbf{k} \cdot \mathbf{x} - \omega_\mathbf{k} t + \alpha_\mathbf{k}) \qquad (1.b.47)$$

where the $B_{\mathbf{k},\hat{\mathbf{e}}}$ are real, and the average energies per mode are given by Eq.(1.b.41) which yields $<E_\ell>(T) \to T$ for $\hbar\omega_\ell \ll T$. Eq(1.b.47) might model waves emitted by Bremsstrahlung. If we calculate the ensemble or spatial average of $|\mathbf{B}|^2(\mathbf{x},t)$ we eliminate phases so that $B_{\mathbf{k},\hat{\mathbf{e}}}$ is real:

$$\langle |\mathbf{B}|^2(\mathbf{x},t)\rangle = \sum_{\mathbf{k},\hat{\mathbf{e}}} B_{\mathbf{k},\hat{\mathbf{e}}}^2 \tag{1.b.48}$$

From Eq.(1.b.46) the wave energy density is then

$$\frac{W_{\mathbf{k},\hat{\mathbf{e}}}}{V} = \frac{<|\mathbf{B}|^2(\mathbf{x},t)>}{4\pi}\frac{\omega_k^2}{k^2 c^2} \to <|\mathbf{B}|^2(\mathbf{x},t)> = \sum_{\mathbf{k},\hat{\mathbf{e}}}\frac{4\pi k^2 c^2}{\omega_k^2}\frac{<W_{\mathbf{k},\hat{\mathbf{e}}}>}{V} \tag{1.b.49}$$

With the assumption of thermal equilibrium and doing statistical averages Eqs.(1.b.43) and (1.b.49) yield

$$\frac{<B^2>}{8\pi} = \frac{1}{2V}\sum_{\mathbf{k},\hat{\mathbf{e}}}\frac{k^2 c^2}{\omega_k^2}\frac{\hbar\omega_k}{e^{\beta\hbar\omega_k}-1} = \int\frac{d^3k}{(2\pi)^3}\frac{k^2 c^2}{\omega_k^2}\frac{\hbar\omega_k}{e^{\beta\hbar\omega_k}-1} = \frac{T^4}{2\pi^2(\hbar c)^3}\int_\alpha^\infty dx\frac{(x^2-\alpha^2)^{3/2}}{e^x-1} \tag{1.b.50}$$

where $\omega_k^2 = k^2 c^2 + \omega_p^2$, $\frac{k^2 c^2}{\omega_k^2} = 1$ in vacuum, $x = \frac{\hbar\omega_k}{T}$, and $\alpha = \frac{\hbar\omega_p}{T}$. For $\alpha = \frac{\hbar\omega_p}{T} \ll 1$ the last integral on the right side of Eq.(1.b.50) yields $\frac{\pi^4}{15}$ which is the result for classical black body radiation, Eq.(1.b.45). For $\alpha = \frac{\hbar\omega_p}{T} \gg 1$ the integral yields $3\sqrt{\frac{\pi}{2}}\alpha^{3/2}e^{-\alpha}$ which implies that the magnetic energy is exponentially small for $T \to 0$ accompanying a coalescence of the photons in the ground state as the entropy likewise goes to zero (Nernst theorem).

Exercise: For a d-dimensional medium supporting normal modes with $\omega_k \sim k^p$ with $p > 0$, e.g., $p = \frac{1}{2}$ for water waves, $p = 1$ for sound waves, and $p = 2$ for a de Broglie matter wave, find the specific heat $C \sim T^q$. The specific heat capacity is defined as $C = T\,\partial\mathcal{S}/\partial T$ and recall that $\mathcal{S}$ is the specific entropy. Use Eq.(1.b.16) to evaluate the entropy in terms of the partition function and the examples in Sec. 1.b.v as a template to calculate the partition function.

We now extend the analysis to an electromagnetic plasma with applied fields. Consider a set of charged particles interacting with a given external field, e.g., $\{\phi_0(\mathbf{x},t), \mathbf{A}_0(\mathbf{x},t)\}$ with Lagrangian given by [ref. (Jackson, 1975), Chapt. 12]

$$L\{\mathbf{r}_i,\mathbf{v}_i;\phi_0,\mathbf{A}_0\} = \sum_{i=1}^N \frac{1}{2}m_i v_i^2 - \sum_{i=1}^N e_i\,\phi_0(\mathbf{r}_i,t) + \sum_{i=1}^N \frac{e_i}{c}\mathbf{v}_i\cdot\mathbf{A}_0(\mathbf{r}_i,t) - \sum_{i<j}\frac{e_i e_j}{r_{ij}} \tag{1.b.51}$$

The equations of motion determined by the Euler-Lagrange equations are

$$m_i\dot{\mathbf{v}}_i = e_i\left[\mathbf{E}_0(\mathbf{r}_i,t) + \frac{1}{c}\mathbf{v}_i\times\mathbf{B}_0(\mathbf{r}_i,t)\right] + e_i\sum_j e_j(-\nabla_i\frac{1}{r_{ij}}) \tag{1.b.52}$$

We can further expand the expressions in Eqs.(1.b.51) and (1.b.52) to include an internal electromagnetic field (also incorporating retarded time). To add a radiation field to the Lagrangian we posit (via guesswork or covariance arguments, $\{\mathbf{E}\cdot\mathbf{B}, E^2 - B^2\}$ ) (Galloway and Kim, 1971)

$$L\{\mathbf{r}_i, \mathbf{v}_i; \mathbf{A}, \dot{\mathbf{A}}; \phi_0, \mathbf{A}_0\} = L\{\mathbf{r}_i, \mathbf{v}_i; \phi_0, \mathbf{A}_0\} + \int \frac{d^3\mathbf{x}}{8\pi}(E_{rad}^2 - B_{rad}^2) + \ldots \quad (1.b.53)$$

where $\mathbf{E}_{rad}(\mathbf{x},t) = -\frac{1}{c}\dot{\mathbf{A}}(\mathbf{x},t)$ and $\mathbf{B}_{rad}(\mathbf{x},t) = -\nabla \times \mathbf{A}(\mathbf{x},t)$ in Coulomb (transverse) gauge, $\nabla \cdot \mathbf{A} = 0$. After a little bit of guess and check, the complete Lagrangian including radiation fields for a plasma in the presence of applied and internal electromagnetic fields is

$$L\{\mathbf{r}_i, \mathbf{v}_i; \mathbf{A}, \dot{\mathbf{A}}; \phi_0, \mathbf{A}_0\} = L\{\mathbf{r}_i, \mathbf{v}_i; \phi_0, \mathbf{A}_0\} + \int \frac{d^3\mathbf{x}}{8\pi}\left(\frac{1}{c^2}|\dot{\mathbf{A}}(\mathbf{x})|^2 - |\nabla \times \mathbf{A}(\mathbf{x})|^2\right)$$
$$+ \frac{1}{c}\int d^3\mathbf{x}\, \mathbf{j}(\mathbf{x}, \{\mathbf{r}_i, \mathbf{v}_i\}) \cdot \mathbf{A} \quad (1.b.54)$$

where

$$\mathbf{j}(\mathbf{x}, \{\mathbf{r}_i, \mathbf{v}_i\}) \equiv \sum_i e_i\, \mathbf{v}_i \delta(\mathbf{x} - \mathbf{r}_i) \quad (1.b.55)$$

The current $\mathbf{j}$ can be decomposed into a sum of longitudinal (curl free) and transverse (divergence free) terms $\mathbf{j} = \mathbf{j}^\ell + \mathbf{j}^t$. Only the transverse part contributes to (1.b.55) in consequence of the following result:

$$\int d^3\mathbf{x}\, \nabla\varphi \cdot \mathbf{A} = 0 = -\int d^3\mathbf{x}\, \varphi \nabla \cdot \mathbf{A} = 0 \text{ where } \nabla \cdot \mathbf{A} = 0$$

We decompose $\mathbf{j}$ into longitudinal and transverse pieces by calculating $\nabla \cdot \mathbf{j}$ and $\nabla \times \mathbf{j}$, and then inverting scalar and vector Poisson equations.

Exercise: Work out the Euler-Lagrange equations for the particles using (1.b.54) to find

$$m_i\dot{\mathbf{v}}_i = e_i\left[\mathbf{E}_0(\mathbf{r}_i, t) + \frac{1}{c}\mathbf{v}_i \times \mathbf{B}_0(\mathbf{r}_i, t)\right] + e_i \sum_j e_j\left(-\nabla_i \frac{1}{r_{ij}}\right) +$$
$$e_i\left[\mathbf{E}_{rad}(\mathbf{r}_i, t) + \frac{1}{c}\mathbf{v}_i \times \mathbf{B}_{rad}(\mathbf{r}_i, t)\right] \quad (1.b.56)$$

Definition: Define the functional derivatives (introducing bars over the partial signs $\bar{\partial}$)

$$\mathbf{\Pi} \equiv \frac{\bar{\partial}L}{\bar{\partial}\dot{\mathbf{A}}(x)} \equiv \frac{1}{4\pi c^2}\dot{\mathbf{A}}(x) = -\frac{1}{4\pi}\mathbf{E}_{rad} \text{ and } \dot{\mathbf{\Pi}}(x) \equiv \frac{\bar{\partial}L}{\bar{\partial}\mathbf{A}(x)} = -\frac{1}{4\pi c}\dot{\mathbf{E}}_{rad}(\mathbf{x}) \quad (1.b.57)$$

These are used in recovering Maxwell's equations from the Euler-Lagrange equations applied to (1.b.54). For example, from the term $\int d^3\mathbf{x}\,|\nabla \times \mathbf{A}|^2$ one forms $2\int d^3\mathbf{x}\,\nabla \times \mathbf{A} \cdot \nabla \times \delta\mathbf{A} \to$ $2 \int d^3x \, \delta \mathbf{A} \cdot \nabla \times (\nabla \times \mathbf{A}) = 2 \int d^3x \, \delta \mathbf{A} \cdot \frac{4\pi}{c} \mathbf{j}$ from which $\dot{\Pi}(\mathbf{x}) = -\frac{1}{4\pi c} \dot{\mathbf{E}}_{rad}(\mathbf{x}) = -\frac{1}{4\pi} \nabla \times \mathbf{B}_{rad} + \frac{1}{c} \mathbf{j}^t$. In summary, the Lagrangian in (1.b.54) recovers the correct Maxwell equations:

$$\mathbf{E}_{rad} = -\frac{1}{c} \dot{\mathbf{A}}_{rad}, \quad \nabla \times \mathbf{B}_{rad} - \frac{1}{c} \dot{\mathbf{E}}_{rad} = \frac{4\pi}{c} \mathbf{j}^t \text{ and}$$

$$\nabla \cdot \left( \nabla \times \mathbf{B} - \frac{1}{c} \dot{\mathbf{E}} = \frac{4\pi}{c} \mathbf{j} \right) \to 0 = \dot{\mathbf{E}} + 4\pi \mathbf{j} \to \nabla \cdot \mathbf{E} = 4\pi \rho \tag{1.b.58}$$

where we have made use of charge continuity: $\dot{\rho} + \nabla \cdot \mathbf{j} = 0$.

<u>Definition</u>: The canonical momentum in an electromagnetic field is defined by

$$\mathbf{p}_i \equiv \frac{\partial L}{\partial \mathbf{v_i}} = m_i \mathbf{v}_i + \frac{e_i}{c} [\mathbf{A}_0(\mathbf{r}_i, t) + \mathbf{A}(\mathbf{r}_i, t)] \tag{1.b.59}$$

and as noted in Eq.(1.b.57) $\Pi(\mathbf{x}) \equiv \frac{\partial L}{\partial \mathbf{A}(\mathbf{x})} = \frac{1}{4\pi c^2} \dot{\mathbf{A}}(\mathbf{x}) = -\frac{1}{4\pi c} \mathbf{E}(\mathbf{x})$.

Recalling the definitions in (1.b.57), the Hamiltonian implied by the Lagrangian in Eq.(1.b.54) is

$$H = \sum_i \mathbf{p}_i \cdot \mathbf{v}_i + \int d^3x \, \Pi(\mathbf{x}) \cdot \dot{\mathbf{A}}(\mathbf{x}) - L =$$

$$\sum_i \frac{1}{2} m_i v_i^2 + \sum_{i<j} \frac{e_i e_j}{r_{ij}} + \sum_i e_i \, \phi_0(\mathbf{r}_i, t) + \int \frac{d^3x}{8\pi} (E_{rad}^2 + B_{rad}^2) = K + C + R \tag{1.b.60}$$

where $K \equiv \sum_i \frac{1}{2} m_i v_i^2$, $C \equiv \sum_{i<j} \frac{e_i e_j}{r_{ij}} + \sum_i e_i \, \phi_0(\mathbf{r}_i, t)$, $R \equiv \int \frac{d^3x}{8\pi} (E_{rad}^2 + B_{rad}^2)$. We note that there is no magnetic interaction energy in the Hamiltonian.

<u>Exercise</u>: Calculate $\dot{p} = -\frac{\partial H}{\partial q}$, $\dot{q} = \frac{\partial H}{\partial p}$ and recover Maxwell's equations. The generalized momentum and Maxwell's equation in Eqs.((1.b.57), (1.b.58), and (1.b.59) have been calculated already from the Lagrangian in Eq.(1.b.54).

The classical partition function for an electromagnetic plasma with applied field is given by

$$Z(\beta, v; N_s; \phi_0, \mathbf{A}_0) \equiv \int \Pi_\mathbf{x} \frac{d\Pi(\mathbf{x}) d\mathbf{A}(\mathbf{x})}{h} \frac{1}{\prod_s N_s!} \prod_{i=1}^{N} \int \frac{d^3 p_i d^3 r_i}{h^3} e^{-\beta H} \tag{1.b.61}$$

It is convenient to transform coordinates from $(\mathbf{p}_i, \mathbf{r}_i, \mathbf{A}, \dot{\mathbf{A}}) \to (m_i \mathbf{v}_i, \mathbf{r}_i, \mathbf{A}, \dot{\mathbf{A}})$. Some coordinate transformations will require the introduction of a non-canonical transformation. In general the new volume element is related to the old volume element by a Jacobian factor, which is one for a canonical transformation and must be evaluated for a non-canonical transformation. The Jacobian matrix and its determinant are generally

$$\mathbf{J}_{ij} \equiv \frac{\partial x_{new,i}}{\partial x_{old,j}} \quad \text{and} \quad J \equiv det[\mathbf{J}_{ij}] \tag{1.b.62}$$

In this case $\mathbf{p}_i \to m_i \mathbf{v}_i$ simply. The partition function can be recast as

$$Z = \left[V^N \frac{\int \Pi_i d^3 \mathbf{v}_i m_i e^{-\beta K}}{h^{3N} N_s!}\right]\left[\int \Pi_i \frac{d^3 \mathbf{r}_i}{V} e^{-\beta \phi}\right]\left[\int \Pi_\mathbf{x} \frac{d\mathbf{A} d\mathbf{\Pi}}{h} e^{-\beta R}\right] \equiv$$

$$Z_{kinetic}(\beta, V) Z_{config.}(\beta, \phi) Z_{rad}(V, \beta) \tag{1.b.63}$$

Note that the dependence on $\mathbf{A}_0$ has vanished. The kinetic component of the partition function becomes (for an ideal gas)

$$Z_{kinetic}(\beta, V) = \Pi_s \frac{1}{N_s!}\left[\frac{V}{\Lambda_s^3(\beta)}\right]^{N_s} \tag{1.b.64}$$

The configuration component of the partition function becomes

$$Z_{config.}(\beta, \phi) = \int \Pi_i \frac{d^3 \mathbf{r}_i}{V} e^{-\beta \left[\sum_i e_i \phi_0(\mathbf{r}_i, t) + \sum_{i<j} \frac{e_i e_j}{r_{ij}}\right]} \tag{1.b.65}$$

Note that $\frac{e_i e_j}{r_{ij}}$ becomes divergent as $r_{ij} \to 0$, which could affect attracting charge pairs, and requires a cutoff at the quantum limit.

Consider a Fourier series representation of the electromagnetic vector potential:

$$\mathbf{A}(\mathbf{x}, t) = \sum_\mathbf{k} \sum_{\hat{\mathbf{e}}} \mathbf{A}_{\mathbf{k},\hat{\mathbf{e}}}(t) e^{i\mathbf{k} \cdot \mathbf{x}} \tag{1.b.66}$$

where $\mathbf{x}$ takes on a continuum of values, the sum over $\mathbf{k}$ is denumerably infinite, $\hat{\mathbf{e}}$ are the two polarization states orthogonal to $\hat{\mathbf{k}}$, and we assume that $\mathbf{A}_{-\mathbf{k},\hat{\mathbf{e}}} = \mathbf{A}^*_{\mathbf{k},\hat{\mathbf{e}}}$ so that the sum over $\mathbf{k}$ is over a half-space (which we will denote $\sum_\mathbf{k}'$). The radiation component of the Lagrangian in (1.b.54) becomes

$$L_{rad.} = \sum_\mathbf{k}' \sum_{\hat{\mathbf{e}}} \frac{V}{4\pi c^2}\left\{\left|\dot{\mathbf{A}}_{\mathbf{k},\hat{\mathbf{e}}}\right|^2 - k^2 c^2 \left|\mathbf{A}_{\mathbf{k},\hat{\mathbf{e}}}\right|^2\right\} \tag{1.b.67}$$

Given (1.b.67) the following two functional derivative expressions are independent:

$$\Pi_{\mathbf{k},\hat{\mathbf{e}}} \equiv \frac{\partial L_{rad}}{\partial \dot{\mathbf{A}}_{\mathbf{k},\hat{\mathbf{e}}}} = \frac{V}{4\pi c^2} \dot{\mathbf{A}}^*_{\mathbf{k},\hat{\mathbf{e}}}, \quad \Pi^*_{\mathbf{k},\hat{\mathbf{e}}} \equiv \frac{\partial L_{rad}}{\partial \dot{\mathbf{A}}^*_{\mathbf{k},\hat{\mathbf{e}}}} = \frac{V}{4\pi c^2} \dot{\mathbf{A}}_{\mathbf{k},\hat{\mathbf{e}}} \tag{1.b.68}$$

The radiation component of the Hamiltonian becomes

$$H_{rad} = R = \sum_{\mathbf{k},\hat{\mathbf{e}}}'(\Pi \dot{A} + \Pi^* \dot{A}^*) - L_{rad} = \sum_{\mathbf{k},\hat{\mathbf{e}}}' \frac{4\pi c^2}{V}|\Pi|^2 + \frac{Vk^2}{4\pi}|A|^2 \tag{1.b.69}$$

We introduce the definition $A \equiv (a + ib)\sqrt{\frac{2\pi c^2}{V}}$ and recast (1.b.69) as

$$H_{rad} = \sum_{\mathbf{k}}' \sum_{\hat{\mathbf{e}}} \left\{ \frac{1}{2} \dot{a}_{\mathbf{k},\hat{\mathbf{e}}}^2 + \frac{k^2 c^2}{2} a_{\mathbf{k},\hat{\mathbf{e}}}^2 \right\} + \sum_{\mathbf{k}}' \sum_{\hat{\mathbf{e}}} \left\{ \frac{1}{2} \dot{b}_{\mathbf{k},\hat{\mathbf{e}}}^2 + \frac{k^2 c^2}{2} b_{\mathbf{k},\hat{\mathbf{e}}}^2 \right\} \quad (1.b.70)$$

We are on our way to calculating the expectation of the statistically averaged Hamiltonian. Recall from (1.b.63) that $Z_{classical} = Z_{kin} Z_{conf} Z_{rad}$ and $e^{-\beta H} = e^{-\beta K} e^{-\beta C} e^{-\beta R}$ based on (1.b.60). We note that there are two harmonic oscillators in $H_{rad}$. Similar to (1.b.43) but accounting for the two harmonic oscillators in the radiation contribution to the Hamiltonian one obtains

$$\langle H_{\mathbf{k}\hat{\mathbf{e}}} \rangle = \frac{2\hbar kc}{e^{\beta \hbar kc} - 1} \to 2T \text{ for } \hbar kc \ll T \quad (1.b.71)$$

and

$$\langle H_{rad} \rangle = \sum'_{\mathbf{k},\hat{\mathbf{e}}} \langle H_{\mathbf{k}\hat{\mathbf{e}}} \rangle = \sum'_{\mathbf{k}\hat{\mathbf{e}}} \frac{2\hbar kc}{e^{\beta \hbar kc} - 1} = \sum_{\mathbf{k}\hat{\mathbf{e}}} \frac{2\hbar kc}{e^{\beta \hbar kc} - 1} \quad (1.b.72)$$

which again recovers the black-body radiation formula. We conclude that the radiation in the volume is independent of the particles except to the extent that they influence the dispersion relation for the electromagnetic normal modes. For example, in a plasma

$$\langle \frac{B^2}{8\pi} \rangle_{\omega_k, k, \hat{\mathbf{e}}} = \frac{T}{2} \frac{k^2 c^2}{k^2 c^2 + \omega_p^2} \quad \text{(in a plasma)} \quad (1.b.73a)$$

and in a vacuum

$$\langle \frac{B^2}{8\pi} \rangle_{\omega_k, k, \hat{\mathbf{e}}} = \frac{T}{2} \quad \text{(in a vacuum)} \quad (1.b.73b)$$

Theorem: In a classical system in thermal equilibrium, statistical mechanics and classical mechanics imply that the average magnetization in response to a finite applied magnetic field **B**₀ vanishes, <**M**>=0. (Bohr-Van Leeuwen)

Definition: $\mathbf{j}(\mathbf{x}(\mathbf{r}_i, \mathbf{v}_i)) \equiv \sum_i e_i \mathbf{v}_i \delta(\mathbf{x} - \mathbf{r}_i)$

To illustrate the Bohr-Van Leeuwen theorem consider the term in the Lagrangian in Eq.(1.b.51) that depends on **A**₀.

$$L = \cdots + \frac{1}{c} \int d^3x \, \mathbf{A}_0 \cdot \mathbf{j}(\mathbf{x}) + \cdots$$

from which follows using the functional derivatives introduced in (1.b.57)

$$\frac{1}{c}\mathbf{j}(\mathbf{x}) = \frac{\overline{\partial}L}{\overline{\partial}\mathbf{A}_0} = -\frac{\overline{\partial}H}{\overline{\partial}\mathbf{A}_0}\bigg|_{\mathbf{p},\mathbf{q}} \tag{1.b.74}$$

Corollary: For small changes $\Delta\lambda$ in $L(q,\dot{q};\lambda)$ and $H(q,\dot{q};\lambda)$

$$\delta L|_{q,\dot{q}} = -\delta H|_{p,q} \tag{1.b.75}$$

Using $Z = \int d\Gamma e^{-\beta H}$ we note

$$\frac{1}{Z}\frac{\overline{\partial}Z}{\overline{\partial}\mathbf{A}_0} = \frac{\overline{\partial}\ln Z(\beta,v,\phi_0)}{\overline{\partial}\mathbf{A}_0} \equiv 0 \tag{1.b.76}$$

and using (1.b.63) $Z = Z_{kinetic}(\beta,V)Z_{config.}(\beta,\phi)Z_{rad}(V,\beta)$, none of which components $Z_i$ depend on $\mathbf{A}_0$, and (1.b.76)

$$\frac{1}{Z}\frac{\overline{\partial}Z}{\overline{\partial}\mathbf{A}_0} = -\frac{\int d\Gamma e^{-\beta H}\frac{\overline{\partial}H}{\overline{\partial}\mathbf{A}_0}}{Z} \rightarrow \langle\frac{\overline{\partial}H}{\overline{\partial}\mathbf{A}_0}\rangle = 0 \tag{1.b.77}$$

Hence, from (1.b.74)

$$\langle\frac{1}{c}\mathbf{j}(\mathbf{x})\rangle = \langle\frac{\overline{\partial}L}{\overline{\partial}\mathbf{A}_0}\rangle = -\langle\frac{\overline{\partial}H}{\overline{\partial}\mathbf{A}_0}\bigg|_{\mathbf{p},\mathbf{q}}\rangle = 0 \tag{1.b.78}$$

and from

$$\langle\mathbf{j}\rangle = c\nabla\times\langle\mathbf{M}\rangle(\mathbf{x}) \rightarrow \langle\mathbf{j}\rangle = 0 \rightarrow \langle\mathbf{M}\rangle = 0 \tag{1.b.79}$$

Thus, the averaged equilibrium current and magnetization are zero in a classical system. Systems governed by quantum mechanics do not have to obey the Bohr-Van Leeuwen theorem, e.g., superconductors, permanent magnets with permanent magnetic dipole moment, etc.

Exercise: In a system with a uniform constant applied magnetic field $\mathbf{B}_0$ and $\mathbf{A}_0 = \frac{1}{2}\mathbf{B}_0\times\mathbf{x}$ with $L_{int} = \boldsymbol{\mu}\cdot\mathbf{B}_0$ and $\boldsymbol{\mu} \equiv \frac{1}{2c}\int d^3\mathbf{x}\,\mathbf{x}\times\mathbf{j}(\mathbf{x})$, show that $<\boldsymbol{\mu}> = 0 = \langle\frac{\partial L}{\partial\mathbf{B}_0}\rangle$ analogous to the arguments and results in Eq.(1.b.74) to (1.b.79).

[*Editor's Note: The Bohr-Van Leeuwen theorem continues to attract attention; and a rich literature exists. Some of the proofs of the BVL theorem reveal subtleties associated with boundary conditions for finite domains.*]

### 1.b.vii Grand canonical ensemble, grand partition function, chemical potential

We next turn to consideration of grand canonical ensembles. A grand canonical ensemble is a macroscopic ensemble of states that is in equilibrium with a reservoir. We assume that the system I is in contact with reservoir system II, and the volume of system I is fixed. System I may exchange particles and heat with system II, but the total energy $E = E_I + E_{II}$ and total number of particles $N_s = N_s^I + N_s^{II}$ are fixed. Typically system II is much larger than system I.

From Eqs.(1.a.38) and (1.a.39), we note the probability of system I being in a particular microstate $n$ with energy $E_I$ and number of particles $N_s^I$ is

$$w^I_{\{N_s^I\},n} \propto \Gamma_{II}[E_{II} = E - E_I(n, \{N_s^I\}'), N_s^{II} = N_s - N_s^I] = e^{S_{II}(\cdots)} =$$

$$e^{S_{II}(E,N_s) - E_I \frac{\partial S_{II}}{\partial E_I}\big|_E - \Sigma_s N_s^I \frac{\partial S_{II}}{\partial N_s^{II}}\big|_{N_s} + \cdots} \tag{1.b.80}$$

Consider the microcanonical entropy $S(E, N_s; \lambda)$ and its properties: $\beta \equiv \frac{\partial S}{\partial E}\big|_{N,\lambda}$ and $\gamma_s \equiv \frac{\partial S}{\partial N_s}\big|_{E,N_{s'},\lambda}$ where $\mathbf{N}_s$ is the vector of particle numbers whose component index $s$ denotes the species. We note that $\gamma_s = O(1)$. For example, the entropy for an ideal gas is

$$S(E, N_s; \lambda) = \Sigma_s N_s \left[\frac{5}{2} - \ln \frac{N_s}{V}\left(\frac{h}{\sqrt{\frac{4\pi}{3} m_s \frac{E}{N_s}}}\right)^3\right] = \Sigma_s N_s \left[\frac{5}{2} - \ln n_s \Lambda_s^3\right], \quad N = \Sigma_s N_s \tag{1.b.81}$$

We next normalize the right side of (1.b.80) to finish evaluating the probability of being in the microstate $n$ at thermal equilibrium:

$$w^I_{\{N_s^I\},n} = \frac{e^{-\beta_{II} E_I - \gamma_{II} \cdot \mathbf{N}_I}}{\Sigma_{\{\mathbf{N}_s^I\}} \Sigma_n e^{-\beta_{II} E_I - \gamma_{II} \cdot \mathbf{N}_I}} \tag{1.b.82}$$

The denominator in (1.b.82) is identified as the grand partition function

$$\mathbb{Z}(\beta_{II}, \gamma_{II}. \lambda) \equiv \Sigma_{\{\mathbf{N}_s^I\}} \Sigma_n e^{-\beta_{II} E_I(n, \mathbf{N}_I) - \gamma_{II} \cdot \mathbf{N}_I} \equiv \Sigma_{\{\mathbf{N}_s^I\}} e^{-\gamma_{II} \cdot \mathbf{N}_I} Z_I(\mathbf{N}_I, \beta_{II}, \lambda) \tag{1.b.83}$$

where

$$Z_I(\mathbf{N}_I, \beta_{II}, \lambda) \equiv \Sigma_n e^{-\beta_{II} E_I(n, \mathbf{N}_I)} \tag{1.b.84}$$

is the "petite" partition function.

A small change in the entropy satisfies the difference equation

$$dS = \beta dE + \boldsymbol{\gamma}_s \cdot d\mathbf{N}_s \rightarrow dE = TdS - T\boldsymbol{\gamma}_s \cdot d\mathbf{N}_s \tag{1.b.85}$$

<u>Definition</u>: $\boldsymbol{\mu}_s \equiv -T\boldsymbol{\gamma}_s$ is a set of chemical potentials. Equivalently $\boldsymbol{\mu}_s \equiv \frac{\partial E(S, \mathbf{N}_s, \lambda)}{\partial \mathbf{N}_s}$.

Recalling the definition of $\gamma_s$, we can express the chemical potential as follows:

$$\mu_s = T \ln n_s \Lambda_s^3 \tag{1.b.86}$$

Keep in mind that $N_s$ is variable.

Example: The grand partition function for an ideal gas (of identical particles) is

$$\mathbb{Z} = \sum_N e^{-\gamma N} \frac{(Z_1)^N}{N!} \approx \sum_{N=0}^{\infty} \frac{(Z_1 e^{-\gamma})^N}{N!} = e^{Z_1 e^{-\gamma}} \tag{1.b.87}$$

where $Z_1$ is the one-particle partition function as defined after (1.b.33), $Z_1 \equiv \sum_{n_i} e^{-\beta \varepsilon_{n_i}} \to \frac{V_I}{\Lambda^3}$ using (1.b.36). We note that (1.b.84) and (1.b.85) yield the simple identities:

$$<N_I> = -\frac{\partial \ln \mathbb{Z}(\beta_{II}, \gamma_{II})}{\partial \gamma_{II}} \quad \text{and} \quad <E_I> = -\frac{\partial \ln \mathbb{Z}(\beta_{II}, \gamma_{II})}{\partial \beta_{II}} \tag{1.b.88}$$

Example: In an ideal gas $<N_I> = -\frac{\partial \ln \mathbb{Z}(\beta_{II}, \gamma_{II})}{\partial \gamma_{II}} = Z_1 e^{-\gamma} = \frac{V_I e^{-\gamma_{II}}}{\Lambda^3(\beta_{II})} = \frac{V_I n_{II} \Lambda^3(\beta_{II})}{\Lambda^3(\beta_{II})} = n_{II} V_I$ and the number densities $n_{II} = n_I$ using (1.b.85), (1.b.86) and the definitions, and assuming that the bath and system I share the same mix of species and can freely exchange particles without disturbing the physics.

Exercise: Show that $\langle (\delta N_I)^2 \rangle \sim \frac{\partial^2 \ln \mathbb{Z}}{\partial \gamma_{II}^2}$ is small if system I is macroscopic, i.e., $N_I \sim O(10^{23})$.

Consider a macroscopic system I whose probability is given by

$$w_{\{N^I\}} \equiv \sum_n w_{n,\{N^I\}} = \frac{1}{\mathbb{Z}} e^{-\gamma \cdot \mathbf{N}} Z(\beta, \mathbf{N}) = \frac{1}{\mathbb{Z}} e^{-\beta(F - \mathbf{\mu} \cdot \mathbf{N})} \equiv \frac{1}{\mathbb{Z}} e^{-\beta \Omega} \tag{1.b.89}$$

where $F$ is the Helmholtz free energy, $\mathbf{\mu} = -T\mathbf{\gamma}$, and the grand potential is defined as

$$\Omega \equiv \Omega(\beta, \mathbf{\mu}; \mathbf{N}) \equiv F(\beta, \mathbf{N}) - \mathbf{\mu} \cdot \mathbf{N} \tag{1.b.90}$$

Recall that $\mathbf{N}$ is the vector representing the set of occupation numbers in different states.

The probability of having a given set of occupation numbers is

$$w_{\mathbf{N}} = \frac{1}{\mathbb{Z}(\beta, \mathbf{\mu})} e^{-\beta \Omega(\beta, \mathbf{\mu}; \mathbf{N})} \tag{1.b.91}$$

The maximum of the probability $w_{\mathbf{N}}$ with respect to $\mathbf{N}$ occurs at $\mathbf{N}^*$, i.e., $<\mathbf{N}> = \mathbf{N}^*$. Given (1.b.91), the maximum of $w_{\mathbf{N}}$ corresponds to minimizing $\Omega$ with respect to $\mathbf{N}$. At the most probable set of occupation numbers $\mathbf{N}^*$ the grand potential satisfies the relation

$$\Omega \cong -\frac{1}{\beta} \ln \mathbb{Z} \tag{1.b.92}$$

plus a constant of smaller order.

Using the relations $F \equiv E - TS$, $\Omega = F - \boldsymbol{\mu} \cdot \mathbf{N}$, (1.b.85), and (1.b.86), we have

$$dE = TdS + \boldsymbol{\mu} \cdot d\mathbf{N} + \Lambda d\lambda \quad \text{and} \quad dF = -SdT + \boldsymbol{\mu} \cdot d\mathbf{N} + \Lambda d\lambda \tag{1.b.93}$$

where, for example, we might choose $\lambda = V$ and $\Lambda = -P$ from Eq.(1.a.59); and we arrive at

$$d\Omega = -SdT - \mathbf{N} \cdot d\boldsymbol{\mu} + \Lambda d\lambda \quad \text{and} \quad \mathbf{N} = -\frac{\partial \Omega}{\partial \boldsymbol{\mu}} = -\frac{\partial \ln \mathbb{Z}}{\partial \boldsymbol{\gamma}} \to \langle \mathbf{N} \rangle = \mathbf{N}^* \tag{1.b.94}$$

for a macroscopic system. From (1.b.94) with $\Lambda = -P$ and $\lambda = V$

$$d\Omega = -SdT - \mathbf{N} \cdot d\boldsymbol{\mu} - PdV \to P = -\frac{\partial \Omega}{\partial V}(T, \boldsymbol{\mu}, V) = -\frac{\Omega}{V} \tag{1.b.95}$$

In deriving $P = -\frac{\Omega}{V}$ we argue that $\Omega$ is extensive (should scale with volume), while $T$ and $\boldsymbol{\mu}$ are intensive. Hence, $\Omega = V \, \text{fn}(T, \boldsymbol{\mu})$; and

$$-PV = \Omega = F - \boldsymbol{\mu} \cdot \mathbf{N} \tag{1.b.96}$$

Example: Grand partition function and grand potential for an ideal classical gas

$$\Omega(T, \mu, V) = -T \ln \mathbb{Z} = -\frac{VT}{\Lambda^3(\beta)} e^{\beta \mu} \tag{1.b.97}$$

$$P = -\frac{\Omega}{V} = \frac{T}{\Lambda^3(\beta)} e^{\beta \mu} \quad N = \frac{V}{\Lambda^3(\beta)} e^{\beta \mu} \quad \frac{P}{N} = \frac{T}{V} \to P = nT \tag{1.b.98}$$

Theorem: (Gibbs-Duhem) Take the differential of (1.b.96), substitute (1.b.95), and divide through by $V$ to obtain

$$dP = \frac{S}{V} dT + \mathbf{n} \cdot d\boldsymbol{\mu} \tag{1.b.99}$$

and $P$ is determined as a function of $T$ and $\boldsymbol{\mu}$, or $T$ and $\mathbf{n} \equiv \mathbf{N}/V$, i.e., the equation of state,

$$P(T, \boldsymbol{\mu}): \quad \mathbf{n}(T, \boldsymbol{\mu}) = \frac{\partial P}{\partial \boldsymbol{\mu}}(T, \boldsymbol{\mu}) \to P(T, \mathbf{n}) \tag{1.b.100}$$

We next consider a few interesting examples.

Example: Quantum ideal gas

Consider a subsystem consisting of a single-particle quantum state *k* for a simple noninteracting electron gas. The energy of a single electron is

$$\mathcal{E}_k = \begin{cases} \frac{p^2}{2m} & \text{(non-relativistic)} \\ \sqrt{p^2 c^2 + m^2 c^4} & \text{(relativistic)} \end{cases} \quad (1.b.101)$$

Note: If the particles are photons (bosons) instead, one must be careful because they are not conserved. Ions, molecules, fermions, molecules, etc., are conserved if non-interacting (no ionization, no recombination, no chemistry). Including a magnetic field *B* and spin, the subsystem energy is

$$E = \sum_k \mathcal{E}_k N_k \pm \hat{\mu} B \quad (1.b.102)$$

where $N_k$ is an occupation number ( $N_k = 0,1$ for fermions due to the Pauli principle; and $N_k = 0,1,2,..,\infty$ for bosons); and $\hat{\mu}$ is the magnetic moment associated with the spin. The probability of the macroscopic state *k* with occupation number $N_k$ is then

$$w_{N_k} = \frac{e^{-\gamma N_k - \beta \mathcal{E}_k N_k}}{\mathbb{Z}_k} \equiv \frac{e^{-\beta(\mathcal{E}_k - \mu_s)N_k}}{\sum_{N_k=0}^{1\ or\ \infty} e^{-\beta(\mathcal{E}_k - \mu_s)N_k}} \quad (1.b.103)$$

where the sum in the denominator for the grand partition function is just two terms for fermions and a geometric series for bosons.

Example: Bosons

In order that $\mathbb{Z}_k$ converges, $\mathcal{E}_k > \mu$ for all *k*. We also note that $\mathcal{E}_0 = 0$, which implies that $\mu < 0$. Hence, $w_{N_k} \propto e^{-\beta(\mathcal{E}_k + |\mu|)N_k}$, i.e., $w_{N_k}$ is a monotonic and exponentially decreasing function of $N_k$. The most probable state is the state $N_k = 0$. From Eq.(1.b.103) one concludes

$$\langle N_k \rangle \equiv \sum_{N_k=0}^{\infty} w_{N_k} N_k = \frac{1}{e^{\beta(\mathcal{E}_k + |\mu|)} - 1} \quad (1.b.104)$$

$<N_k>$ is a monotonically decreasing function of $\mathcal{E}_k$, and Einstein condensation can occur when $<N_k>$ becomes macroscopically large which is possible at $\mathcal{E}_k = 0$.

Example: Fermions

Because of the Pauli principle the occupation number $N_k$ = 0 or 1 for fermions. The probability $w_{N_k}$ is proportional to

$$w_{N_k} \propto e^{-\beta(\mathcal{E}_k - \mu)N_k} \quad (1.b.105)$$

For $\mu > 0$ the argument of the exponential is positive for $\mathcal{E}_k < \mu_s$ and is negative for $\mathcal{E}_k > \mu$. $w_{N_k}$ takes on just two values as a function of $N_k$, at $N_k = 0$ and $1$. The argument of the exponential vanishes for $\mathcal{E}_k = \mu$ the Fermi level. Figure 1.b.2 plots $\langle N_k \rangle = \sum_{N_k=0}^{1} w_{N_k} N_k$ as a function of energy $\mathcal{E}_k$ (Fermi-Dirac distribution function).

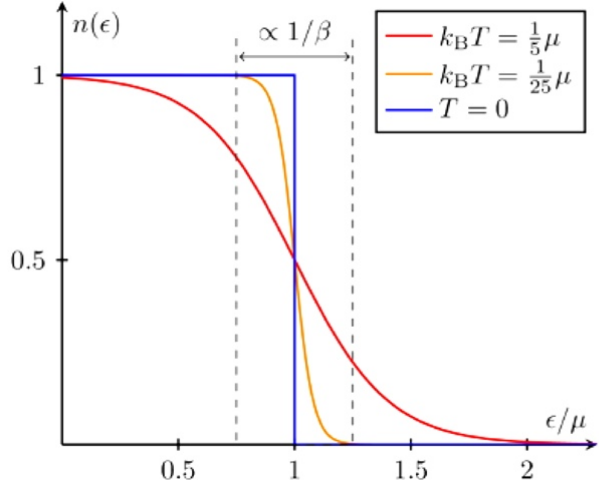

Fig. 1.b.2 Fermi-Dirac distribution function <$N_k$>=$n(\mathcal{E})$. (Riebesell, 2022)

We define $\mathcal{E}' \equiv \mathcal{E}_k - \mu$ and the partition function $\mathbb{Z}_k$ is then

$$\mathbb{Z}_k = \left(1 + \sigma e^{-\beta \mathcal{E}'}\right)^\sigma \quad \sigma \equiv \begin{cases} +1 & \text{(fermions)} \\ -1 & \text{(bosons)} \end{cases} \quad (1.b.106)$$

Then

$$\ln \mathbb{Z}_k = \sigma \ln(1 + \sigma e^{-\beta \mathcal{E}'}) \quad \text{and} \quad \mathbb{Z} = \prod_k \mathbb{Z}_k \rightarrow \ln \mathbb{Z} = \sum_k \ln \mathbb{Z}_k \quad (1.b.107)$$

We recall the analysis leading to (1.b.96) and obtain

$$P = -\frac{\Omega}{V} = \frac{T}{V} \ln \mathbb{Z} = \frac{T}{V} \sigma \sum_k \ln\left(1 + \sigma e^{-\beta(\mathcal{E}_k - \mu)}\right) \quad (1.b.108)$$

We introduce the deBroglie wavenumber $k$ and evaluate $\mathcal{E}_k = \frac{(\hbar k)^2}{2m}$. In Eq.(1.b.108) we replace $\sum_k \rightarrow gV \int \frac{d^3 \mathbf{k}}{(2\pi)^3}$ where the factor $g = 2S + 1$ and the spin factor is $S = \frac{1}{2}$ or an integer. Then (1.b.108) becomes using $\ln(1 + x) = \sum_{\ell=1}^{\infty} (-1)^{\ell-1} \frac{x^\ell}{\ell}$

$$P = \sigma g T \int \frac{d^3 k}{(2\pi)^3} \ln\left(1 + \sigma e^{-\beta(\mathcal{E}_k - \mu)}\right)$$

$$= \sigma g T \sum_{\ell=1}^{\infty} \frac{(-1)^{\ell-1}}{\ell} \sigma^{\ell} \int \frac{d^3 \mathbf{k}}{(2\pi)^3} e^{-\beta \ell (\varepsilon_k - \mu)}$$

$$= g T \sum_{\ell=1}^{\infty} \frac{(-\sigma)^{\ell-1}}{\ell} \int \frac{d^3 \mathbf{k}}{(2\pi)^3} e^{-\beta \ell (\varepsilon_k - \mu)} \qquad (1.b.109)$$

Recalling the definition $\Lambda = \frac{h}{\sqrt{2\pi m T}}$ and introducing the dimensionless fugacity or absolute activity $\xi \equiv e^{\beta \mu}$, (1.b.109) leads to

$$P = \frac{gT}{\Lambda^d(\beta)} \sum_{\ell=1}^{\infty} \frac{(-\sigma)^{\ell-1}}{\ell^{1+\frac{d}{2}}} \xi^{\ell} \to P(T, \xi) = \frac{gT}{\Lambda^3(\beta)} \left[ \xi - \frac{\sigma \xi^2}{2^{5/2}} + \frac{\xi^3}{3^{5/2}} + \cdots \right] \qquad (1.b.110)$$

where the dimensionality $d$ has been set to $d = 3$. We recall (1.b.100) which determines the particle density $n = \frac{\partial P}{\partial \mu_s}$

$$n(T, \xi) = \frac{\partial P(T, \mu)}{\partial \mu} = \frac{\beta \xi \partial P(T, \xi)}{\partial \xi} = \frac{g}{\Lambda^d(\beta)} \left[ \xi \left( 1 - \frac{\sigma}{2^{\frac{d}{2}}} \xi + \cdots \right) \right]$$

$$= \frac{g}{\Lambda^d(\beta)} \sum_{\ell=1}^{\infty} \frac{(-\sigma)^{\ell-1}}{\ell^{d/2}} \xi^{\ell} \qquad (1.b.111)$$

We note that the convergence of the expressions in (1.b.110) and (1.b.111) for $P$ and $n$ requires that the absolute activity $\xi \equiv e^{\beta \mu} < 1$, i.e., $\mu < 0$. (1.b.111) can be inverted and solved iteratively for

$$\xi(T, n) = \frac{n \Lambda^d(\beta)}{g} \left( 1 + \frac{\sigma}{2^{d/2}} \frac{n \Lambda^d(\beta)}{g} + \cdots \right) \qquad (1.b.112)$$

The value of $n\Lambda^3$ yields a measure of how quantum mechanical the gas is: $n\Lambda^3 \ll 1$ is the classical limit.

Using the definition $\xi \equiv e^{\beta \mu}$ and (1.b.112), one obtains

$$\mu = T \ln \left( \frac{n \Lambda^d(\beta)}{g} \right) + \text{corrections} \qquad (1.b.113)$$

and from (1.b.110) and (1.b.112)

$$P(n, T) = nT \left( 1 + \frac{\sigma}{2^{1+d/2}} \frac{n\Lambda^d}{g} + O(n\Lambda^d)^2 \right) \qquad (1.b.114)$$

where 3 is replaced by $d$.

The influence of $\sigma$ on the pressure is clear: $\sigma = +1$ for fermions has a repulsive effect, while $\sigma = -1$ for bosons has an attractive effect (symmetric wave function).

Example: Bose gas

Consider a gas of bosons, $\sigma = -1$. The pressure relation Eq.(1.b.110) becomes

$$P(T,\xi) = \frac{gT}{\Lambda^d(\beta)} \sum_{\ell=1}^{\infty} \frac{1}{\ell^{1+d/2}} \xi^\ell \qquad (1.b.115)$$

and the density relation (1.b.111) becomes

$$n(T,\xi) = \frac{g}{\Lambda^d(\beta)} \sum_{\ell=1}^{\infty} \frac{1}{\ell^{d/2}} \xi^\ell \qquad (1.b.116)$$

The expression

$$\frac{\Lambda^d(\beta)}{g} n(T,\xi) = \sum_{\ell=1}^{\infty} \frac{1}{\ell^{d/2}} \xi^\ell \qquad (1.b.117)$$

is a monotonic increasing function of $\xi$ on [0,1] and takes on larger values for $d = 2$ than for $d = 3$ (it diverges at $\xi = 1$ for $d = 2$). The limiting value of $\xi$ is $\xi = 1$, and the right side of (1.b.117) becomes the Riemann zeta function $R\left(x = \frac{d}{2}\right)$ where $R(x) = \sum_{\ell=1}^{\infty} \frac{1}{\ell^x}$. A few values of $R(x)$ are given in the following list in order of increasing $x$ : $R\left(\frac{1}{2}\right) = \infty$, $R(1) = \infty$, $R\left(\frac{3}{2}\right) = 2.612$, $R(2) = \frac{\pi^2}{6} = 1.645$, $R(4) = \frac{\pi^4}{90} = 1.082$, $R(10) = 1.001$. R is a monotonic decreasing function of its argument and asymptotes to unity for large argument. Thus, for $d = 3$, $(n\Lambda^3)_{max} = 2.612g$, $g = 2S + 1$. Because $\xi$ is less than one, the value of $(n\Lambda^3)_{max}$ is actually less than $2.612g$. Recall the discussion accompanying Eq.(1.b.104) that $<N_k>$ is a monotonically decreasing function of energy $\mathcal{E}_k$ and that a condensate can occur in the ground state. From (1.b.107) the partition function satisfies:

$$\ln \mathbb{Z}_0 = -\ln(1-\xi) \quad \text{and} \quad \ln \mathbb{Z} = \sum_k \sigma \ln(1 + \sigma e^{-\beta \varepsilon'})$$

Hence,

$$\langle N_0 \rangle = -\frac{\partial \ln \mathbb{Z}_0}{\partial \gamma} = \xi \frac{\partial \ln \mathbb{Z}_0}{\partial \xi} = \frac{\xi}{1-\xi} \to \lim_{\xi \to 1} \frac{\xi}{1-\xi} = \frac{1}{1-\xi} \qquad (1.b.118)$$

and this is volume independent. Thus, $\langle N_0 \rangle \to \infty$ as $\xi \to 1$, and $\xi = 1 - 1/\langle N_0 \rangle$. To illustrate the onset of the Bose-Einstein condensation, set $\xi$ to its limiting value $\xi=1$ in Eq.(1.b.117), set $d = 3$, use $R\left(\frac{3}{2}\right) = 2.612$, and evaluate $\Lambda$, $g = 1$, and $n$ in a specific experiment for helium II to obtain

$$T_0(n) = 3.31 \frac{\hbar^2}{m} n^{2/3} \quad \rightarrow \quad T_0 \text{ (theory)} = 3.13° K \tag{1.b.119}$$

where $2\pi/R \left(\frac{3}{2}\right)^{\frac{2}{3}} = 3.3128$ ...This compares with an experimental result of $2.19° K$.

[*Editor's Note: No reference to a specific experiment was given. The boiling point of He is 4.2° K, and He II becomes a superfluid at approximately 2.17 ° K at 1 atmosphere pressure.* ]

There is a problem with applying the grand canonical ensemble to the description of the Bose-Einstein condensate. Consider Eqs.(1.b.115) and (1.b.116) for the pressure $P$ and the density $n$ in the limits $g = 1$ and $d = 3$:

$$P(T, \xi) = \frac{T}{\Lambda^3(\beta)} \sum_{\ell=1}^{\infty} \frac{1}{\ell^{5/2}} \xi^\ell \quad n(T, \xi) = \frac{1}{\Lambda^3(\beta)} \sum_{\ell=1}^{\infty} \frac{1}{\ell^{3/2}} \xi^\ell \tag{1.b.120}$$

where $\xi = e^{-\gamma} = e^{\beta\mu} < 1$ because $\mu < 0$. In the limit that $\xi \to 0$ $\xi = n\Lambda^3$ is the number of particles in a deBroglie cube. However, from the expression for the number density in (1.b.120), as $\xi \to 1$

$$n\Lambda^3 = \Lambda^3 \frac{<N>}{V} = 2.612 \to <N> = 2.612 \frac{V}{\Lambda^3} \tag{1.b.121}$$

while in contrast Eq.(1.b.118) asserts $\langle N_0 \rangle = \frac{1}{1-\xi}$ which diverges as $\xi \to 1$ and is volume independent, while <N> scales with V. So there is a problem here. The difficulty is that in using the grand canonical ensemble, any particular energy state uses all the other states (systems) as the bath at a given temperature. However, when the particular state is the ground state, the bath is not so large in comparison to the number of states occupying the ground state at conditions such that $\xi \to 1$ are approached. The model of the grand canonical ensemble falls apart here for Bose statistics.

A solution for bosons is to use the Gibbs ensemble $Z(\beta, V_I, N_I)$ in which the system (I) is in contact with a heat bath allowing exchange of energy but in which particle exchange is prohibited. In Sec. 9.6 of (Reif,1965) is presented an analysis of Bose-Einstein statistics. There is a clever use of a Lagrange multiplier there. $N_0$ is shown to be proportional to $V$, and the ground state is shown to support large fluctuations. (Landau and Lifshitz, 1969) is another good reference on the Bose gas.

A plot of the density vs temperature where the Bose-Einstein condensate onsets based on Eq.(1.b.119) is shown in Figure 1.b.3 A schematic of the relative fraction in the ground state for the corrected theory is shown in Figure 1.b.4, $\frac{<N_0>}{N}$ vs $T$. Note that in the corrected theory both

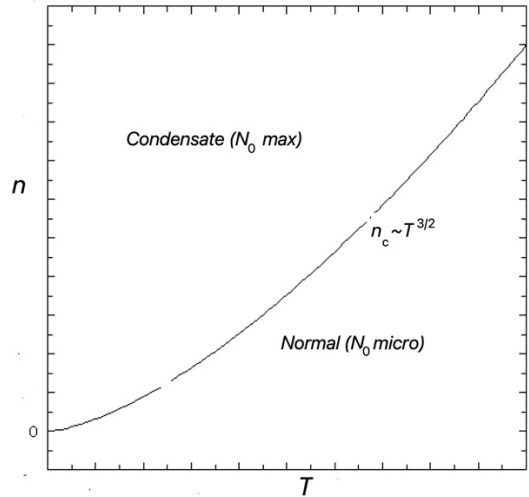

Fig. 1.b.3 Phase diagram for Bose-Einstein condensate, density vs temperature.

$N_0$ and $N$ scale with volume. Regarding the pressure, the ground state has no energy; only the excited states contribute. Equation (1.b.120) gives the correct pressure. Figure 1.b.5 presents a schematic for the pressure vs density for various temperatures. For $n\Lambda^3 \ll 1$ the system satisfies the ideal gas relation $P \sim nT$ while for $n\Lambda^3 > 2.612$ the system begins to fill the ground state.

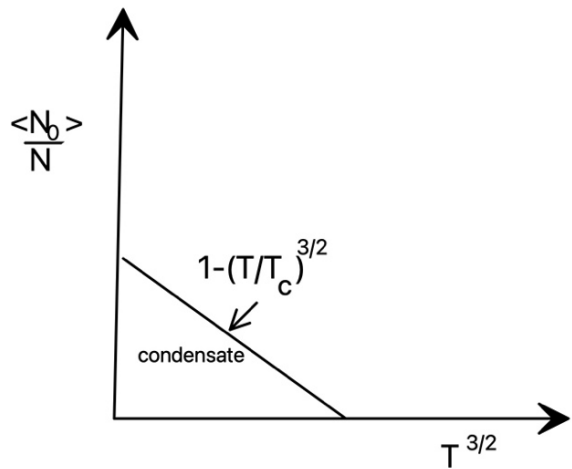

Fig. 1.b.4 Schematic of $\frac{<N_0>}{N}$ vs $T$ for the Bose condensate

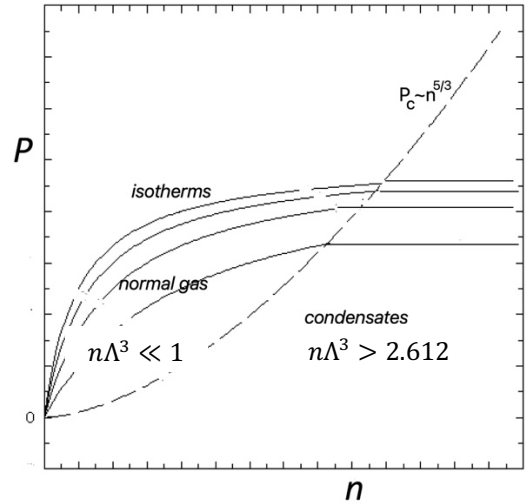

Fig. 1.b.5 Schematic: *P* vs *n* for various temperatures

Figure 1.b.6 presents a schematic of the pressure *P* vs the volume *V*=*N*/*n* for various isotherms. The critical pressure for a given volume above which volume there is no condensate scales as $P_c \sim V^{3/5}$

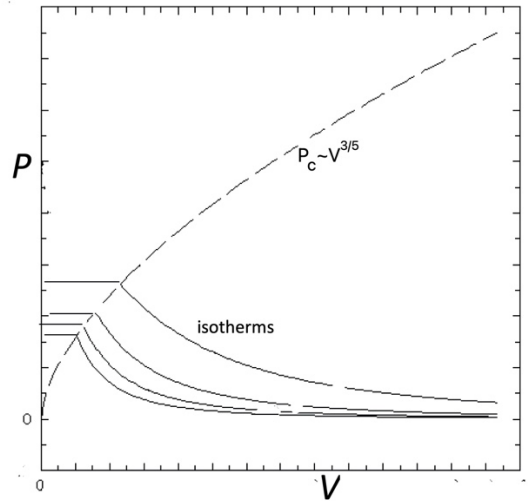

Fig. 1.b.6 Schematic: *P* vs *V* for various temperatures

<u>Example</u>: Bosons in which the number *N* is not conserved, e.g., excitations, photons, …. In such situations *N* is not conserved, $N_k = 0, 1, \ldots, \infty$ and $\mu = 0$. Calculate the properties using the grand canonical ensemble but with $\mu = 0$ (should agree with canonical ensemble). For the special case of photons in vacuum with $\omega_k = kc$, $\mathcal{E}_k = \hbar\omega_k = \hbar kc$, summing over right-hand and left-hand circularly polarized waves in three dimensions, we can calculate the grand potential $\Omega$ from (1.b.97) and (1.b.117):

$$\Omega = -VT\, R(4) \frac{1}{\pi^2} \frac{1}{\lambda_w^3}, \quad \lambda_w \equiv \frac{\hbar c}{T} \tag{1.b.122}$$

where $R(4) = \frac{\pi^4}{90}$.

Exercise: Verify Eq.(1.b.122).

Example: Ideal Fermi gas. Consider an electron gas with $g = 2S + 1 = 2$ and $\sigma = 1$. The pressure from (1.b.109) is

$$P(T,\mu) = 2T \int \frac{d^3k}{(2\pi)^3} \ln\left(1 + \sigma e^{-\beta(\mathcal{E}_k - \mu)}\right) \tag{1.b.123}$$

For $\mathcal{E}_k > \mu$: $e^{-\beta(\mathcal{E}_k - \mu)} \ll 1$ in the limit that $T \to 0$ ($\beta \to \infty$), so that $\ln(1 + 0) = 0$; and the electrons are completely degenerate. Note that for $\mathcal{E}_k \equiv \frac{p^2}{2m} = \frac{\hbar^2 k^2}{2m} < \mu$: $e^{\beta(\mu - \mathcal{E}_k)} \gg 1$ and $\ln e^{\beta(\mu - \mathcal{E}_k)} = \beta(\mu - \mathcal{E}_k)$ in (1.b.123). The pressure receives finite contributions only for $\mathcal{E}_k < \mu$ when $T \to 0$:

$$P(0,\mu) = 2 \int_0^{k_{max}\,(\mathcal{E}_k < \mu)} \frac{d^3k}{(2\pi)^3} (\mu - \mathcal{E}_k) \tag{1.b.124}$$

The density satisfies

$$n(T = 0, \mu) = \frac{\partial P}{\partial \mu} = 2 \int_0^{k_{max}\,(\mathcal{E}_k < \mu)} \frac{d^3k}{(2\pi)^3} = 2 \frac{\frac{4\pi}{3} k_f^3}{(2\pi)^3} = \frac{8\pi}{3}\left(\frac{p_F}{h}\right)^3 \equiv \frac{8\pi}{3} \Lambda_F^{-3} \tag{1.b.125}$$

where we define the Fermi level:

$$\mu = \mathcal{E}_F = \frac{p_F^2}{2m} = \frac{\hbar^2 k_F^2}{2m} \tag{1.b.126}$$

From (1.b.125) and (1.b.126)

$$k_F^3 = \frac{3}{8\pi} \frac{N_0}{V} (2\pi)^3 \implies \Lambda_F = \frac{h}{p_F} \text{ and } n\Lambda_F^3 = \frac{8\pi}{3} \tag{1.b.127}$$

Hence, as $T \to 0$

$$E = \frac{3}{5} \mathcal{E}_F N \text{ and } P = \frac{2 \frac{3}{5}\mathcal{E}_F N}{3\phantom{5} V} = \frac{2}{5} n \mathcal{E}_F \tag{1.b.128}$$

in the non-relativistic limit. Beware that $\mathcal{E}_F$ depends on $n$ via Eqs.(1.b.126) and (1.b.127).

Example: A non-ideal gas ("real gas") in the classical and quantum limits

Consider a simple one-species gas with partition function in the classical limit. From (1.b.36)

$$Z(\beta, N, V) \equiv \sum_n e^{-\beta E_n(V,N)} = \frac{1}{N!}\left(\frac{V}{\Lambda^3}\right)^N Q_N(\beta, V) \quad (1.b.129)$$

where $Q_N(\beta, V) \equiv \int \left(\frac{d^3\mathbf{r}}{V}\right)^N e^{-\beta\Phi(r_i)}$ is the configurational partition function and $\Phi$ is the interaction potential. The grand partition function is then

$$\mathbb{Z}(\gamma, \beta, V) = \sum_N e^{-\gamma N} Z(\beta, N, V) \quad (1.b.130)$$

Substituting (1.b.129) into (1.b.130)

$$\mathbb{Z}(\gamma, \beta, V) = \sum_{N=0}^{\infty} \frac{1}{N!}\left(\frac{Ve^{-\gamma}}{\Lambda^3}\right)^N Q_N(\beta, V)$$

$$= \sum_{N=0}^{\infty} \frac{1}{N!}(Vz)^N Q_N(\beta, V)$$

$$= 1 + VzQ_1 + \frac{1}{2}V^2 z^2 Q_2 + \cdots \quad (1.b.131)$$

where $z \equiv \frac{\xi}{\Lambda^3} = \frac{e^{-\gamma}}{\Lambda^3}$ is the activity in density units. Note that in the Boltzmann gas limit $\xi \to n\Lambda^3$. In (1.b.131), $Q_1 = 1$ (no self-interaction); and

$$Q_2 = \int \frac{d^3 r_1 d^3 r_2}{V^2} e^{-\beta \Phi_{12}} = \int \frac{d^3 r_1 d^3 r_2}{V^2} e^{-\beta \phi(r_{12})} \quad (1.b.132)$$

For large numbers of particles the thermodynamics is independent of the particular ensemble. However, for finite systems, e.g., $N=100$, the grand canonical ensemble is invalid; $N$ is not much larger than $\ln(N)$. Consider applications of using the grand canonical ensemble, Eqs.(1.b.130)-(1.b.132). For $z = \frac{e^{-\gamma}}{\Lambda^3} \to n$ as $n \to 0$ with $\gamma = \frac{\partial S}{\partial N}(E, V, N)$, we have Eq.(1.b.131) for $\mathbb{Z}(\gamma, \beta, V)$ with

$$Q_N(T, V) = \int \frac{d^{3N} r_i}{V^N} e^{-\beta \Phi(r_i)} > 0 \quad (1.b.133)$$

Consider (1.b.132) in more detail:

$$Q_2 = \int \frac{d^3 r_1 d^3 r_2}{V^2} e^{-\beta \phi(r_{12})} = \int \frac{d^3 r_1}{V}\left\{\int \frac{d^3 s}{V}\left[e^{-\beta \phi(s)} - 1\right] + 1\right\} = 1 - \frac{2b_2(T)}{V} \quad (1.b.134)$$

where $b_2(T) \equiv -\frac{1}{2}\int d^3 s \left[e^{-\beta \phi(s)} - 1\right]$ is defined in (1.b.10). We note that the term $\frac{2b_2(T)}{V} = O\left(\frac{1}{V}\right)$. Then

$$\mathbb{Z}(\gamma,\beta,V) = 1 + Vz + \frac{1}{2}V^2z^2Q_2 + O(N^{*3}) + \cdots + O(N^*)^{N^*} \quad (1.b.135)$$

where $Vz = O(N^*)$, $\frac{1}{2}V^2z^2Q_2 = O(N^{*2})$, and so on. The terms increase successively until the $N^*$ term, after which the terms in the series fall off, and the series converges. The expansion in (1.b.135) is not as useless as one might think because the series is monotonic increasing in $z$, the activity. As a function of $z$, $\mathbb{Z}(\gamma,\beta,V)$ increases from unity at $z=0$; and $\ln \mathbb{Z}$ is greater than 0 and increases with $z$. There are some assumptions to keep in mind. For example, one requires that $\Phi > -\infty$ in order that $Q_N$ not go to $\infty$, which excludes point masses and point charges. For the hard core + van der Waals potential diagrammed in Fig. 1.b.1, there is a minimum volume for the hard-sphere particle $\frac{4}{3}\pi r_0^3$ and a maximum number of particles in the volume: $N_{max} \sim \frac{V}{\frac{4}{3}\pi r_0^3}$ For $N > N_{max}$ $Q_N \to 0$. In these circumstances we can terminate the series in (1.b.135) at $N = N_{max}$. The grand partition function is analytic and finite. From Eq.(1.b.108)

$$P = \frac{1}{\beta V}\ln \mathbb{Z}(z,T,V) \quad (1.b.136)$$

and the physical pressure independent of volume satisfies

$$P(z,T)_{physical} = \lim_{V \to \infty} \frac{1}{\beta V}\ln \mathbb{Z}(z,T,V) \quad (1.b.137)$$

$\beta P$ is a non-negative and a monotonic increasing function of $z$. There is a possibility of a discontinuity in the slope of $\beta P$ with respect to $z$. Using (1.b.111)

$$n(z,T,V) = \frac{\partial P(\mu,T,V)}{\partial \mu} = z\frac{\partial}{\partial z}[\beta P(z,T,V] = z\frac{\partial}{\partial z}\frac{\ln\left(1+Vz+\frac{1}{2}V^2z^2Q_2\right)}{V} \quad (1.b.138)$$

The density $n$ is positive and so is $\frac{\partial n}{\partial z} > 0$. Where $\beta P$ has a discontinuity in its slope at $z = z_T$, for $V \to \infty$, a jump discontinuity can develop in the density $n$. The equation of state for $P$ as a function of density $n$ can then exhibit a phase transition at $z_T$. Figure 1.b.7 depicts a schematic of an equation of state and a phase diagram. Phase I might represent a gas, while II might represent a liquid or a solid. In constructing the equation of state and phase diagram in Fig. 1.b.7, we note that $0 < \langle(\delta N)^2\rangle \equiv \langle N^2\rangle - \langle N\rangle^2$ and in consequence of (1.b.138)

$$N - \langle N\rangle = \frac{\partial^2 \ln \mathbb{Z}}{\partial^2 \gamma^2} = Vz\frac{\partial n}{\partial z} = T\langle N\rangle\frac{\partial n}{\partial P} > 0 \implies \frac{\partial n}{\partial P} > 0 \text{ and } \frac{dP}{dn} \geq 0 \quad (1.b.139)$$

using $T>0$, $\langle N\rangle > 0$, and $\frac{\partial n}{\partial z} > 0$. Hence, $P$ either increases with respect to increasing $n$ or has flat intervals.

There was fundamental work on phase transitions in physical systems in the thermodynamic limit based on the properties of small, model systems by Lee and Yang (Yang and Lee, 1952; Lee and Yang, 1952) theory revolves around complex zeros of the partition function in finite-sized systems which may permit the possibility of phase transitions.

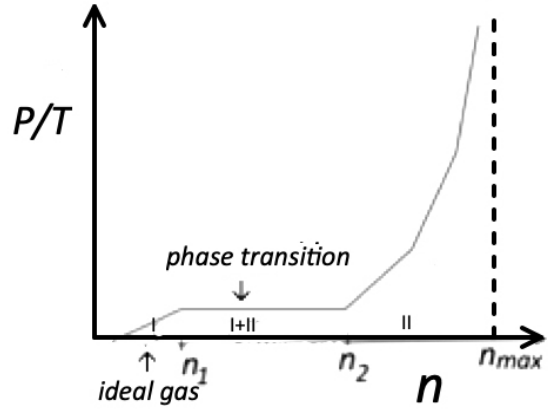

Fig. 1.b.7 Schematic: $\beta P = P/T$ vs $n$ equation of state and phase diagram

**1.b.viii Systems with external fields**

We next analyze a system in the presence of an external field. Consider a system with a downward directed (with respect to $z$) gravitational field with gravitational acceleration $g$. The gravitational potential for a particle in energy state $a$ with occupation number $N_a$ in subsystem II at $z_{II}$ with mass species $s$ is juxtaposed with a subsystem I at $z_I$ above it is given by

$$\psi_a^s = m_s g z_a \tag{1.b.140}$$

and the subsystem total energy summed over the internal energy and the energy in the external field is

$$E_a = E_a^{int} + N_a \psi_a^s \tag{1.b.141}$$

The subsystem II and bath I are assumed to satisfy the conservation laws:

$$N = N_I + N_{II}, \quad E = E_I + E_{II} \tag{1.b.142}$$

We maximize the total entropy $S = S_I + S_{II}$ and deduce the relations:

$$0 = dS = \sum_a dS_a = \sum_a \frac{\partial S_a}{\partial E_a^{int}} dE_a^{int} + \frac{\partial S_a}{\partial N_a} dN_a \tag{1.b.143}$$

with

$$\gamma \equiv \left.\frac{\partial S(E,N)}{\partial N}\right|_E \quad \text{and} \quad \beta_a \equiv \left.\frac{\partial S(E,N)}{\partial E_a}\right|_{N_a} \tag{1.b.144}$$

and $T_I = T_{II}$. The dependence of the entropy $S(E,N)$ on the internal energy in II can be expressed in terms of the internal energy $E_{int}$ and the occupation number $N$ as $S(E,N) = S^0(E_{int}, N) = S^0(E - N\psi, N)$ from which follows

$$-\beta\mu \equiv \gamma = \left.\frac{\partial S(E,N)}{\partial N}\right|_E = \left.\frac{\partial S^0}{\partial N}\right|_{E_{int}} - \psi\left.\frac{\partial S^0}{\partial E_{int}}\right|_N = \gamma^0 - \beta\psi = -\beta(\mu_0 + \psi)$$

$$\Rightarrow \mu = \mu_0 + \psi \qquad (1.b.145)$$

i.e., the potential energy of the subsystem is the sum of the internal chemical potential and the external potential energy. Sub-systems I and II are contiguous and in equilibrium with one another, and they share a common temperature. Thus, (1.b.145) applies to both; and $\gamma$ and, hence, $\mu$ are continuous:

$$\mu_I = \mu_{II} = \mu_0^I + \psi^I = \mu_0^{II} + \psi^{II} \qquad (1.b.146)$$

For the case of an ideal gas,

$$\mu_0 = T\ln(n\Lambda^3) \text{ and } T\ln(n_I\Lambda^3) + mgz_I = T\ln(n_{II}\Lambda^3) + mgz_{II} \qquad (1.b.147)$$

from which follows the result for an isothermal system:

$$\frac{n_I}{n_{II}} = e^{-\beta mg(z_I - z_{II})} \qquad (1.b.148)$$

If there is no net force on the isothermal system then

$$\nabla\mu = 0 \rightarrow \nabla\mu_0 = -\nabla\psi \rightarrow \frac{T\nabla n}{n} = mg \Rightarrow n(z) = n(0)e^{-\beta mgz} \qquad (1.b.149)$$

We can further elaborate the results in Eqs.(1.b.147)-(1.b.149) using the Gibbs-Duhem relation (1.b.100)

$$\mathbf{n}(T, \boldsymbol{\mu}) = \frac{\partial P}{\partial \boldsymbol{\mu}}(T, \boldsymbol{\mu}) \rightarrow P(T, \boldsymbol{\mu})$$

In the absence of the external field,

$$P(T, \mu) \rightarrow P^0(T, \mu_0) = P^0(T, \mu - \psi) \Rightarrow \nabla P = \frac{\partial P^0}{\partial \mu_0}(-\nabla\psi) \qquad (1.b.150)$$

For example, $-\nabla\psi = m\mathbf{g}$ and then $\Delta P = -nmg\Delta z$.

### 1.b.ix Particle interactions: hard discs, pair and triplet correlations

We next return to consideration of correlated particles, the virial expansion, and Eqs.(1.b.131)-(1.b.132):

$$\mathbb{Z}(\gamma, \beta, V) = \sum_{N=0}^{\infty} \frac{1}{N!} (Vz)^N Q_N(\beta, V)$$

$$= 1 + Vz + \frac{1}{2}V^2 z^2 Q_2 + \frac{1}{6}V^3 z^3 Q_3 + \cdots$$

$$Q_N(T, V) = \int \frac{d^{3N} r_i}{V^N} e^{-\beta \Phi(r_i)}$$

$$Q_2 = \int \frac{d^3 r_1 d^3 r_2}{V^2} e^{-\beta \Phi_{12}} = 1 - \frac{2b_2(T)}{V}$$

for the case of studying the transition from a fluid to a solid, and the formation of a close-packed structure. Berni Alder was a pioneer in molecular dynamics simulations and exploited hard-disk models in the interaction potentials. Consider a working model for $Q_3$ which captures the interaction of three particles:

$$Q_3 = \int \frac{d^3 r_1 d^3 r_2 d^3 r_3}{V^3} e^{-\beta[\Phi_{12} + \Phi_{13} + \Phi_{23}]} = \int \frac{d^3 r_1 d^3 r_2 d^3 r_3}{V^3} e^{-\beta \Phi(r_{12})} e^{-\beta \Phi(r_{13})} e^{-\beta \Phi(r_{23})} \quad (1.b.151)$$

<u>Definition</u>: Introduce $f_{ij}$, $e^{-\beta \phi_{ij}} = 1 + f_{ij}$ where $\lim_{r_{ij} \to \infty} f_{ij} \to 0$. Then

$$(1 + f_{12})(1 + f_{23})(1 + f_{13}) = 1 + f_{12} + f_{23} + f_{13} + f_{12}f_{23} + f_{12}f_{13}$$

$$+ f_{23}f_{13} + f_{12}f_{23}f_{13} \quad (1.b.152)$$

Now start performing the integrals in (1.b.151), e.g., $\frac{V}{V^3} \int d^3 r_1 d^3 r_2 f_{12} = \frac{V^2}{V^3} \int d^3 r_{21} f_{12}$; all of the terms are like this. Hence,

$$Q_3 = 1 + \frac{3}{V} \int d^3 r_{21} f_{12} + \frac{3}{V^2} \int d^3 r_{21} \int d^3 r_{31} f_{12} f_{13} + \frac{1}{V^3} \int d^3 r_{23} \int d^3 r_{21} \int d^3 r_{31} f_{12} f_{13} f_{23}$$

$$(1.b.153)$$

The second through fourth terms on the right side of (1.b.153) correspond pictorially to —, ∧, Δ i.e., two-particle interactions among two vertices (particles) or three vertices (particles). $Q_4$ has additional terms like ⊠ and other ways to connect four vertices depicting two-particle interactions among four vertices (particles). From (1.b.131) we can employ a Taylor-series expansion to calculate

$$\ln \mathbb{Z} = Vz + \frac{1}{2}V^2z^2[Q_2 - 1] + \frac{1}{6}V^3z^3[Q_3 - 3Q_2 + 2] \ldots =$$

$$Vz + \frac{1}{2}V^2z^2[\frac{1}{V}(-)] + \frac{1}{6}V^3z^3[\frac{1}{V^2}(3\wedge + \Delta)] + \ldots \quad (1.b.154)$$

in terms of the pictograms (Hill, 1960). Hence,

$$\frac{\ln \mathbb{Z}}{V} = z + \frac{1}{2}z^2(-) + \frac{1}{6}z^3(3\wedge + \Delta) + \ldots = \beta P \quad (1.b.155)$$

Thus, the pressure is volume independent. We can rewrite (1.b.155) in the following form:

$$\beta P(z,T) = \sum_{j=1}^{\infty} z^j C_j(T)$$

$$C_1 = 1 \quad (1.b.156)$$

$$C_2 = \frac{1}{2}(-)$$

$$C_3 = \frac{1}{2}(\wedge) + \frac{1}{6}(\Delta) = \frac{1}{2}(-)^2 + \frac{1}{6}(\Delta)$$

$$C_4 = \cdots$$

From (1.b.138)

$$n(z,T) = z\frac{\partial}{\partial z}(\beta P) = \sum_{j=1}^{\infty} jz^j C_j(T) \quad (1.b.157)$$

Hence, $\quad n = z + 2z^2 C_2 + 3z^3 C_3 + \cdots \quad (1.b.158)$

and $\quad \beta P = z + z^2 C_2 + z^3 C_3 + \cdots \quad (1.b.159)$

from which follows $\quad z = n - 2n^2 C_2 + n^3(8C_2^2 - 3C_3) + \cdots \quad (1.b.160)$

Substituting $z$ from (1.b.160) into (1.b.159) and collecting terms in a power series in $n$, a virial expansion for $\beta P$ can be obtained:

$$\beta P(n,T) = \sum_{j=1}^{\infty} n^j b_j(T) \quad (1.b.161)$$

where $b_1 = 1$, $b_2 = -C_2 = -\frac{1}{2}(-)$, $b_3 = 4C_2^2 - 2C_3 = (-)^2 - (-)^2 - \frac{1}{3}(\Delta) = -\frac{1}{3}(\Delta)$, and $b_4$ contains irreducible forms involving ( ⊠, ...). We discussed the evaluation of $b_2$ earlier for the van der Waals model, Eq.(1.b.11). For $b_3$ we have

$$b_3 = -\frac{1}{3}(\Delta) = -\frac{1}{3}\int d^3r_{32}\int d^3r_{21}\int d^3r_{31} f_{12}\, f_{13} f_{23} \quad (1.b.162)$$

Example: Assume a hard-sphere model

$$f_{12} = \begin{cases} -1, & r_{12} \leq a \\ 0, & r_{12} > a \end{cases} \qquad (1.b.163)$$

which corresponds to $\phi_{12} = \infty$ for $r_{12} \leq a$ and $\phi_{12} = 0$ for $r_{12} > a$, where $a$ is the diameter of the hard sphere or disk. We note that this model is highly simplified and artificial. Actual configurations of three hard spheres generally cannot satisfy $r_{ij} \leq a$ for all three interacting pairs except for one orientation in which the three hard sphere centers correspond to vertices of an equilateral triangle. Configurations of four hard spheres cannot satisfy $r_{ij} \leq a$ for all possible pairs. Nevertheless, use of the hard sphere model in (1.b.163) and ignoring the reality that certain interacting pairs can never satisfy $r_{ij} \leq a$ allow us to push through the calculations determining the virial coefficients. (Frisch, 1964)

The evaluation of the virial coefficients in two dimensions leads to

$$b_1 = 1, \ b_2 = -C_2 = -\tfrac{1}{2}(-) = \tfrac{1}{2}\pi a^2, \ b_3 = 0.782(b_2)^2, \ b_4 = 0.5327(b_2)^3 \qquad (1.b.164)$$

Then using Eq.(1.b.161) it follows

$$\beta P = n + b_2 n^2 + 0.782 b_2^{\,2} n^3 + 0.533 b_2^{\,3} n^4 + O(n^5) \qquad (1.b.165)$$

We note that $\tfrac{1}{2}\pi a^2$ is the area of two hard discs, and $n\tfrac{1}{2}\pi a^2$ is the average number of particles within two discs in two dimensions. The maximum two-dimensional density for close-pack hard discs is $n_{max} = 2/a^2$ from which $n\tfrac{1}{2}\pi a^2 = \pi n/n_{max}$ and (1.b.165) becomes

$$\frac{P}{nT} = 1 + \frac{n}{n_{max}} + \pi^2 0.782 \left(\frac{n}{n_{max}}\right)^2 + \pi^3 0.533 \left(\frac{n}{n_{max}}\right)^3 + \cdots \qquad (1.b.166)$$

$P/nT$ is an increasing function of $n/n_{max}$, and increases faster as $n/n_{max}$ increases in this example. $P/nT$ diverges at a value of $n/n_{max}$ equal to the radius of convergence which is less than or equal to unity. However, the virial coefficients are not necessarily positive in the non-fluid region; and the $P/T$ vs $n/n_{max}$ plot can have a flat region as in Fig. 1.b.7 where there is a phase transition. In the flat region the local number density is the ratio of the sum of the number of particles or molecules in the two phases divided by the sum of the volumes of the two phases. Recall the arguments accompanying (1.b.137)-(1.b.139) regarding the positivity of both $P/T$ and $\frac{dP}{dn}$ for the grand canonical ensemble. By including more physics in $\phi_{12}$, e.g., attractive forces, $P/T$ vs $n$ can acquire more structure.

**1.b.x Simple model of a phase transition**

Example: Hard-disc interaction in a two-dimensional periodic domain (particles leaving the domain re-enter symmetrically). Alder and collaborators computed the pressure using the virial expansion. In two dimensions and taking the time average to simplify

$$P = \frac{<K>}{V} + \frac{1}{2V} \langle \Sigma_{collisions}(\mathbf{r}_i - \mathbf{r}_j) \cdot \mathbf{f}_j^i \rangle \qquad (1.b.167)$$

The collisional interaction can be evaluated as $\mathbf{f}_j^i = \frac{d}{dt} m\mathbf{v}_i|_{collision} = \frac{\Delta m\mathbf{v}_i}{\Delta t}$ and

$$\frac{1}{\Delta t} \Sigma_{coll}^{in\ \Delta t} r_{ij} < |\Delta m\mathbf{v}_i| > = v_{coll} \frac{N}{2} a < |\Delta m\mathbf{v}_i| > \qquad (1.b.168)$$

where $r_{ij} = a$ for hard discs and $v_{coll}$ is the single-particle collision rate. We define an effective temperature although there is no heat bath, $K \equiv NT$. Hence,

$$P = nT + \frac{na}{4} v_{coll} < |\Delta m\mathbf{v}_i| > \qquad (1.b.169)$$

or using $v_{coll} \equiv \frac{\sqrt{<v_i^2>}}{\ell}$ where $\ell$ is the collisional mean-free-path

$$\frac{P}{nT} = 1 + \frac{av_{coll}}{4} \frac{<|\Delta m\mathbf{v}_i|>}{<\frac{1}{2} m v_i^2>} \equiv 1 + \frac{a}{\ell} \frac{<|\Delta \mathbf{v}_i|>}{2\sqrt{<v_i^2>}} \qquad (1.b.170)$$

The right side of (1.b.170) has no explicit temperature dependence, only geometry and density dependence, at least for hard discs.

Hence, $P/T$ vs $n$ is a universal curve independent of isotherm for hard-sphere interactions.

We note that in a dilute medium the collisional mean-free-path scales as $\ell \sim 1/n\sigma \sim 1/na^2$ where $\sigma$ is the collision cross-section. In dense regimes the collisions are pretty much head on. As $n \to n_{max}$ then $\ell \to 0$ in crystalline structures.

In 1957 Alder and Wainwright published the results of numerical Monte Carlo calculations based on a hard-sphere model for the interaction potential leading to equation-of-state results (Alder and Wainwright, 1957; Wood and Jacobson, 1957). Figure 8 taken from Wood and Jacobson shows Alder and Wainwright's equation of state results for systems with 108 and 32 molecules, in which $PV_0/T$ is plotted vs the normalized volume $V/V_0$, which is proportional to the number density $n$, and $V_0$ is the close-pack volume for the system. Alder and Wainwright's results show an overlapping region wherein two distinct pressure states can co-exist for the same volume. The system supports the possibility of a spontaneous transition. In the overlap region, i.e., the two-phase regime, there is a great amount of instability with large fluctuations.

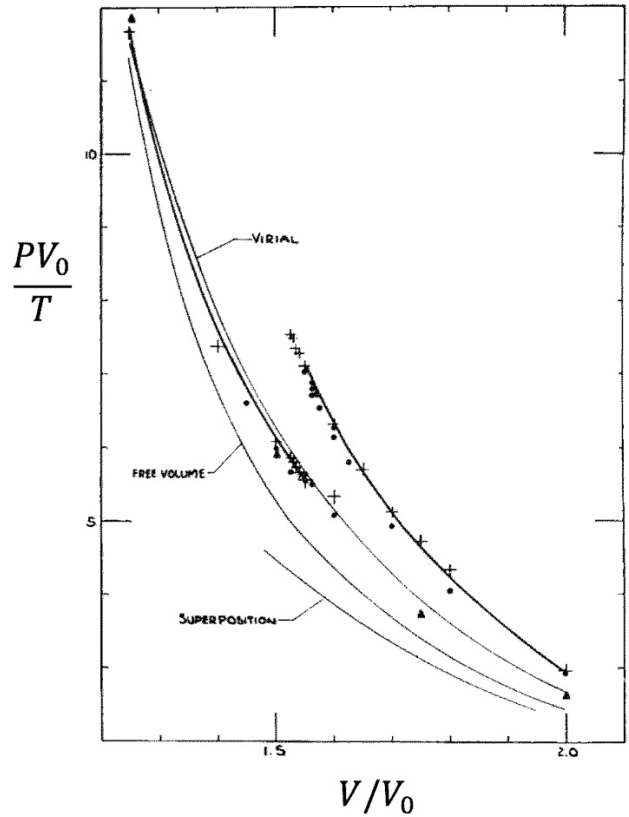

Fig. 1.b.8 Monte Carlo equation of state results from Woods and Jacobson 1957 showing their results and those of Alder and Wainwright 1957 (solid line for 108 molecules and + for 32 molecules).

If we were to plot *P/T* vs *n* for Alder and Wainwright's results in Fig. 1.b.8, *P/T* would grow from 0 at *n*=0 in the fluid region for increasing *n* until it reaches a critical density $n_c$, above which *P/T* might continue to increase if the system remains a fluid, but the system may instead jump to a crystalline solid branch at a lower value of *P/T* whose *P/T* then increases with *n*. Note that we earlier proved $\frac{dP}{dn} > 0$ for the grand canonical ensemble, while the Alder and Wainwright systems do not satisfy this constraint because their systems are microcanonical ensembles with a relatively small number of particles/molecules rather than an infinite number.

Theories of the liquid state are difficult. Liquids are hard-sphere systems with a weak interaction between particles/molecules considered as a perturbation.

$$\frac{P}{T} = \frac{\partial S(E,V,N)}{\partial V} = -n^2 \frac{\partial S(\mathcal{E},n)}{\partial n} \rightarrow \frac{P}{nT} = 1 + h(n) \tag{1.b.171}$$

From Eqs.(1.b.81) and (1.b.171) one obtains

$$S(\mathcal{E}, n) = \frac{5}{2} - \ln n\Lambda^3(\mathcal{E}) - \int_0^n \frac{dn'}{n'} h(n') \tag{1.b.172}$$

Why is the entropy contribution negative as *n* increases? As *n* increases for *N* fixed, V decreases and configuration space is limited so $\Gamma$ decreases and the entropy must decrease.

Consider a solid and its own vapor with strong enough forces such that even at $P \to 0$ there is a solid. Assume there is a solid surrounded by a gas. Further assume that the energy of the solid can be represented by

$$E_{sol} = -N_{sol}\mathcal{E}_b + f(T)(\text{neglible term}) + g(V)(\text{neglible term}), \quad \frac{1}{V}\frac{dV}{dP}P_0 \ll 1 \tag{1.b.173}$$

The specific energy of the solid is $\mathcal{E}_{sol} = -\mathcal{E}_b$, the binding energy. Fixed constants are

$$N = N_{gas} + N_{sol} \text{ and } V = V_{gas} + V_{sol}, \text{ where } V_{sol} = N_{sol}\mathcal{V}_0 \tag{1.b.174}$$

where $\mathcal{V}_0$ is the volume per solid molecule. At equilibrium

$$\mu_{sol} = \mu_{gas} \to \mu_{sol} = \frac{\partial E_{sol}}{\partial N_{sol}} = -\mathcal{E}_b = \mu_{gas} = T \ln n_{gas}\Lambda^3(T)$$

$$\Rightarrow n_{gas}\Lambda^3 = e^{-\mathcal{E}_b/T} \tag{1.b.175}$$

Equation(1.b.175) is a nice formula for the vapor pressure. The classical limit in (1.b.175) is $n_{gas}\Lambda^3 \ll 1$, which implies that for $\mathcal{E}_b \sim 1$ eV, $T \ll 1$ eV = $12000°K$.

Consider increasing $N_{sol}$ and concomitantly $V_{sol} = N_{sol}\mathcal{V}_0$, while holding *T* fixed. The volume $V_{gas}$ must decrease for the total volume *V* fixed. In this case, $N_{gas}\Lambda^3 = V_{gas}e^{-\mathcal{E}_b/T}$ decreases with decreasing $V_{gas}$. $n_{gas} = N_{gas}/V_{gas}$ is just a function of temperature and remains fixed. Hence, $P_{gas} = n_{gas}T = \frac{T}{\Lambda^3}e^{-\mathcal{E}_b/T}$ remains fixed, i.e., the vapor pressure remains constant as we increase $N_{sol}$.

We next include volume dependence in Eq.(1.b.173) so that $V_{sol}$ is allowed to vary around its optimum equilibrium value:

$$E_{sol} = -N_{sol}\mathcal{E}_b + \alpha\mathcal{E}_b \frac{N_{sol}}{2}\frac{(\mathcal{V}_s - \mathcal{V}_0)^2}{\mathcal{V}_0^2} \tag{1.b.176}$$

where $\frac{N_{sol}}{2}$ represents the number of interacting pairs and $\mathcal{V}_s = V_{sol}/N_{sol}$. From (1.b.176) is follows that

$$P_{sol} = -\frac{\partial E_{sol}}{\partial V_{sol}}\bigg|_{N_{sol}} = -\alpha\mathcal{E}_b \frac{(\mathcal{V}_s - \mathcal{V}_0)}{\mathcal{V}_0^2} \tag{1.b.177}$$

which has the form of a Hooke's force law. We can divide through by $N_{sol}$ in (1.b.176) to obtain the specific energy per solid molecule. We note that the system has the following attributes:

$$P_{sol} = P_{gas} \qquad \mu_{sol} = \mu_{gas}$$

$$V = V_{sol} + V_{gas} \qquad N = N_{sol} + N_{gas} \qquad S = S_{sol} + S_{gas} \qquad (1.b.178)$$

The total entropy is the sum of the gas and solid entropies:

$$S = S_{sol}(E_{sol}, V_{sol}, N_{sol}) + S_{gas}(E_{gas}, V_{gas}, N_{gas}) =$$

$$N_{sol} s_{sol}(\mathcal{E}_{sol}, n_{sol}) + N_{gas} s_{gas}(\mathcal{E}_{gas}, n_{gas}) \qquad (1.b.179)$$

Given that the solid and gas are in equilibrium with one another at the same temperature, then $P_{sol} = P_{gas}$ and $\mu_{sol} = \mu_{gas}$ in Eq.(1.b.178). From $\mu_{sol} = \left.\frac{\partial E_{sol}}{\partial N_{sol}}\right|_{S_{sol}, V_{sol}} = -\mathcal{E}_b + \frac{1}{2}\alpha \mathcal{E}_b \frac{(v_s-v_0)^2}{v_0^2}$, $\mu_{gas} = T \ln n_{gas} \Lambda^3(T)$, $\mu_{sol} = \mu_{gas}$, $P_{gas} = n_{gas} T = \frac{T}{\Lambda^3} e^{-\mathcal{E}_b/T} = P_{sol} = -\alpha \mathcal{E}_b \frac{(v_s-v_0)}{v_0^2}$, it then follows

$$\frac{(v_s-v_0)}{v_0} = -\frac{1}{\alpha} \frac{T}{\mathcal{E}_b} \frac{v_0}{\Lambda^3} e^{-\mathcal{E}_b/T} \qquad (1.b.180)$$

where $\frac{1}{\alpha} \sim O(1)$, $\frac{T}{\mathcal{E}_b} < O(1)$, $\frac{v_0}{\Lambda^3} \sim O(1)$, and $e^{-\mathcal{E}_b/T} \ll 1$.

From (1.b.180) we conclude that only negligible deviations (compression or expansion) of $\mathcal{V}_s$ from $\mathcal{V}_0$ are allowed. From (1.b.178) and (1.b.180) we can also derive

$$N_{gas} \cong \frac{V - N_{sol} v_0}{\Lambda^3} e^{-\mathcal{E}_b/T} \qquad (1.b.181)$$

Figure 9 presents of schematic diagram for the *P* vs *V* relation for the gas-solid system with *N* and *T* fixed. For small values of *V* greater than the minimum value $Nv_0$ the system is a solid and the pressure is largest. As *V* increases, *P* decreases while $N_{sol}/N \sim 1$ for a while, until $N_{sol}/N$ begins to decrease and $N_{gas}/N$ increases. Both gas and solid phases occupy the flattish intermediate region of the *P* vs *V* relation. At the largest values of *V* there is only the gas phase. Not shown in Fig. 1.b.9 are trajectories followed from either end of *P* vs *V* where we progress along curves in the complete absence of the other phase, viz., beginning at largest *V* gas only and beginning at smaller *V* solid phase only. Along these curves we can have meta-stable states which require either nonuniformities, e.g., for the super-saturated vapor to collect and precipitate upon, or over a long time cavities will appear (when the solid is subjected to too little pressure) in which there is vapor.

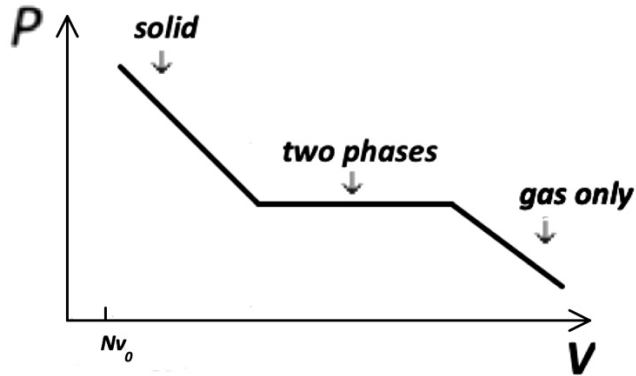

Fig. 1.b.9 Schematic for *P* vs *V* phase diagram for the gas-solid system

### 1.b.xi Quantum virial expansion

Consider hydrogen atoms (no ionization) and H$_2$ formation. In the quantum picture the grand canonical partition function is

$$\mathbb{Z}(\beta,\gamma,V) = \sum_{N=0}^{\infty} e^{-\gamma N} Z_N(\beta,V) = \sum_{N=0}^{\infty} e^{-\gamma N} \sum_n e^{-\beta E_n(V)} = 1 + e^{-\gamma} Z_1 + e^{-2\gamma} Z_2 + \cdots$$

(1.b.182a)

where $\mu \equiv -T\gamma$ and introducing the internal energy $\varepsilon_k^{int}$

$$Z_1 = \sum_k e^{-\beta \varepsilon_k} = \sum_k e^{-\beta\left(\frac{\hbar^2 k^2}{2m} + \varepsilon_k^{int}\right)} = \left(\sum_k e^{-\beta \frac{\hbar^2 k^2}{2m}}\right)\left(\sum_n e^{-\beta \varepsilon_n^{int}}\right) = \frac{V}{\Lambda^3} Z_1^{int}(\beta) \quad (1.b.182b)$$

Recalling the analysis in Eqs.(1.b.106) to (1.b.108) the statistical weight factor associated with the possible quantum states now including angular momentum and spin quantum numbers is $g_0 = (2S+1)(2I+1)$ where *S*=1/2 and *I*=1/2, then $Z_1^{int}(\beta) = \sum_n e^{-\beta \varepsilon_n^{int}} = 4e^{-\beta(-I_H)}$ where $I_H = 13.6$ eV, and we ignore excited states of hydrogen to be consistent with the assumption of no ionization. Recalling the virial expansion in Eqs.(1.b.131)-(1.b.132) we have

$$\mathbb{Z}(\gamma,\beta,V) = \sum_{N=0}^{\infty} \frac{1}{N!}(Vz)^N Q_N(\beta,V) \tag{1.b.183}$$

Define

$$Z_N \equiv \frac{(Z_1)^N}{N!} Q_N \tag{1.b.184}$$

and $Q_N$ are quantum virial coefficients:

$$Q_0 = 1 \quad Q_1 = 1 \quad Q_2 = \frac{2Z_2}{Z_1^2} \ldots \tag{1.b.185}$$

Analogous to (1.b.147), $\mu = T[\ln z\Lambda^3 - \ln g_0] - I_H$ where $g_0 = (2S+1)(2I+1) = O(1)$ compared to $z\Lambda^3$. Note that $z$ as in (1.b.160) is equal to $n$ to lowest order. Analogous to (1.b.161)

$$\beta P(n,T) = \sum_{j=1}^{\infty} n^j b_j(T), \quad b_1 = 1, \quad b_2(T) = -\frac{V}{2}[Q_2(T,V) - 1], \ldots \tag{1.b.186}$$

Example: For an ideal gas we employ classical counting for the possible states and conclude that $z_2 = z_1^2/2$ and $b_2$=0.

Example: For quantum systems we carefully count the possible states. Consider bosons with no internal structure, e.g., the helium atom with two states: $n = (k_1, k_2)$. Then with $E_n = \mathcal{E}_{k_1} + \mathcal{E}_{k_2}$

$$Z_2 = \sum_n e^{-E_n} = \left( \sum_{k_1,k_2,k_1<k_2} + \sum_{k_1=k_2} \right) e^{-\beta(\mathcal{E}_{k_1}+\mathcal{E}_{k_2})} = \left( \frac{1}{2}\sum_{k_1,k_2} + \frac{1}{2}\sum_{k_1=k_2} \right) e^{-\beta(\mathcal{E}_{k_1}+\mathcal{E}_{k_2})} =$$
$$\frac{1}{2}\left(Z_1^2(\beta) + \sum_{k_1} e^{-2\beta \mathcal{E}_k}\right) = \frac{1}{2}\left(Z_1^2(\beta) + Z_1(2\beta)\right) \tag{1.b.187}$$

where from (1.b.36) and (1.b.184) to (1.b.186)

$$Z_1(\beta) = V/\Lambda^3(\beta) \text{ and } b_2(T) = \frac{-V}{2Z_1^2(\beta)} Z_1(2\beta) = -\frac{\Lambda^3(\beta)}{4\sqrt{2}} \tag{1.b.188}$$

Example: For Fermions we exclude the state with $k_1 = k_2$ and derive

$$b_2(T) = \frac{\Lambda^3(\beta)}{4\sqrt{2}} \tag{1.b.189}$$

because Fermions have repulsive interactions.

### 1.b.xii Numerical simulation of equations of state and phase transitions, Berni Alder and molecular dynamics

Berni Alder's work was mentioned earlier in Secs. 1.b.ix and 1.b.x Alder made seminal contributions to molecular dynamics and was a pioneer in demonstrating molecular dynamics as a viable approach to studying the statistical mechanics of many-body interacting systems. (Alder, 1972; Alder, 1973)  By employing Monte Carlo methods in the numerical integration of the Newtonian equations of motion for ensembles of particles, i.e., molecular dynamics simulations, a numerical scheme for solving the Liouville equations was devised; and the partition function and its derivatives were obtained.  This approach was necessarily limited in the number of particles and hence generated a microcanonical ensemble.  The application of Monte Carlo integration methods is not straightforward and is good only when the quasi-ergodic hypothesis is valid.  How does one evaluate Eq.(1.b.133) using Monte Carlo integration methods?

$$Q_N(T,V) = \int \frac{d^{3N}\mathbf{r}_i}{V^N} e^{-\beta \Phi(\mathbf{r}_1,\mathbf{r}_2,\ldots)}$$

In discussing the challenges attendant in numerically calculating $Q_N$ and the partition function there are several points to make.

1. The numerical integration of $Q_N$ involves a multi-dimensional integration with a certain number of Monte Carlo sampling points per dimension.  If there are $l$ points per dimension and $3N$ dimensions, then there could be as many as $l^{3N}$ evaluations of the integrand.  The dimensionality could be a number of order ~$10^{23}$.  Furthermore, $Q$ can be sharply peaked necessitating more numerical resolution locally.  The curse of dimensionality is a formidable aspect in evaluating $Q_N$.

   [*Editor's Note: In recent years researchers dealing with uncertainty quantification and machine learning have introduced systematic approaches to sample a multi-dimensional space in an optimally efficient fashion to mitigate the "curse of dimensionality" problem.  Furthermore, in the 50 years since these lectures computing power has grown enormously; and molecular dynamics simulations have been able to evolve to address systems that are orders of magnitude larger and more complex.*]

   Numerical methods have more success in the case of the virial expansion of the equation of state where the dimensionality and complexity of the successive virial coefficients grows in a very limited fashion, e.g., Eqs.(1.b.165) and (1.b.166).

2. For instantaneous phase-space pictures one can calculate the potential energy by randomly placing particles in a box and computing $V_N = \sum V_{ij}(r_{ij})$, and this is used in $e^{-V_N/kT}$.  Problems emerge at high density where particles get so close together that the interaction potential $V_{ij} \to \infty$ and $e^{-V_N/kT} \to 0$.  In practice, it is unlikely that random loading of particles one by one will get to the critical densities where divergences may occur.  So-called quiet and quasi-random loading algorithms have been developed.  The loading problem becomes a function of the loading history.

3. What succeeds is "importance sampling" in which a modified distribution is sampled instead of the actual distribution in order to reduce the variance in the sampling process.

One generates a Markov chain numerically to obtain results for a canonical ensemble (in reality a micro-canonical ensemble).

Example: Consider hard spheres for which $\phi = 0$ or $\infty$ (not accessible). Load $N$ hard spheres in a defined volume $V$ for a given temperature $T$. Randomly displace one sphere in the list. If this results in no overlap with another hard sphere, then this is a successful new configuration and weight it by $e^{-V_N/kT}$. If instead this results in overlapping another hard sphere, then return the sphere to its original position and count the old configuration in the sum over states weighted by $e^{-V_N/kT}$. Continue through the list. This kind of method works in practice, but in a certain sense it is theoretically incapable of coming up with all configurations for hard spheres. The simulation examples show in Fig. 1.b.8 illustrate phase transition phenomena for relatively small ensembles of simulation test particles. For insufficient numbers of particles it becomes difficult to distinguish the distinct phases, and the relative size of the statistical fluctuations becomes problematic.

### 1.b.xiii Example: structureless particles with an interaction potential

Here we consider structureless particles with an interaction potential represented in the virial expansion (Secs 1.b.ii and 1.b.ix) by

$$b_2(T) = -\frac{V}{Z_1^2}\left(Z_2 - \frac{1}{2}Z_1^2\right) \tag{1.b.190}$$

where

$$Z_1 = \frac{V}{\Lambda^3(\beta,m)}, \quad Z_2 = Z_2^{c.m.} Z_2^{rel}, \quad Z_2^{c.m.} = \frac{V}{\Lambda^3(\beta,2m)} = \frac{V}{\Lambda^3(\beta,m)} 2^{3/2} \tag{1.b.191}$$

We introduce $Z_2^{rel} = \left(Z_2^{rel} - Z_2^{rel(0)}\right) + Z_2^{rel(0)}$ where the terms in the parentheses are the contribution due to the interaction. Then $b_2 = b_2^{(0)} + b_2^{int}$ and

$$b_2^{int}(T) = -\frac{V^2}{\left(\frac{V}{\Lambda^3}\right)^2 \Lambda^3} 2^{\frac{3}{2}}\left(Z_2^{rel} - Z_2^{rel(0)}\right) = -2^{\frac{3}{2}}\Lambda^3(\beta,m)\left[\sum_k e^{-\beta\mathcal{E}_k} - \sum_{k^0} e^{-\beta\mathcal{E}_{k^0}}\right] \tag{1.b.192}$$

The first sum on the right side of (1.b.192) is

$$\sum_k e^{-\beta\mathcal{E}_k} = \sum_{k,bound} e^{+\beta|\mathcal{E}_k|} + \sum_{k,free} e^{-\beta\mathcal{E}_k} \tag{1.b.193}$$

where $\mathcal{E}_k = \frac{1}{2}\frac{\hbar^2 k^2}{\left(\frac{m}{2}\right)}$ (note the reduced mass m/2); and the second sum in (1.b.192) is just over free (or "scattered") states ($\mathcal{E}_{k^0} > 0$). At large distances the asymptotic wave function for free states (positive energy) with angular momentum $\ell$ is (Landau and Lifshitz, 1969; Sec. 77)

$$\psi_\ell \sim \frac{1}{r}\sin\left(kr - \frac{\ell\pi}{2} + \delta_\ell(k)\right), k = \frac{p}{\hbar};\ \delta_\ell(k) = 0 \text{ if no interaction}, \neq 0 \text{ with interaction} \quad (1.b.194)$$

Returning to Eq.(1.b.192) and (1.b.193), in the limit of large volume

$$\sum_{k,free} e^{-\beta\varepsilon_k} - \sum_{k^0,free} e^{-\beta\varepsilon_{k^0}} \to \frac{1}{\pi}\sum_{\ell=0}^\infty (2\ell+1)\int_0^\infty dk \frac{d\delta_\ell}{dk} e^{-\beta\hbar k^2/2m} \quad (1.b.195)$$

and (1.b.192) yields after integrating by parts

$$b_2^{int}(T) = -\Lambda^3\left(\beta, \frac{m}{2}\right)\sum_{\ell=0}^\infty (2\ell+1)\left[\sum_{\varepsilon_k^\ell<0} e^{+\beta|\varepsilon_k^\ell|} - \frac{1}{\pi}\delta_\ell(0) + \frac{\beta}{\pi}\int_0^\infty d\varepsilon\, \delta_\ell(\varepsilon)e^{-\beta\varepsilon}\right]$$

$$= -\Lambda^3\left(\beta, \frac{m}{2}\right)\sum_{\ell=0}^\infty (2\ell+1)\left[\sum_{\varepsilon_k^\ell<0}[e^{+\beta|\varepsilon_k^\ell|} - 1] + \frac{\beta}{\pi}\int_0^\infty d\varepsilon\, \delta_\ell(\varepsilon)e^{-\beta\varepsilon}\right] \quad (1.b.196)$$

using Levinson's theorem from quantum scattering theory: for k=0  $\delta_\ell(0) = \pi \times$ no. of bound states.

[*Editor's Note: Section 77 of (Landau and Lifshitz, 1969) gives a more detailed explanation of the analysis leading to the results in this section. Kaufman's lectures in this section are not self-contained, depend on other sources, and are more of an overview.*]

Example: Bosons – Bosons have symmetric wave functions, and the angular momentum quantum number $\ell$ is even. Note that Fermions have anti-symmetric wave functions. If we assume that the bosons have repulsive interactions then there are no bound states, and the first sum inside the square bracket in (1.b.196) vanishes. Furthermore, from quantum mechanics $\delta_\ell < 0$, which then leads directly to $b_2^{int}(T) > 0$; and the pressure has increased due to the interaction.

Example: Bose gas with hard-sphere repulsive interaction – From the quantum mechanical treatment for the phase shift $\delta_\ell$ in (Schiff, 1968) Sec. 19, there is a treatment for a spherically symmetric interaction potential in the special limit of a hard-sphere interaction, Schiff's Eq.(19.20), which in the low energy limit simplifies to Schiff's Eq.(19.21). As $T \to 0$ only the $\ell = 0$ quantum number significantly contributes, in which limit

$$\delta_0(ka) = -ka \text{ and } b_2^{int}(T \to 0) = \Lambda^2\left(\beta, \frac{m}{2}\right) a \quad (1.b.197)$$

where $r_{ij} = a$ defines the distance between the hard-sphere centers. The pressure increases because $b_2 > 0$. The pressure at low temperatures for the Bose gas with hard-sphere interaction then follows from (1.b.114), (1.b.186), and (1.b.197), and recalling $b_2 = b_2^{(0)} + b_2^{int}$:

$$\frac{P}{nT} = 1 - \frac{n\Lambda^3\left(\frac{m}{2}\right)}{16} + n\Lambda^2 a + O(n^2) \quad (1.b.198)$$

Example: Bose gas with an attractive interaction potential – In this case bound states are possible and $b_2^{int} < 0$, so the pressure is reduced. The term involving $\sum_{\varepsilon_k^\ell < 0}[e^{+\beta|\varepsilon_k^\ell|} - 1]$ in (1.b.196) contributes a net negative contribution to $b_2^{int}$, and from quantum mechanics $\delta_\ell > 0$ so the second term in the bracket (1.b.196) also contributes a net negative contribution. As $T \to 0$ ($\beta \to \infty$) just a single bound state and only $\ell = 0$ are important. Hence, at low temperatures

$$b_2^{int}(T \to 0) = -\Lambda^3 e^{\beta \varepsilon_b} \text{ and } \frac{P}{nT} = 1 - n e^{\beta \varepsilon_b} \Lambda^3 \left(\frac{m}{2}\right) \tag{1.b.199}$$

where $\varepsilon_b$ is the disassociation energy, i.e., the binding energy of the molecule.

## 1.c Chemical equilibrium

### 1.c.i Systems comprised of multiple species allowing for chemical reactions

Here we analyze systems with multiple species which interact with one another through a chemical reaction. Some examples of simple systems are

$$\begin{aligned} H_2 &\leftrightarrow H + H \\ H &\leftrightarrow p^+ + e^- \\ 2H_2O &\leftrightarrow 2H_2 + O_2 \\ H^+ + Cl^- &\leftrightarrow HCl \end{aligned} \tag{1.c.1}$$

As a matter of notation, reactions like $H^+ + Cl^- \leftrightarrow HCl$ can be represented generically as $A + B \leftrightarrow C \equiv AB$. We assume that a system I supporting a reaction like those in (1.c.1) is in contact with system II which acts as a heat bath such that

$$E = E^I + E^{II} = \text{constant} \tag{1.c.2}$$

and the chemical reaction dictates the following conservation laws

$$N_A + N_C = N_a = \text{constant}$$

$$N_B + N_C = N_b = \text{constant} \tag{1.c.3}$$

where $N_s$ is the number of atoms or molecules in the combined system. From the point of view of a grand canonical ensemble, the probability $\rho$ of the system I with $\{N_A^I, N_B^I, N_C^I\}$ is

$$\rho(N_A^I, N_B^I, N_C^I) \sim \sum_{E^I} \Gamma_I(N_A^I, N_B^I, N_C^I, E^I)$$

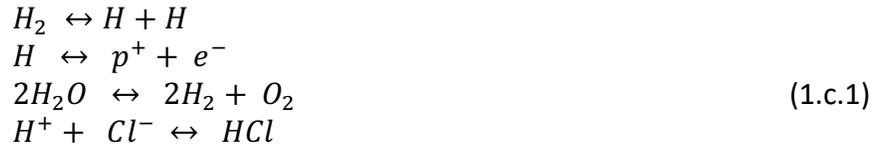

$$\times \Gamma_{II}(N_a^{II} = N_a - [N_A^I + N_C^I], N_B^{II} = N_b - [N_B^I + N_C^I], E^{II} = E - E^I) \tag{1.c.4}$$

and

$$\Gamma_I = e^{S_I} \text{ and } \Gamma_{II} = e^{S_{II}} \tag{1.c.5}$$

We assume that system I is a small perturbation with respect to system II, which allows us to evaluate

$$S_{II} = S_{II}(N_a, N_b, E) - E^I \left(\frac{\partial S_{II}}{\partial E_{II}}\right) - [N_A^I + N_C^I]\left(\frac{\partial S_{II}}{\partial N_a^{II}}\right) - [N_B^I + N_C^I]\left(\frac{\partial S_{II}}{\partial N_b^{II}}\right) =$$

$$S_{II}(N_a, N_b, E) - \beta_{II} E^I - \gamma_a^{II}[N_A^I + N_C^I] - \gamma_b^{II}[N_B^I + N_C^I] \tag{1.c.6}$$

We then use Eqs.(1.c.5) and (1.c.6) to express the probability in (1.c.4) as

$$\rho(N_A^I, N_B^I, N_C^I) \sim \sum_{E^I} e^{-\beta_{II}(E^I - TS^I)} e^{-\gamma_a^{II}(N_A^I + N_C^I) - \gamma_b^{II}(N_B^I + N_C^I)} =$$

$$\sum_{E^I} e^{-\beta(E^I - TS^I)} e^{-\gamma_a(N_A^I + N_C^I) - \gamma_b(N_B^I + N_C^I)} \tag{1.c.7}$$

where $\beta, \gamma_a, \gamma_b$ are determined in (1.c.6) from partial derivatives on $S_{II}$. We note that $F^I = E^I - TS^I$. We next introduce the definitions:

Definition: $\quad\quad \mu_A \equiv -T\gamma_a, \ \mu_B \equiv -T\gamma_b, \ \mu_c \equiv -T(\gamma_a + \gamma_b) \tag{1.c.8}$

We note that the definitions in (1.c.8) dictate $\mu_c = \mu_A + \mu_B$. The probability in (1.c.7) can then be rewritten as

$$\rho(N_A^I, N_B^I, N_C^I) \sim \sum_{E^I} e^{-\beta F^I + \beta \boldsymbol{\mu} \cdot \boldsymbol{N}} = \sum_{E^I} e^{-\beta(E^I - TS(E^I, N)) + \beta \boldsymbol{\mu} \cdot \boldsymbol{N}} \tag{1.c.9}$$

where $\boldsymbol{\mu} \equiv \{\mu_A, \mu_B, \mu_C\}$ and $\boldsymbol{N} \equiv \{N_A, N_B, N_C\}$.

Lemma: From the formalism in Sec. 1.b.vii and the expression in (1.c.9) we deduce

$$\langle \boldsymbol{N} \rangle = T \frac{\partial \ln \mathbb{Z}}{\partial \boldsymbol{\mu}} \tag{1.c.10}$$

where $\mathbb{Z}$ is the grand canonical partition function:

$$\mathbb{Z}(\boldsymbol{\mu}, \beta, V) \equiv \sum_{\{N_s\}} e^{\beta \sum_s \mu_s N_s} Z(\boldsymbol{N}, \beta, V) = \sum_{\{N_s\}} e^{\beta \sum_s \mu_s N_s} \prod_s Z_s(N_s, \beta, V) =$$

$$\prod_s \sum_{\{N_s\}} e^{\beta \mu_s N_s} Z_s(N_s, \beta, V) = \prod_s \mathbb{Z}_s(\beta, \mu_s, V) \tag{1.c.11}$$

extending expressions in Sec.1.b.vii to multiple species, using an ideal-gas approximation for $Z(\mathbf{N}, \beta, V) = \prod_s Z_s(N_s, \beta, V)$, and recalling $\mu_c = \mu_A + \mu_B$. From (1.c.11) it follows that

$$\ln \mathbb{Z} = \sum_s \ln \mathbb{Z}_s(\mu_s, \beta, V) \quad (1.c.12)$$

We recall from Eq.(1.b.88) that $\langle N_s \rangle = T \frac{\partial \ln \mathbb{Z}}{\partial \mu_s} = T \frac{\partial \ln \mathbb{Z}_s}{\partial \mu_s}$ and observe that the existence of the chemical reactions is only felt through $\mu_c = \mu_A + \mu_B$.

### 1.c.ii The law of mass action

From (1.c.12) and (1.b.136) we evaluate the pressure

$$P \equiv \frac{T}{V} \ln \mathbb{Z} = \sum_s P_s \quad (1.c.13)$$

and the partial pressure is

$$P_s \equiv \frac{T}{V} \ln \mathbb{Z}_s = \frac{T}{V} \ln \sum_{N_s=0}^{\infty} e^{\beta \mu_s N_s} \frac{(Z_1^s)^{N_s}}{N_s!} = \frac{T}{V} \ln \exp(e^{\beta \mu_s} Z_1^s) =$$

$$\frac{T}{V} e^{\beta \mu_s} \frac{V}{\Lambda^3} Z_s' = \frac{T}{\Lambda^3} e^{\beta \mu_s} Z_s' \quad (1.c.14)$$

where $Z_1^s = \frac{V}{\Lambda^3} Z_s'$ and $Z_s'$ is the partition function for the internal states and the sum over $N_s$ in (1.c.14) is recognized as the exponential. We recall Eq.(1.b.88) and the Gibbs-Duhem relations Eqs(1.b.99) and (1.b.100) which when used in conjunction with (1.c.14) yield:

$$\frac{\partial P_s}{\partial \mu_s} = \frac{T}{V} \frac{\partial \ln \mathbb{Z}_s}{\partial \mu_s} = \langle n_s \rangle \quad (1.c.15)$$

and note that $P_s = \langle n_s \rangle T$. The statistical average $\langle n_s \rangle$ removes statistical fluctuations in the number density. From (1.c.14) and (1.c.15) we obtain $\langle n_s \rangle \Lambda^3 = e^{\beta \mu_s} Z_s'$ and then with $\langle n_s \rangle \approx n_s$, ignoring fluctuations and taking the logarithm,

$$\mu_s = T[\ln(n_s \Lambda^3) - \ln Z_s'] \quad (1.c.16)$$

From $\mu_c = \mu_A + \mu_B$ and (1.c.16) one derives

$$\ln n_A \Lambda_A^3 + \ln n_B \Lambda_B^3 - \ln n_C \Lambda_C^3 = \ln Z_A' + \ln Z_B' - \ln Z_C'$$

or

$$\frac{n_A \Lambda_A^3 n_B \Lambda_B^3}{n_C \Lambda_C^3} = \frac{Z_A' Z_B'}{Z_C'} \quad (1.c.17)$$

for an ideal gas. It is straightforward to include stochiometric coefficients: $\nu_C \mu_C = \nu_A \mu_A + \nu_B \mu_B$, and then (1.c.17) becomes

$$\frac{(n_A \Lambda_A^3)^{\nu_A}(n_B \Lambda_B^3)^{\nu_B}}{(n_C \Lambda_C^3)^{\nu_C}} = \frac{(z'_A)^{\nu_A}(z'_B)^{\nu_B}}{(z'_C)^{\nu_C}} \tag{1.c.18}$$

We can group all the temperature dependence in (1.c.17) on the right side to obtain:

$$\frac{n_A n_B}{n_C} = \frac{\Lambda_C^3}{\Lambda_A^3 \Lambda_B^3} \frac{z'_A z'_B}{z'_C}(T) \tag{1.c.19}$$

From $P_s = n_s T$ ignoring fluctuations and (1.c.19) one then obtains

$$\frac{P_A P_B}{P_C} = T \frac{\Lambda_C^3}{\Lambda_A^3 \Lambda_B^3} \frac{z'_A z'_B}{z'_C}(T) \equiv K_p(T) \tag{1.c.20}$$

which is Eq.(104.3) in (Landau and Lifshitz, 1969) and is called the law of mass action which is a formula employed in chemistry. With $c_s \equiv P_s/P$ then (1.c.20) is equivalent to $\frac{c_B c_B}{c_C} = P K_p(T) \equiv K_c(P, T)$.

From the point of view of a closed isolated system with energy $E$ and volume $V$, and constraint (1.c.3), we can maximize the entropy $S$ with respect to $(N_A, N_B, N_C, E, V)$ and produce a general derivation of the most probable partition function independent of the assumption of an ideal gas. Then

$$0 = \delta S = \delta N_A \frac{\partial S}{\partial N_A} + \delta N_B \frac{\partial S}{\partial N_B} + \delta N_C \frac{\partial S}{\partial N_C} \tag{1.c.21}$$

with $E$ and $V$ fixed so they don't appear. With $\gamma_s \equiv \frac{\partial S}{\partial N_s}$ and $\delta N_A = \delta N_B = -\delta N_C$, it follows that $\gamma_C = \gamma_A + \gamma_B$. From the definition, $\mu_s = -T\gamma_s$ we then recover $\mu_C = \mu_A + \mu_B$.

### 1.c.iii Derivation of the Saha equation

Consider a system comprised of hydrogen, protons, and electrons, allowing for ionization.

$$H \leftrightarrow p^+ + e^-, \ \mu_H = \mu_p + \mu_e, \ Z'_p = 2 \text{ and } Z'_e = 2 \tag{1.c.22}$$

where the internal partition functions capture the two possible spin states (up and down).

From (1.c.17) assuming negligible excitation of the hydrogen atoms, $T \ll I \sim 13.6 \text{eV}$,

$$\frac{n_p \Lambda_p^3 n_e \Lambda_e^3}{n_H \Lambda_H^3} = \frac{Z'_p Z'_e}{Z'_H} \tag{1.c.23}$$

where $Z'_H = 4e^{-\beta E_0} = 4e^{\beta I}$ and $E_0 = -I$. We further assume that there is no electron/nuclear spin interaction and simplify (1.c.23) using $n_p \sim n_e$, $m_p \sim m_H$, and $\Lambda_p^3 \sim \Lambda_H^3$ to obtain the Saha equation:

$$\frac{n_e^2 \Lambda_e^3}{n_H} = \frac{Z'_p Z'_e}{Z'_H} = \frac{2 \times 2}{4 e^{\beta I}} = e^{-\beta I} \tag{1.c.24}$$

Definition: The degree of ionization is

$$f \equiv \frac{[H^+]}{[H^+]+[H]} = \frac{n_e}{n_e + n_H} \equiv \frac{n_e}{n_0} \tag{1.c.25}$$

The total density is defined $n = n_H + n_e + n_p = n_0 + n_e = n_0(1+f)$ and $P = nT$. We use these definitions and divide Eq.(1.c.24) by $n_0 = n/(1+f)$ to obtain the following.

From (Eq.(106.5) of (Landau and Lifshitz, 1969)

$$\frac{f^2}{1-f^2} = \frac{1}{n \Lambda_e^3} e^{-\beta I} \to f^2(n,T) = \frac{1}{1 + n \Lambda_e^3 e^{\beta I}} \tag{1.c.26}$$

Definition: We define the ionization temperature $T_I$ by setting $n \Lambda_e^3 = e^{-\beta I} \ll 1$ so that $I \gg T_I$ and

$$f(n, T_I) \equiv \frac{1}{\sqrt{2}} \tag{1.c.27}$$

From the ionization temperature relation

$$\ln \frac{1}{n \Lambda_e^3} = \frac{I}{T_I} \quad \text{or} \quad T_I(n) = \frac{1}{\ln \frac{1}{n \Lambda_e^3(T_I)}} \tag{1.c.28}$$

and

$$\ln \frac{1}{n \Lambda_e^3} = 49.3 - 2.3 \left[ \log_{10} n_{cc^{-1}} - \frac{3}{2} \log_{10} T_{eV} \right] \tag{1.c.29}$$

Example: For the interstellar gas, $n \sim 1\ cm^{-3}$, $T_I = \frac{13.6 \text{eV}}{49.3 + 3.5 \ln(T_I \sim \frac{1}{4})} = \frac{13.6}{47.2} \text{eV} = 0.29 \text{eV}$

Example: For atmospheric densities, $n \sim 10^{19}\ cm^{-3}$, $T_I = \frac{13.6}{6.6} \text{eV} \sim 2 \text{eV}$

Note: The recombination temperature for hydrogen is $1000° - 2000°K \sim 0.1 - 0.2$ eV, which sets a lower limit for a physically realistic value of $T_I$.

Equation (1.c.26) can be expressed using (1.c.28)

$$f(n,T) = \frac{1}{\sqrt{1+\left(\frac{T_I(n)}{T}\right)^{3/2} e^{T_i(n)/T}}} \quad (1.c.30)$$

The fractional ionization in (1.c.30) is plotted in Fig. 1.c.1 as a function of $T/T_I(n)$; we note that most of the change in $f$ occurs in the range $1/2 \leq T/T_I \leq 3/2$.

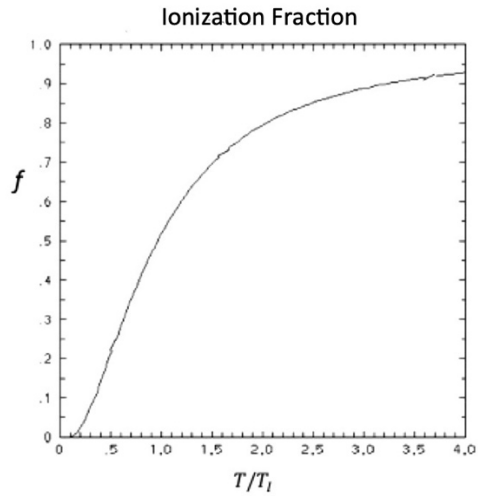

Fig. 1.c.1  Fractional ionization vs. $T/T_I(n)$ based on Eq.(1.c.30)

### 1.c.iv Chemical equilibrium including ionization and excited states

We now extend the analysis in Sec. 1.c.iii to include excitation of the hydrogen atom as a first correction.  The internal partition function becomes

$$Z'_H = \sum_{n=1}^{\infty} n^2 e^{\frac{\beta I}{n^2}} \quad (1.c.31)$$

where $n$ is the principal quantum number and the multiplier $n^2$ account for degeneracy ignoring spin degeneracy.  We note that the maximum atomic radius must scale as $R_{max} \sim \frac{1}{n_0^{1/3}}$ where $n_0$ is the density of atoms, while the atomic radius for an atom in the $n$th excited state scales as $R \approx n^2 a_0$ where $a_0$ is the Bohr radius.  Setting $R \approx R_{max}$ we deduce the maximum quantum number allowed

$$n_{max} = \sqrt{\frac{R_{max}}{a_0}} = \sqrt{\frac{1}{n_0^{1/3} a_0}} = \frac{1}{(n_0 a_0^3)^{1/6}} \gg 1 \quad (1.c.32)$$

Returning to (1.c.31)

$$Z_H' = e^{\beta I} + \sum_{n=2}^{\infty} n^2 e^{\frac{\beta I}{n^2}} \sim e^{\beta I} + \int_0^{n_{max}} n^2 dn = e^{\beta I} + \frac{1}{3} n_{max}^3 = e^{\beta I} + \frac{1}{3} \frac{1}{(n_0 a_0^3)^{1/2}} \quad (1.c.33)$$

The claim is that although $\frac{1}{(n_0 a_0^3)^{1/2}}$ is large compared to unity, it is small compared to $e^{\beta I}$ which scales as $e^{\beta I} \sim \frac{1}{n_0 \Lambda_e^3}$ based on (1.c.24). Hence, the excited state's contribution to the internal partition function is negligible.

### 1.d Long-range interactions

Long-range interactions are important in neutral and non-neutral fluids and gases. Some examples of interactions are Coulomb interactions affecting both non-neutral and neutral systems. Dipole interactions are important when there is an applied electric field. Gravitational interactions are significant in neutral systems at long scales. Some of the topics addressed in this section include self-consistent fields, spatial non-uniformity, quasi-neutrality, Debye shielding, and a virial theorem.

### 1.d.i Classical treatment of interactions: Coulomb, dipole, etc.

We postulate a classical treatment for a system in contact with a heat bath. Then the probability function will have the form

$$\rho(p,q) \sim e^{-\beta(K+\Phi)} \sim e^{-\beta K} e^{-\beta \Phi} \sim \rho_p(p) \rho_q(q) \quad (1.d.1)$$

for a canonical ensemble with prescribed temperature *T*. The configuration space probability function obeys

$$\rho_q(\mathbf{r}^{(N)}) = \frac{e^{-\beta \Phi}}{Q}, \quad Q \equiv \int d^N r \, e^{-\beta \Phi} \quad (1.d.2)$$

Where *N* rolls up the dimensionality of the system and the number of elements. For the Coulomb interaction we have

$$\Phi = \sum_{i<j} \frac{e_i e_j}{r_{ij}} + \sum_i e_i \phi_0(\mathbf{r}_i) \quad (1.d.3)$$

where $\Phi$ is the total electrostatic potential; and the electrostatic electric field satisfies $\mathbf{E} = -\nabla \Phi$. Gauss' law relates the electric field to the charge density.

The one-particle density is defined by

$$n_1(\mathbf{x}|\mathbf{r}_i) \equiv \delta(\mathbf{x} - \mathbf{r}_1) \qquad \int d^3x\, n_1(\mathbf{x}|\mathbf{r}_1) = 1 \qquad (1.d.4)$$

The ensemble average of $n_1$ given a probability density function as in (1.d.1) and (1.d.2) is

$$\langle n_1 \rangle(\mathbf{x}) = \int d^N r\, \rho(\mathbf{r}^N) \delta(\mathbf{x} - \mathbf{r}_1) = \sum_{i=1}^{N} \langle n_i \rangle(\mathbf{x}) \qquad (1.d.5)$$

We note that $\langle n_1 \rangle(\mathbf{x}) = \frac{1}{V}$ if $\phi \neq \phi(\mathbf{x})$, ie., if $\phi$ has no spatial dependence. If we introduce more than one species $s$, we have

$$n^s(\mathbf{x}) = \sum_{i=1}^{N_s} \langle n_{i,s} \rangle(\mathbf{x}) \qquad (1.d.6)$$

We also note that as a matter of economy in notation

$$\rho(\mathbf{r}_1, \mathbf{r}_2, \ldots \mathbf{r}_N) \delta(\mathbf{x} - \mathbf{r}_1) = \rho(\mathbf{r}^N) \delta(\mathbf{x} - \mathbf{r}_1)$$

and

$$\langle n_1 \rangle(\mathbf{x}) = \int d^N r\, \rho(\mathbf{r}^N) \delta(\mathbf{x} - \mathbf{r}_1) \equiv \rho(\mathbf{r}_1 = \mathbf{x}) \qquad (1.d.7)$$

Consider the spatial gradient of the number density in (1.d.6)

$$\frac{\partial}{\partial \mathbf{x}} n^s(\mathbf{x}) \equiv \frac{\partial}{\partial \mathbf{x}} \sum_{i=1}^{N_s} \langle n_{i,s} \rangle(\mathbf{x}) = \sum_{i=1}^{N_s} \int d^N r\, \rho(\mathbf{r}^N) \frac{\partial}{\partial \mathbf{x}} \delta(\mathbf{x} - \mathbf{r}_i)$$

$$= -\sum_{i=1}^{N_s} \int d^N r\, \rho(\mathbf{r}^N) \frac{\partial}{\partial \mathbf{r}_i} \delta(\mathbf{x} - \mathbf{r}_i) =$$

$$= \sum_{i=1}^{N_s} \int d^N r\, \delta(\mathbf{x} - \mathbf{r}_1) \frac{\partial}{\partial \mathbf{r}_i} \rho(\mathbf{r}^N) = -\beta \sum_{i=1}^{N_s} \int d^N r\, \delta(\mathbf{x} - \mathbf{r}_i) \frac{\partial \Phi}{\partial \mathbf{r}_i} \frac{1}{Q} e^{-\Phi}$$

$$= -\beta \sum_{i=1}^{N_s} \langle \delta(\mathbf{x} - \mathbf{r}_i) \frac{\partial \Phi}{\partial \mathbf{r}_i} \rangle \qquad (1.d.8)$$

integrating by parts and identifying the averaging process along the way. We recall (1.d.3) and take its gradient

$$\frac{\partial \Phi}{\partial \mathbf{r}_i} = \sum_{j \neq i} \frac{\partial}{\partial \mathbf{r}_i} \frac{e_i e_j}{r_{ij}} + \sum_j \frac{\partial}{\partial \mathbf{r}_i} e_j \phi_0(\mathbf{r}_i) \equiv \sum_{j \neq i} \frac{\partial}{\partial \mathbf{r}_i} \Phi_{ij} + \sum_j \frac{\partial}{\partial \mathbf{r}_i} \Phi_j(\mathbf{r}_i) \qquad (1.d.9)$$

We return to (1.d.8) to obtain

$$\frac{\partial}{\partial x} n^s(\mathbf{x}) = -\beta \left[ n_s(\mathbf{x}) \frac{\partial \Phi_s}{\partial \mathbf{x}} + \sum_i^S \sum_{j \neq i} \langle n_i(\mathbf{x}) \frac{\partial}{\partial \mathbf{x}} \int d^3 x' \Phi_{ss'} \delta(\mathbf{x}' - \mathbf{r}_j) \rangle \right]$$

$$= -\beta \left[ n_s(\mathbf{x}) \frac{\partial \Phi_s}{\partial \mathbf{x}} + \sum_{s'} \sum_i^S \sum_{j \neq i}^{S'} \int d^3 x' \langle n_i(\mathbf{x}) n_j(\mathbf{x}') \rangle \frac{\partial}{\partial \mathbf{x}} \Phi_{ss'}(\mathbf{x}, \mathbf{x}') \right] \quad (1.d.10)$$

and

$$\langle n_i(\mathbf{x}) n_j(\mathbf{x}') \rangle \equiv \int d^N r \rho(\mathbf{r}^N) \, \delta(\mathbf{x} - \mathbf{r}_i) \delta(\mathbf{x}' - \mathbf{r}_j) \equiv \rho(\mathbf{r}_i = \mathbf{x}, \mathbf{r}_j = \mathbf{x}')$$

$$= \rho(\mathbf{r}_i = \mathbf{x}) \rho(\mathbf{r}_j = \mathbf{x}') = \langle n_i \rangle(\mathbf{x}) \langle n_j \rangle(\mathbf{x}') [1 + g_{ij}(\mathbf{x}, \mathbf{x}')] \quad (1.d.11)$$

where the correlation function is $g_{ij} \ll 1$ for dilute gases and/or for $\frac{\Phi_{ij}}{T} \ll 1$. We shall neglect $g_{ij}$ presently and proceed. We will check on $g_{ij}$ later.

We return to (1.d.10) and note that $\sum_{j \neq i}^{N_{s'}} = \sum_j^{N_{s'}} - \sum_{j=i} = O(N_s) - O(1) \approx \sum_j^{N_{s'}}$. Furthermore, we set $\sum_i^{N_s} \langle n_i \rangle(\mathbf{x}) = n_s(\mathbf{x})$ and $\sum_i^{N_{s'}} \langle n_i \rangle(\mathbf{x}') = n_{s'}(\mathbf{x}')$. Hence, (1.d.10) becomes

$$\frac{\partial}{\partial x} n^s(\mathbf{x}) = -\beta n_s(\mathbf{x}) \frac{\partial}{\partial x} [\Phi_s + \int d^3 x' \sum_{s'} n_{s'}(\mathbf{x}') \, \Phi_{ss'}(\mathbf{x}, \mathbf{x}')] \quad (1.d.12)$$

We introduce the total self-consistent electrostatic potential $\phi(x)$ neglecting correlations

$$\phi(\mathbf{x}) \equiv \phi^0(\mathbf{x}) + \int d^3 x' \sum_{s'} \frac{e_{s'} n_{s'}(x')}{|\mathbf{x} - \mathbf{x}'|} \quad (1.d.13)$$

(1.d.12) then becomes

$$\frac{\partial}{\partial \mathbf{x}} n_s = -\beta e_s n_s(\mathbf{x}) \frac{\partial}{\partial \mathbf{x}} \phi(\mathbf{x}) \quad (1.d.14)$$

or

$$\frac{\partial}{\partial \mathbf{x}} \ln n_s = -\beta e_s \frac{\partial}{\partial \mathbf{x}} \phi(\mathbf{x}) \quad \rightarrow \quad n_s(\mathbf{x}) = n_s(0) e^{-\beta e_s \phi(\mathbf{x})} \quad (1.d.15)$$

<u>Poisson-Boltzmann equation</u>: We apply the Laplacian to (1.d.13) and derive Poisson's equation having identified $\delta(\mathbf{x} - \mathbf{x}')$ from $\nabla^2 \frac{1}{|\mathbf{x}-\mathbf{x}'|}$ inside the volume integral w.r.t. $\mathbf{x}'$:

$$\nabla^2 \phi = -4\pi [\rho^0(\mathbf{x}) + \sum_s e_s n_s(0) e^{-\beta e_s \phi(\mathbf{x})}] \quad (1.d.16)$$

Eq.(1.d.16) is a single quasi-linear partial differential equation for the electric potential.

## 1.d.ii Example of an electron gas and Coulomb interaction

Consider an electron gas with $\phi^0(x) = 0$. Eq.(1.d.16) becomes

$$\nabla^2 \phi = -4\pi e\, n_0 e^{-\beta e\, \phi(\mathbf{x})} \tag{1.d.17}$$

Define $\psi \equiv \beta e\, \phi(\mathbf{x}) = \dfrac{e\phi}{T}$, from which (1.d.17) becomes

$$\nabla^2 \psi = -K^2 e^{-\psi} \tag{1.d.18}$$

where $K^2 \equiv 4\pi\beta e^2 n_0 = 1/\lambda_{Debye}^2$. Now consider a one-dimensional limit of (1.d.18) and solve by standard methods:

$$\frac{d^2}{dx^2}\psi = -K^2 e^{-\psi} \;\rightarrow\; \psi(x) = -2\ln\left[\sec\left(\frac{Kx}{\sqrt{2}}\right)\right] \tag{1.d.19}$$

and

$$n(x) = n(0)\sec^2\left(\frac{Kx}{\sqrt{2}}\right) \tag{1.d.20}$$

We note that the number of electrons *N*, assumed large, can be determined from the integral of (1.d.20) over a domain defined by [-*a*,*a*]:

$$N = \int_{-a}^{a} dx\, n(x) = 2n(x=0)\frac{\sqrt{2}}{K}\tan\left(\frac{Ka}{\sqrt{2}}\right) \tag{1.d.21}$$

Assume that the argument of the tan( ) in (1.d.21) approaches $\pi/2$ so that *N* is large, but not too close. From

$$\frac{Ka}{\sqrt{2}} \approx \frac{\pi}{2} \;\rightarrow\; n_0 \approx \frac{\pi}{8}\frac{T}{e^2}\frac{1}{a^2} \;\rightarrow\; 10^8 \text{ cm}^{-3}$$

for $T\sim 10$ eV and $a\sim 1$ cm. This estimate is independent of any specific value of *N*. Given (1.d.20) which diverges at argument value $\frac{\pi}{2}$, as more particles are added they tend to end up at the edges, which is a consequence of the electron-electron repulsion. This is very different from a neutral system. Suppose $N=10^{12}$ instead of infinity. Then from (1.d.21) and (1.d.19) with $a=1$ cm

$$\frac{Ka}{\sqrt{2}} \sim \tan^{-1}\left(\frac{N}{n_0 a}\right) = \tan^{-1}\left(\frac{N}{10^8}\right) \rightarrow$$

$$\psi(a) = -2\ln\left[\sec\left(\frac{Ka}{\sqrt{2}}\right) \sim \tan\left(\frac{Ka}{\sqrt{2}}\right) = \frac{N}{n_0 a}\right] = -2\ln(N\, 10^{-8}) \approx -20$$

Hence, $|e\phi|(a) \approx 20T = 200$ eV for the parameters of this example.

### 1.d.iii Example of a stellar cluster and gravitational interaction

Consider a system of charge-neutral, finite-mass elements interacting through gravitational forces. Assume that the elements share equal masses. Introduce a gravitational potential $\psi(x)$ such that

$$n(x) = n_0 e^{-\beta m_1 \psi(x)}$$

$$\nabla^2 \psi = 4\pi G m_1 n(x) = 4\pi G m_1 n_0 e^{-\beta m_1 \psi(x)} \tag{1.d.22}$$

Introduce $K^2 \equiv \frac{4\pi G m_1^2 n_0}{T}$ and $\Psi \equiv \frac{m_1 \psi}{T}$ so that (1.d.22) becomes

$$\nabla^2 \Psi = K^2 e^{-\Psi} \tag{1.d.23}$$

It can be shown that (1.d.23) has the solution

$$\Psi(x) = 2\ln \cosh\left(\frac{Kx}{\sqrt{2}}\right) \rightarrow \sqrt{2} K x \text{ for large } x$$

$$n(x) = n_0 \operatorname{sech}^2\left(\frac{Kx}{\sqrt{2}}\right) \rightarrow 4 n_0 e^{-\frac{2Kx}{\sqrt{2}}} \text{ for large } x \tag{1.d.24}$$

In 3D with spherical symmetry the gravitational potential has the asymptotic limit for large $r$, $\psi \rightarrow -\frac{GM_{tot}}{r}$ and $n(r) \rightarrow n_0 e^{-\frac{\beta m_1 GM_{tot}}{r}}$, so that $n(\infty) \rightarrow n_0$ which is a contradiction ($n(\infty)$ needs to vanish)! In 3D with spherical symmetry, one cannot satisfy the equations of thermal equilibrium. Hence, clusters are constantly losing particles; and similarly, terrestrial atmospheres are constantly losing particles.

### 1.d.iv Example with ions and electrons, and Coulomb interaction

Consider a plasma with electrons and singly charge ions (or perhaps positrons). Gauss' law leads to Poisson's equation:

$$\nabla^2 \phi = -4\pi e[n_i(\mathbf{x}) - n_e(\mathbf{x})] \rightarrow n_e(\mathbf{x}) = n_i(\mathbf{x}) + \frac{1}{4\pi e}\nabla^2 \phi \tag{1.d.25}$$

Eq.(1.d.25) becomes a statement characterizing quasi-neutrality if $\frac{1}{4\pi e}\nabla^2 \phi \ll n_i$. In thermal equilibrium

$$n_e(\mathbf{x}) = n_e^0 e^{\beta e \phi}$$

Suppose $L_n$ is the scale length for the density gradient. Then using $\beta e \phi = \ln \frac{n_e}{n_e^0}$ Eq.(1.d.25) leads to

$$n_e(\mathbf{x}) = n_i(\mathbf{x}) + \frac{1}{4\pi\beta e^2}\frac{1}{L_{n_i}^2} = n_i(\mathbf{x})\left[1 + \frac{1}{4\pi\beta n_i e^2}\frac{1}{L_{n_i}^2}\right] \quad (1.d.26)$$

<u>Definition</u>: $\lambda_D \equiv \sqrt{\frac{T}{4\pi n e^2}}$ is the Debye length. Hence, $\lambda_D^2 = \frac{1}{4\pi\beta e^2 n}$

Equation (1.d.26) then becomes

$$n_e(\mathbf{x}) = n_i(\mathbf{x})\left[1 + O\left(\frac{\lambda_D^2}{L_{n_i}^2}\right)\right] \quad (1.d.27)$$

$\lambda_D$ is millimeters in many laboratory plasmas and meters in many space plasmas. Quasi-neutrality corresponds to $L_{n_i}^2 \gg \lambda_D^2$.

**1.d.v Example with ions and electrons, and Coulomb and gravitational interactions**

Next we consider a simple one-dimensional system composed of two species (ions and electrons) with Coulomb and gravitational interactions. In thermal equilibrium we have

$$n_e(\mathbf{x}) = n_e^0 e^{-\beta[-e\phi + m_e gz]} \quad n_i(\mathbf{x}) = n_i^0 e^{-\beta[e\phi + m_i gz]} \quad (1.d.28)$$

and Poisson's equation Eq.(1.d.25) is unmodified. Again $n_e \approx n_i$. At z=0 define $\phi = 0$; then $n_e^0 = n_i^0$. As a consequence of quasi-neutrality and (1.d.28)

$$e\phi + m_i gz = -e\phi + m_e gz \quad (1.d.29)$$

We can then solve (1.d.29) for $\phi$ and take its gradient to determine the electric field:

$$e\phi = -\frac{1}{2}(m_i - m_e)gz \quad (1.d.30)$$

and

$$e\mathbf{E} = \frac{1}{2}(m_i - m_e)\mathbf{g} \quad (1.d.31)$$

The equilibrium number densities are then

$$n_e(\mathbf{x}) \approx n_i(\mathbf{x}) = n_i^0 e^{-\frac{1}{2}\beta(m_i + m_e)gz} \quad (1.d.32)$$

**1.d.vi Example with ions and electrons, and Coulomb interaction with correlations – Debye-Hückel theory and shielding**

Here we assume a plasma in thermal equilibrium with multiple charge species and an imposed test-particle distribution. Poisson's equation becomes

$$\nabla^2 \phi = -4\pi [\sum_s e_s n_s(\mathbf{x}) + \rho_e^0(\mathbf{x})] \qquad (1.d.33)$$

where

$$n_s(\mathbf{x}) = n_s^0 e^{-\beta e_s \phi(\mathbf{x})} \qquad \phi \to \phi^{(0)} + \delta\phi \qquad (1.d.34)$$

The linearly perturbed Poisson equation and charge density satisfy

$$\nabla^2 \delta\phi = -4\pi [\sum_s e_s \delta n_s(\mathbf{x}) + \delta\rho_e^0(\mathbf{x})] \qquad (1.d.35)$$

and

$$\delta n_s(\mathbf{x}) = -n_s^0(\mathbf{x}) \beta e_s \delta\phi(\mathbf{x}) \qquad (1.d.36)$$

Substituting (1.d.36) in (1.d.35) one obtains

$$\nabla^2 \delta\phi = [4\pi\beta \sum_s e_s^2 n_s^0(\mathbf{x})]\delta\phi(\mathbf{x}) - 4\pi\delta\rho_e^0(\mathbf{x}) =$$

$$\frac{1}{\lambda_D^2} \delta\phi(\mathbf{x}) - 4\pi e_0 \delta(\mathbf{x}) \qquad (1.d.37)$$

where the Debye length $\lambda_D$ is defined here by

$$\lambda_D \equiv \sqrt{\frac{T}{4\pi \sum_s e_s^2 n_s^0(\mathbf{x})}} \qquad (1.d.38)$$

The solution of Eq.(1.d.37) with boundary conditions $\delta\phi(r \to \infty) = 0$ and regularity at $r = 0$ is

$$\delta\phi(\mathbf{x}) = \frac{e_0}{r} e^{-(r/\lambda_D)} \qquad (1.d.39)$$

and with the assumption of quasi-neutrality in the unperturbed equilibrium charge densities

$$\delta n_e = \beta e \delta\phi n_e^0 > 0 \quad \text{and} \quad \delta n_i = -\beta e \delta\phi n_e^0 < 0 \qquad (1.d.40)$$

Thus, there is a slight excess of electrons around the test charge at **x**=0 and a slight deficiency in the ions. Both perturbations in the charge densities decay spatially on the scale of the Debye length. This is a manifestation of plasma shielding of the test charge.

We next construct the conditional probability

$$\rho_{ij}(\mathbf{x}, \mathbf{x}') \equiv \rho_i(\mathbf{x})\rho_j(\mathbf{x}'; i\ at\ \mathbf{x}) = \langle n_i \rangle(\mathbf{x})\langle n_j \rangle(\mathbf{x}'; i\ at\ \mathbf{x}) =$$

$$=\langle n_i\rangle(\mathbf{x})\langle n_j\rangle(\mathbf{x}')[1 + \text{correction due to } i \text{ at } \mathbf{x}] \tag{1.d.41}$$

where the correction is due to the perturbation associated with $\delta n_e$:

$$\langle n_j\rangle(\mathbf{x}'; i \text{ at } \mathbf{x}) = \langle n_j\rangle(\mathbf{x}')\left[1 + (-\beta)e_j\delta\phi(\mathbf{x}')\right] =$$

$$\langle n_j\rangle(\mathbf{x}')\left[1 + (-\beta)e_j\frac{e_i}{|\mathbf{x}-\mathbf{x}'|}e^{-\frac{|\mathbf{x}-\mathbf{x}'|}{\lambda_D}}\right] \tag{1.d.42}$$

The conditional probability including the correlation function correction is then obtained from (1.d.41) and (1.d.42):

$$\rho_{ij}(\mathbf{x},\mathbf{x}') \equiv \rho_i(\mathbf{x})\rho_j(\mathbf{x}')\left[1 + \left(-\beta\frac{e_je_i}{|\mathbf{x}-\mathbf{x}'|}e^{-\frac{|\mathbf{x}-\mathbf{x}'|}{\lambda_D}}\right)\right] \tag{1.d.43}$$

Is the correlation $g_{ij}(\mathbf{x},\mathbf{x}') \equiv -\beta\frac{e_je_i}{|x-x'|}e^{-\frac{|\mathbf{x}-\mathbf{x}'|}{\lambda_D}} \ll 1$ ?  We require for the consistency of the theory for the self-consistent field solution that at least the effect of $g_{ij}$, if not $g_{ij}$ itself, is small. If $\lambda_D$ is very small, then long-range $1/r^2$ forces are much reduced, and only short-range effects survive. Thus, electrically charged particles and their systems are uncorrelated in space if $\lambda_D$ is small; and all sorts of quantities like the entropy and energies are additive.

Consider the internal energy:

$$\langle U\rangle = \frac{3}{2}NT + \langle\sum_{i<j}\frac{e_ie_j}{r_{ij}}\rangle = \frac{3}{2}NT + \langle\sum_{i<j}e_ie_j\int d^3\mathbf{x}\int d^3\mathbf{x}'\frac{\delta(\mathbf{x}-\mathbf{x}_i)\delta(\mathbf{x}'-\mathbf{x}_j)}{|\mathbf{x}-\mathbf{x}'|}\rangle =$$

$$\frac{3}{2}NT + \frac{1}{2}\int d^3\mathbf{x}\int d^3\mathbf{x}'\frac{\rho(\mathbf{x})\rho_j(\mathbf{x}')}{|\mathbf{x}-\mathbf{x}'|} + \langle U_{correlation}\rangle =$$

$$\frac{3}{2}NT + \int d^3\mathbf{x}\frac{|\mathbf{E}(\mathbf{x})|^2}{8\pi} - \frac{1}{2}\int d^3\mathbf{x}\sum_s n_s(\mathbf{x})\frac{e_s^2}{\lambda_D} \tag{1.d.44}$$

where the second term is the field energy due to the macro field associated with not having exact charge neutrality.  Where does this result for $\langle U_{correlation}\rangle$ come from? One can expand the expression for $\delta\phi$

$$\delta\phi = \frac{e_i}{r}e^{-\frac{r}{\lambda_D}} \approx \frac{e_i}{r}\left(1 - \frac{r}{\lambda_D}\right) + O(r) \tag{1.d.45}$$

Thus, the correction to the electric potential due to the shielding cloud is $\delta\phi_{cloud} = -\frac{e_i}{\lambda_D} + O(r)$. The associated energy is $e_i \delta\phi_{cloud}(0)$=energy of the test particle interacting with the cloud= $-\frac{e_i^2}{\lambda_D} + O(r)|_{r=0}$. Hence,

$$U_{correlation} = \frac{1}{2}\int d^3\mathrm{x}\sum_s n_s(x)\frac{(-e_s^2)}{\lambda_D} \qquad (1.d.46)$$

Note that the factors of ½ that appear in the intermediate expression in (1.d.44) and in $U_{correlation}$ arise from counting only the unique pairs from the double sums over particles.

The condition for weak correlations on which the validity of the Debye-Hückel theory depends is

$$\frac{1}{2}\frac{Ne^2}{\lambda_D} \ll \frac{3}{2}NT \;\to\; \frac{e^2}{\lambda_D} \ll T \qquad (1.d.47)$$

Define the radius $R_T \equiv \frac{e^2}{T}$. For $T \sim 10$ eV, $R_T \sim 1$ Å. The condition for weak correlations can be rewritten as

$$nR_T^3 \ll 1 \qquad (1.d.48)$$

i.e., if there are many particles within the strong interaction distance $R_T$ the correlations are *not* weak. For $T \sim 10$ eV and $n \ll 10^{24} \text{cm}^{-3}$ the interactions are weakly correlated, i.e., at anything less than solid densities the interactions are weakly correlated. We note that the ratio of $U_{correlation}$ to the total kinetic energy $\frac{3}{2}NT$ scales as

$$\frac{U_{correlation}}{\frac{3}{2}NT} \sim \sqrt{nR_T^3} \qquad (1.d.49)$$

Now that we know $<U>$ in (1.d.44), define it simply as $U$ the total internal energy; and we can derive the other thermodynamic quantities from the relations

$$U = -\frac{\partial \ln Z}{\partial \beta}, \quad F = -T\ln Z, \quad \mathcal{S} = \ln Z + \beta U, \quad \mu_s = \frac{\partial F}{\partial N_s}\bigg|_{V,T} = T\ln n_s \Lambda^3 + \frac{1}{2}\epsilon_s + e_s\phi \quad (1.d.50)$$

where $\epsilon_s \equiv -\frac{e_s^2}{\lambda_D}$ is due to the interaction of the particle with its own shielding cloud and the intrinsic electro-chemical potential is identified as

$$\mu_s^0 \equiv T\ln n_s \Lambda^3 + \frac{1}{2}\epsilon_s \qquad (1.d.51)$$

We note as a warning that if we violate the validity condition in (1.d.47) and (1.d.48) we can obtain a large correlation energy, but it is still short-range in its effect.

For chemical equilibrium in a nonuniform medium one requires

$$0 = \nabla \mu_s = \nabla \mu_s^0 + \nabla e_s \phi = \sum_{s'} \frac{\partial \mu_s^0(n_{s'})}{\partial n_{s'}} \nabla n_{s'} + e_s \nabla \phi, \quad \mathbf{E} = -\nabla \phi \qquad (1.d.52)$$

from which

$$\mathbf{E} = \frac{1}{e_s} \sum_{s'} \frac{\partial \mu_s^0(n_{s'})}{\partial n_{s'}} \nabla n_{s'} \qquad (1.d.53)$$

Example: Correlations are negligible in an ideal gas, $T \ln n_s \Lambda^3 + \frac{1}{2}\epsilon_s \approx T \ln n_s \Lambda^3$ and

$$\mathbf{E} \approx \frac{1}{e_s} \sum_{s'} \frac{\partial T \ln n_s \Lambda^3}{\partial n_{s'}} \nabla n_{s'} \approx \frac{1}{e_s} \sum_{s'} \frac{T}{n_{s'}} \nabla n_{s'} \rightarrow n_s \sim e^{-\beta e_s \phi} \qquad (1.d.54)$$

All this has been classical.

Example: Suppose we have a degenerate electron gas. Recall Eqs.(1.b.125-128) in Sec. 1.b.vii from which we have

$$n \Lambda_f^3 = \frac{8\pi}{3}, \quad \Lambda_f \equiv \frac{h}{p_f}, \quad \frac{p_f^2}{2m} = \mu \qquad (1.d.55)$$

if there is no external electric potential. The effect of an external field is

$$e_1 \phi + \frac{p_f^2}{2m} = \mu = e_1 \phi + \mu_0, \quad \Lambda_f \equiv \frac{h}{\sqrt{2m(\mu - e_1\phi)}}, \quad n \frac{h^3}{(2m(\mu - e_1\phi))^{\frac{3}{2}}} = \frac{8\pi}{3} \qquad (1.d.56)$$

Eq.(1.d.56) gives the lowest-order number density as a function of $\phi$. Performing an analysis similar to that in Sec. 1.d.vi involving the solution of the linearized Poisson equation we recover shielding but with

$$\lambda_D = \sqrt{\frac{\frac{2}{3}\mu_0}{4\pi n_0 e^2}} \qquad (1.d.57)$$

and the condition that the correlations are weak is

$$U_{corr} \ll K = \frac{3}{5} N \mu_0 \quad \text{or} \quad n \left(\frac{e^2}{\mu_0}\right)^3 \ll 1 \qquad (1.d.58)$$

For energies corresponding to 10 eV, the weak correlation condition (1.d.58) fails for $n \sim 10^{24} cm^{-3}$, i.e., for solid state conditions. With considerations similar to those leading to (1.d.52) and (1.d.53) we can deduce the equilibrium electric field in a non-uniform metal by setting $\nabla \mu = 0$ given Eq.(1.d.56).

[*Reviewer Dominique Escande's Comment:* *- By treating n as a continuous function, the analysis implicitly assumes that there are many particles in the shielding cloud: so it is for weakly coupled plasmas only. In reality the theory works for a number of particles larger than 40 https://www.scielo.org.mx/scielo.php?script=sci_arttext&pid=S1870-35422017000100063 .*
*- A Vlasov calculation of shielding can be performed for a weakly perturbative test charge moving very slowly. Shielding does not exist for a fast particle. See for instance section 9.2 of (Nicholson, 1983).*
*- A derivation of shielding avoiding the assumption of Boltzmann equilibrium is provided in (Meyer-Vernet, 1993). It relies upon Gauss' theorem and the Coulomb deflection of particles.*
*- The shielded Coulomb potential is a basic example of a renormalized potential. See section 3.2 of (McComb, 2004).*
*- In order to go further on long-range interacting systems, (Campa, Dauxois, Fanelli, and Ruffo, 2014) is a useful reference.*]

## 2. Non-equilibrium statistical mechanics

[*Editor's Note: Physics 212B addressed non-equilibrium statistical mechanics. As in Physics 212A there was no textbook, and use was made of many of the same references.*]

### 2.a Fundamentals

#### 2.a.i Definitions of a realization, moments, characteristic function, and discrete variables

As a vehicle to introduce a number of fundamental concepts and definitions we consider a system in which a large particle with mass *M* is immersed in a collection of smaller particles with mass *m* that constitutes a gas or fluid. The Hamiltonian for such a system is

$$H = \tfrac{1}{2} M V^2 + \sum_i \tfrac{1}{2} m v_i^2 + \sum_i \Phi(|\mathbf{r}_i - \mathbf{R}|) + \sum_{i<j} \phi(r_{ij}) \qquad (2.a.1)$$

The large particle has velocity **V** and position **R**, while the smaller particles have velocities $v_i$ and positions $r_i$. We posit that the motion of the large particle consists of fast variations due to fluid particles colliding with the large particle and slow variations due to the net diffusion of the large particle. An example of this is the motion of pollen grains as in Robert Brown's famous observations in 1827 that subsequently was named Brownian motion. In this system, the kinetic energy of the large particle satisfies

$$\tfrac{1}{2} M \langle V_x^2 + V_y^2 + V_z^2 \rangle = \tfrac{3}{2} T \qquad (2.a.2)$$

and $\langle V_x^2 \rangle = T/M$.

Definition: A realization of this system is a single instance of the system with defined initial conditions.

A realization of a random process has a specific initial condition and time history. For a sufficiently long time we can compute the time average $\langle V_x^2 \rangle_t = T/M$. Now consider an ensemble of identical systems possessing different initial conditions within the domain of the accessible phase space of the system. Then we can compute the ensemble average $\langle V_x^2 \rangle_{ensemble}(t)$ at a specified time $t$, averaging over the different initial conditions. We expect the same result $T/M$ if the ensemble average over initial conditions is equal to the time average of a single realization of the system. That is to say we expect the ergodic hypothesis to be valid: the dynamics should spend equal times in equal volumes of phase space for a random process. Furthermore, we expect that every degree of freedom of a weakly coupled system should have $T/2$ energy associated with it as a consequence of ergodicity.

However, the results of experiments indicate that ergodicity is not always occurring in systems that we might think are random. An example of this is found in the numerical integration of the equations of motion of a chain of nonlinear oscillators with Lenard-Jones interactions between nearest neighbors reported in (Galgani and Scott, 1972). Instead of an equipartition scaling of the time-average kinetic energy as in the one-dimensional version of (2.a.2), the numerical integration exhibited a Planck-like scaling for the mean energy levels:

$$\bar{E}_n \sim \frac{1}{e^{\alpha \omega_n} - 1} \tag{2.a.3}$$

where $\alpha \equiv \beta \hbar$. We return to the introduction of fundamental concepts that we will make use of in the course of the subsequent discussion.

Definition: Let $x$ be a random variable, whose probability $\rho(x)$ is normalized on the domain of $x$:

$$\int dx\, \rho(x) = 1 \tag{2.a.4}$$

The average of any function $f$ of $x$ is defined as

$$\langle f(x) \rangle = \int dx\, \rho(x) f(x) \tag{2.a.5}$$

There is a 1:1 relation between $f$ and $x$ such that

$$\rho(f) df = \rho(x) dx \qquad \rho(f) = \rho(x) \left|\frac{df}{dx}\right|^{-1} \qquad \int df\, \rho(f) = 1 \tag{2.a.6}$$

Example: Suppose $K = \frac{1}{2} MV^2$, then

$$\rho(V) \sim e^{-\beta\frac{1}{2}MV^2} \quad \rightarrow \quad \rho(K) \sim \frac{e^{-\beta K}}{MV} \sim \frac{e^{-\beta K}}{\sqrt{K}} \tag{2.a.7}$$

If there is absolute certainty that the random variable $x$ has the value $x_0$, then $\rho(x) = \delta(x - x_0)$. If instead we have knowledge of the relative probabilities that $x$ is equal to a set of discrete values, then

$$\rho(x) = \sum_i \rho_i \delta(x - x_i) \qquad \sum_i \rho_i = 1 \tag{2.a.8}$$

Definition: Moments of a distribution of random variables are defined by

$$\langle x^\ell \rangle \equiv \int dx \, x^\ell \rho(x) \tag{2.a.9}$$

The mean value of $x$ corresponds to $\ell = 1$.

Definition: Fluctuations are defined by $\delta x \equiv x - \langle x \rangle$ and $\langle \delta x \rangle \equiv 0$. The standard deviation $\sigma$ is the square root of the variance defined by $\sigma^2 \equiv \langle (\delta x)^2 \rangle = \langle x^2 \rangle - \langle x \rangle^2$

Definition: The characteristic function is defined by

$$Z_x(k) \equiv \int dx \, e^{-ikx} \rho(x) \tag{2.a.10}$$

It then follows

$$Z_x(k=0) = 1, \quad e^{-ikx} = \sum_{\ell=0}^\infty \frac{(-ikx)^\ell}{\ell!}, \quad Z_x(k) = \sum_{\ell=0}^\infty \frac{(-ik)^\ell}{\ell!} \langle x^\ell \rangle \tag{2.a.11}$$

and

$$\left.\frac{d^\ell Z_x}{dk^\ell}\right|_{k=0} = \int dx \, e^{-ikx}\big|_{k=0} (-ix)^\ell \rho(x) = (-i)^\ell \langle x^\ell \rangle \tag{2.a.12}$$

We conclude that the probability function of the random variable and the moments of the random variable completely determine one another:

$$\rho(x) \rightleftharpoons \{\langle x^\ell \rangle, \ \ell = 0, 1, 2, \ldots\}$$

The inverse transform of (2.a.10) yields

$$\rho(x) = \int \frac{dk}{2\pi} e^{ikx} Z_x(k) \tag{2.a.13}$$

Central limit theorem: A statement of the central limit theorem is that when the sample size of the discrete random variables is large enough, the probability distribution tends towards a Gaussian:

$$\rho(x) = \frac{1}{\sqrt{2\pi\sigma^2}} e^{-\frac{[x-\langle x \rangle]^2}{2\sigma^2}} \tag{2.a.14}$$

The characteristic function is obtained from (2.a.10) and (2.a.14):

$$Z_x(k) = e^{-ik\langle x \rangle - \frac{1}{2}k^2\sigma^2} \rightarrow \ln Z_x(k) = -ik\langle x \rangle - \frac{1}{2}k^2\sigma^2 \tag{2.a.15}$$

Taking $\ln Z_x(k)$ has separated <x> from its shape with respect to its mean value.

More generally $\ln Z_x(k)$ is determined by a series expansion of (2.a.10)

$$\ln Z_x(k) = -ik\langle x \rangle - \frac{1}{2}k^2\sigma^2 + \frac{(-i)^3}{3!}k^3\chi_3 + \cdots + \frac{(-i)^n}{n!}k^n\chi_n \tag{2.a.16}$$

where $\chi_n \equiv K_n$ are cumulants of the probability distribution and were described by Danish astronomer T. N. Thiele as semi-invariants more than a century ago. Two examples of the cumulants are the skewness and the kurtosis:

$$\text{skewness: } \frac{K_3}{\sigma^3} \quad K_3 = \langle x^3 \rangle - 3\langle x^2 \rangle \langle x \rangle + 2\langle x \rangle^3 = \langle \delta x^3 \rangle \tag{2.a.17}$$

$$\text{kurtosis: } \frac{K_4}{\sigma^4} \quad K_4 = \langle \delta x^4 \rangle - 3\sigma^4 \tag{2.a.18}$$

<u>Definition</u>: Given two random variables *x* and *y*, the probability $\rho(x,y)$ is defined by

$$\rho(x,y) \equiv \rho(x|y)\rho(y) \equiv \rho(y|x)\rho(x) \tag{2.a.19}$$

where $\rho(x|y)$ is the conditional probability of *x* given *y*; and

$$\rho(x) = \int dy \, \rho(x,y) = \int dy \, \rho(y)\rho(x|y) \tag{2.a.20}$$

<u>Definition</u>: Given the function $f(x,y)$ the probability function $\rho(f)$ is

$$\rho(f) = \int dxdy \, \rho(x,y)\delta(f - f(x,y)) \tag{2.a.21}$$

<u>Definition</u>: *x* and *y* are statistically independent if $\rho(x,y) = \rho(x)\rho(y)$.

If *x* and *y* are statistically independent then

$$\rho(x|y) = \rho(x) \tag{2.a.22}$$

<u>Definition</u>: The correlation of *x* and *y* can be inferred from $\langle \delta x \delta y \rangle \equiv \langle xy \rangle - \langle x \rangle \langle y \rangle$ using the definitions $\delta x \equiv x - \langle x \rangle$, $\delta y \equiv y - \langle y \rangle$. If *x* and *y* are independent then $\langle \delta x \delta y \rangle = 0$, but not the converse. If $\langle \delta x \delta y \rangle \neq 0$ then *x* and *y* are dependent.

<u>Definition</u>: A set of *N* random variables can be represented by

$$\{x_i\} \equiv \mathbf{x}, \quad i = 1, 2, \ldots, N \tag{2.a.23}$$

The probability distribution $\rho(\mathbf{x})$ satisfies the normalization condition $\int d^N x \, \rho(\mathbf{x}) = 1$. If the set of random variables is statistically independent, then $\rho(\mathbf{x}) = \prod_{i=1}^{N} \rho_i(x_i)$. The generalization of (2.a.10) to a *N*-dimensional vector of random variables is

$$Z_x(\mathbf{k}) \equiv \int d^N x \, e^{-i\mathbf{k} \cdot \mathbf{x}} \rho(\mathbf{x}) \tag{2.a.24}$$

and (2.a.13) generalizes to

$$\rho(\mathbf{x}) = \int \frac{d^N k}{2\pi} e^{i\mathbf{k} \cdot \mathbf{x}} Z_x(\mathbf{k}) \tag{2.a.25}$$

The generalization of (2.a.16) is

$$\ln Z_x(\mathbf{k}) = -i\mathbf{k} \cdot \langle \mathbf{x} \rangle - \frac{1}{2} \mathbf{k} \cdot \sigma^2 \cdot \mathbf{k} + O(\mathbf{kkk}) \tag{2.a.26}$$

<u>Example</u>: For a Gaussian probability distribution (2.a.25) becomes

$$\rho(\mathbf{x}) = \frac{1}{(2\pi)^{N/2} \sqrt{|\sigma^2|}} e^{-\frac{1}{2} \mathbf{x} \cdot \sigma^{-2} \cdot \mathbf{x}} \tag{2.a.27}$$

## 2.a.ii Derivation of the central limit theorem

Assume the set $\{x_i\}$ is statistically independent. Define $x \equiv \sum_i x_i$. Then

$$\rho(x) \equiv \int d^N x \, \rho(\{x_i\}) \delta(x - \sum_i x_i) = \int d^N x \, \prod_i \rho_i(x_i) \, \delta(x - \sum_i x_i) \tag{2.a.28}$$

and

$$Z_x(k) \equiv \int dx \, e^{-ikx} \rho(x) = \int d^N x \, \prod_i \rho_i(x_i) \, e^{-ik \sum_i x_i} = \prod_i \int dx_i \, \rho_i(x_i) \, e^{-ikx_i} = \prod_i Z_i(k)$$

$$\tag{2.a.29}$$

then $\ln Z_x(\mathrm{k}) = \sum_i \ln Z_i(\mathrm{k})$

Example: For the special case wherein $\rho_i = \rho_j$ then $\ln Z_x(k) = \sum_i \ln Z_i(k) \to N \ln Z_1(k)$. As a matter of definition, $\langle x \rangle = \sum_i \langle x_i \rangle \to N \langle x_1 \rangle$, $\sigma_x^2 = \sum_i \sigma_i^2 \to N\sigma_1^2$, and $\sigma_x \to \sqrt{N}\sigma_1$. In consequence of these relations $\frac{\sigma_x}{\langle x \rangle} = \frac{1}{\sqrt{N}}\frac{\sigma_1}{\langle x_1 \rangle}$ and

$$\ln Z_x(k) = -ik\langle x \rangle - \frac{1}{2}k^2\sigma_x^2 + \cdots + \frac{(-i)^n}{n!}k^n K_n(x) \qquad (2.a.30)$$

and $K_n(x) = NK_n(x_1)$ where $K_n$ are Thiele's cumulants. Note that the inner products $\mathbf{k} \cdot \langle \mathbf{x} \rangle$ and $\mathbf{k} \cdot \boldsymbol{\sigma}^2 \cdot \mathbf{k}$ in (2.a.26) and use of $\langle x \rangle = \sum_i \langle x_i \rangle$ and $\sigma_x^2 = \sum_i \sigma_i^2$ have led to (2.a.30).

Hence, as $N \to \infty$ the distribution function becomes Gaussian:

$$\ln Z_x(k) = N \ln Z_1(k) = -ikN\langle x_1 \rangle - \frac{1}{2}k^2 N\sigma_1^2 + \cdots + \frac{(-i)^n}{n!}k^n NK_n(x_1) \qquad (2.a.31)$$

Using (2.a.25) and (2.a.31) we obtain

$$\rho(x) = \int \frac{dk}{2\pi} e^{ikx} Z_x(k) = \int \frac{dk}{2\pi} e^{ikx} \exp\left\{-ik\langle x \rangle - \frac{1}{2}k^2\sigma_x^2 + \cdots\right\}$$

$$= \int \frac{dk}{2\pi} e^{ik(x-\langle x \rangle) - \frac{1}{2}k^2 N\sigma_1^2 + \cdots + \frac{(-i)^n}{n!}k^n NK_n(x_1)} \qquad (2.a.32)$$

We evaluate (2.a.32) in the limit of large $N$. Assume that the largest contributions to the integral over $k$ derive from $k \sim N^{1/2}\sigma_1^{-1}$ so that $\frac{1}{2}k^2 N\sigma_1^2 \sim O(1)$. What then happens to the general term?

$$\frac{(-i)^n}{n!}k^n NK_n(x_1) \to \frac{(-i)^n}{n!}\frac{K_n}{\sigma_1^n}N^{1-\frac{n}{2}} \to \frac{(-i)^n}{n!}N^{1-\frac{n}{2}} \qquad (2.a.33)$$

with $K_n \sim O(\sigma_1^n)$. Hence, for $n > 2$, $N^{1-\frac{n}{2}} \to 0$ as $N \to \infty$; and the general term for $n > 2$ is vanishingly small as $N \to \infty$. Then for large $N$ the probability distribution tends toward a Gaussian:

$$\rho(x) = \int \frac{dk}{2\pi} e^{ik(x-\langle x \rangle) - \frac{1}{2}k^2 N\sigma_1^2} = \frac{1}{\sqrt{2\pi N\sigma_1^2}} e^{-\frac{1}{2}\frac{[x-<x>]^2}{N\sigma_1^2}} \qquad (2.a.34)$$

Example: Consider the non-Gaussian probability distribution $\rho(x_1) \sim x_1^{m-1} e^{-x_1}$ $x_1 \geq 0$ For this distribution $\langle x_1 \rangle = m$ and $\sigma_1 = \sqrt{m}$ Based on $\rho(x_1)$, $\rho(x) = x^{Nm-1} e^{-x}$ exactly, which derives from considering the sums of gamma-distributed variables and is not a trivial result; and $\langle x \rangle = Nm$ and $\sigma_x = \sqrt{Nm}$.

Exercise: Take the limit as $N \to \infty$ for $\rho(x) = x^{Nm-1} e^{-x}$ and recover $\rho(x) \sim e^{-\frac{[x-<x>]^2}{2N\sigma_x^2}}$ to verify the central limit theorem.

Example: Consider the non-Gaussian probability distribution $\rho(x_1) \sim \frac{1}{x_1^2+1}$ for positive and negative $x_1$. Then $\langle x_1 \rangle = 0$ and $\sigma_1^2 = \langle x_1^2 \rangle \sim \int_{-\infty}^{\infty} dx \frac{x^2}{1+x^2} = \infty$. Moreover, $\langle x_1^{2n} \rangle = \infty$. The characteristic function is the Fourier transform of the Lorentzian in this example and is not analytic: $Z(k) \sim e^{-|k|}$. The central limit theorem is invalid for this probability distribution because the moments are infinite.

## 2.a.iii Random processes, spectral density, and correlation function

Consider a random process for a particle velocity as a function of time $V(t)$ with probability distribution $\rho(V_{t_1}, V_{t_2}, V_{t_3}, \ldots, V_{t_N})$. If one knows $\rho$, one can calculate all of the moments of $\{V_{t_i}\}$. For example,

$$\langle V_{t_1} V_{t_2}^2 V_{t_3} \rangle = \int d^N V_t \, V_{t_1} V_{t_2}^2 V_{t_3} \rho(V_{t_1}, V_{t_2}, V_{t_3}, \ldots, V_{t_N}) \tag{2.a.35}$$

Definition: (Stationary random process) If $\rho(V_{t+\tau_1}, V_{t+\tau_2}, V_{t+\tau_3}, \ldots, V_{t+\tau_N})$ is independent of $t$ then the system or process is stationary.

Definition: (Ergodicity) In order for a system or process to be ergodic, it must be stationary; and the time average of any moment must equal the ensemble average of the same moment, e.g.,

$$\langle V_{t+\tau_1} V_{t+\tau_2}^2 V_{t+\tau_3} \rangle_t = \langle V_{t+\tau_1} V_{t+\tau_2}^2 V_{t+\tau_3} \rangle_{ensemble} \tag{2.a.36}$$

The systems comprising the ensemble on the right side of (2.a.36) are not prepared necessarily the same uniquely, but are macroscopically identical. The systems must be stationary in order to calculate the time average in (2.a.36) sensibly.

Definition: (Spectral density) Assume that $\langle V \rangle_{ensemble}(t) = 0$ or $\langle V(t) \rangle_{time} = 0$ (equivalent if ergodic and stationary). The spectrum is determined by

$$V(\omega) \equiv \int dt \, e^{i\omega t} V(t) \tag{2.a.37}$$

The two-time correlation is defined by

$$C(\tau) \equiv \langle V(t) V(t-\tau) \rangle_{t \text{ or ensemble}} \tag{2.a.38}$$

The spectral density is defined by

$$S(\omega) \equiv \int d\tau \, e^{i\omega \tau} C(\tau) \tag{2.a.39}$$

That $S(\omega)$ is the power spectral density of the $V(t)$ field is a consequence of the convolution theorem. Consider a process $V(t)$ that satisfies the stationary and ergodic assumptions. Then the Fourier transform satisfies:

$$V(\omega) = \int_{-\infty}^{\infty} dt\, V(t) e^{i\omega t}, \quad \omega = \text{real} \tag{2.a.40}$$

with reality condition

$$V(-\omega) = V^*(\omega) \tag{2.a.41}$$

We note that $\langle V(\omega) \rangle \to 0$ due to symmetry. We want to calculate $\langle |V(\omega)|^2 \rangle$, but to evaluate this will lead to the auto-correlation:

$$\langle V(t_1) V(t_2) \rangle \to \langle V(t_1) V(t_1 + \tau) \rangle_{ensemble} = C(\tau) = \langle V(t) V(t + \tau) \rangle_t \tag{2.a.42}$$

Then $C(\tau) = \langle V(t) V(t - \tau) \rangle_t = C(-\tau)$, i.e., $C(\tau)$ is an even function. Furthermore,

$$C(0) = \langle V(t) V(t) \rangle_{ensemble} = \langle V(t) V(t) \rangle_t = \frac{T}{M}, \quad C(\tau \to \infty) \to 0 \tag{2.a.43}$$

where $T$ in (2.a.43) is the temperature if $V$ is the velocity.

Definition: The normalized correlation function is

$$R(\tau) \equiv C(\tau)/C(0) \tag{2.a.44}$$

Example: The normalized correlation function for a Lorentz model looks like

$$R(\tau) = e^{-\nu|\tau|} \cos \omega_0 \tau \tag{2.a.45}$$

As a consequence of (2.a.39) and since $C(\tau)$ is even

$$S(\omega) \equiv \int_{-\infty}^{\infty} d\tau\, e^{i\omega\tau} C(\tau) \quad \text{and} \quad C(\tau) = \int_{-\infty}^{\infty} \frac{d\omega}{2\pi} e^{-i\omega\tau} S(\omega)$$

Then $S(\omega) = S(-\omega) = S^*(\omega)$, i.e., $S(\omega)$ is an even function and is real also.

$$\langle V^2 \rangle = S(0) = \int_{-\infty}^{\infty} \frac{d\omega}{2\pi} S(\omega) \tag{2.a.46}$$

Definition: It is convenient to introduce subscripts

$$C_\chi \equiv \langle \chi(t) \chi(t - \tau) \rangle_{ensemble} = \langle \chi(t) \chi(t - \tau) \rangle_t \tag{2.a.47}$$

to define the correlation function and spectral density for the general case.

Consider

$$\langle V(\omega)V^*(\omega')\rangle \equiv \int_{-\infty}^{\infty} dt\, e^{i\omega t} \int_{-\infty}^{\infty} dt'\, e^{-i\omega' t'} \langle V(t)V(t')\rangle =$$

$$\int_{-\infty}^{\infty} dt\, e^{i(\omega-\omega')t} \int_{-\infty}^{\infty} d\tau e^{i\omega'\tau}\, C(\tau) = 2\pi\delta(\omega-\omega')S(\omega') \quad (2.a.48)$$

using $t' = t - \tau$. (2.a.48) tells us that there are no correlations between Fourier components of the random field at different frequencies.

<u>Lemma</u>: Given that $\langle V(\omega)V^*(\omega)\rangle = \langle |V(\omega)|^2\rangle = 2\pi\delta(0)S(\omega)$ we conclude that $S(\omega) \geq 0$.

Now we replace the time integrals in (2.a.48) with the limiting forms

$$\int_{-\infty}^{\infty} dt \to \lim_{T\to\infty} \int_{-T/2}^{T/2} dt$$

(where now *T* represents a time interval) so that

$$V(\omega) = \lim_{T\to\infty} \int_{-T/2}^{T/2} dt\, V(t)e^{i\omega t} \text{ and } 2\pi\delta(0)S(\omega) \to 2\pi T S(\omega) \text{ for } \omega = \omega'$$

and we obtain the result of the Wiener-Khinchin theorem:

$$\lim_{T\to\infty} \frac{\langle |V_T(\omega)|^2\rangle}{T} = S(\omega) \quad (2.a.49)$$

<u>Example</u>: Return to consideration of the Lorentz model correlation function

$$R(\tau) = e^{-\nu|\tau|}\cos\omega_0\tau \to S(\omega) = \sum_{\pm} \frac{\nu}{\nu^2+(\omega\pm\omega_0)^2} \quad (2.a.50)$$

where $\nu$ is the inverse correlation time.

<u>Example</u>: Exponential correlation function $R(\tau) = e^{-\nu|\tau|} \to S(\omega) = \frac{2\nu}{\nu^2+\omega^2}$

## 2.b Brownian motion

We return to the consideration of random processes and Brownian motion.

### 2.b.i Brownian motion – Langevin equation

The equation of motion for Brownian motion can be cast in the general form

$$M\dot{V} = F(t) \equiv \langle F\rangle(t) + \delta F(t) \quad (2.b.1)$$

Given V(t) for a Brownian particle, we expect a viscous force for the mean force on the particle, i.e., a viscous drag force:

$$\langle F \rangle(V) = \langle F \rangle(0) + V \frac{d\langle F \rangle}{dV}\bigg|_{V=0} + \frac{1}{2}V^2 \frac{d^2\langle F \rangle}{dV^2}\bigg|_{V=0} = V \frac{d\langle F \rangle}{dV}\bigg|_{V=0} + \frac{1}{2}V^2 \frac{d^2\langle F \rangle}{dV^2}\bigg|_{V=0} \quad (2.b.2)$$

where $\langle F \rangle(0) = 0$ as we assume that there is no external nonzero fields influencing the particle at rest. Hence, to lowest order

$$M\dot{V} = F(t) = -\gamma V + \delta F(t) \quad (2.b.3)$$

From hydrodynamics, Stokes' law gives

$$\gamma = 6\pi\eta R \quad (2.b.4)$$

where $\eta$ is the specific viscosity and $R$ is the radius of the Brownian particle if it is a spherical object.

Definition: Eq.(2.b.3) can be rewritten in the form of a Langevin equation:

$$\left(M\frac{d}{dt} + \gamma\right)V(t) = \delta F(t) \text{ or } (-i\omega M + \gamma)V(\omega) = \delta F(\omega) \quad (2.b.5)$$

Hence,

$$V(\omega) = \frac{\delta F(\omega)}{-i\omega M + \gamma} \rightarrow \langle |V(\omega)|^2 \rangle = \frac{\langle |\delta F(\omega)|^2 \rangle}{|-i\omega M + \gamma|^2} \rightarrow S_V(\omega) = \frac{S_F(\omega)}{\omega^2 M^2 + \gamma^2} \quad (2.b.6)$$

We assume that the Brownian particle has a much larger mass $M$ than the particles in the surrounding fluid. This leads to much lower characteristic frequencies in $S_V(\omega)$ than in $S_F(\omega)$. The fluid forces give rise to very rapid fluctuations in $F$:

$$\omega_{\delta F} \sim \frac{\sqrt{\langle V^2 \rangle}}{a} \sim \frac{\sqrt{T/m}}{a} \equiv \nu_{\delta F} \quad R_F(\tau) \sim e^{-\nu_{\delta F}|\tau|} \quad (2.b.7)$$

while the response of the Brownian particle velocity is

$$S_V(\omega) = \frac{\frac{1}{M^2}S_F(\omega)}{\omega^2 + \nu_V^2} \quad \nu_V \equiv \frac{\gamma}{M} \quad R_V(\tau) \sim e^{-\nu_V|\tau|} \quad (2.b.8)$$

Because $M$ is large, $\nu_V \ll \nu_{\delta F}$; and the power spectrum $S_V(\omega)$ decays with respect to frequency at much lower frequency values than does $S_F(\omega)$. We can estimate $\nu_F \sim n\langle V^2 \rangle^{\frac{1}{2}}\pi a^2$ where $a$ is approximately the atomic radius of the fluid particle and $n$ is the fluid density. We assume that the mass density of the Brownian and fluid particles are the same. The specific viscosity is $\eta \sim \rho \ell \langle V^2 \rangle^{\frac{1}{2}}$ where $\ell \sim \frac{1}{n\pi a^2}$, and $M = \frac{4\pi}{3}\rho R^3$. Hence,

$\nu_V \sim \frac{6\pi\eta R}{M} \sim \frac{6\pi\rho\ell\langle V^2\rangle^{\frac{1}{2}}R}{M} \sim 6\pi\rho \frac{1}{n\pi a^2}\langle V^2\rangle^{\frac{1}{2}}R/(\frac{4\pi}{3}\rho R^3)$. Note that $\eta \sim \frac{m\langle V^2\rangle^{\frac{1}{2}}}{a^2}$ is independent of the density. For a gas, $\langle V^2\rangle^{1/2} \sim \sqrt{T/M}$, and $\eta \sim 10^{-3}$ in cgs units, while for water $\eta \sim 10^{-2}$ in cgs units; so we can introduce a fudge factor O(1-10) in the viscosity. Finally, our estimates of $\nu_F$ and $\nu_V$ lead to

$$\frac{\nu_F}{\nu_V} = \frac{1}{6\pi}\frac{4\pi}{3}\rho R^3 \frac{n\langle V^2\rangle^{1/2}\pi a^2}{\rho \frac{1}{n\pi a^2}\langle V^2\rangle^{1/2}R} = \frac{1}{6\pi}\frac{4\pi}{3}\pi^2(na^3)^2 \frac{R^2}{a^2} \sim O(1)(na^3)^2 \frac{R^2}{a^2} \gg 1 \qquad (2.b.9)$$

If $R \sim 10^{-5}$cm and $a \sim 10^{-8}$cm there is significant margin for satisfying the inequality in (2.b.9). Note that the origin of the $R^2$ factor in the numerator of (2.b.10) is the $1/M$ in the $\nu_V$ expression.

## 2.b.ii Fluctuation-dissipation theorem

We return to consideration of the spectral densities $S_F(\omega)$ and $S_V(\omega)$ in (2.b.8):

$$S_F(\omega) \sim \frac{1}{\omega^2 + \nu_F^2}, \quad \omega \sim \nu_V \ll \nu_F \qquad (2.b.10)$$

for frequencies $\omega$ relevant to the velocity response. Then $\frac{\omega^2}{\nu_F^2} \sim 10^{-12}$ or $10^{-8}$ and $S_F(\omega \ll \nu_F) = S_F(0)$, and from (2.b.9)

$$S_V(\omega) \cong \frac{\frac{1}{M^2}S_F(0)}{\omega^2 + \nu_V^2} \qquad (2.b.11)$$

From (2.a.39) the inverse Fourier transform of $S_V(\omega)$ yields the correlation function:

$$C_V(\tau) = \frac{S_F(0)}{M^2 2\nu_V}e^{-\nu_V|\tau|} \qquad (2.b.12)$$

However, we know from the fundamental definition of $C_V(\tau)$ in (2.a.38) that $C_V(0) = \langle V^2\rangle = T/M$. Hence,

$$C_V(\tau) = \frac{T}{M}e^{-\nu_V|\tau|} \quad \text{and} \quad S_F(\omega) = 2\gamma T \qquad (2.b.13)$$

good for $\omega \ll \nu_F$. Now consider the integral of $S_F(\omega)$ over a frequency interval $[-\Delta\omega, \Delta\omega]$ where $\Delta\omega \ll \nu_F$.

The fluctuation-dissipation relation is

$$\langle(\delta F)^2\rangle_{\Delta\omega} = \int_{-\Delta\omega}^{\Delta\omega} \frac{d\omega}{2\pi} S_F(\omega) = 2\frac{\Delta\omega}{2\pi} 2\gamma T = 4\frac{\Delta\omega}{2\pi}\gamma T \qquad (2.b.14)$$

This is the fluctuation-dissipation theorem or Nyquist theorem due to Einstein in his work on Brownian motion.

## 2.b.iii Spatial diffusion and diffusivity

If the velocity field in Brownian motion is a random process, the particle displacement inherits randomness from the velocity.

Definition: Let $x$ be the position of the particle and $V$ its velocity.

For a specified time interval $\Delta t$ there is an accrued displacement:

$$\Delta x = \int_{t}^{t+\Delta t} dt' V(t') \qquad (2.b.15)$$

The ensemble-averaged displacement inherits its value from the ensemble-averaged velocity:

$$\langle\Delta x\rangle = \int_{t}^{t+\Delta t} dt' \langle V(t')\rangle = 0 \qquad (2.b.16)$$

if $\langle V(t')\rangle = 0$. We can then calculate the ensemble-averaged variance of the displacement:

$$\langle(\Delta x)^2\rangle = \int_{t}^{t+\Delta t} dt' \{\int dt'' \langle V(t')V(t'')\rangle\} =$$

$$= \int_{t}^{t+\Delta t} dt' \{\int dt'' \, C_V(|t'-t''|)\} = 2\frac{\langle V^2\rangle}{\nu_V^2}[\nu_V\Delta t - 1 + e^{-\nu_V\Delta t}] \qquad (2.b.17)$$

where (2.b.13) is used to evaluate $C_V(|t'-t''|) = \langle V^2\rangle e^{-\nu_V|t'-t''|}$ The variance has the two limiting values:

$$\langle(\Delta x)^2\rangle = \begin{cases} \langle V^2\rangle(\Delta t)^2 & \nu_V\Delta t \ll 1 \\ 2\langle V^2\rangle\frac{\Delta t}{\nu_V} & \nu_V\Delta t \gg 1 \end{cases} \qquad (2.b.18)$$

For short times the variance in the displacement grows quadratically in time as if the velocity is constant, while for long times the variance in the displacement grows linearly in time apropros of a diffusion process!

Definition: The diffusivity is defined by

$$D \equiv \lim_{\nu_V\Delta t\to\infty} \frac{\langle(\Delta x)^2\rangle}{2\Delta t} \qquad (2.b.19)$$

The diffusivity is then
$$D = \lim_{\nu_V \Delta t \to \infty} \frac{\langle (\Delta x)^2 \rangle}{2\Delta t} = \frac{\langle V^2 \rangle}{\nu_V} = \frac{T}{M}\frac{M}{\gamma} = \frac{T}{\gamma} \qquad (2.b.20)$$

This result allows us to define and evaluate the steady velocity response to a steady external force, i.e., the mobility:

$$\langle M\dot{V} \rangle + \langle \gamma V \rangle = \langle \delta F \rangle + \langle F^{ext} \rangle \;\to\; \langle \gamma V \rangle = \langle F^{ext} \rangle = F^{ext} \qquad (2.b.21)$$

<u>Definition</u>: The mobility $\mu$ is then defined as

$$\mu \equiv \frac{\langle V \rangle}{F^{ext}} = \frac{1}{\gamma} \qquad (2.b.22)$$

Using (2.b.20) and (2.b.22) we arrive at the Einstein relation:

$$D = \mu T \qquad (2.b.23)$$

The diffusivity, which characterizes random spatial diffusion, is related directly to the mobility, i.e., the response to an external steady force, and the temperature.

We return to the consideration of Eq.(2.b.18) in more detail.

$$\langle (\Delta x)^2 \rangle = \int_{t_0}^{t_0+\Delta t} dt' \left\{ \int_{t_0}^{t_0+\Delta t} dt'' \, \langle V(t')V(t'') \rangle \right\} =$$

$$= \int_{t_0}^{t_0+\Delta t} dt' \left\{ \int d\tau \, \langle V(t')V(t'-\tau) \rangle \right\} \qquad (2.b.24)$$

Note that the limits of integration in the second integral in (2.b.24) are implied based on the original limits of the double integral which corresponded to the boundaries of a rectangle in the $t'$ and $t''$ domain, $t_0$ to $t_0 + \Delta t$ in each direction. We recognize that $C_V(\tau) = \langle V(t')V(t'-\tau) \rangle$ from (2.a.47) and $C_V(|\tau|) = \langle V^2 \rangle e^{-\nu_V |\tau|}$. The correlation function $C_V(|\tau|)$ falls off sharply over a time $\tau_c \sim O(1)\nu_V^{-1}$. We assume that $\Delta t \gg \tau_c$ The direct implication of the sharp fall off of $C_V(|\tau|)$ is that the dominant contributions to the double integrals in Eqs.(2.b.17) and (2.b.24) are over a narrow region surrounding the diagonal $t' = t''$ in the original rectangle in the $t'$ and $t''$ domain. This allows us to extend the limits of integration in the $\int d\tau$ integral in (2.b.24) to $[-\infty, \infty]$ with no loss of precision. Evaluation of (2.b.24) is then straightforward:

$$\int_{t_0}^{t_0+\Delta t} dt' \left\{ \int_{-\infty}^{\infty} d\tau \, C_V(\tau) \right\} = \Delta t \left\{ \int_{-\infty}^{\infty} d\tau \, C_V(\tau) \right\} = 2\Delta t \left\{ \int_{0}^{\infty} d\tau \, C_V(\tau) \right\} \qquad (2.b.25)$$

From (2.b.20) and (2.b.25)

$$D_x \equiv \frac{\langle(\Delta x)^2\rangle}{2\Delta t} = \int_0^\infty d\tau\, C_V(\tau) = \frac{\langle V^2 \rangle}{\nu_V} \qquad (2.b.26)$$

and we recover the result in (2.b.20).

Example: Using the evaluation of $\nu_V$ and the specific viscosity $\eta$ in the last section we have

$$D_x = \frac{T}{6\pi\eta R} \qquad (2.b.27)$$

Assume $T$=20-25 °C, $R \sim 10^{-5}$cm, and $a \sim 10^{-8}$cm, then $D_x \sim 2 \times 10^{-8} \frac{cm^2}{s}$. Hence,

$$\langle(\Delta x)^2\rangle = 2D_x \Delta t \sim 4 \times 10^{-8} \Delta t \text{ cm}^2 \text{ and } \sigma_x \sim 2 \times 10^{-4}\sqrt{\Delta t(\text{sec})} \text{ cm.} \qquad (2.b.28)$$

Because of the $1/R$ dependence in $D_x$ smaller particles diffuse considerably faster.

Next we consider the probability of a particle having a displacement $x_t$ relative to a reference or initial displacement $x_0$, and we appeal to the central limit theorem:

$$\rho(x_t|x_0) = \frac{e^{-\frac{(x_t-x_0)^2}{4Dt}}}{\sqrt{4\pi Dt}} \qquad (2.b.29)$$

Now multiply both sides of equation (2.b.29) by the probability $\rho(x_0)$. Then

$$\rho(x_t, x_0) \equiv \rho(x_0)\rho(x_t|x_0) = \frac{e^{-\frac{(x_t-x_0)^2}{4Dt}}}{\sqrt{4\pi Dt}} \rho(x_0) \qquad (2.b.30)$$

These probabilities are not statistically independent! The displacement at time $t$ is very dependent on the displacement at $t$=0. The probability of the displacement $\rho(x_t)$ without specifying $x_0$ is given by the integral over $x_0$:

$$\rho(x_t) = \int dx_0 \rho(x_t, x_0) \equiv \rho(x; t) \qquad (2.b.31)$$

and $\rho(x_0) = \rho(x; 0)$.

Then $\quad \rho(x; t) = \int dx' \frac{e^{-\frac{(x-x')^2}{4Dt}}}{\sqrt{4\pi Dt}} \rho(x'; 0) \quad$ or $\quad \rho(x; t+\tau) = \int dx' \frac{e^{-\frac{(x-x')^2}{4D\tau}}}{\sqrt{4\pi D\tau}} \rho(x'; t) \qquad (2.b.32)$

The constructed solution $\rho(x; t)$ satisfies the diffusion equation:

$$\frac{\partial}{\partial t}\rho(x; t) = D\frac{\partial^2 \rho}{\partial x^2} \qquad (2.b.33)$$

This all carries over to three dimensions: $\frac{\partial}{\partial t}\rho(\mathbf{x};t) = D\nabla^2\rho$   We note that going from one to three dimensions in the diffusion equation comes with extending the boundary conditions from one to three dimensions, alters the Green's functions for integral solutions of the diffusion equation, and changes how theorems are proven.

### 2.b.iv Boltzmann's H theorem

In this section we consider example calculations derived from (2.b.33) which lead us to Boltzmann's H theorem.

Examples:
1. Given the initial condition $\rho(x,0) = \rho_0(1 + \epsilon \sin kx)$ the solution to the p.d.e. in (2.b.33) is straightforward: $\rho(x,t) = \rho_0\left(1 + \epsilon e^{-k^2 Dt}\sin kx\right)$ for $t > 0$.
2. Consider the integrated form of (2.b.33) with reflecting wall boundary conditions $\frac{\partial \rho}{\partial x} = 0$:

$$0 = \frac{d}{dt}\int dx\, \rho(x,t) = \int dx\, \frac{\partial \rho}{\partial t} = \int dx\, D\frac{\partial^2 \rho}{\partial x^2} = D\frac{\partial \rho}{\partial x}\Big|_{boundary}$$

    Now solve the boundary value problem given the initial conditions in the first example and defining a relation between $k$ and the length of the box $L$, e.g., $k = 2\pi/L$. Given that the definition of the flux is $\Gamma = -D\nabla n$ and the boundary conditions, there should be no flux across the bounding surfaces.

3. H-theorem (Boltzmann) – Introduce the entropy:

$$S(t) = -\int d^3x\, \rho(\mathbf{x};t) \ln \rho(\mathbf{x};t) \tag{2.b.34}$$

    Then from its time derivative

$$\frac{dS}{dt} = -\int d^3\mathbf{x}\left[\frac{\partial \rho}{\partial t}\ln \rho + \frac{\partial \rho}{\partial t}\right] = -\int d^3\mathbf{x}\frac{\partial \rho}{\partial t}\ln \rho - \frac{d}{dt}\int d^3\mathbf{x}\, \rho = -\int d^3\mathbf{x}\frac{\partial \rho}{\partial t}\ln \rho$$

$$= -\int d^3\mathbf{x} \ln\rho\, D\nabla^2\rho = \int d^3\mathbf{x}\frac{D}{\rho}\nabla\rho\cdot\nabla\rho = \int d^3\mathbf{x}\frac{D}{\rho}|\nabla\rho|^2 \geq 0 \tag{2.b.35}$$

    and we note that surface terms in the integration by parts vanish since the flux across bounding surfaces is assumed to be zero, $\rho$ is non-negative, and the volume integral of $\rho$ is conserved with the zero flux boundary conditions. Thus, only if $\rho$ is perfectly flat will $S$ stop growing; and there is irreversible growth until the entropy achieves a maximum.

4. $S$ has an upper bound if $\rho$=constant corresponding to uniformity, which is the asymptotic limit corresponding to $\nabla\rho = 0$ yielding $\frac{dS}{dt} = 0$.

Note: Boltzmann used the Boltzmann equation rather than the diffusion equation used here to derive the H theorem and obtained the Boltzmann distribution as an asymptotic state. (Boltzmann, 1872)

## 2.c Liouville and Klimontovich Equations

In this section kinetic equations are introduced to describe the evolution in phase space of a deterministic system when initial conditions might not be precisely known.

### 2.c.i Liouville equation

We postulate a vector function $\Gamma(t)$ that describes the state of a system of particles, i.e., its momenta $\{p_i\}$ and positions $\{q_i\}$, and assume further that there exists a Hamiltonian $H(p,q;t)$ that determines the evolution of the system:

$$\dot{q}_i = \frac{\partial H}{\partial p_i} \quad \dot{p}_i = -\frac{\partial H}{\partial q_i} \tag{2.c.1}$$

Example: Charged particle motion in electromagnetic fields is described by the equations

$$\{\mathbf{r},\mathbf{v}\}: \quad \dot{\mathbf{r}} = \mathbf{v} \quad \dot{\mathbf{v}} = \frac{e}{m}\left[\mathbf{E}(\mathbf{r},t) + \frac{1}{c}\mathbf{v}(t) \times \mathbf{B}(\mathbf{r},t)\right] \tag{2.c.2}$$

Example: Viscous system

$$\{\mathbf{r},\mathbf{v}\}: \quad \dot{\mathbf{r}} = \mathbf{v} \quad \dot{\mathbf{v}} = -\frac{\gamma}{m}\mathbf{v}(t) \tag{2.c.3}$$

$\Gamma(t)$ has 2$f$ dimensions, where $f$ is the number of degrees of freedom. The domain of $\Gamma(t)$ is sometimes called phase space. The evolution of $\Gamma(t)$ is formally expressed as

$$\frac{d}{dt}\Gamma(t) = \dot{\Gamma}(\Gamma,t) \tag{2.c.4}$$

Given $\Gamma(\Gamma_0) \to \Gamma_t \equiv \Gamma(t|\Gamma_0)$ is determined. Keep in mind that $\Gamma$ represents the phase-space independent variable $\{\mathbf{r},\mathbf{v}\}$. For three spatial dimensions $\Gamma-$ space has 6$f$ dimensions. For a specified fixed time $t$, $\Gamma \to \delta(\mathbf{x}-\mathbf{x}_i, \mathbf{p}-\mathbf{p}_i)$ in probability. Hence,

$$\rho(\Gamma; t|\Gamma_0) = \delta\big(\Gamma - \Gamma(t|\Gamma_0)\big) \tag{2.c.5a}$$

$$\rho(\Gamma; t|\Gamma_0)\rho(\Gamma_0) = \rho(\Gamma_t, \Gamma_0) \quad \rho(\Gamma;t) \equiv \int d\Gamma_0\, \rho(\Gamma_t, \Gamma_0) \tag{2.c.5b}$$

How does $\rho(\Gamma;t)$ evolve with time $t$? This is the fundamental question of non-equilibrium statistical mechanics.

For a state function $A(\Gamma)$,

$$\langle A \rangle(t) \equiv \int d\Gamma \, A(\Gamma) \rho(\Gamma; t) \qquad \frac{d}{dt}\langle A \rangle = \int d\Gamma \, A(\Gamma) \frac{d\rho}{dt}(\Gamma; t) \qquad (2.c.6)$$

So we evaluate

$$\frac{\partial}{\partial t} \rho(\Gamma; t|\Gamma_0) = \frac{\partial}{\partial \mathbf{x}} \delta(\mathbf{x}) \cdot \frac{\partial \mathbf{x}}{\partial t} = -\frac{\partial}{\partial \Gamma} \delta(\mathbf{x}) \cdot \dot{\Gamma}(\Gamma, t) \qquad \frac{\partial}{\partial \Gamma} \delta(\mathbf{x}) = \frac{\partial}{\partial \mathbf{x}} \delta(\mathbf{x}) \cdot \frac{\partial \mathbf{x}}{\partial \Gamma} = \frac{\partial}{\partial \mathbf{x}} \delta(\mathbf{x}) \cdot \mathbf{I} \quad (2.c.7)$$

where $\mathbf{X} \equiv \Gamma - \Gamma(t|\Gamma_0)$, $\frac{\partial}{\partial \mathbf{x}} \delta(\mathbf{x}) = \frac{\partial}{\partial \Gamma} \delta(\mathbf{x})$, and $\frac{\partial \mathbf{X}}{\partial t} = -\dot{\Gamma}(\Gamma, t)$. Hence,

$$\frac{\partial}{\partial t} \rho(\Gamma; t|\Gamma_0) = -\frac{\partial}{\partial t} \Gamma(t|\Gamma_0) \cdot \frac{\partial}{\partial \Gamma} \rho(\Gamma; t|\Gamma_0) = -\frac{\partial}{\partial \Gamma} \cdot \left( \frac{\partial}{\partial t} \Gamma(t|\Gamma_0) \rho(\Gamma; t|\Gamma_0) \right) \qquad (2.c.8)$$

Note that $\Gamma(t|\Gamma_0)$ is not a function of $\Gamma$ and $\Gamma(t|\Gamma_0) \neq \Gamma$. However, $\frac{\partial}{\partial t} \Gamma(t|\Gamma_0) = \dot{\Gamma}(\Gamma(t|\Gamma_0), t)$ ; and from (2.c.5a) $\rho(\Gamma; t|\Gamma_0) = \delta(\Gamma - \Gamma(t|\Gamma_0))$; hence, (2.c.8) becomes

$$\frac{\partial}{\partial t} \rho(\Gamma; t|\Gamma_0) = -\frac{\partial}{\partial \Gamma} \cdot \left( \frac{\partial \Gamma(\Gamma, t)}{\partial t} \rho \right) = -\frac{\partial}{\partial \Gamma} \cdot \left( \dot{\Gamma}(\Gamma, t) \rho(\Gamma; t|\Gamma_0) \right) \qquad (2.c.9)$$

Example: Consider a one-dimensional, field-free, viscous system to illustrate (2.c.9)

$$\frac{d}{dt} V = -\gamma V \qquad \frac{\partial}{\partial t} \rho(V; t|V_0) = -\frac{\partial}{\partial v} \cdot (-\gamma V \rho) \qquad (2.c.10)$$

where $\Gamma \to \mathbf{V} \to V$ in one dimension.

Now we return to (2.c.8) and integrate over the $\Gamma_0$ domain:

$$\frac{\partial}{\partial t} \rho(\Gamma; t) = \int d\Gamma_0 \rho(\Gamma_0) \frac{\partial}{\partial t} \rho(\Gamma; t|\Gamma_0) = -\int d\Gamma_0 \rho(\Gamma_0) \frac{\partial}{\partial \Gamma} \cdot \left( \dot{\Gamma}(\Gamma, t) \rho(\Gamma; t|\Gamma_0) \right) =$$

$$-\frac{\partial}{\partial \Gamma} \cdot \int d\Gamma_0 \rho(\Gamma_0) \dot{\Gamma}(\Gamma, t) \rho(\Gamma; t|\Gamma_0) = -\frac{\partial}{\partial \Gamma} \cdot \left( \dot{\Gamma}(\Gamma, t) \rho(\Gamma; t) \right) \qquad (2.c.11)$$

which is the continuity equation for the probability in phase space.

Example: Return to the one-dimensional, field-free, viscous system in Eq.(2.c.10). We guess a solution

$$\rho(V; t) = \frac{1}{\sqrt{2\pi\sigma^2(t)}} e^{-\frac{V^2}{2\sigma^2(t)}} \quad \text{and} \quad \sigma(t) = \sigma_0 e^{-\gamma t} \qquad (2.c.12)$$

Exercise: Check whether (2.c.12) is a solution of (2.c.10) and (2.c.11).

We return to (2.c.11) and expand the right side of the continuity equation:

$$\frac{\partial}{\partial t}\rho(\mathbf{\Gamma};t) = -\rho\frac{\partial}{\partial \mathbf{\Gamma}}\cdot\dot{\mathbf{\Gamma}}(\mathbf{\Gamma},t) - \dot{\mathbf{\Gamma}}\cdot\frac{\partial}{\partial \mathbf{\Gamma}}\rho(\mathbf{\Gamma};t) \rightarrow$$

$$\left(\frac{\partial}{\partial t} + \dot{\mathbf{\Gamma}}\cdot\frac{\partial}{\partial \mathbf{\Gamma}}\right)\rho(\mathbf{\Gamma};t) = -\rho\frac{\partial}{\partial \mathbf{\Gamma}}\cdot\dot{\mathbf{\Gamma}}(\mathbf{\Gamma},t) \quad (2.c.13)$$

We define the convective derivative $\frac{D}{Dt} \equiv \frac{\partial}{\partial t} + \dot{\mathbf{\Gamma}}\cdot\frac{\partial}{\partial \mathbf{\Gamma}}$ and (2.c.13) becomes

$$\frac{D}{Dt}\rho(\mathbf{\Gamma};t) = -\rho\frac{\partial}{\partial \mathbf{\Gamma}}\cdot\dot{\mathbf{\Gamma}}(\mathbf{\Gamma},t) \quad (2.c.14)$$

<u>Example</u>: With $\dot{V} = -\gamma V$ then $\frac{D}{Dt}\rho = \gamma V$ and $\rho$ increases following the orbit.

For a Hamiltonian system $\frac{\partial}{\partial \mathbf{\Gamma}}\cdot\dot{\mathbf{\Gamma}}(\mathbf{\Gamma},t) = 0$

<u>Proof</u>: $\quad \frac{\partial}{\partial \mathbf{\Gamma}}\cdot\dot{\mathbf{\Gamma}}(\mathbf{\Gamma},t) = \sum_i \frac{\partial}{\partial q_i}\dot{q}_i + \frac{\partial}{\partial p_i}\dot{p}_i = \sum_i \frac{\partial}{\partial q_i}\frac{\partial H}{\partial p_i} - \frac{\partial}{\partial p_i}\frac{\partial H}{\partial q_i} \equiv 0 \quad (2.c.15)$

Thus, for a Hamiltonian system $\frac{D}{Dt}\rho(\mathbf{\Gamma};t) = 0$; or more formally

$$\frac{\partial}{\partial t}\rho(\mathbf{\Gamma};t) + \{\rho,H\} = 0 \quad (2.c.16)$$

where $\{\rho,H\}$ is the Poisson bracket. (2.c.16) is a statement of Liouville's theorem.

<u>Example</u>: Consider $\dot{\mathbf{V}} = \frac{e}{mc}\mathbf{V}\times\mathbf{B}(\mathbf{r},t) \rightarrow \dot{V}_i = \frac{e}{mc}\varepsilon_{ijk}V_j B_k$ We note that $\sum_i \frac{\partial \dot{x}_i}{\partial x_i} + \frac{\partial \dot{V}_i}{\partial V_i} = \cdots = 0$
With this force law the system is not a Hamiltonian system; nevertheless, the divergence of the flow field is zero. Thus, the flow can be incompressible independent of whether the system has a Hamiltonian or not.

### 2.c.ii Klimontovich phase space and distribution function

Consider the velocity (or momentum) and position phase space for *N* particles. For three spatial dimensions and three velocity dimensions, this is a six-dimensional by *N* points phase space.

<u>Definition</u>: This phase space corresponds to the Klimontovich space $\equiv \mu$ and the corresponding phase-space density distribution function is the Klimontovich distribution.

The equations of motion (assumed non-relativistic) are

$$\frac{d\mathbf{x}_i}{dt} = \mathbf{V}_i \quad (2.c.17a)$$

$$\frac{d\mathbf{v}_i}{dt} = \frac{1}{m_i}\left[f_i^{ext}(\mathbf{x}_i, \mathbf{v}_i, t) + \sum_{j(\neq i)} f_{i,j}\right] \quad (2.c.17b)$$

For $\xi_i$ defined as the vector defining the $i^{th}$ particle in the 6-D $\mu$ space, (2.c.17) can be written as

$$\frac{d\boldsymbol{\xi}_i(t)}{dt} = \dot{\boldsymbol{\xi}}_i^{ext}(\boldsymbol{\xi}_i, t) + \sum_{j(\neq i)}^{N} \dot{\boldsymbol{\xi}}^{in}(\boldsymbol{\xi}_i, \boldsymbol{\xi}_j) \quad (2.c.18)$$

<u>Definition</u>: Let $F \equiv$ density, $\quad F \equiv \sum_{i=1}^{N} \delta(\boldsymbol{\xi} - \boldsymbol{\xi}_i(t)) \quad (2.c.19)$

With the use of the density, the second term on the right side of (2.c.18) can be expressed as

$$\sum_{j(\neq i)}^{N} \dot{\boldsymbol{\xi}}^{in}(\boldsymbol{\xi}_i, \boldsymbol{\xi}_j) \rightarrow \int d\boldsymbol{\xi}' F(\boldsymbol{\xi}') \dot{\boldsymbol{\xi}}^{in}(\boldsymbol{\xi}_i, \boldsymbol{\xi}') \quad (2.c.20)$$

$$\frac{d\boldsymbol{\xi}(t)}{dt} = \dot{\boldsymbol{\xi}}^{ext}(\boldsymbol{\xi}, t) + \int d\boldsymbol{\xi}' F(\boldsymbol{\xi}') \dot{\boldsymbol{\xi}}^{in}(\boldsymbol{\xi}, \boldsymbol{\xi}') = \dot{\boldsymbol{\xi}}(\boldsymbol{\xi}, t, F) \quad (2.c.21)$$

We can take the partial derivative of (2.c.19) with respect to time

$$\frac{\partial F(\boldsymbol{\xi},t)}{\partial t} = -\sum_i \frac{d\boldsymbol{\xi}(t)}{dt} \cdot \frac{\partial}{\partial \boldsymbol{\xi}} \delta(\boldsymbol{\xi} - \boldsymbol{\xi}_i(t)) =$$

$$-\frac{\partial}{\partial \boldsymbol{\xi}} \cdot \sum_i \frac{d\boldsymbol{\xi}}{dt}(\boldsymbol{\xi}_i, t, F)\, \delta(\boldsymbol{\xi} - \boldsymbol{\xi}_i(t)) =$$

$$-\frac{\partial}{\partial \boldsymbol{\xi}} \cdot \{\frac{d\boldsymbol{\xi}}{dt}(\boldsymbol{\xi}, t, F) \sum_i \delta(\boldsymbol{\xi} - \boldsymbol{\xi}_i(t))\} =$$

$$-\frac{\partial}{\partial \boldsymbol{\xi}} \cdot \{\dot{\boldsymbol{\xi}}(\boldsymbol{\xi}, t, F)\, F(\boldsymbol{\xi}; t)\} \quad (2.c.22)$$

Equation (2.c.22) is a nonlinear p.d.e. in seven variables, the Klimontovich equation.

<u>Lemma 1</u>: Usually (unless the physical system is pathological) $\frac{\partial}{\partial \boldsymbol{\xi}} \cdot \dot{\boldsymbol{\xi}} = 0 \quad (2.c.23)$

<u>Lemma 2</u>: Following from the previous lemma (if true) and (2.c.22):

$$\left(\frac{\partial}{\partial t} + \dot{\boldsymbol{\xi}} \cdot \frac{\partial}{\partial \boldsymbol{\xi}}\right) F = 0 \quad (2.c.24)$$

which is a Liouville-type equation.

<u>Definition</u>: We introduce the ensemble average of $F$ over initial conditions for $\boldsymbol{\xi}_i$:

$$f_1(\boldsymbol{\xi}, t) \equiv \langle F \rangle(\boldsymbol{\xi}, t) = \int d\Gamma_0 \rho(\Gamma_0) \sum_i \delta(\boldsymbol{\xi} - \boldsymbol{\xi}_i(t|\Gamma_0)) \quad (2.c.25)$$

The mean value of the Klimontovich equation (2.c.22) is then

$$\frac{\partial}{\partial t} f_1(\pmb{\xi}, t) = -\frac{\partial}{\partial \pmb{\xi}} \cdot \langle \dot{\pmb{\xi}} \, F \rangle =$$

$$-\frac{\partial}{\partial \pmb{\xi}} \cdot \left\{ \dot{\pmb{\xi}}^{ext}(\pmb{\xi}, t) f_1(\pmb{\xi}; t) \right\} - \frac{\partial}{\partial \pmb{\xi}} \cdot \int d\pmb{\xi}' \, \dot{\pmb{\xi}}^{in}(\pmb{\xi}, \pmb{\xi}') \langle F(\pmb{\xi}', \mathbf{t}) F(\pmb{\xi}, \mathbf{t}) \rangle \qquad (2.c.26)$$

<u>Definition</u>: In (2.c.26) we have introduced the two-position correlation function

$$\langle F(\pmb{\xi}', \mathbf{t}) F(\pmb{\xi}, \mathbf{t}) \rangle \equiv \sum_i \sum_j < \delta(\pmb{\xi} - \pmb{\xi}_i) \delta(\pmb{\xi}' - \pmb{\xi}_j) >$$

$$= \delta(\pmb{\xi} - \pmb{\xi}') f_1(\pmb{\xi}, t) + \sum_{i \neq j} \langle F_i(\pmb{\xi}; t) F_j(\pmb{\xi}'; t) \rangle \qquad (2.c.27)$$

and $F_i(\pmb{\xi}; t) \equiv \delta(\pmb{\xi} - \pmb{\xi}_i)$. The first term on the right side of (2.c.27) is zero to preclude self-forces.

[*Editor's Note: A slightly cleaner notation in which the double sum has $i \neq j$ to exclude self-forces could have been employed*.]

Now expand the ensemble average bracket after introducing $\delta F_i = F_i - \langle F_i \rangle$:

$$\langle F_i F_j \rangle = \langle (\langle F_i \rangle + \delta F_i)(\langle F_j \rangle + \delta F_j) \rangle = \langle F_i \rangle \langle F_j \rangle + \langle \delta F_i \delta F_j \rangle \qquad (2.c.28)$$

using the identity $\langle \delta F_i \rangle = 0$. We note that $F_i$ is a single term out of the *N* terms in the sum over *i* leading to $f_1$; hence, $F_i \rightarrow O(\frac{1}{N}) f_1(\pmb{\xi})$ and $F_j \rightarrow O(\frac{1}{N}) f_1(\pmb{\xi}')$; and the sum $\sum_{i \neq j} \rightarrow N(N-1)$ pairs. These arguments lead to

$$\langle F(\pmb{\xi}') F(\pmb{\xi}) \rangle = \delta(\pmb{\xi} - \pmb{\xi}') f_1(\pmb{\xi}) + (1 - \frac{1}{N}) f_1(\pmb{\xi}) f_1(\pmb{\xi}') + \sum_{i \neq j} \langle \delta F_i(\pmb{\xi}) \delta F_j(\pmb{\xi}') \rangle \qquad (2.c.29)$$

*N* is large; so we drop the 1/*N* term in (2.c.29). We identify the last term

$$h_2 \equiv \sum_{i \neq j} \langle \delta F_i(\pmb{\xi}) \delta F_j(\pmb{\xi}') \rangle \qquad (2.c.30)$$

as the two-particle correlation which accounts for the forces between particles. We also drop the first term on the right side of (2.c.29) $\delta(\pmb{\xi} - \pmb{\xi}') f_1(\pmb{\xi})$ because the contribution to $\dot{\pmb{\xi}}^{in}$ from this term must be zero because self-forces are disallowed. If we assume the lemma (2.c.23) is valid and can commute $\frac{\partial}{\partial \pmb{\xi}} \cdot \dot{\pmb{\xi}}$ leading to (2.c.24), then (2.c.26) becomes

$$\left[ \frac{\partial}{\partial t} + \dot{\pmb{\xi}}^{ext}(\pmb{\xi}, t) \cdot \frac{\partial}{\partial \pmb{\xi}} + \int d\pmb{\xi}' \, \dot{\pmb{\xi}}^{in}(\pmb{\xi}, \pmb{\xi}') f_1(\pmb{\xi}') \cdot \frac{\partial}{\partial \pmb{\xi}} \right] f_1(\pmb{\xi}; t) =$$

$$= -\frac{\partial}{\partial \pmb{\xi}} \cdot \int d\pmb{\xi}' \, \dot{\pmb{\xi}}^{in}(\pmb{\xi}, \pmb{\xi}') h_2(\pmb{\xi}, \pmb{\xi}'; t) \qquad (2.c.31)$$

We note that the second term on the left side of (2.c.31) derives from the external force(s), and the third term contains the mean internal force. Equation (2.c.31) is not a closed equation determining the evolution of the distribution $f_1(\boldsymbol{\xi};t)$ because $h_2$ the two-particle correlation appears. The next step will be to get an equation for $h_2$, but this will drag in $h_3$, etc., i.e., the Bogoliubov–Born–Green–Kirkwood–Yvon (BBGKY) hierarchy emerges.

If the correlations vanish, the right side of (2.c.31) is zero; and the Vlasov equation results:

$$\left[\frac{\partial}{\partial t} + \dot{\boldsymbol{\xi}}^{ext}(\boldsymbol{\xi},t) \cdot \frac{\partial}{\partial \boldsymbol{\xi}} + \int d\boldsymbol{\xi}'\, \dot{\boldsymbol{\xi}}^{in}(\boldsymbol{\xi},\boldsymbol{\xi}') f_1(\boldsymbol{\xi}') \cdot \frac{\partial}{\partial \boldsymbol{\xi}}\right] f_1(\boldsymbol{\xi};t) = 0 \qquad (2.c.32)$$

Definition: Introduce $f_2(\boldsymbol{\xi},\boldsymbol{\xi}')$ defined by

$$f_2(\boldsymbol{\xi},\boldsymbol{\xi}') \equiv \sum_{i \neq j} \langle F_i(\boldsymbol{\xi}) F_j(\boldsymbol{\xi}') \rangle \qquad (2.c.33)$$

Lemma: Given (2.c.28) and the definition of $h_2$ in (2.c.30) then

$$f_2(\boldsymbol{\xi},\boldsymbol{\xi}') = f_1(\boldsymbol{\xi}) f_1(\boldsymbol{\xi}') + h_2(\boldsymbol{\xi},\boldsymbol{\xi}') \qquad (2.c.34)$$

From (2.c.34) we note that

$$h_2 \equiv \langle FF \rangle - f_1 f_1 \quad \rightarrow \quad \frac{\partial h_2}{\partial t} = \langle \frac{\partial F}{\partial t} F \rangle + \langle F \frac{\partial F}{\partial t} \rangle - \frac{\partial f_1}{\partial t} f_1 - f_1 \frac{\partial f_1}{\partial t} \qquad (2.c.35)$$

## 2.d Landau equation

In the previous section we derived (2.c.32) as the collisionless limit of (2.c.31). Progress can be made evaluating the right side of (2.c.31) in particular limits. If the gas is sufficiently dilute, then the Boltzmann parameter is small, $na_0^3 \ll 1$. A simple representation of the collision operator can be derived if the particle interaction potential is small, $\frac{\phi}{T} \ll 1$, leading to the Landau equation. Another limit useful for plasmas is obtained when $(e^2/\lambda_D)/T \ll 1$, and the Lenard-Balescu-Guernsey equation can be derived.

### 2.d.i Derivation of the Landau equation

To derive the Landau equation we assume $\frac{\phi}{T} \ll 1$ and consider a plasma which is weakly coupled, i.e., the interaction potential and collisions are weak. In this limit, $h_2 \sim \varepsilon \sim \phi_{ij}$; and $h_3 \sim \text{force} \times h_2 \sim \varepsilon^2$, which is higher order in $\varepsilon$ and will be discarded. This truncates the hierarchy and allows the set of equations to be closed. Recall from the previous section, the most severe truncation of the hierarchy arises when $h_2$ and all higher interaction terms are discarded, in which limit the Vlasov equation (2.c.32) is obtained. From 2.c.35 we have

$$\left[\frac{\partial}{\partial t} + \mathbf{v}\cdot\frac{\partial}{\partial \mathbf{r}} + \mathbf{v}'\cdot\frac{\partial}{\partial \mathbf{r}'} + \dot{\mathbf{p}}^{ext}(\mathbf{p})\cdot\frac{\partial}{\partial \mathbf{p}} + \dot{\mathbf{p}}^{ext}(\mathbf{p}')\cdot\frac{\partial}{\partial \mathbf{p}'}\right] h_2(\mathbf{r},\mathbf{r}',\mathbf{p},\mathbf{p}';t) =$$

$$\mathbf{f}(\mathbf{r},\mathbf{r}')\cdot\left(\frac{\partial}{\partial \mathbf{p}'} - \frac{\partial}{\partial \mathbf{p}}\right) f_1(\mathbf{r},\mathbf{p};t) f_1(\mathbf{r}',\mathbf{p}';t) \qquad (2.\text{d}.1)$$

In (2.d.1) $\mathbf{f}(\mathbf{r},\mathbf{r}')$ is the electric field force.

Next consider a system in which there is no external forces and to further simplify we assume that system is uniform so that $f \to f(\mathbf{p};t)$. With these simplifications and defining $\mathbf{s} = \mathbf{r} - \mathbf{r}'$ (2.d.1) becomes

$$\left[\frac{\partial}{\partial t} + (\mathbf{v}-\mathbf{v}')\cdot\frac{\partial}{\partial \mathbf{s}}\right] h_2(\mathbf{s},\mathbf{p},\mathbf{p}';t) = -\frac{\partial \phi(\mathbf{s})}{\partial \mathbf{s}}\cdot\left(\frac{\partial}{\partial \mathbf{p}'} - \frac{\partial}{\partial \mathbf{p}}\right) f_1(\mathbf{p};t) f_1(\mathbf{p}';t) \qquad (2.\text{d}.2)$$

<u>Definition</u>: The Fourier transform of $g(\mathbf{s})$ is $g(\mathbf{k}) = \int d^3 s\, g(\mathbf{s}) e^{-i\mathbf{k}\cdot\mathbf{s}}$

The Fourier transform of (2.d.2) is then

$$\left[\frac{\partial}{\partial t} + (\mathbf{v}-\mathbf{v}')\cdot i\mathbf{k}\right] h_2(\mathbf{k},\mathbf{p},\mathbf{p}';t) = -i\mathbf{k}\phi(\mathbf{k})\cdot\left(\frac{\partial}{\partial \mathbf{p}'} - \frac{\partial}{\partial \mathbf{p}}\right) f_1(\mathbf{p};t) f_1(\mathbf{p}';t) \qquad (2.\text{d}.3)$$

(2.d.3) is a quasi-linear differential equation that is first order in time. The corresponding limit of Eq.(2.c.31) for $f_1$ is

$$\frac{\partial}{\partial t} f_1(\mathbf{p};t) = -\frac{\partial}{\partial \mathbf{p}}\cdot \int d^3 r' d^3 p' \left(-\frac{\partial \phi}{\partial \mathbf{r}}(\mathbf{s}) h_2(\mathbf{s},\mathbf{p},\mathbf{p}';t)\right) \qquad (2.\text{d}.4)$$

Fourier transforming (2.d.4) and using the convolution theorem, (2.d.4) becomes

$$\frac{\partial}{\partial t} f_1(\mathbf{p};t) = \frac{\partial}{\partial \mathbf{p}}\cdot \int \frac{d^3 k}{(2\pi)^3} d^3 p' \left([i\mathbf{k}\phi(\mathbf{k})]^* h_2(\mathbf{k},\mathbf{p},\mathbf{p}';t)\right) \qquad (2.\text{d}.5)$$

We note that for $\frac{\partial}{\partial t} f_1 \to -\frac{\partial}{\partial t} f_1$ and $\mathbf{p} \to -\mathbf{p}$ in (2.d.5) the sign changes in both left and right sides of (2.d.5); hence, (2.d.5) appears to be fully time reversible. Thus far there is no "H-theorem" in Eqs.(2.d.1) to (2.d.5).

With respect to the reversibility or irreversibility of (2.d.5) consider the following.

$$\left(\frac{\partial}{\partial t} + \alpha\right) h_2 = g(t) \quad \to \quad \alpha = 0 \quad \frac{\partial}{\partial t} h(t) = g(t) \qquad (2.\text{d}.6)$$

which has solutions:

$$h(t) = \int_{-\infty}^{t} dt' g(t') + h(-\infty) \tag{2.d.7a}$$

or based on future values

$$h(t) = h(\infty) - \int_{t}^{\infty} dt' g(t') \tag{2.d.7b}$$

The solution in (2.d.7b) is disturbing because of causality considerations. It is irreversibility that provides a philosophical basis for not solving history problems backwards. Macroscopic variables lead to equations that do not tolerate "backward" or "final-value" problems. However, the basic equations (2.d.1) to (2.d.5) at the moment are fully time reversible. More work is needed to derive an irreversible kinetic equation.

We return to the solutions of (2.d.2) and consider first the homogeneous equation in the non-relativistic limit. With the masses $m = m' \equiv 1$ nd $\mathbf{w} \equiv \mathbf{v} - \mathbf{v}'$, the solution to the homogeneous limit (right side equals zero) of (2.d.2) is simply

$$h_2(\mathbf{k}, \mathbf{p}; \mathbf{p}'; t) = h_2(\mathbf{k}, \mathbf{p}; \mathbf{p}'; 0) e^{-i\mathbf{k} \cdot \mathbf{w} t} \tag{2.d.8}$$

The particular solution of (2.d.2) is

$$h_2(\mathbf{k}, \mathbf{p}; \mathbf{p}'; t) = h_2(\mathbf{k}, \mathbf{p}; \mathbf{p}'; 0) e^{-i\mathbf{k} \cdot \mathbf{w} t}$$
$$+ \int_0^t dt' e^{-i\mathbf{k} \cdot \mathbf{w}(t-t')} \phi(\mathbf{k}) i\mathbf{k} \cdot \left(\frac{\partial}{\partial \mathbf{p}} - \frac{\partial}{\partial \mathbf{p}'}\right) (f_1(\mathbf{p}; t') f_1(\mathbf{p}'; t)) \tag{2.d.9}$$

The solution in (2.d.9) is the causal solution to the initial-value problem. The interaction embodied in $\phi$ has been identified as the cause of the correlation.

<u>Definition</u>: $\tau \equiv t - t'$

Eq.(2.d.9) becomes

$$h_2(\mathbf{k}, \mathbf{p}; \mathbf{p}'; t) = h_2(\mathbf{k}, \mathbf{p}; \mathbf{p}'; 0) e^{-i\mathbf{k} \cdot \mathbf{w} t} + \int_0^t d\tau e^{-i\mathbf{k} \cdot \mathbf{w} \tau} \phi(\mathbf{k}) i\mathbf{k} \cdot \left(\frac{\partial}{\partial \mathbf{p}} - \frac{\partial}{\partial \mathbf{p}'}\right) (f_1(\mathbf{p}, t - \tau) f_1(\mathbf{p}', t))$$
$$\tag{2.d.10}$$

$\tau > 0$ is always causal. With (2.d.10) used for $h_2$ Eq.(2.d.5) becomes

$$\frac{\partial}{\partial t} f_1(\mathbf{p}; t) = \frac{\partial}{\partial \mathbf{p}} \cdot \int \frac{d^3 \mathbf{k}}{(2\pi)^3} d^3 \mathbf{p}' (-i\mathbf{k}) \phi^*(\mathbf{k}) h_2(\mathbf{k}, \mathbf{p}, \mathbf{p}'; 0) e^{-i\mathbf{k} \cdot \mathbf{w} t} + \cdots \tag{2.d.11}$$

We wish to show that the effect of initial correlations (the first term for $h_2$ in Eq.(2.d.10) used in (2.d.11)) falls off rapidly and only the second term in (2.d.10), i.e., recent collisions, persists.

Lemma: (Riemann-Lebesgue) The Fourier transform of an $L^1$ function vanishes at infinity. The Fourier transform of a smooth function falls off rapidly in transform space, e.g., the transform of $h_2$ falls off for large $\mathbf{w}t$.

For a system at or near thermal equilibrium $h_2 \to \beta\, \phi(\mathbf{k}) f_1(\mathbf{p}, 0) f_1(\mathbf{p}', 0)$. So we need only show

$$\frac{\partial}{\partial \mathbf{w}t}\left\{ \int \frac{d^3\mathbf{k}}{(2\pi)^3} |\phi(k)|^2\, e^{-i\mathbf{k}\cdot\mathbf{w}t} \right\} \to 0 \text{ rapidly} \qquad (2.d.12)$$

Example: For $\phi(s) \sim e^{-\frac{s^2}{2a^2}} \to |\phi(k)|^2 \sim e^{-k^2 a^2}$ and $\int d^3 k\, e^{-\frac{k^2 a^2}{2} - i\mathbf{k}\cdot\mathbf{w}t} \to e^{-\frac{|\mathbf{w}t|^2}{4a^2}}$ From this last expression we conclude that the characteristic time $t$ in which the correlation falls off is set by $t \sim a/|\mathbf{w}| \sim a/\bar{v}$ where $a$ is the range of the interaction, which is typically a small microscopic distance. Hence, the initial correlations disappear rapidly. For particles with $\mathbf{w} \sim 0$, co-traveling with the test particle, their contribution to $f_1(\mathbf{v})$ is small. Furthermore, three-particle correlations will destroy this special case.

We conclude that the contribution to $h_2$ from the initial correlation, the first term on the right side of (2.d.9) is subdominant to the second term in contributing to the right side of (2.d.11); hence, (2.d.11) becomes

$$\frac{\partial}{\partial t} f_1(\mathbf{p}; t) = \frac{\partial}{\partial \mathbf{p}} \cdot \int d^3 \mathbf{p}' \int_0^t d\tau \int \frac{d^3\mathbf{k}}{(2\pi)^3} \mathbf{k} |\phi(k)|^2 e^{-i\mathbf{k}\cdot\mathbf{w}\tau} \mathbf{k} \cdot \left( \frac{\partial}{\partial \mathbf{p}} - \frac{\partial}{\partial \mathbf{p}'} \right) (f_1(\mathbf{p}; t-\tau) f_1(\mathbf{p}', t))$$

(2.d.13)

The integrand in (2.d.13) falls off for $\tau \gg a/\bar{v}$, which allows us to extend the integral $\int_0^t d\tau \to \int_0^\infty d\tau$. (2.d.13) is a closed kinetic equation for $f_1$. It is irreversible and depends only on earlier times. (2.d.13) with the time integral extended to $\infty$ is

$$\frac{\partial}{\partial t} f_1(\mathbf{p}; t) = \frac{\partial}{\partial \mathbf{p}} \cdot \int d^3 \mathbf{p}' \int_0^\infty d\tau \int \frac{d^3\mathbf{k}}{(2\pi)^3} \mathbf{k} |\phi(k)|^2 e^{-i\mathbf{k}\cdot\mathbf{w}\tau} \mathbf{k} \cdot \left( \frac{\partial}{\partial \mathbf{p}} - \frac{\partial}{\partial \mathbf{p}'} \right) (f_1(\mathbf{p}; t-\tau) f_1(\mathbf{p}', t))$$

(2.d.14)

We argue that only present times matter in $f_1 f_1$ and $f_1$ can be expanded in time: $f_1(\mathbf{p}; t-\tau) = f_1(\mathbf{p}; t) - \tau \frac{\partial f_1}{\partial t}(\mathbf{p}; t)$ subject to $\tau < \tau_{coll} \sim a/\bar{v}$ which is a small collision duration. We note that $\frac{\partial f_1}{\partial t} \sim \frac{f_1}{\tau_f}$ where $\tau_f$ is a very large time corresponding to the time between collisions, $\tau_f \sim \ell/\bar{v}$. Once again there are two time scales present and $\frac{\tau_{coll}}{\tau_f} \ll 1$. Hence, $f_1(\mathbf{p}; t-\tau) f_1(\mathbf{p}', t-\tau) \to f_1(\mathbf{p}; t) f_1(\mathbf{p}', t)$

Lemma: $\int d\tau\, e^{-i\mathbf{k}\cdot\mathbf{w}\tau} = \pi \delta(\mathbf{k}\cdot\mathbf{w}) + \frac{P}{i\mathbf{k}\cdot\mathbf{w}}$ where $P$ denotes the principal value. We use this lemma to evaluate the right side of (2.d.14). We note that the second term in the lemma is an odd function of $\mathbf{k}$ and will never lead to a contribution.

We arrive at the Landau equation

$$\frac{\partial}{\partial t}f_1(\mathbf{p};t) = \frac{\partial}{\partial \mathbf{p}} \cdot \int d^3\mathbf{p}' \int \frac{d^3\mathbf{k}}{(2\pi)^3} \mathbf{k}\mathbf{k}|\phi(k)|^2\, \pi\delta(\mathbf{k}\cdot\mathbf{w}) \cdot \left(\frac{\partial}{\partial \mathbf{p}} - \frac{\partial}{\partial \mathbf{p}'}\right)(f_1(\mathbf{p};t)f_1(\mathbf{p}',t)) \quad (2.\text{d}.15)$$

Let us recap the assumptions in arriving at the Landau equation. The simplifying assumptions are spatial uniformity, no external forces, and a single species. The essential assumptions are weak coupling ($\phi << T$) leading to two disparate time scales (fast collision time and slower relaxation time), and the kinetic equation describes an initial-value problem.

### 2.d.ii Elaboration of the Landau equation and derivation of an H-theorem

In Sec. 2.c we analyzed the continuity equation for the phase-space density:

$$\frac{\partial f}{\partial t}(\mathbf{p};t) = -\frac{\partial}{\partial \mathbf{p}} \cdot \widetilde{\boldsymbol{\Gamma}}(\mathbf{p};t) \quad (2.\text{d}.16)$$

where $\widetilde{\boldsymbol{\Gamma}}$ is the particle flux. The ordinary number density is

$$n(t) \equiv \int d^3\mathbf{p}\, f(\mathbf{p};t) \quad (2.\text{d}.17)$$

Conservation of particles in phase space derives from the time derivative of (2.d.17) and the application of boundary conditions:

$$\frac{dn}{dt} = \int d^3\mathbf{p}\, \frac{\partial f}{\partial t}(\mathbf{p};t) = -\int d^3\mathbf{p}\, \frac{\partial}{\partial \mathbf{p}} \cdot \widetilde{\boldsymbol{\Gamma}}(\mathbf{p};t) = -\oint d\boldsymbol{\sigma} \cdot \widetilde{\boldsymbol{\Gamma}}(\mathbf{p};t) = 0 \quad (2.\text{d}.18)$$

assuming $\widetilde{\boldsymbol{\Gamma}}\big|_{\delta\boldsymbol{\sigma}=\infty} = 0$. In this case the flux has been assumed to vanish on the system boundaries. If instead $\widetilde{\boldsymbol{\Gamma}}\big|_{\delta\boldsymbol{\sigma}} \neq 0$, then there is a flow into or out of the region bounded by $\delta\boldsymbol{\sigma}$. The continuity equation (2.d.16) is a statement of continuous flow in momentum space. In the Boltzmann collision equation there are large-angle scatters ($\equiv$ strong interactions), in which case there is no continuous flow in momentum space. In a collision a particle will disappear from one position in momentum space and appear instantaneously elsewhere (unless the collision process is time-resolved on a microscopic time scale). From Sec. 2.d.i

$$\widetilde{\boldsymbol{\Gamma}}(\mathbf{p};t) = \int d^3\mathbf{p}'\, \mathbf{Q}(\mathbf{w}) \cdot \left(\frac{\partial}{\partial \mathbf{p}'} - \frac{\partial}{\partial \mathbf{p}}\right) f(\mathbf{p};t)f(\mathbf{p}';t), \quad \mathbf{w} \equiv \mathbf{v} - \mathbf{v}' \quad (2.\text{d}.19)$$

$$\mathbf{Q}(\mathbf{w}) = \int \frac{d^3\mathbf{k}}{(2\pi)^3} |\phi(\mathbf{k})|^2 \mathbf{k}\mathbf{k}\pi\delta(\mathbf{k}\cdot\mathbf{w}) \quad (2.\text{d}.20)$$

In (2.d.20) $\phi(\mathbf{k})$ is the Fourier transform of the interaction potential. Note: An alternative derivation using the Born approximation and the Fermi golden rule yields exactly the same results.

Lemma: $\mathbf{Q}(\mathbf{w})$ is manifestly symmetric, $\mathbf{w} \cdot \mathbf{Q}(\mathbf{w}) = 0$, and is positive-semi-definite ($\equiv \mathbf{a} \cdot \mathbf{Q}(\mathbf{w}) \cdot \mathbf{a} \geq 0$).

We now derive an explicit expression for $\mathbf{Q}(\mathbf{w})$. We begin with the decomposition $\mathbf{k} = \mathbf{k} \cdot \widehat{\mathbf{w}}\widehat{\mathbf{w}} + \mathbf{k} \cdot (\vec{\mathbf{I}} - \widehat{\mathbf{w}}\widehat{\mathbf{w}})$. By choosing $\widehat{\mathbf{w}} = \widehat{\mathbf{z}}$ and using spherical coordinates $(k, \theta, \phi)$, then $\mathbf{k}(k, \theta, \phi) \cdot \mathbf{w} = kw\cos\theta$, so that $\delta(\mathbf{k} \cdot \mathbf{w}) = \delta(\cos\theta)/(kw)$, and $\mathbf{k}(k, \theta = \pi/2, \phi) = k\widehat{\boldsymbol{\rho}}$, where $\widehat{\boldsymbol{\rho}} = \cos\phi\,\widehat{\mathbf{x}} + \sin\phi\,\widehat{\mathbf{y}}$. Equation (2.d.20), therefore, becomes

$$\mathbf{Q}(\mathbf{w}) = \int \frac{d^3\mathbf{k}}{(2\pi)^3} |\tilde{\phi}(\mathbf{k})|^2 \mathbf{k}\mathbf{k}\pi\delta(\mathbf{k} \cdot \mathbf{w})$$

$$= \int_0^\infty \frac{k^2 dk}{(2\pi)^3} |\tilde{\phi}(\mathbf{k})|^2 \frac{k^2}{kw} \int_0^\pi \pi\delta(\cos\theta)\sin\theta\,d\theta \int_0^{2\pi} \widehat{\boldsymbol{\rho}}\widehat{\boldsymbol{\rho}}\,d\phi$$

$$= \left[\frac{1}{8\pi w}\int_0^\infty k^3 |\tilde{\phi}(\mathbf{k})|^2 dk\right](\widehat{\mathbf{x}}\widehat{\mathbf{x}} + \widehat{\mathbf{y}}\widehat{\mathbf{y}}) \equiv \frac{Q}{w}(\vec{\mathbf{I}} - \widehat{\mathbf{w}}\widehat{\mathbf{w}}) \quad (2.d.21)$$

where $\int_0^{2\pi} \widehat{\boldsymbol{\rho}}\widehat{\boldsymbol{\rho}}\,d\phi = \pi(\widehat{\mathbf{x}}\widehat{\mathbf{x}} + \widehat{\mathbf{y}}\widehat{\mathbf{y}})$.

Aside on initial-value vs. final-value problem: Consider the initial-value problem for a non-singular $f$ ($t > 0$) given $h_2(t=0)$. We note that we might instead wish to solve a final-value problem for $f$ ($t < 0$) given $h_2(t=0)$. Then the only change would be that for the initial-value problem $\boldsymbol{\Gamma}(\mathbf{p}; t) = \int d^3\mathbf{p}' \ldots$ vs. $\boldsymbol{\Gamma}(\mathbf{p}; t) = -\int d^3\mathbf{p}' \ldots$ for the final-value problem.

At this point we have laid the groundwork for deriving an H-theorem for the Landau equation. Consider the expression for the entropy:

$$S(t) \equiv -\int d^3\mathbf{p}\, f(\mathbf{p}; t) \ln f(\mathbf{p}; t) \quad (2.d.22)$$

Using (2.d.16) and (2.d.19), the time derivative of (2.d.22) is

$$\frac{dS}{dt} = -\int d^3\mathbf{p} \left[\frac{\partial f}{\partial t}\ln f + \frac{\partial f}{\partial t}\right] =$$

$$-\int d^3\mathbf{p}\,\frac{\partial f}{\partial t}\ln f = \int d^3\mathbf{p}\,\ln f \frac{\partial}{\partial \mathbf{p}} \cdot \widetilde{\boldsymbol{\Gamma}} = -\int d^3\mathbf{p}\,\frac{1}{f}\frac{\partial f}{\partial \mathbf{p}} \cdot \widetilde{\boldsymbol{\Gamma}}(\mathbf{p}; t) =$$

$$-\frac{1}{2}\int d^3\mathbf{p} \int d^3\mathbf{p}' \left[\frac{1}{f(\mathbf{p})}\frac{\partial f(\mathbf{p})}{\partial \mathbf{p}} - \frac{1}{f(\mathbf{p}')}\frac{\partial f(\mathbf{p}')}{\partial \mathbf{p}'}\right] \cdot \mathbf{Q}(\mathbf{w}) \cdot \left(\frac{\partial}{\partial \mathbf{p}'} - \frac{\partial}{\partial \mathbf{p}}\right) f(\mathbf{p}; t) f(\mathbf{p}'; t) =$$

$$\frac{1}{2}\int d^3\mathbf{p} \int d^3\mathbf{p}' \frac{1}{f(\mathbf{p})f(\mathbf{p}')}\left[\left(\frac{\partial}{\partial \mathbf{p}'} - \frac{\partial}{\partial \mathbf{p}}\right)f(\mathbf{p})f(\mathbf{p}')\right] \cdot \mathbf{Q}(\mathbf{w}) \cdot \left[\left(\frac{\partial}{\partial \mathbf{p}'} - \frac{\partial}{\partial \mathbf{p}}\right)f(\mathbf{p}; t)f(\mathbf{p}'; t)\right] =$$

$$= \frac{1}{2}\int d^3\mathbf{p} \int d^3\mathbf{p}' \frac{1}{f(\mathbf{p})f(\mathbf{p}')} \mathbf{a} \cdot \mathbf{Q}(\mathbf{w}) \cdot \mathbf{a} \geq 0 \quad (2.d.23)$$

where

$$\mathbf{a} \equiv \left(\frac{\partial}{\partial \mathbf{p}'} - \frac{\partial}{\partial \mathbf{p}}\right) f(\mathbf{p}; t) f(\mathbf{p}'; t) \tag{2.d.24}$$

(2.d.23) demonstrates the H-theorem for the Landau equation.

Theorem: $\quad \frac{dS}{dt} = 0$ iff $\mathbf{a} = \mathbf{w} g(\mathbf{p}, \mathbf{p}')$ (2.d.25)

where $g$ is any smooth function of **p** and **p'**.

Now instead of **a** as defined in (2.d.24) consider

$$\mathbf{a}' \equiv \left(\frac{\partial}{\partial \mathbf{p}'} - \frac{\partial}{\partial \mathbf{p}}\right) \ln[f(\mathbf{p}; t) f(\mathbf{p}'; t)] \tag{2.d.26}$$

**a'** can be recast in the same form **a'**= $\mathbf{w} g(\mathbf{p}, \mathbf{p}')$ as in (2.d.25), i.e.,

$$\mathbf{a}' = (\mathbf{v} - \mathbf{v}') g = \frac{\partial \ln f(\mathbf{p}'; t)}{\partial \mathbf{p}'} - \frac{\partial \ln f(\mathbf{p}; t)}{\partial \mathbf{p}} \tag{2.d.27}$$

with which $\frac{dS}{dt} = 0$.

Theorem: The only solution of (2.d.25) has the form $\ln f(\mathbf{p}) = C_1 + \mathbf{C}_2 \cdot \mathbf{p} + C_3 p^2$ (2.d.28)

We can assign $\mathbf{C}_2 = \mathbf{u}$, a mean drift of the velocity distribution, and invert (2.d.28) to obtain the solution for the velocity distribution $f$ that satisfies $\frac{dS}{dt} = 0$, i.e., the equilibrium distribution:

$$f(\mathbf{p}) = \frac{n}{\sqrt{2\pi mT}} e^{-\beta \frac{1}{2} m(\mathbf{v} - \mathbf{u})^2} \tag{2.d.29}$$

(2.d.29) is the formula for a drifting Maxwellian distribution. $\frac{dS}{dt} > 0$ if $f$ is not a Maxwellian, and $f$ will relax to an asymptotic equilibrium that is a Maxwellian.

Definition: Define the kinetic energy

$$K(t) \equiv \int d^3 p \, \frac{p^2}{2m} f(\mathbf{p}; t) \tag{2.d.30}$$

since the interaction energy is higher order.

Exercise: Using the Landau equation (2.d.15) show that the kinetic energy is conserved, i.e., $\frac{dK}{dt} = 0$.

Definition: Define the momentum moment of $f$

$$\mathbf{g}(t) \equiv \int d^3\mathrm{p}\, \mathbf{p} f(\mathbf{p}; t) \tag{2.d.31}$$

Exercise: Using the Landau equation show that $\frac{d\mathbf{g}}{dt} = 0$.

## 2.d.ii Irreversibility

Here we present a discussion and precise definition of irreversibility.

Reversibility

Definition (Reversibility): If $\mathbf{v} \to -\mathbf{v}$ and $t \to -t$ without changing the physics except for merely duplicating the trajectory of the process $\chi(t) \to \chi(-t)$, ending up with the initial conditions defines a reversible process.

From the perspective of the BBGKY hierarchy consider $\big((\mathbf{p}; t), h_2(\mathbf{s}, \mathbf{p}, \mathbf{p}'; t)\big)$ defined at $t = 0$. Solve the equations in 2.d.i to obtain $(f_1, h_2)$ at $t = t_1 > 0$. Now instead introduce

$$\tilde{h}_2(\mathbf{s}, \mathbf{p}, \mathbf{p}'; t_1) \equiv h_2(\mathbf{s}, -\mathbf{p}, -\mathbf{p}'; t_1) \tag{2.d.32}$$

and solve for $(f_1, \tilde{h}_2)$ for $t > t_1$ up to $t \to t_2 = 2t_1$ which yields

$$\tilde{h}_2(\mathbf{s}, \mathbf{p}, \mathbf{p}'; t_2) \equiv h_2(\mathbf{s}, -\mathbf{p}, -\mathbf{p}'; 0) \text{ and } f_1(t_2) = f_1(0) \tag{2.d.33}$$

if the system is reversible. We have assumed weak coupling and no time ordering.

Alternatively we could integrate the kinetic equations forward in time, use the solutions at $t_1$ for initial conditions, then integrate backwards in time to recover the initial conditions at $t = 0$ once again, if the system is reversible.

Irreversibility

Given $(f_1(\mathbf{p}; t), h_2(\mathbf{s}, \mathbf{p}, \mathbf{p}'; t))$ defined at $t = 0$, solve the kinetic equations for $(f_1, h_2)$ at $t > 0$. Given the solutions, calculate the entropy (2.d.22): $S(t) \equiv -\int d^3\mathrm{p}\, f(\mathbf{p}; t)\ln f(\mathbf{p}; t)$. The system is *irreversible* if for any $t_1$, $S(t)$ is asymmetric about $t_1$, i.e., $S(t)$ is growing for increasing $t$. In making these arguments, there must arise a distinction between the microscopic evolution of the system which includes fluctuations and the macroscopic evolution as dictated by a kinetic equation like the Landau equation in which ensemble averages have smoothed over the microscopic fluctuations. The Landau macroscopic evolutionary equation has a discontinuity in slope at $t = 0$. Of course this is not a problem because the Landau equation only applies for $t > 0$.

Example: Consider the simple 1D diffusion equation as an example of an irreversible process.

$$\frac{\partial}{\partial t}\rho(x;t) = D\frac{\partial^2}{\partial x^2}\rho(x;t) \qquad (2.d.34)$$

It $t \to -t$ the left side of (2.d.34) changes sign but the right side does not. Given $\rho(x;0)$ we can find $\rho(x;t)$ for $t \gtrless 0$ by separating variables and Fourier analyzing:

$$\rho(x;t) \equiv \int \frac{dk}{2\pi} e^{ikx}\rho(k;t) \qquad (2.d.35a)$$

$$\frac{\partial}{\partial t}\rho(k;t) = -Dk^2\rho(k;t) \qquad (2.d.35b)$$

$$\rho(k;t) = \rho(k;0)e^{-Dk^2 t} \qquad (2.d.35c)$$

$$\rho(x;t) = \int_{-\infty}^{\infty} \frac{dk}{2\pi} \rho(k;0) e^{ikx - Dk^2 t} \qquad (2.d.35d)$$

We see in (2.d.35c) and (2.d.35d) that the solution for $\rho$ decays for $t > 0$ and blows up for $t < 0$. Moreover, the integral over $k$ in (2.d.35d) does not exist for $t < 0$ as $k \to \pm\infty$ because the integral diverges; so there is no solution for $\rho$ for $t < 0$.

Example: Suppose the initial condition for $\rho$ in the preceding example is $\rho(x;0) = \frac{1}{\sqrt{2\pi\sigma^2}} e^{-\frac{x^2}{2\sigma^2}}$. Then the solution of (2.d.35d) is given by

$$\rho(x;t) = \int_{-\infty}^{\infty} \frac{dk}{2\pi} \rho(k;0) e^{ikx - \frac{1}{2}\sigma^2 k^2 - Dk^2 t} \qquad (2.d.36)$$

We observe that the integral in (2.d.36) converges as long as $-Dt < \frac{1}{2}\sigma^2$, i.e., there is a non-singular solution for $\rho(x;t)$ for a finite interval of negative times. At $t = -\frac{\sigma^2}{2D}$ we have a $\delta$-function solution for $\rho$. In terms of a Green's function we find that

$$\rho(x;t) = \int dx' \rho(x';0) \frac{e^{-\frac{(x-x')^2}{4Dt}}}{\sqrt{4\pi Dt}} \qquad (2.d.37)$$

For $t < 0$, the $\sqrt{4\pi Dt}$ in the denominator is imaginary; and the exponential in the numerator blows up with large $|x-x'|$. However, the integral in (2.d.37) may still converge for a finite interval of negative times if $\rho(x';0)$ falls off with $|x'|$ fast enough. That $\sqrt{4\pi Dt}$ is imaginary is not fatal for obtaining a solution for negative times in the interval where the integral in (2.d.37) converges.

Exercise: For the Gaussian initial condition used in obtaining (2.d.36) show that the Green's function method in (2.d.37) can recover the same solution as in (2.d.36).

## 2.e Markov processes and the Fokker-Planck equation

Definition: A Markov process has no memory.

[*Editor's Note: The definition in Wikipedia is "A Markov chain or Markov process is a stochastic model describing a sequence of possible events in which the probability of each event depends only on the state attained in the previous event."*]

Processes, random, stochastic or otherwise, fall into a few categories. There are Markov and non-Markov processes. Within Markov processes there are continuous and discontinuous processes. An example of a continuous Markov process is Brownian motion with Gaussian statistics. Large-angle collisions described by the Boltzmann equation fall into the discontinuous Markov process category. The Landau equation can describe a continuous Markov process with non-Gaussian statistics. There are examples of generalized Brownian motion that are non-Markov processes. Processes can also be characterized as ergodic, stationary, Gaussian, and so on.

Suppose there is a random process with probability distribution:

$$\rho(x_1, x_2, \ldots, x_n) = \rho(x_1, x_2, \ldots, x_n)\rho(x_n | x_{n-1}, x_{n-2}, \ldots) \quad (2.e.1)$$

where $x(t_i) \equiv x_i$ measured at successive times. A Markov process corresponds to the condition

$$\rho(x_n | x_{n-1}, x_{n-2}, \ldots) = \rho(x_n, x_{n-1}) \quad (2.e.2)$$

A classical random walk is an example of a Markov process.

Definition: Define $\Delta_n \equiv x_n - x_{n-1}$.

We note $\rho(\Delta_n | x_{n-1}, x_{n-2}, \ldots) = \rho(\Delta_n | x_{n-1})$ also defines a Markov process.

Consider

$$\rho(x_n, x_{n-1}, \ldots, x_0) = \rho(x_n | x_{n-1})\rho(x_{n-1} | x_{n-2}) \ldots \rho(x_3 | x_2)\rho(x_2 | x_1)\rho(x_1 | x_0)\rho(x_0) \quad (2.e.3)$$

Divide both sides of (2.e.3) by $\rho(x_0)$ to obtain

$$\rho(x_n, x_{n-1}, \ldots | x_0) = \rho(x_n | x_{n-1})\rho(x_{n-1} | x_{n-2}) \ldots \rho(x_3 | x_2)\rho(x_2 | x_1)\rho(x_1 | x_0)$$

Chapman-Kolmogorov equation: $\quad \rho(x_2, x_1 | x_0) = \rho(x_2 | x_1)\rho(x_1 | x_0) \quad (2.e.4)$

(2.e.4) is the Chapman-Kolmogorov equation for any three times. We can integrate (2.e.4) $\int dx_1$ to obtain

$$\rho(x_2|x_0) = \int dx_1\, \rho(x_2|x_1)\rho(x_1|x_0) \tag{2.e.5}$$

which is true only for Markov processes. More generally $x_0$ would appear in $\rho(x_2, x_1|x_0)$.

Lemma:
$$\langle x_2 \rangle_{|x_0} \equiv \int dx_1\, \langle x_2 \rangle_{|x_1} \rho(x_1|x_0) \tag{2.e.6}$$

Lemma: We recall the definition of the normalized correlation function $R(\tau)$ from (2.a.44) and use (2.e.5) and (2.e.6) to obtain

$$\langle x_n \rangle_{|x_m} = x_m\, R(|t_n - t_m|) \tag{2.e.7}$$

to represent the average value of <x> at time $t_n$ following the precise value $x_m$ at $t_m$, and

$$x_0\, R(|t_2 - t_0|) = \int dx_1\, x_1\, R(|t_2 - t_1|)\, \rho(x_1|x_0) = R(|t_2 - t_1|)\, \langle x_1 \rangle_{|x_0} \tag{2.e.8}$$

For a stationary Gaussian process the correlation function can be built up in multiplicative pieces

$$R(t_2 - t_0) = R(t_2 - t_1) R(t_1 - t_0) \tag{2.e.9}$$

For a stationary Gaussian Markov process the correlation function has the form

$$R(\tau) = e^{-\lambda|\tau|} \tag{2.e.10}$$

to be consistent with (2.e.9), and with Gaussian statistics and stationarity.

## 2.e.i Expansion of the Chapman-Kolmogorov equation to derive the Fokker-Planck equation

Consider a continuous Markov process. Initial conditions become implicit, and we change the notation. For the probability distribution as a function of $x$ at time $t_n$, $\rho(x; t_n)$ given $\rho(x; t_0) = \delta(x - x_0)$ is

$$\rho(x_n; t_n) = \int dx_{n-1} \rho(x_n|x_{n-1})\, \rho(x_{n-1}; t_{n-1}) = \int dx' \rho(x|x')\, \rho(x'; t_{n-1}) \tag{2.e.11}$$

Definition: We introduce the transition probability

$$\psi(\Delta x | x - \Delta x; t - \Delta t, \Delta t) = \rho(x \leftarrow x'; t - \Delta t, \Delta t) \tag{2.e.12}$$

Then

$$\rho(x; t) = \int dx' \rho(x \leftarrow x'; t - \Delta t, \Delta t)\, \rho(x'; t - \Delta t) =$$

$$= \int d(\Delta x)\, \psi(\Delta x | x - \Delta x; t - \Delta t, \Delta t)\, \rho(x - \Delta x; t - \Delta t) \tag{2.e.13}$$

where $\Delta x = x - x'$. Note that $t$ is discrete and $x$ is continuous. Assume $\Delta x$ the step size is small so we can Taylor-series expand:

$$\rho(x;t) = \int d(\Delta x) \sum_{\ell=0}^{\infty} \frac{(-\Delta x)^{\ell}}{\ell!} \frac{\partial^{\ell}}{\partial x^{\ell}} \psi(\Delta x|x; t - \Delta t, \Delta t) \rho(x; t - \Delta t) \quad (2.\text{e}.14)$$

Now we change the notation by shifting $t$, $t \to t + \Delta t$, so that

$$\rho(x; t + \Delta t) = \int d(\Delta x) \, \psi(\Delta x|x - \Delta x; t, \Delta t) \, \rho(x - \Delta x; t) =$$

$$\int d(\Delta x) \sum_{\ell=0}^{\infty} \frac{(-\Delta x)^{\ell}}{\ell!} \frac{\partial^{\ell}}{\partial x^{\ell}} \psi(\Delta x|x; t, \Delta t) \, \rho(x; t) \quad (2.\text{e}.15)$$

If the series expansion in (2.e.15) is uniformly convergent then we can commute the integration and series summation to obtain

$$\rho(x; t + \Delta t) = \sum_{\ell=0}^{\infty} \frac{(-1)^{\ell}}{\ell!} \frac{\partial^{\ell}}{\partial x^{\ell}} \int d(\Delta x) \, (\Delta x)^{\ell} \, \psi(\Delta x|x; t, \Delta t) \, \rho(x; t)$$

$$= \sum_{\ell=0}^{\infty} \frac{(-1)^{\ell}}{\ell!} \frac{\partial^{\ell}}{\partial x^{\ell}} \langle (\Delta x)^{\ell} \rangle (x; t, \Delta t) \, \rho(x; t) \quad (2.\text{e}.16)$$

Next we subtract $\rho(x; t)$, which is just the $\ell = 0$ term on the right side, from both sides of (2.e.16) and then divide both sides of the resulting equation by $\Delta t$ to obtain:

$$\frac{\partial}{\partial t} \rho(x; t) = \sum_{\ell=1}^{\infty} (-1)^{\ell} \frac{\partial^{\ell}}{\partial x^{\ell}} \left[ \lim_{\Delta t \to 0} \frac{\langle (\Delta x)^{\ell} \rangle (x; t, \Delta t)}{\ell! \, \Delta t} \rho(x; t) \right]$$

$$\equiv \sum_{\ell=1}^{\infty} (-1)^{\ell} \frac{\partial^{\ell}}{\partial x^{\ell}} \left[ D^{(\ell)}(x, t) \rho(x; t) \right]$$

$$= -\frac{\partial}{\partial x} \left[ D^{(1)}(x, t) \rho(x; t) \right] + \frac{\partial^2}{\partial x^2} \left[ D^{(2)}(x, t) \rho(x; t) \right]$$

$$+ \sum_{\ell=3}^{\infty} (-1)^{\ell} \frac{\partial^{\ell}}{\partial x^{\ell}} \left[ D^{(\ell)}(x, t) \rho(x; t) \right] \quad (2.\text{e}.17\text{a})$$

$$D^{(\ell)}(x, t) \equiv \lim_{\Delta t \to 0} \frac{\langle (\Delta x)^{\ell} \rangle (x; t, \Delta t)}{\ell! \, \Delta t} \quad (2.\text{e}.17\text{b})$$

Eq.(2.e.17a) is the generalized Fokker-Planck equation. It is useful if we can truncate the equation after the first two terms on the right side: $D^{(1)}, D^{(2)} \neq 0$; $D^{(\ell)} \equiv 0, \ell > 2$. We then rewrite the truncated version of (2.e.17a) in the conventional form:

Fokker-Planck Equation: $\frac{\partial}{\partial t} \rho(x; t) = -\frac{\partial}{\partial x} \left[ \frac{\langle \Delta x \rangle}{\Delta t} \rho(x; t) \right] + \frac{\partial^2}{\partial x^2} \left[ D(x, t) \rho(x; t) \right]$ (2.e.18)

The first term on the right side of (2.e.18) is defined as the dynamic friction. In the following two examples we show how the Langevin equation model for Brownian motion and the Landau equation can lead to the Fokker-Planck equation.

[*Reviewer Dominique Escande's Comment: - The passage to the derivative in (2.e.17a) can be further discussed. See for instance (Ryskin, 1997)*
*- The generalized Fokker-Planck equation Eq.(2.e.17a) is called Kramers–Moyal expansion, or van Kampen's system-size expansion (see the corresponding Wikipedia articles for original references).*
*- Equation (2.e.18) is not a theorem by Fokker-Planck, but by Ryskin (above reference).*
*- The Pawula theorem, (Pawula, 1967) might be referenced here, since it shows that the only truncation of the expansion, which ensures solutions to be physically meaningful (e.g., positive everywhere) is that to the second order.*]

Example: Brownian motion – Consider the Langevin equation for a particle with unit mass (M=1):

$$\dot{v} = -\gamma v + \delta F \qquad (2.e.19)$$

Here the velocity $v(t)$ is the random variable of interest in the Langevin equation. Integrate in time over $\Delta t \ll \gamma^{-1}$ but $\Delta t \gg \tau_{\delta F}$ the characteristic time for fluctuations in the forces. Here $\nu_V \equiv \gamma$ using our previously introduced notation:

$$\Delta v = -\gamma v \Delta t + \int_0^{\Delta t} dt\, \delta F(t) \qquad (2.e.20)$$

Taking the ensemble average over fluctuations in $\delta F$, (2.e.20) becomes

$$\langle \Delta v \rangle = -\gamma v \Delta t + \int_0^{\Delta t} dt\, \langle \delta F \rangle(t) \qquad (2.e.21)$$

assuming there is no correlation between the velocity and the fluctuating force, i.e., $\langle \delta F \rangle = 0$. Hence,

$$\langle \Delta v \rangle = -\gamma v \Delta t \qquad (2.e.22)$$

and

$$\frac{\partial}{\partial t} \rho(v; t) = -\frac{\partial}{\partial v}[-\gamma v \rho(x; t)] + \frac{\partial^2}{\partial v^2}[D(v, t)\rho(v; t)] \qquad (2.e.23)$$

Now calculate the ensemble average of the square of (2.e.20):

$$\langle (\Delta v)^2 \rangle = \gamma^2 v^2 \Delta t^2 + \int_0^{\Delta t}\int dt\, dt'\, \langle \delta F(t) \delta F(t') \rangle - 2\gamma v \Delta t \int_0^{\Delta t} dt \langle \delta F \rangle(t) \qquad (2.e.24)$$

but $\langle \delta F \rangle = 0$ and $\langle \delta F(t)\delta F(t') \rangle = C_F(|t - t'|)$. With $\Delta t \gg \tau_{\delta F}$ a(2.e.24) becomes

$$\langle (\Delta v)^2 \rangle = \gamma^2 v^2 \Delta t^2 + \Delta t \int_{-\infty}^{\infty} d\tau \, C_F(\tau) \tag{2.e.25}$$

We divide (2.e.25) by $2\Delta t$ and take the limit as $\Delta t \to 0$ to obtain

$$\lim_{\Delta t \to 0} \frac{\langle (\Delta v)^2 \rangle}{2\Delta t} = \frac{\gamma^2 v^2 \Delta t}{2} + \frac{1}{2} \int_{-\infty}^{\infty} d\tau \, C_F(\tau) \tag{2.e.26}$$

At this point we recall that $\Delta t \ll \gamma^{-1}$ and $\Delta t \gg \tau_{\delta F}$ which allows us to argue that the first term on the right side of (2.e.26) is small compared to the second term and is negligible.

From (2.e.26) we conclude

$$D_v \equiv \lim_{\Delta t \to 0} \frac{\langle (\Delta v)^2 \rangle}{2\Delta t} = \frac{1}{2} \int_{-\infty}^{\infty} d\tau \, C_F(\tau) \tag{2.e.27}$$

From expressions we derived in Sec. 2.b.i for Brownian motion,

$$D_v \equiv \frac{1}{2} \int_{-\infty}^{\infty} d\tau \, C_F(\tau) = S(\omega = 0) = \gamma T \tag{2.e.28}$$

which recovers the Einstein relation. Note that $\Delta t \to 0$ only on the slow time scale.

[*Reviewer Dominique Escande's Comment: (Ryskin, 1997) can be referenced with respect to $\Delta t \to 0$ only on the slow time scale.*]

We can now identify terms in the Fokker-Planck equation (2.e.23):

$$\frac{\partial}{\partial t} \rho(v; t) = -\frac{\partial}{\partial v}[-\gamma v \rho] + \frac{\partial^2}{\partial v^2}[\gamma T \rho] = \gamma \frac{\partial}{\partial v}\left(v\rho + T \frac{\partial \rho}{\partial v}\right) \tag{2.e.29}$$

This is a universal Fokker-Planck equation, a property of any one-dimensional Gaussian Markov process.

Example: Landau equation – In this example $x \to \boldsymbol{p}(t)$ and $\rho \to f$. The Fokker-Planck equation is

$$\frac{\partial}{\partial t} f(\boldsymbol{p}; t) = -\frac{\partial}{\partial \boldsymbol{p}} \cdot \left[ \lim_{\Delta t \to 0} \frac{\langle \Delta \boldsymbol{p} \rangle}{\Delta t} f(\boldsymbol{p}; t) - \frac{\partial}{\partial \boldsymbol{p}} \cdot \left( \boldsymbol{D}(\boldsymbol{p}, t) f(\boldsymbol{p}; t) \right) \right] \tag{2.e.30}$$

where

$$\boldsymbol{D}(\boldsymbol{p}, t) = \lim_{\Delta t \to 0} \frac{\langle \Delta \boldsymbol{p} \Delta \boldsymbol{p} \rangle}{2\Delta t} \tag{2.e.31}$$

The Landau equation asserts

$$\frac{\partial}{\partial t} f(\mathbf{p};t) = -\frac{\partial}{\partial \mathbf{p}} \cdot \widetilde{\boldsymbol{\Gamma}}(\mathbf{p};t) \qquad (2.e.32)$$

Can we show that the right side of (2.e.30) is equal to $-\frac{\partial}{\partial \mathbf{p}} \cdot \widetilde{\boldsymbol{\Gamma}}(\mathbf{p};t)$? In Sec. 2.d.i we derived (2.d.19)

$$\widetilde{\boldsymbol{\Gamma}}(\mathbf{p};t) = \int d^3\mathbf{p}'\, \mathbf{Q}(\mathbf{w}) \cdot \left(\frac{\partial}{\partial \mathbf{p}'} - \frac{\partial}{\partial \mathbf{p}}\right) f(\mathbf{p};t) f(\mathbf{p}';t), \quad \mathbf{w} \equiv \mathbf{v} - \mathbf{v}'$$

Inside the square bracket in Eq.(2.e.30) the two terms can be expressed as

$$\lim_{\Delta t \to 0} \frac{\langle \Delta \mathbf{p} \rangle}{\Delta t} f + f \frac{\partial}{\partial \mathbf{p}} \cdot \mathbf{D} - \frac{\partial}{\partial \mathbf{p}} \cdot (\mathbf{D} f) - f \frac{\partial}{\partial \mathbf{p}} \cdot \mathbf{D} \qquad (2.e.33)$$

The first two terms in (2.e.33) can be identified with the $\frac{\partial}{\partial \mathbf{p}'}$ in the right side of (2.d.19) and the third and fourth terms in (2.e.33) can be identified with the $\frac{\partial}{\partial \mathbf{p}}$:

$$\mathbf{D}(\mathbf{p},t) = \int d^3\mathbf{p}'\, \mathbf{Q}(\mathbf{w}) f(\mathbf{p}';t) \qquad (2.e.34)$$

Definition: $\qquad \lim_{\Delta t \to 0} \frac{\langle \Delta \mathbf{p} \rangle}{\Delta t} \equiv \langle \mathbf{F} \rangle(\mathbf{p};t) = 2 \frac{\partial}{\partial \mathbf{p}} \cdot \mathbf{D}(\mathbf{p};t) \qquad (2.e.35)$

With the use of (2.e.33), (2.d.19), (2.e.34), and (2.e.35) the Fokker-Planck equation (2.e.30) is recovered. The Landau equation is a particular Fokker-Planck equation for a Markov process. On an appropriate time scale the transition probability $\psi(\Delta \mathbf{p}|\mathbf{p};t,\Delta t)$ has no dependence of the jump $\Delta \mathbf{p}$ on the past history. Weak coupling has been assumed, and the effects of large-angle collisions are neglected. The Boltzmann equation can accommodate large-angle scatters.

[*Reviewer Dominique Escande's Comment: The consequences of large-angle collisions are overlooked. The latter may have a large effect: see (Shoub, 1987).*]

To summarize, the Landau equation written in the form of the Fokker-Planck equation is

$$\frac{\partial}{\partial t} f(\mathbf{p};t) = \frac{\partial}{\partial \mathbf{p}} \cdot \left[ -\langle \mathbf{F} \rangle(\mathbf{p};t) f + \frac{\partial}{\partial \mathbf{p}} \cdot (\mathbf{D}(\mathbf{p},t) f) \right] \qquad (2.e.36)$$

where from Eq.(2.e.34)

$$\mathbf{D}(\mathbf{p},t) = \int d^3\mathbf{p}'\, \mathbf{Q}(\mathbf{w}) f(\mathbf{p}';t), \quad \mathbf{w} \equiv \mathbf{v} - \mathbf{v}'$$

For $v \ll \bar{v}$, $\mathbf{D}(\mathbf{p}, t) \to D\mathbf{I}$, i.e., the diffusion is isotropic in certain situations. The friction or drag term is

$$\langle \mathbf{F} \rangle (\mathbf{p}; t) = 2 \frac{\partial}{\partial \mathbf{p}} \cdot \mathbf{D}(\mathbf{p}; t) \tag{2.e.37}$$

From Sec. 2.d.i

$$\mathbf{Q}(\mathbf{w}) = \frac{Q}{w}(\mathbf{I} - \hat{\mathbf{w}}\hat{\mathbf{w}}), \quad \hat{\mathbf{w}} \cdot \mathbf{Q} = 0 \tag{2.e.38}$$

where from (2.d.21)

$$Q = \frac{1}{8\pi} \int dk \, k^3 |\tilde{\phi}(k)|^2, \quad \mathbf{a} \cdot \mathbf{Q} \cdot \mathbf{a} = \frac{Q}{w}(a^2 - (\mathbf{a} \cdot \hat{\mathbf{w}})^2) \geq 0 \tag{2.e.39}$$

Example: Consider $\phi(r) = \pm \phi_0 e^{-\frac{r^2}{2a^2}}$ attractive or repulsive. Calculate the Fourier transform to obtain $\tilde{\phi}(k) = \pm(\sqrt{2\pi}a)^3 \phi_0 e^{-k^2 a^2/2}$, from which

$$Q = \frac{1}{8\pi} \int dk \, k^3 |\tilde{\phi}(k)|^2 = \frac{(2\pi)^3 a^6}{8\pi} \phi_0^2 \int_0^\infty k^3 e^{-k^2 a^2} dk$$

$$= \pi^2 \phi_0^2 a^2 \left(\frac{1}{2} \int_0^\infty dt \, t e^{-t}\right) = \frac{\pi^2}{2} \phi_0^2 a^2$$

We can now find an expression for the Fokker-Planck equation:

$$\frac{\partial}{\partial t} f = \frac{\partial}{\partial \mathbf{p}} \cdot \left[ -\mathbf{F} f + \frac{\partial}{\partial \mathbf{p}} \cdot (\mathbf{D} f) \right] = \frac{\partial}{\partial \mathbf{p}} \cdot \left[ -\frac{1}{2} \mathbf{F} f + \mathbf{D} \cdot \frac{\partial}{\partial \mathbf{p}} f \right]$$

$$= -\frac{1}{2} \left( \frac{\partial}{\partial \mathbf{p}} \cdot \mathbf{F} \right) f + \mathbf{D} : \frac{\partial^2 f}{\partial \mathbf{p} \, \partial \mathbf{p}} = -\left[ \frac{\partial}{\partial \mathbf{p}} \cdot \left( \frac{\partial}{\partial \mathbf{p}} \cdot \mathbf{D} \right) \right] f + \mathbf{D} : \frac{\partial^2 f}{\partial \mathbf{p} \, \partial \mathbf{p}},$$

where we have used the relation (valid for a single-species plasma)

$$\mathbf{F}(\mathbf{p}) \equiv \int d^3 p' \left[ \frac{\partial}{\partial \mathbf{p}} \cdot \mathbf{Q}(\mathbf{w}) - \frac{\partial}{\partial \mathbf{p'}} \cdot \mathbf{Q}(\mathbf{w}) \right] f(\mathbf{p'}) = 2 \frac{\partial}{\partial \mathbf{p}} \cdot \mathbf{D}(\mathbf{p})$$

between the friction vector and the momentum diffusion (dyadic) tensor. Since Eq.(2.d.34) yields the definition

$$\mathbf{D}(\mathbf{p}) = \int d^3 p' \, \mathbf{Q}(\mathbf{w}) f(\mathbf{p'}) = Q \int d^3 p' \left( \frac{\vec{\mathbf{I}}}{w} - \frac{\hat{\mathbf{w}}\hat{\mathbf{w}}}{w^3} \right) f(\mathbf{p'})$$

we obtain

$$-\left[\frac{\partial}{\partial \mathbf{p}}\cdot\left(\frac{\partial}{\partial \mathbf{p}}\cdot \mathbf{D}\right)\right] = -\int d^3\mathbf{p}'\left[\frac{\partial}{\partial \mathbf{p}}\cdot\left(\frac{\partial}{\partial \mathbf{p}}\cdot \mathbf{Q}(\mathbf{w})\right)\right]f(\mathbf{p}')$$

Next, using Eq.(2.d.21) and **w=v-v'**, we first calculate (not that $Q$ is a constant)

$$\frac{\partial}{\partial \mathbf{p}}\cdot \mathbf{Q}(\mathbf{w}) = \frac{1}{m}\frac{\partial}{\partial \mathbf{v}}\cdot \mathbf{Q}(\mathbf{w}) = \frac{Q}{m}\frac{\partial}{\partial \mathbf{v}}\cdot\left(\frac{\vec{\mathbf{I}}}{w} - \frac{\hat{\mathbf{w}}\hat{\mathbf{w}}}{w^3}\right) = -\frac{2Q}{m}\frac{\hat{\mathbf{w}}}{w^3} = \frac{2Q}{m}\frac{\partial}{\partial \mathbf{v}}\left(\frac{1}{w}\right),$$

so that we obtain the friction vector from Eq.(2.e.37):

$$\mathbf{F} = 2\frac{\partial}{\partial \mathbf{p}}\cdot \mathbf{D} = \frac{2}{m}\int d^3\mathbf{p}'\left[\frac{\partial}{\partial \mathbf{v}}\cdot \mathbf{Q}(\mathbf{w})\right]f(\mathbf{p}') = -\frac{2Q}{m}\int d^3\mathbf{p}'\left[\frac{\mathbf{v}-\mathbf{v}'}{|\mathbf{v}-\mathbf{v}'|^3}\right]f(\mathbf{p}')$$

which naturally satisfies Newton' Third Law. Lastly, using the definition of the Dirac delta function, $\nabla^2|\mathbf{r}-\mathbf{r}'|^{-1} \equiv -4\pi\delta^3(\mathbf{r}-\mathbf{r}')$, we find

$$\frac{\partial}{\partial \mathbf{p}}\cdot\left(\frac{\partial}{\partial \mathbf{p}}\cdot \mathbf{Q}(\mathbf{w})\right) = \frac{2Q}{m^2}\frac{\partial}{\partial \mathbf{v}}\cdot\frac{\partial}{\partial \mathbf{v}}(|\mathbf{v}-\mathbf{v}'|^{-1}) \equiv -\frac{8\pi Q}{m^2}\delta^3(\mathbf{v}-\mathbf{v}') = -8\pi Qm\delta^3(\mathbf{p}-\mathbf{p}')$$

Hence, we conclude

$$-\left[\frac{\partial}{\partial \mathbf{p}}\cdot\left(\frac{\partial}{\partial \mathbf{p}}\cdot \mathbf{D}\right)\right] = -\int d^3\mathbf{p}'[-8\pi Qm\delta^3(\mathbf{p}-\mathbf{p}')]f(\mathbf{p}') = 8\pi Qmf(\mathbf{p}).$$

Example: Despite the long-range interactions in a plasma, assume the plasma is sufficiently tenuous so that weak coupling prevails. Debye shielding affects the interactions:

$$\phi(s) = \begin{cases} \pm\frac{e^2}{s_0} & (s \leq s_0) \\ \pm\frac{e^2}{s}e^{-\frac{s}{\lambda_D}} & (s > s_0) \end{cases}, \quad \lambda_D = \sqrt{\frac{T}{4\pi n_e e^2}} \quad (2.e.40)$$

where $s_0$ determines a cutoff of the potential at short distances such that $\frac{e^2}{s_0} \lesssim T$ and $\Lambda \equiv \frac{\lambda_D}{\left(\frac{e^2}{T}\right)} \gg 1$ to be consistent with the weak coupling assumption. In this example, $Q = 2\pi e^4 \ln\Lambda$; $\ln\Lambda =$ 3 to 15. At distances $s > \lambda_D$ the plasma screens the potential, and the potential decays exponentially in $s/\lambda_D$. Hence, $\tau_{relax} \sim 1/Q \sim 1/\phi_0^2$. $Q$ only affects the time scale for the relaxation of the distribution function, not the form of the solution.

Example: Multi-species Landau equation – Having two or more species only appears in the interaction potential. The Fokker-Planck equation becomes

$$\frac{\partial}{\partial t}f^s(\mathbf{p};t) = \frac{\partial}{\partial \mathbf{p}}\cdot\left[-\langle\mathbf{F}\rangle^s(\mathbf{p};t)f^s + \frac{\partial}{\partial \mathbf{p}}\cdot(\mathbf{D}^s(\mathbf{p},t)f^s)\right] \quad (2.e.41)$$

where

$$\mathbf{D}^s(\mathbf{p}, t) = \sum_{s\prime} \int d^3 \mathrm{p}' \, \mathbf{Q}^{ss\prime}(\mathbf{w}) f^{s\prime}(\mathbf{p}, t), \quad \mathbf{w} = \mathbf{v} - \mathbf{v}' = \frac{\mathbf{p}}{m_s} - \frac{\mathbf{p}'}{m_{s\prime}} \qquad (2.\text{e}.42)$$

and

$$\langle \mathbf{F} \rangle^s(\mathbf{p}; t) = 2 \frac{\partial}{\partial \mathbf{p}} \cdot \mathbf{D}^s(\mathbf{p}; t) \qquad (2.\text{e}.43)$$

where

$$\mathbf{Q}^{ss\prime}(\mathbf{w}) = \frac{Q^{ss\prime}}{w} (\mathbf{I} - \hat{\mathbf{w}}\hat{\mathbf{w}}) \qquad (2.\text{e}.44)$$

**2.e.ii Discontinuous Markov process and derivation of a master equation**

We reconsider the Chapman-Kolmogorov equation in the context of discrete steps in *x*:

$$\rho(x_n|x_0) = \int dx_{n-1} \, \rho(x_n|x_{n-1}) \rho(x_{n-1}|x_0) \qquad (2.\text{e}.45)$$

which we recast in the form

$$\rho(x; t|x_0) = \int dx' \, \rho(x; t|x', t') \rho(x'; t'|x_0) \qquad (2.\text{e}.46)$$

where *x* is discrete with index *m* and $t \to t + \Delta t$. The Chapman-Kolmogorov equation becomes

$$\rho_m(t + \Delta t) = \sum_{m'} \rho_{m \leftarrow m'}(t, \Delta t) \rho_{m'}(t) \qquad (2.\text{e}.47)$$

and we have suppressed the initial condition $x_0$ in the notation. Define the transition probability $\psi_{mm'} \equiv \rho_{m \leftarrow m'}(t, \Delta t), \sum_m \psi_{mm'} = 1$. Using the identity $\rho_m(t) = \sum_{m'} \psi_{mm'} \rho_{m'}(t)$ and (2.e.47)

$$\rho_m(t + \Delta t) - \rho_m(t) = \sum_{m'} \left[ \psi_{mm'}^{(\Delta t)} \rho_{m'}(t) - \psi_{m'm}^{(\Delta t)} \rho_m(t) \right] =$$

$$\sum_{m' \neq m} \left[ \psi_{mm'}^{(\Delta t)} \rho_{m'}(t) - \psi_{m'm}^{(\Delta t)} \rho_m(t) \right] \qquad (2.\text{e}.48)$$

since the *m=m'* term cancels. We next divide (2.e.48) by $\Delta t$ and take the limit $\Delta t \to 0$:

$$\lim_{\Delta t \to 0} \frac{\rho_m(t + \Delta t) - \rho_m(t)}{\Delta t} = \sum_{m' \neq m} \left[ \frac{\psi_{mm'}^{(\Delta t)}}{\Delta t} \rho_{m'}(t) - \frac{\psi_{m'm}^{(\Delta t)}}{\Delta t} \rho_m(t) \right] \to$$

$$\frac{\partial}{\partial t}\rho_m(t) = \sum_{m' \neq m}[a_{mm'}\rho_{m'}(t) - a_{m'm}\rho_m(t)] = \sum_{m'}[a_{mm'}\rho_{m'}(t) - a_{m'm}\rho_m(t)]$$

(2.e.49)

where $a_{mm'} \equiv \lim_{\Delta t \to 0} \frac{\psi_{mm'}^{(\Delta t)}}{\Delta t}$ the transition probability per unit time which is non-negative. Thus, the rate of change of probability for a discrete state *m* is just a function of the present time, due to $\Delta t \ll \tau_{evolution}$ The process becomes explicitly Markovian. The master equation (2.e.49) has not been derived from first principles: the derivation has used the Chapman-Kolmogorov equation.

Definition: A master equation as in (2.e.49) is a set of first-order differential equations describing the time evolution of the probability of a system to occupy each one of a discrete set of states with regard to a continuous time variable *t*. Pauli, Tolman, and Van Hove are among those credited with presenting master equations.

Rate equation for probability:

$$\frac{\partial}{\partial t}\rho_m(t) = \sum_{m'}[a_{mm'}\rho_{m'}(t) - a_{m'm}\rho_m(t)] \qquad (2.e.50)$$

Using the lemmas $\sum_m \rho_m(t) = 1$ and $\frac{d}{dt}\sum_m \rho_m(t) = 0$, and the definition for the entropy $S(t) = -\sum_m \rho_m(t) \ln \rho_m(t)$, we calculate the time derivative of the entropy:

$$\frac{d}{dt}S = \frac{1}{2}\sum_{mm'}(\ln \rho_m - \ln \rho_{m'})(a_{m'm}\rho_m - a_{mm'}\rho_{m'}) \qquad (2.e.51)$$

At this point we must assume something about $a_{m'm}$ vs. $a_{mm'}$.

Postulate: Assume detailed balance (à la Boltzmann), but not microscopic reversibility:

$$a_{m'm} = a_{mm'} \qquad (2.e.52)$$

Detailed balance means that the probability of the process has the same probability as does the inverse process. In some instances detailed balance does not hold, but states can be grouped into super-states where detailed balance does hold.

Using (2.e.52), Eq.(2.e.51) yields an "H-theorem":

$$\frac{d}{dt}S = \frac{1}{2}\sum_{mm'} a_{m'm}(\ln \rho_m - \ln \rho_{m'})(\rho_m - \rho_{m'}) \geq 0 \qquad (2.e.53)$$

Theorem: $\frac{d}{dt}S = 0$ iff $\rho_m = \rho_{m\prime}$, i.e., all states are equally probable, which defines equilibrium. The concept of equal *a priori* probabilities is synonymous with equilibrium. Moreover, as $t \to \infty$, $\frac{d}{dt}S \to 0$ and $\rho_m \to \rho_{m\prime}$.

## 2.f Linear response theory, linear Boltzmann equation, and transport theory

### 2.f.i Evolution of velocity angle probability distribution due to scattering

Consider a system of scatterers, e.g., neutrons being scattered by point scatterers or light being scattered. For specificity, consider a neutron in a uniform system of scatterers (at least statistically uniform).

Definition: Define a scattering direction $\boldsymbol{\Omega}$, $\int d\boldsymbol{\Omega} = 4\pi$, and the velocity direction probability function $\rho(\boldsymbol{\Omega}; t)$. We assume that the magnitude of the velocity $|\mathbf{v}|$ is unaffected by the scattering off recoilless particles.

$$\frac{\partial}{\partial t}\rho(\boldsymbol{\Omega}; t) = \int_{4\pi} d\boldsymbol{\Omega}' \nu(\boldsymbol{\Omega} \leftrightarrow \boldsymbol{\Omega}')[\rho(\boldsymbol{\Omega}'; t) - \rho(\boldsymbol{\Omega}; t)] \qquad (2.f.1)$$

where $\nu$ is the probability per unit time for scattering from $\boldsymbol{\Omega}$ to $\boldsymbol{\Omega}'$ or the reverse $\boldsymbol{\Omega}'$ to $\boldsymbol{\Omega}$: $\nu \sim n_0 v \sigma(\Theta)$ where $\sigma$ is the differential cross-section through the angle $\Theta$ between scattering directions $\boldsymbol{\Omega}$ and $\boldsymbol{\Omega}'$, $n_0$ is the number density of scatterers, and v is the relative velocity between the neutron and the scatterer.

The differential cross-section can be decomposed into a series expansion separating its angular dependence from its dependence on speed:

$$\sigma(\Theta) = \sum_{\ell=0}^{\infty} P_\ell(\cos\Theta) \frac{2\ell+1}{4\pi} \sigma_\ell(v) \qquad (2.f.2)$$

where $P_\ell$ are Legendre polynomials. We decompose (2.f.1) into spherical harmonics and solve the linear equation to obtain:

$$\rho(\boldsymbol{\Omega}; t) = \sum_{\ell,m} Y_\ell^m(\boldsymbol{\Omega})\rho_\ell^m(t=0)e^{-(\nu_0 - \nu_\ell)t} \qquad (2.f.3)$$

where $\nu_\ell \equiv n_0 v \sigma_\ell$. In going from (2.f.2) to (2.f.3) it is useful to employ the addition theorem for spherical harmonics:

$$P_\ell(\cos\gamma) = \frac{4\pi}{2\ell+1}\sum_{m=-\ell}^{m=\ell} Y_\ell^m(\theta_1, \phi_1) Y_\ell^{m*}(\theta_2, \phi_2)$$

where $\cos\gamma = \cos\theta_1 \cos\theta_2 + \sin\theta_1 \sin\theta_2 \cos(\phi_1 - \phi_2)$. We note that $\frac{1}{\sigma_{\ell\neq 0}} > \frac{1}{\sigma_0}$ and the $\ell = 0$ term in right side of (2.f.3) does not vanish as $t \to \infty$.

Suppose that the scattering is isotropic, i.e., $\sigma(\Theta) = \sigma_0$ and $\sigma_{\ell\neq 0} = 0$. In this case (2.f.3) becomes

$$\rho(\mathbf{\Omega}; t) = \sum_{\ell\neq 0, m} Y_\ell^m(\mathbf{\Omega}) \rho_\ell^m(t=0) e^{-\nu_0 t} + \rho_0^0(t=0) \tag{2.f.4}$$

Thus, any initial anisotropy will decay away exponentially. The physical mechanism is that random scattering causes a loss of order and structure. These results require $\sigma_{\ell\neq 0} < \sigma_0$ to be physical; otherwise any initial anisotropy would grow. (2.f.3) can be rewritten as

$$\rho(\mathbf{\Omega}; t) = \frac{1}{4\pi} + \sum_{\ell=1}^{\infty} \sum_{m=-\ell}^{m=\ell} Y_\ell^m(\mathbf{\Omega}) \rho_\ell^m(0) e^{-(\nu_0 - \nu_\ell)t} \tag{2.f.5}$$

with $\nu_\ell < \nu_0$ for all $\ell > 0$, $\nu_\ell = n_0 v \sigma_\ell$ and $\sigma(\Theta)$ given in (2.f.2).

### 2.f.ii Linear response fundamentals

Before discussing the analysis of the linear Boltzmann equation we first introduce some necessary definitions and properties associated with linear response functions. Consider the linear response *J(t)* of a system due to an external agency *F(t)*.

Postulate: Assume linearity, causality, and stationarity.

Given the postulated assumptions, quite generally one can write

$$J(t) = \int_{-\infty}^{t} dt' R(t, t') F(t') = \int_{0}^{\infty} d\tau R(\tau = t - t') F(t - \tau) \tag{2.f.6}$$

Definition: The response or transfer function *R* satisfies $R(\tau) = \begin{cases} R(\tau) & \tau > 0 \\ 0 & \tau < 0 \end{cases}$ (2.f.7)

With (2.f.7), the integral in (2.f.6) can be extended:

$$J(t) = \int_{-\infty}^{\infty} d\tau R(\tau) F(t - \tau) \tag{2.f.8}$$

Definition: Define the Fourier transform $g(\omega) = \int_{-\infty}^{\infty} dt\, g(t) e^{i\omega t}$

We use the convolution theorem and (2.f.8) to obtain

$$J(\omega) = R(\omega) F(\omega) \tag{2.f.9}$$

where $F(\omega) = \int_{-\infty}^{\infty} dt\, F(t)e^{i\omega t}$ with $F(t)|_{-\infty} = 0$ ($F$ is turned on at a finite time) and $F(t)|^{\infty} =$ finite; $R(\omega) = \int_{0}^{\infty} dt\, R(t)e^{i\omega t}$; and $J(\omega) = \int_{-\infty}^{\infty} dt\, J(t)e^{i\omega t} = \int_{-\infty}^{\infty} dt\, J(t)e^{i(\Omega+i\gamma)t}$, $\gamma > 0$ for convergence.

We note $R(\tau)$ is a causal function and is analytic in the upper half of the complex $\omega$ plane, which follows from $R(\tau) = 0$ for $\tau < 0$; and $R(-\omega) = R^*(\omega)$ for real $\omega$.

<u>Example 1</u>: $R(\tau) = e^{-\nu\tau} \to R(\omega) = \frac{1}{\nu - i\omega}$ which has a simple pole at $\omega = -i\nu$.

<u>Example 2</u>: $R(\tau) = \sin \omega_0 \tau \to R(\omega) = \frac{\omega_0}{\omega_0^2 - \omega^2}$ ($\omega \neq \omega_0$)

<u>Example 3</u>: $R(\tau) = \int_{-\infty}^{\infty} d\nu\, e^{-\nu(v)\tau} g(v)$, $0 \leq \nu_0 < \nu(v) < \nu_1$, $\to R(\omega) = \int_{-\infty}^{\infty} d\nu\, \frac{g(v)}{\nu(v) - i\omega}$
There is a branch cut for $-i\nu_0 < \omega < -i\nu_1$.

If we evaluate $R(\omega)$ on the real $\omega$ axis we note that $R(\omega) = R'(\omega) + iR''(\omega)$ where $R'(\omega)$ is an even function of $\omega$ and $iR''(\omega)$ is an odd function. The Kramers-Kronig relations assert that $R'(\omega)$ and $R''(\omega)$ are Hilbert transforms of one another.

Kramers-Kronig relations: $R'(\omega)$ and $R''(\omega)$ satisfy

$$R'(\omega) = \frac{1}{\pi} p.v. \int_{-\infty}^{\infty} d\xi\, \frac{R''(\xi)}{\xi - \omega} \qquad R''(\omega) = -\frac{1}{\pi} p.v. \int_{-\infty}^{\infty} d\xi\, \frac{R'(\xi)}{\xi - \omega} \qquad (2.f.10)$$

where $p.v. \int_{-\infty}^{\infty} d\xi$ is the Cauchy principal-value integral.

*[Editor's Addendum: The Kramers-Kronig relations are derived as an application of the Cauchy residue formula for a function $f(z) \equiv u(z) + iv(z)$ that is analytic in the upper-half plane (Im $z \geq 0$). Under this assumption, the contour integral $\oint_C f(z)dz/(z - \zeta) = 0$ vanishes for any real variable $\zeta$ along a closed contour C that is composed of 4 segments: two segments along the real axis (from -R to $\zeta - \epsilon$ and $\zeta + \epsilon$ to +R), a semi-circular segment in the clockwise direction (from $\zeta - \epsilon$ to $\zeta + \epsilon$), and a semi-circular segment in the counter-clockwise direction (from +R to -R). In the limits $R \to \infty$ and $\epsilon \to 0$, we therefore obtain*

$$0 = p.v. \int_{-\infty}^{\infty} dx\, \frac{u(x) + iv(x)}{x - \zeta} - i\pi[u(\zeta) + iv(\zeta)]$$

*where*

$$p.v. \int_{-\infty}^{\infty} dx\, \frac{f(x)}{(x - \zeta)} \equiv \lim_{\epsilon \to 0} \left[ \int_{\zeta + \epsilon}^{\infty} dx\, \frac{f(x)}{(x - \zeta)} + \int_{-\infty}^{\zeta - \epsilon} dx\, \frac{f(x)}{(x - \zeta)} \right]$$

*denotes the Cauchy principal-value integral. Hence, we obtain the dual relations*

$$u(\zeta) = \frac{1}{\pi} p.v. \int_{-\infty}^{\infty} dx \, \frac{v(x)}{x-\zeta} \equiv H[v](\zeta) \text{ and } v(\zeta) = -\frac{1}{\pi} p.v. \int_{-\infty}^{\infty} dx \, \frac{u(x)}{x-\zeta} \equiv -H[u](\zeta)$$

which are expressed in terms of the Hilbert transform $H[f](\zeta) \equiv \frac{1}{\pi} p.v. \int_{-\infty}^{\infty} dx \, f(x)/(x-\zeta)$. We note that these relations are completely general for the real and imaginary parts of an analytic function in the upper-half complex plane.]

Example: For atomic spectra, absorption can occur at a particular frequency:

$$R''(\omega) = \delta(\omega - \omega_0) - \delta(\omega + \omega_0) \qquad R'(\omega) = \frac{2}{\pi} \frac{\omega_0}{\omega_0^2 - \omega^2}, \omega \neq \omega_0 \qquad (2.f.11)$$

Thus, dissipation implies dispersion; and dispersion implies absorption.

Example: $R(\tau) = \sin(\omega_0 \tau) \rightarrow R(\omega) = \frac{\omega_0}{\omega_0^2 - \omega^2}$ which is wrong! The correct result is

$$R(\omega) = \frac{\omega_0}{\omega_0^2 - \omega^2} + i \frac{\pi}{2} (\delta(\omega - \omega_0) - \delta(\omega + \omega_0)) \qquad (2.f.12)$$

which we should have caught when we performed the Cauchy integral carefully. Thus, the Kramers-Kronig relations provide a valuable check.

Exercise: Include damping in a model response function *R*(t). Use Kramers-Kronig to determine *R″*.

[*Editor's Solution: Consider the model response function which includes damping*

$$R(t) = A \exp(-\nu t) \sin(\Omega t + \alpha)$$

*that is a solution (for $t > 0$) of the damped oscillator equation $\ddot{R}(t) + 2\nu \dot{R}(t) + \omega_0^2 R(t) = 0$, where $\Omega^2 \equiv \omega_0^2 - \nu^2 > 0$ and the constants ($A, \alpha$) are determined from the initial conditions $R(0)$ and $\dot{R}(0)$. The Fourier transform $\mathcal{R}(\omega) \equiv \int_0^\infty dt R(t) \exp(i\omega t)$ is expressed as a complex-valued function*

$$\mathcal{R}(\omega) = \frac{A}{2} \left[ \frac{\exp(i\alpha)}{(\omega + \Omega + i\nu)} - \frac{\exp(-i\alpha)}{(\omega - \Omega + i\nu)} \right] \equiv \mathcal{R}'(\omega) + i\mathcal{R}''(\omega),$$

*which has poles in the lower-half complex-$\omega$ plane at $\omega = \pm \Omega - i\nu$, with $\mathcal{R}'(\omega) \equiv \mathrm{Re}[\mathcal{R}](\omega)$ and $\mathcal{R}''(\omega) \equiv \mathrm{Im}[\mathcal{R}](\omega)$ for real $\omega$. Hence, the real and imaginary parts of $\mathcal{R}(\omega)$ are guaranteed to be related by the Kramers-Kronig relations Eq.(2.f.10)*

$$R'(\omega) = \frac{1}{\pi} p.v. \int_{-\infty}^{\infty} d\xi \, \frac{R''(\xi)}{\xi - \omega} \qquad R''(\omega) = -\frac{1}{\pi} p.v. \int_{-\infty}^{\infty} d\xi \, \frac{R'(\xi)}{\xi - \omega}$$

*which hold for arbitrary constants (A,α). For example, we consider the case (A, α) = (1,0), which yields the double Lorentzian distributions (centered at $\pm\omega_0$)*

$$\mathcal{R}'(\omega) = \frac{\Omega(\omega_0^2 - \omega^2)}{(\omega^2 - \omega_0^2)^2 + 4\nu^2\omega^2}$$

$$\mathcal{R}''(\omega) = \frac{2\Omega\omega\nu}{(\omega^2 - \omega_0^2)^2 + 4\nu^2\omega^2}$$

*where $\mathcal{R}'(\omega)$ and $\mathcal{R}''(\omega)$ are even and odd functions of ω, respectively. At resonance $\omega = \pm\omega_0$ (for $\nu \neq 0$), we find $\mathcal{R}'(\pm\omega_0) = 0$ and $\mathcal{R}''(\pm\omega_0) = \pm\Omega/(2\nu\omega_0)$. Lastly, in the limit $\nu \to 0$, we find*

$$R(\omega) = \frac{\omega_0}{\omega_0^2 - \omega^2} + i\frac{\pi}{2}(\delta(\omega - \omega_0) - \delta(\omega + \omega_0))$$

*which is given by Eq.(2.f.12).*]

### 2.f.iii Linear Boltzmann equation

Consider a system comprised of an electron gas with electron charge *e*, immersed in a gas of neutrals in which the electrons scatter. If there is an externally applied electric field *E(t)*, the current in response to the electric field is

$$\mathbf{j}(t) = \int_0^\infty d\tau\, \overleftrightarrow{\boldsymbol{\sigma}}(\tau) \cdot \mathbf{E}(t - \tau) \quad (2.f.13)$$

and

$$\mathbf{j}(\omega) = \overleftrightarrow{\boldsymbol{\sigma}}(\omega) \cdot \mathbf{E}(\omega) \quad (2.f.14)$$

We use the linear Boltzmann equation to describe the scattering of the electrons by the surrounding neutrals.

Postulate: Assume conditions such that $\nu_{e,neutrals} \gg \nu_{ee}$ with Debye shielding, i.e., the electron-neutral collisions are dominant.

Before linearization the scattering equation for the electron velocity probability distribution is

$$\frac{\partial}{\partial t}\rho(\mathbf{v};t) + \dot{\mathbf{v}} \cdot \frac{\partial}{\partial \mathbf{v}}\rho(\mathbf{v};t) = \int_{4\pi} d\,\Omega'\nu(\Theta,\mathbf{v})[\rho(\mathbf{v}';t) - \rho(\mathbf{v};t)] \quad (2.f.15)$$

where $\dot{\mathbf{v}} = \frac{e}{m}\mathbf{E}$ and $\nu \sim n_0\sigma(\Theta,v)v$ which depends on the particular neutral atoms. Eq.(2.f.15) is not linear in $\rho, \mathbf{E}$, etc. To justify linearization we require that **E** is weak and produces a small perturbation in $\rho$:

$$\rho(\mathbf{v}; t) = \rho^{(0)}(v) + \delta\rho(\mathbf{v}; t), \quad \delta\rho \ll \rho^{(0)} \tag{2.f.16}$$

where $\rho^{(0)}(v)$ is isotropic and only depends on the electron speed, e.g., a Maxwellian. To first order (2.f.15) becomes

$$\frac{\partial}{\partial t}\delta\rho(\mathbf{v}; t) + \frac{e}{m}\mathbf{E}\cdot\frac{\partial}{\partial \mathbf{v}}\rho^0(\mathbf{v}; t) = \int_{4\pi} d\,\mathbf{\Omega}'\nu(\Theta, \mathbf{v})[\delta\rho(\mathbf{v}'; t) - \delta\rho(\mathbf{v}; t)] \tag{2.f.17}$$

(2.f.17) is a linear Boltzmann equation. Now solve the linear integro-differential equation.

<u>Definition</u>: Introduce the notation $\int_{4\pi} d\,\mathbf{\Omega}'\nu(\Theta, \mathbf{v})[\delta\rho(\mathbf{v}'; t) - \delta\rho(\mathbf{v}; t)] \equiv -\tilde{\nu}\delta\rho(\mathbf{v}; t)$ (2.f.18)
where $\tilde{\nu}$ is a positive-definite operator operating on $\delta\rho$ inside the integral on the right side of (2.f.17). The operator $\tilde{\nu}$ is elaborated in the rest of this section and in particular (2.f.26) through (2.f.29).

With the use of (2.f.18), (2.f.17) becomes

$$-i\omega\delta\rho(\mathbf{v}; \omega) + \frac{e}{m}\mathbf{E}(\omega)\cdot\frac{\partial\rho^{(0)}}{\partial \mathbf{v}} = -\tilde{\nu}\delta\rho(\mathbf{v}; \omega)$$
$$\to (\omega + i\tilde{\nu})\delta\rho(\mathbf{v}; \omega) = -i\frac{e}{m}\mathbf{E}(\omega)\cdot\frac{\partial\rho^{(0)}}{\partial \mathbf{v}} \tag{2.f.19}$$

The current is related to the electron charge density and fluid velocity:

$$\mathbf{j}(t) \equiv n_e e \langle\mathbf{v}\rangle(t) = e n_e \int d^3\mathbf{v}\,\mathbf{v}\left(\rho^{(0)}(v) + \delta\rho(\mathbf{v}; t)\right) \tag{2.f.20}$$

where $\rho$ is normalized to unity, $n_e$ is the unperturbed electron density normalization factor, and only the $\delta\rho$ term on the right side leads to a finite current. Using (2.f.19) and (2.f.20) the Fourier transformed linearized current density is

$$\mathbf{j}(\omega) = \left[\frac{n_e e^2}{m}(-i)\int d^3\mathbf{v}\,\frac{\mathbf{v}}{(\omega+i\tilde{\nu})}\frac{\partial\rho^{(0)}}{\partial \mathbf{v}}\right]\cdot\mathbf{E}(\omega) = \sigma(\omega)\,\mathbf{E}(\omega) \tag{2.f.21}$$

The quantity inside the square bracket in (2.f.21) is the conductivity tensor $\boldsymbol{\sigma} = \sigma\overleftrightarrow{\mathbf{I}}$ which is isotropic due to the assumed isotropy of $\rho^{(0)}$ and has positive-definite eigenvalues:

$$\boldsymbol{\sigma}(\omega) = -\left[\frac{n_e e^2}{m}\int d^3\mathbf{v}\,\mathbf{v}\frac{1}{(\tilde{\nu}-i\omega)}\mathbf{v}\frac{\partial\rho^{(0)}}{v\partial v}\right] \tag{2.f.22}$$

This is a universal result.

For $\nu(v; \Theta) \equiv n_0 v\sigma(v; \Theta)$ (here $\sigma$ is the scattering cross-section, not to be confused with other definitions of $\sigma$) the linear Boltzmann equation derived from (2.f.1) is

$$\tilde{\nu}f(\mathbf{\Omega}; t) = \int d^2\,\mathbf{\Omega}'\nu(v; \Theta)[f(\mathbf{\Omega}; t) - f(\mathbf{\Omega}'; t)] \tag{2.f.23}$$

From the Landau and Boltzmann equations one can write down a generic kinetic equation for the electron velocity distribution in a spatially uniform medium:

$$\frac{\partial}{\partial t} f^e(\mathbf{v};t) + \frac{e}{m}\mathbf{E}(t)\cdot\frac{\partial}{\partial \mathbf{v}} f^e(\mathbf{v};t) = C^{ee}(f^e,f^e) + C^{ei}(f^e,f^i) \tag{2.f.24}$$

where $f^e(\mathbf{v};t) = f^{e(0)}(\mathbf{v}) + \delta f^e(\mathbf{v};t)$. In (2.f.24) $f^e(\mathbf{v};t) = n^e \rho(\mathbf{v};t)$, and the right side can be linearized

$$C^{ee}(f^e,f^e) + C^{ei}(f^e,f^i) = -\tilde{\nu}^{ee}\delta f^e - \tilde{\nu}^{ei}\delta f^e \equiv -\tilde{\nu}\delta f^e \tag{2.f.25}$$

so that (2.f.24) has the same form as the kinetic equation (2.f.17) after linearization.

The eigenfunctions of $\tilde{\nu}$ are the $Y_\ell^m(\Omega)$:

$$\nu(v;\Theta) = \sum_{\ell=0}^{\infty} P_\ell(\cos\Theta)\frac{2\ell+1}{4\pi}\nu_\ell(v) \tag{2.f.26}$$

$$\nu_\ell(v) \equiv \int d^2\Omega\, P_\ell(\cos\Theta)\,\nu(v;\Theta) \leq \int d^2\Omega\, \nu(v;\Theta) = \nu_0(v) \tag{2.f.27}$$

where $\nu_0(v) = \nu(v)$ is the "total" rate, $\nu_0(v) = n_0\sigma(v)v$, and we note that $\nu_\ell > 0$. Using (2.f.23), (2.f.26), (2.f.27) and the addition theorem mentioned after Eq. (2.f.3), it can be shown

$$\tilde{\nu} Y_\ell^m(\Omega) = (\nu_0 - \nu_\ell) Y_\ell^m(\Omega) \tag{2.f.28}$$

where on the left side of (2.f.28) $\tilde{\nu}$ operates on $Y_\ell^m$. Thus, $\nu_0 - \nu_\ell$ is the positive eigenvalue of $\tilde{\nu}$ operating on the eigenvector $Y_\ell^m$. Hence, any function of the operator $F(\tilde{\nu})$ operating on $Y_\ell^m$, $F(\tilde{\nu})Y_\ell^m$ will yield $F(\nu_0 - \nu_\ell)Y_\ell^m$. To illustrate the operator $\tilde{\nu}$, for $\ell = 1$ these relations imply

$$\nu_0 - \nu_1 \equiv \int d^2\Omega\, \nu(v;\Theta)(1-\cos\theta) = \int d^2\Omega\, n_0\, v\sigma(v;\Theta)(1-\cos\theta) \tag{2.f.29}$$

where $\sigma(v;\Theta)$ is the differential cross section and $\sigma_{tr}(v) = \int d^2\Omega\, \sigma(v;\Theta)(1-\cos\theta) = (\nu_0 - \nu_1)/n_0 v$ is the transport cross-section.

### 2.f.iv Collision models and conductivity

We can now apply (2.f.29) to the conductivity tensor in (2.f.22) for the $\ell = 1, m = \pm 1, 0$ terms.

In the expression for the conductivity $(\tilde{\nu} - i\omega)^{-1} \to (\nu_0 - \nu_1 - i\omega)^{-1}$ and using (2.f.22) and (2.f.29) we obtain

$$\boldsymbol{\sigma}(\omega) = \sigma\overleftrightarrow{\mathbf{I}} = -\frac{n_e e^2}{m}\int d^3\mathbf{v}\, \mathbf{v}\frac{1}{(\nu_0-\nu_1-i\omega)}\mathbf{v}\frac{\partial \rho^{(0)}}{v\partial v} \tag{2.f.30}$$

With $\rho^{(0)} \sim e^{-\beta \frac{1}{2}mv^2}$ (2.f.30) becomes

$$\boldsymbol{\sigma}(\omega) = \sigma(\omega)\overleftrightarrow{\mathbf{I}} = \beta n_e e^2 \int d^3\mathbf{v}\, \mathbf{v}\mathbf{v}\, \frac{1}{(\nu_{tr} - i\omega)} \rho^{(0)} \qquad (2.f.31)$$

Using the definition of the average over the velocity distribution and $tr\,\overleftrightarrow{\mathbf{I}}=3$, (2.f.31) becomes

$$\sigma(\omega) = \frac{1}{3}\beta n_e e^2 \langle v^2(\nu_{tr}(v) - i\omega)^{-1}\rangle_0 \qquad (2.f.32)$$

Example: Consider a single electron in the presence of an atom. The atom feels the electric field of the electron $\mathbf{E} = \frac{e}{r^2}\hat{\mathbf{r}}$ which induces a dipole moment in the atom, $\boldsymbol{\Pi} = \alpha \mathbf{E}$. The interaction energy of the atom's dipole with the electron is $e\phi = e\boldsymbol{\Pi}\cdot\nabla\frac{1}{r} \to \alpha\frac{e^2}{r^4}$. A model for the interaction of the electron with the dipole might be $\dot{\mathbf{v}} = -\frac{\partial}{\partial\mathbf{r}}\left(-\frac{(\frac{\alpha e^2}{m})}{r^4}\right) = -\nabla_r e\phi/m$, with $\alpha \sim O(\text{volume})$. The interaction of an electron with an ion is $\dot{\mathbf{v}} = -\frac{\partial}{\partial\mathbf{r}}\left(-\frac{(\frac{Ze^2}{m})}{r}\right)$. In general, the interaction of the electron with scattering center can be modeled as

$$\dot{\mathbf{v}} = -\frac{\partial}{\partial\mathbf{r}}\left(\frac{a_s}{r^s}\right) \qquad (2.f.33)$$

where $s=4$ for the dipole interaction and $s=1$ for the interaction with the ion. The quantity $a_s$ has units $a_s \sim \frac{velocity \cdot L}{time} L^s$. The transport cross-section has units of $L^2$ with functional dependence

$$\sigma_{tr}^{(s)}(v, a_s) = \frac{C_s |a_s|^q}{v^n} \qquad (2.f.34)$$

Hence, from (2.f.34) $L^2 = \frac{1}{velocity^n}(velocity^2 L^s)^q$; and $2q=n$ and $2=sq$ from which we conclude that $q=2/s$ and $n=4/s$.

$$\sigma_{tr}^{(s)}(v, a_s) = \frac{C_s |a_s|^{2/s}}{v^{4/s}} \qquad (2.f.35)$$

Example: For $s=1$ and interaction of electrons with an ion,

$$\sigma_{tr}^{(1)} = \frac{C_1}{v^4}\left(\frac{Ze^2}{m}\right)^2 = C_1 \frac{Z^2 e^4}{m^2 v^4} \qquad (2.f.36)$$

which we recognize as $C_1$ multiplied by the Rutherford cross section for the Coulomb interaction.

Exercise: Derive the expression for $\sigma_{tr}$ for a Debye-shielded electron-ion interaction based on the analysis in Sec. 1.d.vi

Example: Electron interaction with a neutral atom

From the literature we find the transport cross section

$$\sigma_{tr}(v) = \frac{c_4}{v}\left(\frac{\alpha e^2}{m}\right)^{\frac{1}{2}}, \quad c_4 = 1.1052 \tag{2.f.37}$$

and $\alpha$ is the polarizability. From (2.f.37) the scattering rate is

$$\nu_{tr}(v) = n_0 \sigma_{tr}(v) v = n_0 c_4 \left(\frac{\alpha e^2}{m}\right)^{\frac{1}{2}} \tag{2.f.38}$$

which is independent of speed. From (2.f.32) we solve for the conductivity

$$\sigma(\omega) = \frac{n_e e^2/m}{\nu_{tr} - i\omega} = \frac{\omega_{pe}^2}{4\pi(\nu_{tr}-i\omega)} \tag{2.f.39}$$

and there is a pole at $\omega = -i\nu_{tr}$. After Fourier transforming,

$$\sigma(\tau) = \frac{n_e e^2}{m} e^{-\nu_{tr}\tau} \quad \mathbf{j}(t) = \int_0^\infty d\tau\, \sigma(\tau)\mathbf{E}(t-\tau) \tag{2.f.40}$$

Exercise: Sketch $\sigma'(\omega) = \text{Re}\,\frac{\omega_{pe}^2}{4\pi(\nu_{tr}-i\omega)}$ and $\sigma''(\omega) = \text{Im}\,\frac{\omega_{pe}^2}{4\pi(\nu_{tr}-i\omega)}$ using the estimate

$$\sigma(\omega=0) = \frac{n_e}{n_0}\left(\frac{e^2}{m}\right)^{1/2}\frac{1}{\alpha^{1/2}c_4} \sim 10^{16}\frac{n_e}{n_0}\text{s}^{-1} \quad n_0 \sim 10^{19}\text{cm}^{-3} \quad \nu_{tr}\sim n_0 10^{-8}\text{cm}^3\text{s}^{-1}\sim 10^{11}\text{s}^{-1}$$

Using Maxwell's equation (electrostatic limit) $\frac{4\pi\mathbf{j}}{c} + \frac{1}{c}\frac{\partial \mathbf{E}}{\partial t} = 0$ and $\mathbf{j}(\omega) = \overleftrightarrow{\sigma}(\omega)\cdot\mathbf{E}(\omega)$ we obtain the dispersion relation for electron plasma waves with collisional damping:

$$\omega(\omega + i\nu_{tr}) = \frac{4\pi n_e e^2}{m} \equiv \omega_{pe}^2 \tag{2.f.41a}$$

$$\omega = \omega' + i\omega'' \sim \omega_{pe} - i\frac{\nu_{tr}}{2} \tag{2.f.41b}$$

The generalization of (2.f.40) for $\sigma(\tau)$ when $\nu_{tr}(v)$ is a function of v is

$$\sigma(\tau) = \frac{n_e e^2}{m}\frac{\langle e^{-\nu_{tr}(v)\tau}v^2\rangle}{\langle v^2\rangle} \tag{2.f.42}$$

Example: Useful formulae for a linear Boltzmann model of a plasma:

$$\sigma'(\omega = 0) = \frac{8}{\sqrt{\pi}} \frac{n_e e^2}{m\bar{v}} \quad \text{where} \quad \bar{v} \equiv \frac{n_0}{(2mT^3)^{1/3}} Q \text{ and } Q \cong 2\pi e^4 \ln \Lambda \tag{2.f.43}$$

$\sigma'(\omega = 0)$ is the dc plasma conductivity. The linear Boltzmann model for plasma collisions is equivalent to a Lorentz model. A linear Landau equation plasma model yields $\sigma'_{Landau}(\omega = 0) = 1.98 \sigma'_{linear\ Boltz.}(\omega = 0)$.

## 2.f.v Linear response theory and Kubo formulae

M. Green (Green, 1954) and R. Kubo (Kubo, 1957) derived relations that give exact mathematical expressions for transport coefficients in terms of integrals of time correlation functions. These relations lead to a powerful fluctuation-dissipation theorem.

We posit a system that can be described with a Hamiltonian and obeys the Liouville equation. We further assume that the system possesses a small parameter that allows one to expand the Hamiltonian and the probability distribution to first order in the time-dependent perturbations:

$$H = H_0(\boldsymbol{\Gamma}) + \delta H(\boldsymbol{\Gamma}; t) \qquad \boldsymbol{\Gamma} = (\boldsymbol{p}_i, \boldsymbol{q}_i) \tag{2.f.44}$$

The system being Hamiltonian can be represented by the Liouville equation. For

$$\rho(\boldsymbol{\Gamma}; t) = \rho_0(\boldsymbol{\Gamma}) + \delta\rho(\boldsymbol{\Gamma}; t) \tag{2.f.45}$$

the Liouville equation is

$$\frac{d\rho}{dt} = \frac{\partial \rho}{\partial t} + \{\rho, H\} = 0 \tag{2.f.46}$$

where $\{f,g\}$ is the Poisson bracket. The equilibrium distribution $\rho_0$ is time independent and satisfies

$$\{\rho_0, H_0\} = 0 \tag{2.f.47}$$

and the ergodic hypothesis applies:

$$\rho_0(H_0) = \frac{e^{-\beta H_0(\boldsymbol{\Gamma})}}{Z} \tag{2.f.48}$$

The linearized version of (2.f.46) is

$$\frac{d\delta\rho}{dt} = \frac{\partial \delta\rho}{\partial t} + \{\delta\rho, H_0\} + \{\rho_0, \delta H\} = 0 \tag{2.f.49}$$

Definitions:

$$\mathcal{L}_0 = \{,H_0\} \equiv \dot{q}^{(0)}\frac{\partial}{\partial q} + \dot{p}^{(0)}\frac{\partial}{\partial p} \qquad (2.f.50a)$$

$$\delta\mathcal{L}(t) \equiv \{,\delta H(t)\} \equiv \delta\dot{q}\frac{\partial}{\partial q} + \delta\dot{p}\frac{\partial}{\partial p} \qquad (2.f.50b)$$

Equation (2.f.49) can be integrated by introducing an integrating factor as

$$e^{-t\mathcal{L}_0}\frac{d}{dt}\left(e^{t\mathcal{L}_0}\delta\rho(t)\right) = -\delta\mathcal{L}(t)\rho_0$$

which can be integrated from $-\infty$ to $t$:

$$\delta\rho(t')e^{t'\mathcal{L}_0}\Big|_{t'=-\infty}^{t'=t} = \delta\rho(t)e^{t\mathcal{L}_0} = -\int_{-\infty}^{t}dt'e^{-t'\mathcal{L}_0}\,\delta\mathcal{L}(t')\rho_0$$

$$= -\int_0^{\infty}d\tau e^{-(t-\tau)\mathcal{L}_0}\,\delta\mathcal{L}(t-\tau)\rho_0$$

where the last expression is obtained with the substitution $t' = t - \tau$. Hence, the solution for $\delta\rho(t)$ is given by

$$\delta\rho(\Gamma,t) = -\int_0^{\infty}d\tau e^{-\tau\mathcal{L}_0}\,\delta\mathcal{L}(t-\tau)\rho_0(\Gamma) \qquad (2.f.51)$$

Lemma: Using (2.f.50), Hamilton's equations of motion, and (2.f.48),

$$\delta\mathcal{L}(t)\,\rho_0 = -\beta\rho_0\delta\mathcal{L}(t)H_0 = -\beta\rho_0\{H_0,\delta H(t)\} = \beta\rho_0\{\delta H(t),H_0\} = \beta\rho_0\mathcal{L}_0\,\delta H(t) \qquad (2.f.52)$$

Hence, $\delta\mathcal{L}(t)\,\rho_0 = \beta\rho_0\mathcal{L}_0\,\delta H(t)$.

Given (2.f.52), $\delta\mathcal{L}(t-\tau)\,\rho_0(\Gamma) = \beta\rho_0(\Gamma)\mathcal{L}_0\,\delta H(t-\tau)$, (2.f.51) is equivalent to

$$\delta\rho(\Gamma,t) = -\beta\rho_0(\Gamma)\int_0^{\infty}d\tau e^{-\tau\mathcal{L}_0}\,\mathcal{L}_0\delta H(t-\tau) \qquad (2.f.53)$$

We assume that the linear perturbation to the Hamiltonian can be expressed as a sum over terms that can be decomposed into products of spatial and temporal phase factors:

$$\delta H(\Gamma,t) \equiv -\sum_\mu A_\mu(\Gamma)\delta F_\mu(t) \qquad (2.f.54)$$

Example: Electrons subject to externally imposed fields

$$\delta H = \int d^3x\left\{\frac{1}{c}\mathbf{j}(\mathbf{x}|\Gamma)\cdot\delta\mathbf{A}^{ext}(\mathbf{x},t) - \rho_{elec}(\mathbf{x},t)\delta\phi^{ext}(\mathbf{x},t)\right\} \qquad (2.f.55)$$

Here **x** denotes the field position in configuration space, distinct from $\Gamma$ which is the phase space of all the particle positions and momenta. This is the classical perturbed Hamiltonian for electromagnetic forces. Here

$$\rho_{elec} = \sum_i e_i \delta(\mathbf{x} - \mathbf{r}_i) \qquad \mathbf{j} = \sum_i e_i \mathbf{v}_i \delta(\mathbf{x} - \mathbf{r}_i) \qquad (2.f.56)$$

Using (2.f.54) in (2.f.53) the perturbed probability distribution becomes

$$\delta\rho(\Gamma, t) = \beta \rho_0(\Gamma) \sum_\mu \int_0^\infty d\tau \delta F_\mu(t-\tau) e^{-\tau \mathcal{L}_0} \mathcal{L}_0 A_\mu(\Gamma) \qquad (2.f.57)$$

Since $A_\mu(\Gamma)$ has no explicit time dependence, its total time derivative along the zero order particle trajectory is given by

$$\frac{d^{(0)} A_\mu(\Gamma)}{dt} = \frac{\partial A_\mu}{\partial q} \dot{q}^{(0)} + \frac{\partial A_\mu}{\partial p} \dot{p}^{(0)} = \{A_\mu, H_0\} \equiv \dot{A}_\mu(\Gamma) \qquad (2.f.58)$$

<u>Example</u>: $A_\mu = \mathbf{r}_i, \quad \dot{\mathbf{r}}_i = \{\mathbf{r}_i, H_0\} \equiv \mathbf{v}_i$

We introduce 
$$B(\Gamma, t) \equiv e^{t\mathcal{L}_0} B(\Gamma) = \sum_{n=0}^\infty \frac{(t\mathcal{L}_0)^n}{n!} B(\Gamma) \qquad (2.f.60)$$

<u>Example</u>: $B(\Gamma, t) = \mathbf{r}_i$ Hence, $\mathbf{r}_i(t)$ is the position $\mathbf{r}_i(t)$ at time $t$ given $\mathbf{r}_i(0)$ at $t=0$.

<u>Corollary</u>: $e^{-\tau \mathcal{L}_0} \mathcal{L}_0 A_\mu(\Gamma) = e^{-\tau \mathcal{L}_0} \dot{A}_\mu(\Gamma) = \dot{A}_\mu(\Gamma, -\tau)$ and thus (2.f.57) becomes

$$\delta\rho(\Gamma, t) = \beta \rho_0(\Gamma) \sum_\mu \int_0^\infty d\tau \delta F_\mu(t-\tau) \dot{A}_\mu(\Gamma, -\tau) \qquad (2.f.61)$$

Consider a set $\{B_\nu(\Gamma)\}$ from which we calculate an average value at a given time:

$$\langle B_\nu \rangle(t) \equiv \int d\Gamma B_\nu(\Gamma) \rho(\Gamma, t) = \int d\Gamma B_\nu(\Gamma)(\rho_0 + \delta\rho) \qquad (2.f.62)$$

and

$$\delta \langle B_\nu \rangle(t) \equiv \int d\Gamma B_\nu(\Gamma) \delta\rho(\Gamma, t) =$$

$$\beta \sum_\mu \int_0^\infty d\tau \delta F_\mu(t-\tau) \int d\Gamma \rho_0(\Gamma) B_\nu(\Gamma) \dot{A}_\mu(\Gamma, -\tau) =$$

$$\beta \sum_\mu \int_0^\infty d\tau \delta F_\mu(t-\tau) \langle B_\nu \dot{A}_\mu(-\tau) \rangle_0 =$$

$$\beta \sum_\mu \int_0^\infty d\tau \delta F_\mu(t-\tau) \langle B_\nu(t) \dot{A}_\mu(t-\tau) \rangle_0 \qquad (2.f.63)$$

$\langle B_\nu \dot{A}_\mu(-\tau) \rangle_0$ is a correlation function and can be shown to be stationary $\langle B_\nu \dot{A}_\mu(-\tau) \rangle_0 = \langle B_\nu(t) \dot{A}_\mu(t-\tau) \rangle_0$, which is a "cross-correlation function."

<u>Definition</u>: 
$$C_{\nu\mu}^{B\dot{A}}(\tau) \equiv \langle B_\nu(t) \dot{A}_\mu(t-\tau) \rangle_0 \qquad (2.f.64)$$

Hence
$$\delta\langle B_\nu\rangle(t) \equiv \beta \sum_\mu \int_0^\infty d\tau\, C^{B\dot{A}}_{\nu\mu}(\tau)\delta F_\mu(t-\tau) \tag{2.f.65}$$

The integrals over time in Eqs.(2.f.61), (2.f.63), (2.f.64) and (2.f.65) follow the same convention as in (2.f.51) with respect to the limits of the integrations given the actual initial conditions.

(2.f.65) gives the response of the system at thermal equilibrium to an external field based just on the unperturbed system. This is a fluctuation response theorem describing the linear response of a system in thermal equilibrium to a small-amplitude external (non-equilibrium) field.

Because (2.f.65) is a convolution, we can Fourier analyze and use the convolution theorem to obtain:
$$\delta\langle B_\nu\rangle(\omega) \equiv \beta \sum_\mu C^{B\dot{A}}_{\nu\mu}(\omega)\delta F_\mu(\omega) \tag{2.f.66}$$

where
$$C^{B\dot{A}}_{\nu\mu}(\omega) \equiv \int_0^\infty d\tau\, e^{i\omega\tau} C^{B\dot{A}}_{\nu\mu}(\tau) \tag{2.f.67}$$

is complex and satisfies the Kramers-Kronig relations. However,
$$S(\omega) \equiv \int_{-\infty}^\infty d\tau\, e^{i\omega\tau} C(\tau) \tag{2.f.68}$$

is real, positive, and an even function of $\omega$; and

$$C(\omega) = \int_0^\infty d\tau\, e^{i\omega\tau} \int_{-\infty}^\infty \frac{d\omega'}{2\pi} e^{-i\omega'\tau}\, S(\omega') =$$

$$= \int_{-\infty}^\infty \frac{d\omega'}{2\pi} S(\omega') \int_0^\infty d\tau\, e^{i(\omega-\omega')\tau} =$$

$$= \frac{1}{2}S(\omega) - \frac{i}{2\pi}\oint d\xi\, \frac{S(\xi)}{\xi-\omega} \tag{2.f.69}$$

using $\frac{1}{2\pi}\int_0^\infty d\tau\, e^{i(\omega-i\omega')\tau} = \frac{1}{2}\delta(\omega-\omega') - \frac{i}{2\pi}\frac{P}{\omega'-\omega}$. Thus, $2\mathrm{Re}\, C(\omega) = S(\omega)$ and $C''(\omega) = \mathrm{Im}\, C(\omega) = -\frac{1}{\pi}\oint d\xi\, \frac{C'(\xi)}{\xi-\omega}$, which verifies the Kramers-Kronig relations.

The generalization of the linear response of the system at thermal equilibrium to an external field in three dimensions is

$$\delta H = -\mathbf{A}\cdot\delta\mathbf{F} \tag{2.f.70}$$

$$\delta\langle \mathbf{B}_\nu\rangle(t) \equiv \beta \int_0^\infty d\tau\, \mathbf{C}(\tau)\cdot\delta\mathbf{F}(t-\tau) \tag{2.f.71}$$

$$\langle \mathbf{B}\rangle(\omega) = \beta \mathbf{C}(\omega) \cdot \mathbf{F}(\omega) \tag{2.f.72}$$

$$\mathbf{C}(\omega) \equiv \int_0^\infty d\tau e^{i\omega\tau}\mathbf{C}(\tau) \tag{2.f.73}$$

$$2\mathbf{C}'(\omega) = \mathbf{S}(\omega) = \int_{-\infty}^\infty d\tau e^{i\omega\tau}\mathbf{C}(\tau) \tag{2.f.74}$$

where $\mathbf{S}(\omega)$ is Hermitian (real, symmetric) and positive-definite, and $\mathbf{C}(\omega)$ is complex and non-Hermitian.

Hence $\qquad (\mathbf{C}')_{\mu\nu} \equiv \frac{C_{\mu\nu}+C^*_{\nu\mu}}{2} \quad (\mathbf{C}'')_{\mu\nu} \equiv \frac{C_{\mu\nu}-C^*_{\nu\mu}}{2i}$ $\qquad\qquad$ (2.f.75)

The real part is Hermitian, and the imaginary part is anti-Hermitian.

Exercise: Prove (2.f.75) using $\mathbf{C}(\tau)$ is real.

Example: Consider Brownian motion with the perturbed Hamiltonian

$$\delta H = e\delta\phi(X,t) = -Xe\delta E(t) \tag{2.f.76}$$

Here $A(\Gamma) = -X$ and $\delta F = e\delta E(t)$ to touch base with (2.f.70), and $B \equiv \mathrm{v}$ to connect with (2.f.71) and (2.f.72). Hence,

$$\langle \mathrm{v}\rangle(\omega) = \beta C^{\mathrm{v}}(\omega) e\delta E(\omega) \tag{2.f.77}$$

We will use knowledge of the fluctuations $\delta E(\omega)$ to obtain the response $\langle \mathrm{v}\rangle(\omega)$, or vice versa. Recall from Sec. 2.B that

$$C^{\mathrm{v}}(\tau) = \langle \mathrm{v}^2\rangle e^{-\nu|\tau|}, \ \nu = \frac{\gamma}{M} \ \rightarrow \ C^{\mathrm{v}}(\omega) = \frac{T/M}{\nu-i\omega} \tag{2.f.78}$$

From (2.f.77) and (2.f.78) it follows that

$$\langle \mathrm{v}\rangle(\omega) = \frac{e\delta E(\omega)}{\gamma-iM\omega} \tag{2.f.79}$$

This agrees with the Langevin equation (2.b.5) suitably ensemble averaged:

$$\left(M\frac{d}{dt}+\gamma\right)\langle V\rangle(t) = \langle \delta F\rangle(t) = e\,\delta E(t) \text{ or } (-i\omega M + \gamma)\langle V\rangle(\omega) = \langle \delta F\rangle(\omega) = e\delta E(\omega)$$

Example: Consider the current response to an external electric field turned on from zero.

We choose the gauge $\mathbf{E} = -\nabla\phi$. The perturbed Hamiltonian is

$$\delta H(\Gamma, t) = \int d^3x \, \rho^{elec}(\mathbf{x}|\Gamma) \delta\phi^{ext}(\mathbf{x}, t) \tag{2.f.80}$$

Relative to our previous notation for the linear response:

$$\mu \to x' \quad \rho \to A_\mu \quad \delta\phi^{ext} \to -\delta F \quad B_\mu \to \mathbf{j}(\mathbf{x}') \quad B_\nu \to \mathbf{j}(\mathbf{x})$$

Then using earlier results

$$\delta\langle \mathbf{j}\rangle(x, t) = -\beta \int d^3x' \int_0^\infty d\tau \, C^{j\dot{\rho}}_{\mathbf{x},\mathbf{x}'}(\tau) \delta\phi^{ext}(\mathbf{x}', t-\tau) \tag{2.f.81}$$

where

$$C^{j\dot{\rho}}_{\mathbf{x},\mathbf{x}'}(\tau) = \langle \mathbf{j}(\mathbf{x}|\Gamma, t)\dot{\rho}^{elec}(\mathbf{x}'|\Gamma, t-\tau)\rangle_0 \tag{2.f.82}$$

$$\dot{\rho}^{elec}(\mathbf{x}'|\Gamma, t-\tau) = -\frac{\partial}{\partial \mathbf{x}'} \cdot \mathbf{j}(\mathbf{x}'|\Gamma, t-\tau) \tag{2.f.83}$$

The ensemble average of Eq. (2.f.81) for the internal current removes the fluctuations from the current response. The internal current is the current carried by the charged particles within the system in response to fields but not including externally imposed currents in wires, say.

We use (2.f.83) and (2.f.82) in (2.f.81) and integrate wrt $d^3x'$ by parts so that $\frac{\partial}{\partial \mathbf{x}'} \cdot$ operates on $\delta\phi^{ext}$ with a sign change to obtain

$$\delta\langle \mathbf{j}\rangle(\mathbf{x}, t) = \int d^3x' \int_0^\infty d\tau \, \boldsymbol{\sigma}^{ext}(\tau) \cdot \mathbf{E}^{ext}(\mathbf{x}', t-\tau) \tag{2.f.84}$$

where $\boldsymbol{\sigma}^{ext}$ is the response tensor

$$\boldsymbol{\sigma}^{ext}(\mathbf{x}, \mathbf{x}', \tau) = \beta \langle \mathbf{j}(\mathbf{x}|\Gamma, t)\mathbf{j}(\mathbf{x}'|\Gamma, t-\tau)\rangle_0 \tag{2.f.85}$$

Eq.(2.f.84) is nonlocal and involves a two-point correlation function for the fluctuating current density $\mathbf{j}$.

Discussion of special cases:

-- If $\boldsymbol{\sigma}^{ext}(\mathbf{x}, \mathbf{x}', \tau) = \boldsymbol{\sigma}^{ext}(\mathbf{x} - \mathbf{x}', \tau)$; perhaps this is good only for a uniform infinite medium based on a translational invariance argument.

-- If $\boldsymbol{\sigma}^{ext}(\mathbf{x}, \mathbf{x}', \tau) = \boldsymbol{\sigma}^{ext}(|\mathbf{x} - \mathbf{x}'|, \tau)$, i.e., an isotropic conductivity, then

$$\boldsymbol{\sigma}^{ext} = \sigma_1(s, \tau)\mathbf{I} + \sigma_2(s, \tau)\hat{\mathbf{s}}\hat{\mathbf{s}}, \quad \mathbf{s} = \mathbf{x} - \mathbf{x}' \tag{2.f.86}$$

-- If $\sigma^{ext}(\mathbf{x}, \mathbf{x}', \tau)$ is rotationally symmetric about a preferred direction, e.g., with respect to an applied magnetic field, the equation $\mathbf{J}(\omega) = \boldsymbol{\sigma}(\omega) \cdot \mathbf{E}(\omega)$ can be written as

$$\mathbf{J}(\omega) = \boldsymbol{\sigma}(\omega) \cdot \mathbf{E}(\omega) = \sigma_{\parallel}(\omega)\mathbf{E}_{\parallel}(\omega) + \sigma_{\perp}(\omega)\mathbf{E}_{\perp}(\omega) + \sigma_{\wedge}(\omega)\mathbf{E}_{\wedge}(\omega) \qquad (2.f.87)$$

where $\mathbf{E}_{\parallel} = (\mathbf{E} \cdot \hat{\mathbf{b}})\hat{\mathbf{b}}$, $\mathbf{E}_{\perp} = \mathbf{E} - \mathbf{E}_{\parallel}$, and $\mathbf{E}_{\wedge} = \mathbf{E} \times \hat{\mathbf{b}}$.

Example: The most general external electromagnetic field can be expressed as

$$\mathbf{E}^{ext} = -\nabla \phi^{ext} - \frac{1}{c}\frac{\partial \mathbf{A}^{ext}}{\partial t} \qquad (2.f.88)$$

Example: The inclusion of thermal fluctuations in a system with non-trivial boundary conditions is considered in (Landau and Lifshitz, 1963). This is a very difficult calculation.

Note that the linear response calculations presented here insist that the unperturbed system is in thermal equilibrium before the external field is turned on, and the theory is linear.

Example: Consider the linear response of the current in a uniform medium. Let $\mathbf{s} \equiv \mathbf{x} - \mathbf{x}'$

$$\delta\langle \mathbf{j} \rangle(\mathbf{x}, t) = \int d^3 x' \int_0^\infty d\tau \, \boldsymbol{\sigma}^{ext}(\mathbf{s}, \tau) \cdot \mathbf{E}^{ext}(\mathbf{x} - \mathbf{s}, t - \tau) \qquad (2.f.89)$$

$$\delta\langle \mathbf{j} \rangle(\mathbf{k}, \omega) = \boldsymbol{\sigma}^{ext}(\mathbf{k}, \omega) \cdot \mathbf{E}^{ext}(\mathbf{k}, \omega) \qquad (2.f.90)$$

$$g(\mathbf{k}, \omega) \equiv \int_{-\infty}^{\infty} dt \int d^3 x \, g(\mathbf{x}, t) e^{i\omega t - i\mathbf{k} \cdot \mathbf{x}} \qquad (2.f.91)$$

$$\boldsymbol{\sigma}^{ext}(\mathbf{k}, \omega) \equiv \int_{-\infty}^{\infty} d\tau \int d^3 s \, \boldsymbol{\sigma}^{ext}(\mathbf{s}, \tau) e^{i\omega \tau - i\mathbf{k} \cdot \mathbf{s}} \qquad (2.f.92)$$

and $\delta\langle \mathbf{j} \rangle \to \langle \mathbf{j} \rangle$ subsequently in the notation. Now what about the conventional conductivity?

Definition: $\langle \mathbf{j} \rangle(\mathbf{k}, \omega) = \boldsymbol{\sigma}(\mathbf{k}, \omega) \cdot \mathbf{E}^{tot}(\mathbf{k}, \omega)$ and $\mathbf{E}^{tot} = \mathbf{E}^{ext} + \langle \mathbf{E}^{int} \rangle$ $\qquad (2.f.93)$

This includes the internal electric field with the fluctuation field averaged out. In the limit that $\mathbf{k} \to 0$, $\langle \mathbf{j} \rangle(\omega) = \boldsymbol{\sigma}(\omega) \cdot \mathbf{E}^{tot}(\omega)$ for $\lambda \gg d$ usually. Maxwell's equations tell us

$$\nabla \times \langle \mathbf{B}^{int} \rangle - \frac{1}{c}\langle \frac{\partial \mathbf{E}^{int}}{\partial t} \rangle = \frac{4\pi}{c}\langle \mathbf{j}^{int} \rangle \to i\mathbf{k} \times \mathbf{B}^{int} + \frac{i\omega}{c}\mathbf{E}^{int} = \frac{4\pi}{c}\mathbf{j}^{int} \qquad (2.f.94)$$

$$\nabla \times \mathbf{E}^{int} = -\frac{1}{c}\frac{\partial \mathbf{B}^{int}}{\partial t} \to \mathbf{k} \times \mathbf{E}^{int} = \frac{\omega}{c}\mathbf{B}^{int} \qquad (2.f.95)$$

The Maxwell equations being linear, one can use the superposition principle and decompose.

<u>Definition</u>: Define $\mathbf{I}' \equiv \mathbf{I} - \frac{k^2 c^2}{\omega^2}(\mathbf{I} - \hat{\mathbf{k}}\hat{\mathbf{k}})$

Then given (2.f.90) and (2.f.93), and recognizing that $\delta\langle\mathbf{j}\rangle(\mathbf{k}, \omega)$ and $\langle\mathbf{j}\rangle(\mathbf{k}, \omega)$ are equal to the internal current

$$\frac{i\omega}{c}\mathbf{I}' \cdot \mathbf{E}^{int} = \frac{4\pi}{c}\mathbf{j}^{int} \rightarrow \frac{i\omega}{4\pi}\mathbf{I}' \cdot (\mathbf{E}^{tot} - \mathbf{E}^{ext}) = \boldsymbol{\sigma} \cdot \mathbf{E}^{tot} = \boldsymbol{\sigma}^{ext} \cdot \mathbf{E}^{ext} \quad (2.\text{f}.96)$$

where $\boldsymbol{\sigma}^{ext}$ is the Kubo conductivity given by the fluctuations and

$$\frac{i\omega}{c}\mathbf{I}' \cdot \mathbf{E}^{ext} = (\boldsymbol{\sigma} - \frac{i\omega}{4\pi}\mathbf{I}') \cdot \mathbf{E}^{tot} \quad (2.\text{f}.97)$$

Then using $\nabla \times \langle\mathbf{B}^{tot}\rangle - \frac{1}{c}\langle\frac{\partial \mathbf{E}^{tot}}{\partial t}\rangle = \frac{4\pi}{c}\langle\mathbf{j}^{tot} = \mathbf{j}^{int} + \mathbf{j}^{ext}\rangle$ and (2.f.96)

$$i\mathbf{k} \times \mathbf{B}^{tot} + \frac{i\omega}{c}\left(\mathbf{I} - \frac{4\pi}{i\omega}\boldsymbol{\sigma}\right) \cdot \mathbf{E}^{tot} = \frac{4\pi}{c}\mathbf{j}^{ext} \quad (2.\text{f}.98)$$

We note that the dielectric function is

$$\boldsymbol{\varepsilon} \equiv \left(\mathbf{I} - \frac{4\pi}{i\omega}\boldsymbol{\sigma}\right) \quad (2.\text{f}.99)$$

and (2.f.96) can be rewritten as

$$\boldsymbol{\sigma}^{ext} \cdot \mathbf{E}^{ext} = \boldsymbol{\sigma} \cdot \mathbf{E}^{tot} = \boldsymbol{\sigma}\left(\boldsymbol{\sigma} - \frac{i\omega}{4\pi}\mathbf{I}'\right)^{-1} \cdot \left(-\frac{i\omega}{4\pi}\right)\mathbf{I}' \cdot \mathbf{E}^{ext} \quad (2.\text{f}.100)$$

Solving for $\boldsymbol{\sigma}^{ext}$ in (2.f.100) one obtains the relation between the Kubo $\boldsymbol{\sigma}^{ext}$ and the conventional conductivity $\boldsymbol{\sigma}$

$$\boldsymbol{\sigma}^{ext} = \boldsymbol{\sigma} \cdot \left(\boldsymbol{\varepsilon} - \frac{k^2 c^2}{\omega^2}(\mathbf{I} - \hat{\mathbf{k}}\hat{\mathbf{k}})\right)^{-1} \cdot \mathbf{I}' \quad (2.\text{f}.101)$$

<u>Example</u>: For an isotropic uniform medium we can separate longitudinal and transverse components of the conductivity tensor:

$$\boldsymbol{\sigma}(\mathbf{k}, \omega) = \sigma^\ell(\mathbf{k}, \omega)\hat{\mathbf{k}}\hat{\mathbf{k}} + \sigma^t(\mathbf{k}, \omega)(\mathbf{I} - \hat{\mathbf{k}}\hat{\mathbf{k}}) \quad (2.\text{f}.102)$$

From (2.f.99) $\boldsymbol{\varepsilon}^{\ell,t} = \left(\mathbf{I} - \frac{4\pi}{i\omega}\boldsymbol{\sigma}^{\ell,t}\right)$ which is used on the right side of (2.f.101) to obtain

$$\sigma^\ell_{ext}(k, \omega) = \frac{i\omega}{4\pi}\left(\frac{1}{\varepsilon^\ell(k,\omega)} - 1\right) \qquad \sigma^t_{ext}(k, \omega) = \frac{i\omega}{4\pi}\left(1 - \frac{k^2 c^2}{\omega^2}\right)\left(\frac{1 - \frac{k^2 c^2}{\omega^2}}{\varepsilon^{tr} - \frac{k^2 c^2}{\omega^2}} - 1\right) \quad (2.\text{f}.103)$$

It follows that

$$\text{Re } \sigma^\ell_{ext}(k,\omega) = -\frac{\omega}{4\pi}\text{Im}\frac{1}{\varepsilon^\ell(k,\omega)} \quad \text{and} \quad \text{Re } \sigma^t_{ext}(k,\omega) = -\frac{\omega}{4\pi}\left(1-\frac{k^2c^2}{\omega^2}\right)\text{Im}\frac{1}{\varepsilon^t - \frac{k^2c^2}{\omega^2}} \qquad (2.\text{f}.104)$$

We recall from (2.f.82) and (2.f.85) that

$$\boldsymbol{\sigma}^{ext}(\mathbf{s},\tau) \equiv \beta\langle \mathbf{j}(\mathbf{x}|\Gamma,t)\mathbf{j}(\mathbf{x}-\mathbf{s}|\Gamma,t-\tau)\rangle_0 \equiv \beta\mathbf{C}^j(\mathbf{s},\tau),\ \tau>0 \qquad (2.\text{f}.105)$$

where the "0" subscript denotes thermal equilibrium. We take the Fourier transforms of the longitudinal part of (2.f.105) to obtain

$$\boldsymbol{\sigma}^{ext}_\ell(\mathbf{k},\omega) \equiv \beta\mathbf{C}^j_\ell(\mathbf{k},\omega) \qquad (2.\text{f}.106)$$

Next take the real Hermitian part of (2.f.106)

$$\mathcal{H}e\,\boldsymbol{\sigma}^{ext}_\ell(\mathbf{k},\omega) \equiv \beta\,\mathcal{H}e\,C^j_\ell(\mathbf{k},\omega) = \frac{\beta}{2}S^j_\ell(\mathbf{k},\omega) \qquad (2.\text{f}.107)$$

From (2.f.104) and (2.f.107)

$$\langle jj\rangle^\ell_{k,\omega} = 2T\left(-\frac{\omega}{4\pi}\text{Im}\frac{1}{\varepsilon^\ell(k,\omega)}\right) \qquad (2.\text{f}.108)$$

Charge conservation asserts that $\dot{\rho} = -\nabla\cdot\mathbf{j} \to -i\omega\rho = -i\mathbf{k}\cdot\mathbf{j} = -ikj^\ell$, which we use in conjunction with (2.f.108) to obtain the following result.

The fluctuation-dissipation relation obtained (Kubo, 1966) is

$$\left(\frac{\omega}{k}\right)^2\langle\rho\rho\rangle_{k\omega} = \langle jj\rangle^\ell_{k,\omega} = 2T\left(-\frac{\omega}{4\pi}\text{Im}\frac{1}{\varepsilon^\ell(k,\omega)}\right)$$

$$\to S^{\rho el.}(k,\omega) \equiv \langle\rho\rho\rangle_{k\omega} = -\frac{T}{2\pi}\frac{k^2}{\omega}\text{Im}\frac{1}{\varepsilon^\ell(k,\omega)} \qquad (2.\text{f}.109)$$

Example: Scattering of radio frequency waves off the ionosphere – To understand the scattering experiments people calculated the conductivity and then inferred the $S^{\rho el.}(k,\omega)$.

Example: In the very long wavelength limit, we claim that for $k\to 0$, $\sigma^t = \sigma^\ell = \sigma$ from (2.f.103) and (2.f.104). Using $\varepsilon = 1 - \frac{4\pi}{i\omega}\sigma$ and $\sigma(\omega) = \frac{n_e e^2}{m}\frac{1}{\nu_{tr}-i\omega}$ from (2.f.109) one obtains

$$S^{\rho el.}(k,\omega) = \frac{\omega_p^2}{2\pi}Tk^2\frac{\nu_{tr}}{\nu_{tr}^2+\omega^2}\frac{1}{\left|\omega - \frac{\omega_p^2}{\omega+i\nu_{tr}}\right|^2} \qquad (2.\text{f}.110)$$

We note that $S^{\rho el.}(k \to 0, \omega = 0) = \frac{T}{2\pi}k^2 \frac{\nu_{tr}}{\omega_p^2}$ and $\sigma(\omega = 0) = \frac{\omega_p^2}{4\pi\nu_{tr}} \equiv \frac{1}{\eta(\omega=0)}$; hence, $S^{\rho el.}(k \to 0, \omega \approx 0) = \frac{1}{8\pi^2} T k^2 \eta(\omega = 0)$, which is the Johnson-Nyquist noise spectrum result. The resistivity at $\omega = 0$ has been introduced here as the inverse of $\sigma(\omega = 0)$. As a function of frequency $\omega$, $S^{\rho el.}$ has a peak at $\omega_p$ with a width of order $\nu_{tr}$ with the implicit assumption $\nu_{tr} \ll \omega_p$.

There are symmetry conditions pertinent to the two-time correlation function (recall the relations introduced in Eqs.(2.f.73), (2.f.74), and (2.f.75)). Associated with stationarity in time we have

$$C_{\mu\nu}(\tau) \equiv \langle A_\mu(t) A_\nu(t-\tau) \rangle = \langle A_\mu(t+\tau) A_\nu(t) \rangle = \langle A_\nu(t) A_\mu(t+\tau) \rangle = C_{\nu\mu}(-\tau) \quad (2.f.111)$$

and

$$S_{\mu\nu}(\omega) \equiv \int_{-\infty}^{\infty} d\tau e^{i\omega\tau} C_{\mu\nu}(\tau)$$

$$= \int_{-\infty}^{\infty} d\tau' e^{-i\omega\tau} C_{\mu\nu}(-\tau'), \quad \tau = -\tau'$$

$$= S_{\nu\mu}(-\omega) = S_{\nu\mu}^*(\omega) \quad (2.f.112)$$

due to the reality of $C_{\mu\nu}(\tau)$. From (2.f.112) we deduce that $S_{\mu\nu}(\omega)$ is Hermitian.

Now we assume time reversibility, and we will prove that $C_{\mu\nu}$ is symmetric. Consider the phase space $\Gamma(r_i, v_i)$ which becomes $\tilde{\Gamma}(r_i, -v_i)$ under time reversal. Assume a model Hamiltonian with unit mass, $m=1$, and no magnetic effects:

$$H = \sum_i v_i^2 + \phi(\mathbf{r}_i - \mathbf{r}) + e_i \phi_\xi(\mathbf{r}_i) \quad (2.f.113)$$

Assume further: $\rho_0(H): \rho_0(\Gamma) = \rho_0(\tilde{\Gamma})$. Given the assumptions, we have

$$A_\mu(\Gamma) = \pm A_\mu(\tilde{\Gamma}) \qquad A_\mu(\Gamma, \tau) = \pm A_\mu(\tilde{\Gamma}, -\tau) \quad (2.f.114)$$

For example, $\mathbf{j}(\mathbf{x}) = \sum_i \mathbf{v}_i \delta(\mathbf{x} - \mathbf{r}_i) \to -\sum_i \mathbf{v}_i \delta(\mathbf{x} - \mathbf{r}_i)$ under time reversal. Under time reversal the correlation function becomes

$$C_{\mu\nu}(\tau) \equiv \int d\Gamma \rho_0(\Gamma) A_\mu(\Gamma, t) A_\nu(\Gamma, t-\tau) = \int d\tilde{\Gamma} \rho_0(\tilde{\Gamma}) A_\mu(\tilde{\Gamma}, -t) A_\nu(\tilde{\Gamma}, -t+\tau) \quad (2.f.115)$$

We note that with $\mathbf{r}_i(r_i, \mathbf{v}_i, \tau) = \mathbf{r}_i + \mathbf{v}_i \tau \to \mathbf{r}_i(r_i, -\mathbf{v}_i, -\tau) = \mathbf{r}_i + (-\mathbf{v}_i)(-\tau)$ under time reversal.

From (2.f.111) and (2.f.115) we conclude the proof of the symmetry of $C_{\mu\nu}(\tau)$

$$C_{\mu\nu}(\tau) = C_{\mu\nu}(-\tau) \quad \text{and} \quad C_{\mu\nu}(\tau) = C_{\nu\mu}(\tau) \tag{2.f.116}$$

Thus, $C_{\mu\nu}(\tau)$ is symmetric as is $S_{\mu\nu}(\omega)$; and both are real. Hence, both $C_{\mu\nu}(\tau)$ and $S_{\mu\nu}(\omega)$ are Hermitian.

We now return to the electrical conductivity. Recall (2.f.84)

$$\delta\langle \mathbf{j}\rangle(\mathbf{x},t) = \int d^3x' \int_0^\infty d\tau \boldsymbol{\sigma}^{ext}(\mathbf{x},\mathbf{x}';\tau) \cdot \mathbf{E}^{ext}(\mathbf{x}',t-\tau)$$

Using the symmetry relations for $\boldsymbol{\sigma}^{ext}$ we have

$$\sigma_{ij}^{ext}(\mathbf{x},\mathbf{x}';\tau) = \sigma_{ji}^{ext}(\mathbf{x}',\mathbf{x};\tau) \tag{2.f.117}$$

In (2.f.117) the $(\mathbf{x},\mathbf{x}')$ dependence is $(\mathbf{x}-\mathbf{x}')$. For no magnetic field $B_0 = 0$, (2.f.117) leads to

$$\sigma_{ij}^{ext}(\mathbf{k};\omega) = \sigma_{ji}^{ext}(-\mathbf{k};\omega) \tag{2.f.118}$$

With a magnetic field, Onsager in a 1932 publication, making no assumption about isotropy, showed (Onsager and Fuoss, 1932)

$$\sigma_{ij}(\mathbf{k},\omega;\mathbf{B}_0) = \sigma_{ij}(-\mathbf{k},\omega;-\mathbf{B}_0) \tag{2.f.119}$$

if $\mathbf{B}_0 \to -\mathbf{B}_0$ under time reversal. (2.f.119) was discovered experimentally around 1900.

For future use, keep in mind that $C_{\mu\nu}(\tau)$ and $S_{\mu\nu}(\omega)$ are real and symmetric, but $C_{\mu\nu}(\omega)$ is not real.

### 2.f.vi Relation of entropy production to electrical conductivity

In this section we will derive a relation between the entropy production and the electrical conductivity tensor. First consider a general form of the system Hamiltonian:

$$H(\mathbf{\Gamma},t) = H_0(\mathbf{\Gamma}) + \delta H(\mathbf{\Gamma},t) \tag{2.f.120}$$

from which we deduce

$$\dot{H} = \frac{\partial}{\partial t}\delta H(\mathbf{\Gamma},t) = -\sum_\mu A_\mu(\mathbf{\Gamma})\frac{d}{dt}F_\mu(t) \tag{2.f.121}$$

using the notation in (2.f.54) to separate phase factor components. We form the ensemble average and perform the time integral of (2.f.121)

$$\Delta \langle H \rangle = \int_{-\infty}^{\infty} dt \, \langle \dot{H}(t) \rangle = -\sum_\mu \int_{-\infty}^{\infty} dt \, \langle A_\mu(\Gamma) \rangle \frac{d}{dt} F_\mu(t) = -\sum_\mu \int_{-\infty}^{\infty} dt \, \langle A_\mu \rangle (t) \frac{d}{dt} F_\mu(t)$$

$$= \sum_\mu \int_{-\infty}^{\infty} dt \, F_\mu(t) \frac{d}{dt} \langle A_\mu \rangle (t) = \sum_\mu \int_{-\infty}^{\infty} dt \, F_\mu(t) \langle \dot{A}_\mu \rangle (t) \quad (2.f.122)$$

integrating by parts with vanishing contributions at the endpoints $t = \pm \infty$ and using earlier results. (2.f.122) can be expressed in alternative form using Parseval's theorem:

$$\Delta \langle H \rangle = \int_{-\infty}^{\infty} \frac{d\omega}{2\pi} \sum_\mu F_\mu^*(\omega) \langle \dot{A}_\mu \rangle (\omega) = \beta \int_{-\infty}^{\infty} \frac{d\omega}{2\pi} \sum_\mu F_\mu^*(\omega) \sum_\nu C_{\mu\nu}^{\dot{A}\dot{A}} F_\nu(\omega)$$

$$= \beta \int_{-\infty}^{\infty} \frac{d\omega}{2\pi} \mathbf{F}^*(\omega) \cdot \mathbf{C}^{\dot{A}}(\omega) \cdot \mathbf{F}(\omega)$$

$$= \beta \int_{-\infty}^{\infty} \frac{d\omega}{2\pi} \mathbf{F}(\omega) \cdot \mathbf{C}^{\dot{A}*}(\omega) \cdot \mathbf{F}^*(\omega)$$

$$= \beta \int_{-\infty}^{\infty} \frac{d\omega}{2\pi} \mathbf{F}^*(\omega) \cdot \tilde{\mathbf{C}}^{\dot{A}*}(\omega) \cdot \mathbf{F}(\omega)$$

$$= \beta \int_{-\infty}^{\infty} \frac{d\omega}{2\pi} \mathbf{F}^*(\omega) \cdot \mathbf{C}^{\dot{A}T}(\omega) \cdot \mathbf{F}(\omega) \quad (2.f.123)$$

We comment that Parseval's theorem involves integrals over the infinite domains in both time and frequency. Hence, from the different forms of the right side of (2.f.123) it follows

$$\Delta \langle H \rangle = \beta \int_{-\infty}^{\infty} \frac{d\omega}{2\pi} \mathbf{F}^*(\omega) \cdot \mathbf{C}_{\dot{A}}^{\mathcal{H}e}(\omega) \cdot \mathbf{F}(\omega)$$

$$= \frac{\beta}{2} \int_{-\infty}^{\infty} \frac{d\omega}{2\pi} \mathbf{F}^*(\omega) \cdot \mathbf{S}^{\dot{A}}(\omega) \cdot \mathbf{F}(\omega)$$

$$= \frac{\beta}{2} \int_{-\infty}^{\infty} \frac{d\omega}{2\pi} \mathbf{F}^*(\omega) \cdot \omega^2 \mathbf{S}^A(\omega) \cdot \mathbf{F}(\omega) \quad (2.f.124)$$

where $\mathbf{C}_{\dot{A}}^{\mathcal{H}e}$ denotes the Hermitian part of the tensor and the Wiener-Khinchin theorem (2.a.49) has been used.

<u>Definition</u>: The entropy is given by $\Delta S \equiv \beta \Delta \langle H \rangle$ from (1.b.29).

For the specific example of the electrical conductivity (2.f.124) becomes

$$\Delta \langle H \rangle = \int_{-\infty}^{\infty} \frac{d\omega}{2\pi} \int d^3 x \, \langle \dot{\rho}_{elec} \rangle (\mathbf{x}, \omega) \phi^{ex*}(\mathbf{x}, \omega) =$$

$$= -\int_{-\infty}^{\infty} \frac{d\omega}{2\pi} \int d^3 x \, \nabla \cdot \langle \mathbf{j} \rangle (\mathbf{x}, \omega) \phi^{ex*}(\mathbf{x}, \omega)$$

$$= \int_{-\infty}^{\infty} \frac{d\omega}{2\pi} \int d^3 x \, \langle \mathbf{j} \rangle (\mathbf{x}, \omega) \cdot \mathbf{E}^{ext*}(\mathbf{x}, \omega) \quad (2.f.125)$$

using charge continuity and integrating by parts.  We use

$$\langle \mathbf{j}\rangle(\mathbf{x},\omega) \cdot \mathbf{E}^{ext}(\mathbf{x},\omega) = \int d^3x' \, \boldsymbol{\sigma} \cdot \langle \mathbf{E}^{tot}\rangle(\mathbf{x}',\omega) \cdot \langle \mathbf{E}^{tot} - \mathbf{E}^{int}\rangle^*(\mathbf{x},\omega)$$

to express (2.f.125) as

$$\Delta\langle H\rangle = \int_{-\infty}^{\infty} \frac{d\omega}{2\pi} \int d^3x \, \langle \mathbf{j}\rangle(\mathbf{x},\omega) \cdot \mathbf{E}^{ext*}(\mathbf{x},\omega) =$$

$$\int_{-\infty}^{\infty} \frac{d\omega}{2\pi} \int d^3x \int d^3x' \langle \mathbf{E}^{tot}\rangle(\mathbf{x},\omega) \cdot \boldsymbol{\sigma}(\mathbf{x},\mathbf{x}',\omega) \cdot \langle \mathbf{E}^{tot}\rangle^*(\mathbf{x}',\omega)$$

$$- \int_{-\infty}^{\infty} \frac{d\omega}{2\pi} \int d^3x \, \langle \mathbf{j}\rangle(\mathbf{x},\omega) \cdot \langle \mathbf{E}^{int}\rangle^*(\mathbf{x},\omega) \qquad (2.f.126)$$

However, conservation of energy derived for Maxwell's equations using Parseval's theorem for the $\int dvol \, \mathbf{j} \cdot \mathbf{E}$ term yields:

$$\int_{-\infty}^{\infty} \frac{d\omega}{2\pi} \int d^3x \, \langle \mathbf{j}\rangle(\mathbf{x},\omega) \cdot \langle \mathbf{E}^{int}\rangle^*(\mathbf{x},\omega) = \int_{-\infty}^{\infty} dt \int d^3x \, \frac{\partial}{\partial t}\left(\frac{E^2 + B^2}{8\pi}\right) - \nabla \cdot \frac{c}{4\pi}(\mathbf{E} \times \mathbf{B}) = 0$$

with no sources and suitable boundary conditions, so that (2.f.126) becomes

$$\Delta\langle H\rangle = \int_{-\infty}^{\infty} \frac{d\omega}{2\pi} \int d^3x \int d^3x' \langle \mathbf{E}^{tot}\rangle(\mathbf{x},\omega) \cdot \boldsymbol{\sigma}(\mathbf{x},\mathbf{x}',\omega) \cdot \langle \mathbf{E}^{tot}\rangle^*(\mathbf{x}',\omega) =$$

$$= \int_{-\infty}^{\infty} \frac{d\omega}{2\pi} \int d^3x \, \langle \mathbf{E}^{tot}\rangle(\mathbf{x},\omega) \cdot \boldsymbol{\sigma}(\mathbf{x},\omega) \cdot \langle \mathbf{E}^{tot}\rangle^*(\mathbf{x},\omega) \qquad (2.f.127)$$

If the media is isotropic, then (2.f.127) simplifies:

$$\Delta\langle H\rangle = \int_{-\infty}^{\infty} \frac{d\omega}{2\pi} \int d^3x \, \sigma(\mathbf{x},\omega)|\mathbf{E}^{tot}(\mathbf{x},\omega)|^2 = \int_{-\infty}^{\infty} \frac{d\omega}{2\pi} \int d^3x \, \text{Re}\, \sigma(\mathbf{x},\omega)|\langle \mathbf{E}^{tot}\rangle(\mathbf{x},\omega)|^2$$

(2.f.128)

because there is no contribution to the integral from Im $\sigma(\mathbf{x},\omega)$.  We note that

$$\text{Re}\, \sigma(\mathbf{x},\omega)|\langle \mathbf{E}^{tot}\rangle(\mathbf{x},\omega)|^2 \to \eta(\mathbf{x},\omega)|\langle \mathbf{j}\rangle(\mathbf{x},\omega)|^2 \qquad (2.f.129)$$

which is just the $i^2 R$ resistive heating source for the entropy production ( $\Delta S \equiv \beta\Delta\langle H\rangle$ ).

**2.f.vii Transport relations and coefficients**

What if there is a temperature gradient $\nabla T$? Temperature gradients tend to be accompanied by a heat flow (Ref. Newton, Maxwell, etc.): $\mathbf{Q} = -\mathbf{K} \cdot \nabla T$, where $\mathbf{K}$ is the heat-flow tensor, i.e., the thermal conductivity (Mori, 1965; Kubo, 1966). In the presence of a gradient in flow velocity there can arise a viscous stress: $\mathbf{\Pi} = -\mu \nabla \mathbf{u}$, where $\mu$ is the viscosity coefficient in this equation. In the presence of a density gradient there can arise a particle-flux density as a result of diffusion: $\mathbf{\Gamma} = -\mathbf{D} \cdot \nabla n$, or more generally a diffusive flux driven by the gradient in the chemical potential: $\mathbf{\Gamma} = -\mathbf{D} \cdot \nabla \mu$ where $\mu$ is the chemical potential. The current density is another example of diffusive transport: $\mathbf{j} = -\boldsymbol{\sigma} \cdot \nabla \phi = \boldsymbol{\sigma} \cdot \mathbf{E}$.

Definition: Thermodynamic forces $\mathbf{E}$, $\nabla \mathbf{u}$, $\nabla n$, $\nabla T$, $\nabla \mu$ all give rise to thermodynamic fluxes.

Example: Onsager showed $\mathbf{Q} = -\mathbf{K} \cdot \nabla T \Leftrightarrow \mathbf{j} = \boldsymbol{\sigma} \cdot \mathbf{E}$, i.e., in appropriate units these are the same relation.

In this section we will present a few derivations of transport equations. An example of a model system that leads to kinetic transport equations is the derivation due to Chapman and Enskog (1911) (Reif, 1965; Liboff, 1969). We will derive kinetic transport equations from the linear Boltzmann equation. We will also start from the Liouville equation and identify a small parameter, e.g., the magnitude of the gradient relative to the inverse of a characteristic length in the system, to facilitate the derivation of kinetic transport equations; Mori in (Mori, 1965) used this approach.

Consider the linear Boltzmann equation for the probability density:

$$\frac{\partial}{\partial t}\rho(\mathbf{r},\mathbf{v},t) + \mathbf{v} \cdot \frac{\partial}{\partial \mathbf{r}}\rho(\mathbf{r},\mathbf{v},t) = -\tilde{\nu}\rho(\mathbf{r},\mathbf{v},t) \qquad (2.\text{f}.130)$$

where $\mathbf{v} = v\mathbf{\Omega}$ where $\mathbf{\Omega}$ is a unit vector defining the direction of $\mathbf{v}$. We define the $\tilde{\nu}$ operator:

$$\tilde{\nu} f(\mathbf{\Omega}) \equiv \nu f(\mathbf{\Omega}) - \int d^3\Omega' f(\mathbf{\Omega}') \nu(\Theta = (\mathbf{\Omega}; \mathbf{\Omega}')) \qquad (2.\text{f}.131)$$

For elastic scattering of the particles the speed v remains constant. We drop $\mathbf{v}$ as an independent variable for constant speed and use $\mathbf{\Omega}$ instead. We Fourier transform

$$\int d^3\mathbf{r}\, e^{-i\mathbf{k}\cdot\mathbf{r}} \rho(\mathbf{r},\mathbf{v};t) = \rho(\mathbf{k},\mathbf{v};t)$$

to obtain

$$\frac{\partial}{\partial t}\rho(\mathbf{k},\mathbf{v},t) + i\mathbf{k}\cdot\mathbf{v}\, \rho(\mathbf{k},\mathbf{v},t) = -\tilde{\nu}\rho(\mathbf{k},\mathbf{v},t) \qquad (2.\text{f}.132)$$

Representing the time dependence as $\rho(\mathbf{k},\mathbf{v},t) = e^{-i\omega t}\rho_\omega(\mathbf{k},\mathbf{v})$ we get

$$i(-\omega + \mathbf{k}\cdot\mathbf{v})\, \rho_{\omega,k}(\mathbf{\Omega}) = -\tilde{\nu}\rho_{\omega,k}(\mathbf{\Omega}) \qquad (2.\text{f}.133)$$

with $\mathbf{v} = v\mathbf{\Omega}$. With use of (2.f.131), (2.f.133) becomes an eigenfunction-eigenvalue problem in time where $\omega$ is complex:

$$i(-\omega + \mathbf{k} \cdot \mathbf{v})\, \rho_{\omega,k}(\mathbf{\Omega}) = -\nu \rho_{\omega,k}(\mathbf{\Omega}) + \int d^3\Omega'\, \rho_{\omega,k}(\mathbf{\Omega}')\nu(\mathbf{\Theta}) \qquad (2.f.134)$$

Next we simplify by assuming that the scattering is isotropic to obtain $\nu(\mathbf{\Theta}) = \nu/4\pi$ so that

$$\int d^3\Omega'\, \rho_{\omega,k}(\mathbf{\Omega}')\nu(\mathbf{\Theta}) = \int \frac{d\Omega'}{4\pi} \nu \rho_{\omega,k}(\mathbf{\Omega}') = \nu \bar{\rho} \qquad (2.f.135)$$

Hence, (2.f.134) becomes

$$i(-\omega + \mathbf{k} \cdot \mathbf{v})\, \rho_{\omega,k}(\mathbf{\Omega}) = \nu(\bar{\rho} - \rho_{\omega,k}(\mathbf{\Omega})) \qquad (2.f.136)$$

With $\mathbf{k} \cdot \mathbf{v} = kv \cos\theta$, (2.f.136) is

$$-i(\omega + i\nu - kv\cos\theta)\, \rho_{\omega,k}(\theta,\phi) = \nu\bar{\rho} \qquad (2.f.137a)$$

or

$$\rho_{\omega,k}(\theta,\phi) = \frac{i\nu\bar{\rho}}{\omega + i\nu - kv\cos\theta} \qquad (2.f.137b)$$

Let $\mu \equiv \cos\theta$ and integrate (2.f.137b) over $\theta$ and $\phi$ to obtain the average $\bar{\rho}$:

$$\bar{\rho} = i\nu\bar{\rho} \int_{-1}^{1} d\mu\, \frac{1}{\omega + i\nu - kv\mu} \frac{2\pi}{4\pi} = \frac{i\nu\bar{\rho}}{2} \int_{-1}^{1} d\mu\, \frac{1}{\omega + i\nu - kv\mu} \qquad (2.f.138)$$

We obtain a dispersion relation from (2.f.138) by dividing out $\bar{\rho}$ from both sides:

$$1 = \frac{i\nu}{2} \int_{-1}^{1} d\mu\, \frac{1}{\omega + i\nu - kv\mu} \qquad (2.f.139)$$

The denominator vanishes at $\omega = -i\nu + kv\mu$. In the complex $\omega$ plane there are branch points at $\omega = -i\nu \pm kv$ and a branch cut between. We do the integral in (2.f.139) carefully to obtain the result:

$$\ln\left(\frac{kv - (\omega + i\nu)}{-kv - (\omega + i\nu)}\right) = 2i\left(\frac{kv}{\nu}\right) \equiv 2ik\ell$$

which yields

$$\omega = -i\nu(1 - k\ell \cot k\ell), \quad k\ell < \pi/2 \qquad (2.f.140)$$

where $\ell \equiv v/\nu$ is the mean free path. (2.f.140) is valid iff $k\ell < \pi/2$ and there is no solution of (2.f.139) for $k\ell > \pi/2$. Thus, there is the one solution (2.f.140) for the dispersion relation given by (2.f.139). We note that $k\ell \cot k\ell$ equals 1 for $k\ell = 0$, decreases as $1 - (k\ell)^2/3$ for small $k\ell$, and equals 0 at $k\ell = \pi/2$. Using (2.f.140) in (2.f.137b), we obtain

$$\rho_{\omega,k}(\theta,\phi) = \frac{const}{\cot k\ell + i\cos\theta} \tag{2.f.141}$$

where the constant in the numerator is a normalization constant. (2.f.141) reflects that there is only one eigenvalue given in (2.f.140) as all the other possible eigenvalues are on the branch cut.

We return to Eq.(2.f.137b) to consider the eigenvalues on the branch cut, i.e., the contributions to the probability density from the frequencies on the branch cut. (2.f.137b) has the form $\rho(y) = a/y$ if $y \neq 0$. If $y = 0$ is possible, then we must include $\lambda\delta(y)$, i.e.,

$$\rho(y) = \mathcal{P}\left(\frac{a}{y}\right) + \lambda\delta(y) \tag{2.f.142}$$

where the first term is the principal value and the second involves a $\delta$-function. We are reminded of Van Kampen's analysis of the linearized Vlasov equation in which he obtained singular eigenfunctions.

Equation (2.f.137b) becomes

$$\rho_{\omega,k}(\theta,\phi) = \mathcal{P}\left(\frac{iv\bar{\rho}}{\omega+iv-kv\cos\theta}\right) + \lambda(\phi)\delta(\omega+iv-kv\cos\theta) \equiv \rho_{\mu\omega}(\Omega) \tag{2.f.143}$$

The relation (2.f.138) becomes

$$\bar{\rho} = \int_0^{2\pi}\frac{d\phi}{2\pi}\lambda(\phi)\int_{-1}^1\frac{d\mu}{2}\delta(\omega+iv-kv\mu) + \frac{iv\bar{\rho}}{2}\oint\frac{d\mu}{\omega+iv-kv\mu} =$$

$$\frac{\bar{\lambda}}{2kv} + \frac{iv\bar{\rho}}{2}\frac{1}{kv}\ln\frac{-1-\mu_\omega}{1-\mu_\omega}, \quad \mu_\omega \equiv \frac{\omega+iv}{kv}, \quad -1 < \mu_\omega < 1 \tag{2.f.144}$$

Note that if we substitute $\mu_\omega \equiv \frac{\omega+iv}{kv} = \cot(k\ell)$ consistent with the dispersion relation in Eq.(2.f.140), then the solution to Eq.(2.f.144) dictates that $\bar{\lambda} = 0$; and Eq.(2.f.140) is recovered.

Let $\rho_0(\mathbf{k},\mathbf{v}) \equiv \frac{1}{\cot k\ell + i\mu}$ for the $\omega = 0$ mode, where $\mu = \cos\theta$ and $\ell \equiv v/\nu$. We transform (2.f.143) back from the $\omega$ domain to the time domain separating out the $\omega = 0$ mode contribution:

$$\rho(\mathbf{k},\mathbf{v};t) = e^{-\nu(1-k\ell\cot k\ell)t}\frac{a_0(\mathbf{k},\mathbf{v})}{\cot k\ell+i\mu} + \int_{-1}^1 d\mu_\omega e^{-\nu t}e^{-ikv\mu_\omega t}\rho_{\mu\omega}^{\mathbf{k},\mathbf{v}}(\Omega)a_{\mu\omega}(\mathbf{k},\mathbf{v}) \tag{2.f.145}$$

where $a_0(\mathbf{k},\mathrm{v})$ and $a_{\mu\omega}(\mathbf{k},\mathbf{v})$ are Fourier coefficients associated with a normalization constant and initial conditions. (2.f.145) is the solution of the linearized Boltzmann equation for the probability distribution in phase space. We note that the first term in (2.f.145) dominates at long time because $1 - k\ell \cot k\ell < 1$, so it does not drop off as rapidly as the second term.

For long times $tv/\ell \gg 1$, the dominant solution for the probability distribution is then

$$\rho(\mathbf{k}, \mathbf{v}; t) = e^{-v(1-k\ell \cot k\ell)t} \frac{a_0(\mathbf{k}, \mathbf{v})}{\cot k\ell + i\mu} \qquad (2.f.146)$$

Next we Fourier transform $\rho(\mathbf{k}, \mathbf{v}; t)$ back to $\rho(\mathbf{r}, \mathbf{v}; t)$:

$$\rho(\mathbf{r}, \mathbf{v}; t) = \int_{k\ell < \frac{\pi}{2}} \frac{d^3k}{(2\pi)^3} e^{i\mathbf{k}\cdot\mathbf{r}} e^{-v(1-k\ell \cot k\ell)t} \frac{a_0(\mathbf{k}, \mathbf{v})}{\cot k\ell + i\mu} \approx$$

$$\int_{k\ell < \pi/2} \frac{d^3k}{(2\pi)^3} e^{i\mathbf{k}\cdot\mathbf{r}} e^{-\frac{1}{3}\ell v k^2 t} \frac{a_0(\mathbf{k}, \mathbf{v})}{\frac{1}{k\ell} + i\mu} \qquad (2.f.147)$$

where we have used $k\ell \cot k\ell \approx 1 - (k\ell)^2/3$ for $k\ell \ll 1$, $a_0(\mathbf{k}, \mathbf{v}) \to a_0(\mathbf{k}, v)$ cannot depend on the direction of **v**, v is constant in time, and $tv/\ell \gg 1$. We next make use of

$$\frac{1}{\frac{1}{k\ell} + i\mu} = \frac{k\ell}{1 + i\mu k\ell} \approx k\ell\left(1 - i\frac{\mathbf{k}\cdot\mathbf{v}}{v}\right) \qquad (2.f.148)$$

in (2.f.147) to obtain

$$\rho(\mathbf{r}, \mathbf{v}; t) \approx \int_{k\ell < \pi/2} \frac{d^3k}{(2\pi)^3} e^{i\mathbf{k}\cdot\mathbf{r}} e^{-\frac{1}{3}\ell v k^2 t} k\ell\left(1 - i\frac{\mathbf{k}\cdot\mathbf{v}}{v}\right) a_0(\mathbf{k}, v) \qquad (2.f.149)$$

Now we calculate the average flux density:

$$\widetilde{\Gamma}(\mathbf{r}; t) = \rho(\mathbf{r}; t)\langle\mathbf{v}\rangle(r, t) = \int d^3v\, \rho(\mathbf{r}, \mathbf{v}; t)\mathbf{v} \qquad (2.f.150)$$

where we note that $\mathbf{v} = v\mathbf{\Omega}$ and v is constant. If we take the time derivative of Eq.(1.f.149), we easily obtain

$$\frac{\partial \rho}{\partial t} = \frac{\partial}{\partial t}\left[\int \frac{d^3k}{(2\pi)^3} e^{i\mathbf{k}\cdot\mathbf{r}} e^{-Dk^2 t}(\ldots)\right] = -D\int \frac{d^3k}{(2\pi)^3} e^{i\mathbf{k}\cdot\mathbf{r}} e^{-Dk^2 t}(\ldots) = D\nabla^2 \rho \qquad (2.f.151)$$

in which we identify $D = \ell v/3$ from inside the exponential in (2.f.149), and $\rho(\mathbf{r}; t) = \int d^3\mathbf{v}\, \rho(\mathbf{r}, \mathbf{v}; t)$. Given that $\frac{\partial \rho}{\partial t} = -\nabla \cdot \widetilde{\Gamma} = D\nabla^2 \rho$, it follows that

$$\widetilde{\Gamma} = -D\nabla \rho(\mathbf{r}; t) \qquad (2.f.152)$$

No knowledge of (…) in Eq.(2.f.151) is required to obtain Eq.(2.f.152).

Example: Anisotropic scattering – In the limit of small $k\ell$ the eigenfunction can be shown to have the same form as the isotropic scattering limit. The other terms that are not retained are different in the anisotropic case, but it does not matter. In the anisotropic scattering case

$$\nu \to \nu_{tr} \equiv \int \frac{d\Omega}{4\pi} \, \nu(\Theta)(1 - \cos\theta) = n_0 \sigma_{tr} v \qquad (2.f.153)$$

We can generalize the analysis leading to (2.f.150) and (2.f.151) to all transport phenomena. Consider any density field: $\mathcal{A}(\mathbf{x}|\Gamma)$ where $\mathcal{A}$ is the momentum, energy, current, etc., scalar or vector field and $\Gamma$ is the total system "phase."

Example: Let $K \equiv \sum_i K_i$, $K_i \equiv \frac{1}{2} m_i v_i^2 = p_i^2/2m_i$. Then the kinetic energy density is

$$K(\mathbf{x}|\Gamma) \equiv \sum_i K_i \delta(\mathbf{x} - \mathbf{x}_i) \qquad (2.f.154)$$

and the number density is

$$n(\mathbf{x}|\Gamma) \equiv \sum_i \delta(\mathbf{x} - \mathbf{x}_i) \qquad (2.f.155)$$

We assert that the general conservation law for any density field : $\mathcal{A}(\mathbf{x}|\Gamma)$ is

$$\dot{\mathcal{A}}(\mathbf{x}|\Gamma) = -\frac{\partial}{\partial x} \cdot \boldsymbol{\Gamma}^{\mathcal{A}}(\mathbf{x}|\Gamma) + \Sigma^{\mathcal{A}}(\mathbf{x}|\Gamma) \qquad (2.f.156)$$

where the first term on the right side is minus the divergence of the flux associated with the density field $\mathcal{A}$ and the second term is a source/sink term (if finite). Remember from (2.c.16) that

$$\frac{\partial}{\partial t}\rho(\Gamma;t) + \{\rho, H\} = 0$$

involving the Poisson bracket of $\rho$ with $H$ the Hamiltonian, as a consequence of Liouville's theorem in the absence of explicit sources and sinks.

Lemma: Any function $A(\Gamma)$ satisfies

$$\langle \dot{A} \rangle \equiv \int d\Gamma \, \rho(\Gamma, t)\{A, H\} = -\int d\Gamma \, A(\Gamma)\{\rho, H\} = \int d\Gamma \, A(\Gamma)\frac{\partial}{\partial t}\rho(\Gamma, t)$$

$$= \frac{d}{dt}\int d\Gamma \, A(\Gamma)\rho(\Gamma, t) = \frac{d}{dt}\langle A \rangle \qquad (2.f.157)$$

using (2.c.16) and integrating by parts.

Hence, from (2.f.156) and (2.f.157), the volume integral of the ensemble average of (2.f.156) is

$$\frac{d}{dt}\int d^3x \, \langle \mathcal{A} \rangle (\mathbf{x}, t) = -\oint d\boldsymbol{\sigma} \cdot \langle \boldsymbol{\Gamma}^{\mathcal{A}} \rangle (\mathbf{x}, t) + \int d^3x \, \langle \Sigma^{\mathcal{A}} \rangle (\mathbf{x}, t) \tag{2.f.158}$$

Example: Returning to our example of the kinetic energy density (2.f.154), (2.f.156) becomes

$$\dot{K}(x|\Gamma) = -\nabla \cdot \sum_i \mathbf{v}_i \frac{1}{2} m_i v_i^2 \, \delta(\mathbf{x} - \mathbf{x}_i) + \sum_i \mathbf{v}_i \cdot \mathbf{f}^i \delta(\mathbf{x} - \mathbf{x}_i) \tag{2.f.159}$$

where the second term on the right side of (2.f.158) is the power due to the work done on the particle by the force $\mathbf{f}^i$ on it. In the absence of external forces, the force on particle i is just the sum over j of the force of particle j on particle i:

$$\mathbf{f}^i = \sum_{j \neq i} \mathbf{f}^i_j = -\sum_{j \neq i} \frac{\partial}{\partial r_j} \phi(r_{ij}) \tag{2.f.160}$$

where $\phi(r_{ij})$ is a potential energy. The first term on the right side of (2.f.158) is just minus the divergence of the kinetic energy flux.

In the potential introduced in Eq.(2.f.160) where does the potential energy reside, at what location? If the potential energy resides in the particles, then an arbitrary designation is necessary. For example, is the potential energy shared equally by the two particles that define $r_{ij}$ or perhaps can the potential energy be assigned to the midpoint of the two particle locations along $r_{ij}$? The latter implies:

$$\mathbf{r}_{i \text{ or } j} \equiv \mathbf{R}_{ij} \pm \frac{1}{2} \mathbf{r}_{ij} \tag{2.f.161}$$

which defines $\mathbf{R}_{ij}$ the midpoint between particles i and j. We take the midpoint and write

$$\Phi(\mathbf{x}|\Gamma) \equiv \sum_{i<j} \phi(\mathbf{r}_{ij}) \delta(\mathbf{x} - \mathbf{R}_{ij}) \tag{2.f.162}$$

and

$$\dot{\Phi}(\mathbf{x}|\Gamma) \equiv -\nabla \cdot \sum_{i<j} \mathbf{V}_{ij} \, \phi(\mathbf{r}_{ij}) \delta(\mathbf{x} - \mathbf{R}_{ij}) + \sum_{i<j} \dot{\phi}(\mathbf{r}_{ij}) \delta(\mathbf{x} - \mathbf{R}_{ij}) \tag{2.f.163}$$

where $\mathbf{V}_{ij} = \dot{\mathbf{r}}_{ij}$. In the absence of any explicit source or sink of particle energy, we expect relations of the form

$$\mathcal{E}(\mathbf{x}|\Gamma) \equiv K + \Phi \qquad \dot{\mathcal{E}}(\mathbf{x}|\Gamma) \equiv -\nabla \cdot \{\overline{\mathbf{V}}[K + \Phi]\} \tag{2.f.164}$$

We flesh out (2.f.164) in the following. In the analysis we symmetrize

$$\sum_{ij} \mathbf{V}_{ij} \, \mathbf{f}^i_j \delta(\mathbf{x} - \mathbf{r}_i) \rightarrow \frac{1}{2} \sum_{ij} [\mathbf{V}_i \delta(\mathbf{x} - \mathbf{r}_i) + \mathbf{V}_j \delta(\mathbf{x} - \mathbf{r}_j)] \mathbf{f}^i_j \tag{2.f.165}$$

and

$$\delta(\mathbf{x} - \mathbf{r}_i) \approx \delta(\mathbf{x} - \mathbf{R}_{ij}) - \frac{1}{2}\mathbf{r}_{ij}\nabla \cdot \delta(\mathbf{x} - \mathbf{R}_{ij}) + \cdots \quad (2.f.166a)$$

$$\delta(\mathbf{x} - \mathbf{r}_j) \approx \delta(\mathbf{x} - \mathbf{R}_{ij}) + \frac{1}{2}\mathbf{r}_{ij}\nabla \cdot \delta(\mathbf{x} - \mathbf{R}_{ij}) + \cdots \quad (2.f.166b)$$

(Irving and Kirkwood, 1950).

We use Eqs.(2.f.159)-(2.f.166) to obtain, after cancellations,

$$\dot{\mathcal{E}} = -\nabla \cdot \left[\boldsymbol{\Gamma}_K + \boldsymbol{\Gamma}_\Phi + \sum_{i<j} \mathbf{r}_{ij} \mathbf{V}_{ij} \cdot \mathbf{f}_j^i \delta(\mathbf{x} - \mathbf{R}_{ij})\right] \quad (2.f.167)$$

where $\boldsymbol{\Gamma}_K$ is the kinetic energy flux density [first term on right side of (2.f.159)], $\boldsymbol{\Gamma}_\Phi$ is the potential energy flux density of the moving midpoint [first term on right side of (2.f.163)], and the third term on the right side of (2.f.167) is the work done on the moving midpoint.

Definition: (2.f.167) is in the form

$$\dot{\mathcal{E}} = -\nabla \cdot \mathbf{S}(\mathbf{x}|\boldsymbol{\Gamma}) \quad (2.f.168)$$

where $\mathbf{S}$ is the total energy flux density given by the sum of the three terms on the right side of (2.f.167) and is analogous to the Poynting vector in electromagnetic theory.

Example: Temperature evolution – Given $T(\mathbf{x})$ at $t = 0$, find $\frac{\partial}{\partial t}T(\mathbf{x}, t)$. Recall that the temperature is related to the entropy by the relation $\beta = \frac{\partial S}{\partial E}$ which can be used to measure $T$ in a very small volume. We note that matter may not flow, but energy can and will. The probability distribution function in phase space consistent with the definition of a local temperature can be expressed as

$$\rho(\boldsymbol{\Gamma})|_{t=0} \sim e^{-\beta E} \to \frac{1}{Z}e^{-\int d^3\mathbf{x}\beta(\mathbf{x})\mathcal{E}(\mathbf{x}|\boldsymbol{\Gamma})} \quad Z \equiv \int d\boldsymbol{\Gamma}\, e^{-\int d^3\mathbf{x}\beta(\mathbf{x})\mathcal{E}(\mathbf{x}|\boldsymbol{\Gamma})} \quad (2.f.169)$$

Following Mori

$$Z(t) \equiv e^{-\int d^3\mathbf{x}\beta(\mathbf{x},t)[<\mathcal{E}>(\mathbf{x},t) - T(\mathbf{x},t)S(\mathbf{x},t)]} \quad (2.f.170)$$

$<\mathcal{E}>(\mathbf{x}, t)$ is not really a function of time. We expand $\rho(\boldsymbol{\Gamma}, t)$ around a local thermal equilibrium for which the entropy is a maximum with respect to the internal energy

$$\rho(\boldsymbol{\Gamma}, t) = \rho_0(\boldsymbol{\Gamma}, t) + \delta\rho(\boldsymbol{\Gamma}, t) \quad \rho_0(\boldsymbol{\Gamma}, t) = \frac{1}{Z(t)}e^{-\int d^3\mathbf{x}\beta(\mathbf{x},t)\mathcal{E}(\mathbf{x}|\boldsymbol{\Gamma})} \quad (2.f.171)$$

We use the Liouville equation $\frac{d\rho}{dt} = 0$, then using (2.f.168) and (2.f.169), integrate by parts, and use the chain rule $\frac{d}{dt}$ applied to $\rho_0(\boldsymbol{\Gamma}, t)$ to obtain

$$\frac{d\delta\rho}{dt} = -\frac{d\rho_0}{dt} = \rho_0 \int d^3x \beta(\mathbf{x},t)\dot{\mathcal{E}}(\mathbf{x}|\mathbf{\Gamma}) = -\rho_0 \int d^3x \beta(\mathbf{x},t)\nabla \cdot \mathbf{S}(\mathbf{x}|\mathbf{\Gamma})$$

$$= \rho_0 \int d^3x\, \mathbf{S}(\mathbf{x}|\mathbf{\Gamma}) \cdot \nabla\beta(\mathbf{x},t) \qquad (2.f.172)$$

We integrate (2.f.172) with respect to time noting that only **S** varies rapidly with time so one can set everything else to its value at *t* = 0:

$$\delta\rho(\mathbf{\Gamma},t) \approx \rho_0 \int d^3x \int_0^t d\tau\, \mathbf{S}(\mathbf{x}|\mathbf{\Gamma},\tau) \cdot \nabla\beta(\mathbf{x},0) \qquad (2.f.173)$$

Going back to (2.f.168) we can take the ensemble average $\langle\dot{\mathcal{E}}\rangle(\mathbf{x},t) = -\nabla \cdot \langle \mathbf{S}\rangle(\mathbf{x},t)$, but the average energy flux density is

$$\mathbf{Q} \equiv \langle \mathbf{S}\rangle(\mathbf{x},t) \equiv \int d\mathbf{\Gamma}\rho(\mathbf{\Gamma},t)\mathbf{S}(\mathbf{x},\mathbf{\Gamma}) = \int d\mathbf{\Gamma}[\rho_0(\mathbf{\Gamma},t) + \delta\rho(\mathbf{\Gamma},t)]\mathbf{S}(\mathbf{x},\mathbf{\Gamma}) =$$

$$\int d\mathbf{\Gamma}\delta\rho(\mathbf{\Gamma},t)\mathbf{S}(\mathbf{x},\mathbf{\Gamma}) = \int d^3x \int_0^t d\tau \int d\mathbf{\Gamma}\rho_0(\mathbf{\Gamma})\mathbf{S}(\mathbf{x}|\mathbf{\Gamma})\mathbf{S}(\mathbf{x}'|\mathbf{\Gamma},\tau) \cdot \nabla\beta(\mathbf{x}',\mathbf{x}) =$$

$$\int d^3x' \int_0^t d\tau\, \mathbf{C}^S(\mathbf{x},\mathbf{x}';\tau) \cdot (-\beta^2 \nabla T(\mathbf{x},t)) \qquad (2.f.174)$$

where we have used that **S** is odd in $\mathbf{\Gamma}$ while $\rho_0$ is even, have assumed that $t \gg \frac{1}{\nu_{coll}}$, have let $\mathbf{x} \to \mathbf{x}'$ so no macroscopic correlations for large $|\mathbf{x}-\mathbf{x}'|$, and used the analysis and notation in Sec. 2.f.v The result in (2.f.174) can be rewritten as (ref. Kubo (1966), Mori (1965))

$$\mathbf{Q} = -\mathbf{K} \cdot \nabla T \qquad (2.f.175a)$$

with the thermal conduction **K** defined by (ref. Mori)

$$\mathbf{K}(\mathbf{x},t) \equiv \beta^2 \int d^3s' \int_0^\infty d\tau\, \mathbf{C}^S(|\mathbf{x}' - \mathbf{x}|; \mathbf{x}; t)\, e^{-i\mathbf{k}\cdot\mathbf{s}' + i\omega\tau}$$

$$= \beta^2\, \mathbf{C}^S(\mathbf{x}, \omega = 0, \mathbf{k} = 0; t) \qquad (2.f.175b)$$

where $\mathbf{s}' = \mathbf{x}' - \mathbf{x}$, and $k \equiv \omega \equiv 0$ in the complex exponential.

[*Editor's Note: The inclusion of the complex exponential in (2.f.175b) is artificial, because with $k \equiv \omega \equiv 0$ the complex exponential is equal to unity. The presence of the complex exponential foreshadows the Fourier transform introduced subsequently in Eqs.(2.f.177b) and (2.f.178).*]

From time reversibility, there is an Onsager symmetry: **K** is symmetric, $K_{xy}=K_{yx}$. To recap the results in (2.f.174) and (2.f.175)

$$\mathbf{Q} \equiv \langle \mathbf{S}\rangle(\mathbf{x},t) = -\mathbf{K}(\mathbf{x},t) \cdot \nabla T(\mathbf{x},t) \qquad (2.f.176)$$

and

$$\mathbf{K}(\mathbf{x}, t) = \beta^2 \int_0^\infty d\tau \int d^3 s \langle \mathbf{S}(\mathbf{x}, t) \mathbf{S}(\mathbf{x} - \mathbf{s}; t - \tau) \rangle \qquad (2.\text{f}.177)$$

due to Mori. $\mathbf{S}(\mathbf{x}, t)$ is the microscopic heat flux in the absence of $\nabla T$, while $\langle \mathbf{S} \rangle(\mathbf{x}, t)$ is the macroscopic heat flux. The correlation $\mathbf{C}^S$ does not fall off exponentially; it only obeys a power law, and convergence is marginally obtained. The process is not Markov which puts the Onsager approach in trouble.

[*Editor's Note: Professor Kaufman's remarks about the fall off of $\mathbf{C}^S$ were not explained nor was a reference provided. However, these remarks have no bearing on the subsequent analysis.*]

Suppose we insert the exponential phase factor $e^{i\omega\tau - i\mathbf{k}\cdot\mathbf{s}}$ inside the two integrals on the right side of (2.f.177) which then yields the Fourier transform $\mathbf{K}(\mathbf{k}, \omega)$ and

$$\langle \mathbf{S} \rangle(\mathbf{k}, \omega) = - \mathbf{K}(\mathbf{k}, \omega) \cdot (\nabla T)(\mathbf{k}, \omega) \qquad (2.\text{f}.178)$$

which is good for a stationary, uniform medium. From (2.f.176) and (2.f.168)

$$\langle \mathbf{S} \rangle(\mathbf{x}, t) = - \mathbf{K} \cdot \nabla T(\mathbf{x}, t) \qquad \langle \dot{\varepsilon} \rangle(\mathbf{x}, t) = -\nabla \cdot \langle \mathbf{S} \rangle(\mathbf{x}, t)$$

$$\rightarrow \frac{\partial \langle \varepsilon \rangle}{\partial t}(\mathbf{x}, t) = \nabla \cdot (\mathbf{K} \cdot \nabla T) = \frac{\partial \langle \varepsilon \rangle}{\partial T}\bigg|_V \frac{\partial T}{\partial t}(\mathbf{x}, t) = K \nabla^2 T \qquad (2.\text{f}.179)$$

assuming $\nabla K = 0$ and isotropy.

<u>Definition</u>: $C_V \equiv \frac{\partial \langle \varepsilon \rangle}{\partial T}\big|_V$ and $D_T \equiv \frac{K}{C_V} \sim \ell \bar{v}$ (ref. Reif or Liboff).

From the definitions of the heat capacity $C_V$ and $D_T$, and (2.f.179), we obtain the diffusion equation for the temperature:

$$\frac{\partial T}{\partial t} = D_T \nabla^2 T \qquad (2.\text{f}.180)$$

<u>Example</u>: Momentum density evolution – Consider the evolution of the momentum density in a one-species system with point particles and central forces. We define the momentum density as

$$\mathbf{g}(\mathbf{x}|\Gamma) \equiv \sum_i \mathbf{p}_i \delta(\mathbf{x} - \mathbf{r}_i), \quad \mathbf{p}_i \equiv m \mathbf{v}_i \qquad (2.\text{f}.181)$$

with evolution equation

$$\dot{\mathbf{g}}(\mathbf{x}|\Gamma) = -\nabla \cdot \Pi^K + \mathbf{F} \qquad (2.\text{f}.182)$$

where the force density **F** is defined by the sum of forces on the particles

$$\mathbf{F}(\mathbf{x}|\Gamma) \equiv \sum_i \mathbf{f}^i \, \delta(\mathbf{x} - \mathbf{r}_i) \tag{2.f.183}$$

and the momentum flux density $\mathbf{\Pi}^K$ is defined by

$$\mathbf{\Pi}^K(\mathbf{x}|\Gamma) = \sum_i \mathbf{v}_i \, \mathbf{p}_i \delta(\mathbf{x} - \mathbf{r}_i) \tag{2.f.184}$$

The force **F** can be related to an interaction stress tensor $\mathbf{\Pi}^F$ using (2.f.183):

$$\mathbf{F}(\mathbf{x}|\Gamma) = -\nabla \cdot \mathbf{\Pi}^F \tag{2.f.185a}$$

$$\mathbf{\Pi}^F(\mathbf{x}|\Gamma) = \sum_{i,j} \boldsymbol{r}_{ij} \mathbf{f}_j^i \frac{1}{2} \, \delta(\boldsymbol{x} - \boldsymbol{r}_{ij}) \tag{2.f.185b}$$

We define a total stress tensor as follows

$$\mathbf{\Pi} \equiv \mathbf{\Pi}^K + \mathbf{\Pi}^F \tag{2.f.186}$$

The evolution equation (2.f.182) then becomes

$$\dot{\mathbf{g}} = -\nabla \cdot \mathbf{\Pi} \tag{2.f.187}$$

and from (2.f.168) we have $\dot{\mathcal{E}} = -\nabla \cdot \mathbf{S}(\mathbf{x}|\Gamma)$. In addition to the momentum and energy relations we also include conservation of particle number density as expressed in the continuity equation:

$$\dot{n} = -\nabla \cdot \widetilde{\boldsymbol{\Gamma}} \tag{2.f.188a}$$

where the flux density $\widetilde{\boldsymbol{\Gamma}}$ is defined by

$$\widetilde{\boldsymbol{\Gamma}}(\mathbf{x}, t) = \sum_i \mathbf{v}_i \, \delta(\mathbf{x} - \mathbf{r}_i) \tag{2.f.188b}$$

From (2.f.188) and the definitions it follows that the fluid flow velocity is given by

$$\mathbf{u}(\mathbf{x}, t) \equiv \frac{\langle \widetilde{\boldsymbol{\Gamma}} \rangle(\mathbf{x}, t)}{\langle n \rangle(\mathbf{x}, t)} \tag{2.f.189}$$

and the continuity equation for the macroscopic fluid quantities can then be expressed as

$$\frac{\partial \langle n \rangle}{\partial t}(\mathbf{x}, t) = -\nabla \cdot \langle \widetilde{\boldsymbol{\Gamma}} \rangle = -\boldsymbol{\nabla} \cdot (\langle n \rangle \mathbf{u}) \tag{2.f.190}$$

From Eqs.(2.f.181) – (2.f.190) the fluid equation of motion is obtained:

$$m\langle n \rangle \left(\frac{\partial}{\partial t} + \mathbf{u} \cdot \nabla\right) \mathbf{u}(\mathbf{x}, t) = -\nabla \cdot \mathbf{P} \tag{2.f.191}$$

where the pressure tensor **P** is

$$\mathbf{P} = \langle \mathbf{\Pi} \rangle - m \langle n \rangle \mathbf{u}\mathbf{u} = \mathbf{P}^K + \mathbf{P}^F \tag{2.f.192}$$

with the kinetic stress tensor $\mathbf{P}^K$

$$\mathbf{P}^K \equiv \sum_i m(\mathbf{v}_i - \mathbf{u}_i)(\mathbf{v}_i - \mathbf{u}_i)\,\delta(\mathbf{x} - \mathbf{r}_i) \tag{2.f.193}$$

and the interaction stress tensor $\mathbf{P}^F$

$$\mathbf{P}^F = \langle \mathbf{\Pi}^F \rangle \tag{2.f.194}$$

Exercise: Fill in the steps in deriving (2.f.192) – (2.f.194).

In thermal equilibrium for an isotropic medium, the pressure tensor **P** simplifies: $\mathbf{P} = \mathrm{P}\mathbf{I}$:

$$\mathbf{P}^K = \langle n \rangle T \mathbf{I} \qquad \mathbf{P}^F = -\frac{2\pi}{3} \langle n \rangle^2 \int_0^\infty s^3 ds \frac{d\phi}{ds} g_2(s) \mathbf{I} \tag{2.f.195}$$

where $g_2(s)$ is the two-particle correlation function (see Sec. 1.b.i).

In a non-equilibrium system one has

$$\mathbf{P} = \mathbf{I}\mathrm{P}\big(\langle n \rangle(\mathbf{x}, t), T(\mathbf{x}, t)\big) + \mathbf{P}^{visc}(\mathbf{x}, t) \tag{2.f.196}$$

if one can define a *local* temperature. In the non-equilibrium system (2.f.191) becomes

$$m \langle n \rangle \left( \frac{\partial}{\partial t} + \mathbf{u} \cdot \nabla \right) \mathbf{u}(\mathbf{x}, t) = -\nabla \mathrm{P} - \nabla \mathbf{P}^{visc} \tag{2.f.197}$$

Using group theory or Mori's approach, one can show for an isotropic fluid:

$$\mathbf{P}^{visc} = -\zeta \mathbf{I} \nabla \cdot \mathbf{u} - \mu (\nabla \mathbf{u})' \tag{2.f.198}$$

where $\zeta$ is the bulk viscosity, $\mu$ or $\eta$ is the shear viscosity, and we define

$$(\nabla \mathbf{u})' \equiv \nabla \mathbf{u} + \mathbf{u}\overleftarrow{\nabla} - \frac{2}{3} \nabla \cdot \mathbf{u} \equiv 2(\nabla \mathbf{u})^{shear} \tag{2.f.199}$$

associated with the shear and note that $(\nabla \mathbf{u})'$ is traceless. The shear stress is associated with a change in shape due to a change in volume or a rotation. The bulk stress is associated with a change volume without a change in shape.

[*Editor's Note: We assume that Professor Kaufman deemed the detailed consideration of the viscous stress tensor was more appropriate for lectures on fluid mechanics and did not have the* *time to take it up in detail here. A good reference for the viscous stress tensor is (Landau and Lifshitz, 1987)]*

Example: $(\nabla \mathbf{u})'$ in Cartesian two dimensions is $(\nabla \mathbf{u})' = \frac{\partial}{\partial x} u_y \hat{\mathbf{y}} + \frac{\partial}{\partial y} u_x \hat{\mathbf{x}}$

Now we pause and make some order of magnitude estimates of the transport coefficients in which we relate them to other quantities:

thermal conduction $K \sim \frac{\langle \mathbf{v} \rangle}{\sigma_{coll}} \sim nD$ where the diffusion coefficient $D \sim \ell \langle \mathbf{v} \rangle$ shear viscosity $\mu \sim \frac{P}{\nu} \sim \frac{T}{\sigma \langle \mathbf{v} \rangle} \sim \frac{m \langle \mathbf{v} \rangle}{\sigma} \sim mK$ bulk viscosity $\zeta \sim \begin{cases} 0 & \text{for a dilute gas of particles} \\ \frac{P}{\nu_{relax}} & \text{for a dense gas or liquid with internal degrees of freedom} \end{cases}$

Note: The bulk viscosity coefficient for a dense gas or liquid depends on the degrees of freedom (Landau and Lifshitz, 1987).

We return to (2.f.191) rewritten as

$$m \langle n \rangle \frac{D}{Dt} \mathbf{u}(\mathbf{x}, t) = -\nabla P(n, T) - \nabla \cdot [-\zeta(n, T) \mathbf{I} \nabla \cdot \mathbf{u} - 2\mu(n, T)(\nabla \mathbf{u})^{shear}] \qquad (2.f.200)$$

where $\frac{D}{Dt} \equiv \frac{\partial}{\partial t} + \mathbf{u} \cdot \nabla$ From $\dot{\mathcal{E}} = -\nabla \cdot \mathbf{S}(\mathbf{x}|\Gamma)$ (2.f.168) and the definition of the specific internal energy $\mathcal{U}_m \equiv \frac{energy}{mass}$,

$$m \langle n \rangle \mathcal{U}_m \equiv \langle \mathcal{E} \rangle - \frac{1}{2} m \langle n \rangle |\mathbf{u}|^2 \qquad (2.f.201)$$

which is just the energy in the moving frame. The internal heat flow is the heat flow in the moving frame of the flow, *i.e.*, the internal heat flow is the heat flow in the laboratory frame with flow terms subtracted off:

$$\mathbf{Q}^{int} = \langle \mathbf{S} \rangle - \mathbf{u} \langle \mathcal{E} \rangle - \mathbf{P} \cdot \mathbf{u} = -K \nabla T \qquad (2.f.202)$$

It then follows $\qquad m \langle n \rangle \frac{D}{Dt} \mathcal{U}_m = -\nabla \cdot \mathbf{Q}^{int} - \mathbf{P} : \nabla \mathbf{u} \qquad (2.f.203)$

and
$$-\mathbf{P} : \nabla \mathbf{u} = -P \nabla \cdot \mathbf{u} + \zeta (\nabla \cdot \mathbf{u})^2 + 2\mu (\nabla \mathbf{u})^{shear} : (\nabla \mathbf{u})^{shear} \qquad (2.f.204)$$

where the first term on the right side of (2.f.204) is the adiabatic cooling or heating, the second term is the heating due to bulk viscosity, and the third term is the heating due to shear viscosity (: denotes the sum of squares of all components, $\nabla u_{ij} \nabla u_{ij}$).

The specific entropy is $\mathcal{S}_m(\mathcal{U}_m, V_m)$ where $V_m \equiv \frac{1}{\rho} = \frac{1}{m\langle n \rangle}$ satisfies an equation

$$d\mathcal{S}_m = \beta d\mathcal{U}_m + \beta P dV_m \quad (2.f.205)$$

from which follows

$$\frac{D}{Dt}\mathcal{S}_m = \beta \frac{D}{Dt}\mathcal{U}_m - \frac{\beta P}{m\langle n \rangle^2} \frac{D}{Dt}\langle n \rangle \quad (2.f.206)$$

and with use of (2.f.202) and (2.f.203)

$$\frac{D}{Dt}\mathcal{S}_m = -\beta \nabla \cdot \mathbf{Q} - \beta \mathbf{P}^{visc} : \nabla \mathbf{u} \quad (2.f.207)$$

where A:B=$A_{ij}B_{ij}$. Defining the entropy density $\mathcal{S}_V \equiv m\langle n \rangle \mathcal{S}_m$ one can show

$$\frac{\partial}{\partial t}\mathcal{S}_V(\mathbf{x}, t) = -\nabla \cdot (\mathbf{u}\mathcal{S}_V + \beta \mathbf{Q}) + \dot{\mathcal{S}}_V \quad (2.f.208)$$

where

$$\mathbf{\Gamma}^{\mathcal{S}} \equiv \mathbf{u}\mathcal{S}_V + \beta \mathbf{Q} \quad \text{and} \quad \dot{\mathcal{S}}_V = \beta^2 K (\nabla T)^2 + \beta \zeta (\nabla \cdot u)^2 + 2\beta \mu (|\nabla u|^{shear})^2 \quad (2.f.209)$$

To be consistent with the 2$^{nd}$ law of thermodynamics the source term $\dot{\mathcal{S}}_V \geq 0$; thus $\zeta, \mu$, and $K$ are all non-negative. Necessarily $\beta > 0$ in this classical theory.

### 2.f.viii Normal mode solutions of the transport equations

We next analyze the linear normal modes supported by the transport equations. For this purpose we assume that the system is uniform and isotropic in space stretching to infinity. We assume infinitesimal-amplitude perturbations and examine both microscopic and macroscopic modes. Perturbed amplitudes will all have the form $A(x,t) = \tilde{A}e^{i\mathbf{k}\cdot\mathbf{x} - i\omega t}$.

<u>Definitions</u>: Polarizations – Longitudinal modes have **k** and velocity perturbation **u** parallel to one another. Moreover, a purely longitudinal mode is irrotational (curl free) and has a divergence. A compressional wave is a longitudinal wave and has a finite density perturbation. A transverse wave has **k** and velocity perturbation **u** perpendicular to one another. A purely transverse wave has a finite curl and is divergence free. In a uniform, isotropic medium there is no coupling between longitudinal and transverse waves.

In the macroscopic theory all equations are for mean values (ensemble averages have removed random fluctuations). An example of a simple linear equations set is as follows.

$$\mathbf{g} = mn_0\mathbf{u} = \rho_0\mathbf{u} \quad \mathbf{u}_0 = 0 \quad \frac{\partial \mathbf{g}}{\partial t} = -\nabla \cdot \delta\mathbf{\Pi} \quad \mathbf{\Pi} = \mathbf{\Pi}_0 + \delta\mathbf{\Pi} \tag{2.f.210}$$

After Fourier analyzing in time and space, (2.f.210) becomes

$$-i\omega\mathbf{g} = -i\mathbf{k} \cdot \delta\mathbf{\Pi} \;\rightarrow\; \omega\rho_0\mathbf{u} = \mathbf{k} \cdot \delta\mathbf{\Pi} \;\rightarrow\; \omega\rho_0\mathbf{u} = \mathbf{k} \cdot \delta\mathbf{P} \tag{2.f.211}$$

using $\mathbf{P} = \mathbf{\Pi} - \rho\mathbf{uu} \approx P\mathbf{I} + \mathbf{P}^{visc}$ because $\rho\mathbf{uu}$ is higher order.

<u>Example</u>: Shear mode – A shear mode is a transverse wave, for which $\mathbf{k} \cdot$ all of the perturbed quantities in (2.f.211) vanish. However, $\mathbf{k} \times$ on (2.f.211) yields

$$\omega\rho_0 \mathbf{k} \times \mathbf{u} = \mathbf{k} \times \delta\mathbf{P}^{visc} \cdot \mathbf{k} \tag{2.f.212}$$

However, from (2.f.198) $\delta\mathbf{P}^{visc} = -\zeta\mathbf{I}\nabla \cdot \mathbf{u} - \mu(\nabla\mathbf{u})'$. The bulk viscosity term does not contribute because $\mathbf{k} \times \mathbf{k} = 0$, which leaves the shear viscosity term:

$$\omega\rho_0 \mathbf{k} \times \mathbf{u} = -i\mu_0 \mathbf{k} \times \mathbf{u}\, k^2 \tag{2.f.213}$$

Hence, the dispersion relation for the shear wave is

$$\omega = -i\frac{\mu_0}{\rho_0}k^2 = -iD_{sh}k^2 \tag{2.f.214}$$

where we have introduce $D_{sh} = \frac{\mu_0}{\rho_0}$ which is called the kinematic viscosity and has units of a spatial diffusivity. Thus, the shear mode just decays. The vorticity $\nabla \times \mathbf{u} = \mathbf{\Omega}$ is a shear mode which just decays in a liquid: $\frac{\partial \mathbf{\Omega}}{\partial t} = D_{sh}\nabla^2\mathbf{\Omega}$ (ref. H. Helmholtz). In a solid, vorticity may propagate. Note that the transport coefficients here have been assumed to be frequency independent. The correction for shear viscosity that has dependence on frequency and wavenumber is $\mu_0(\mathbf{k},\omega) \rightarrow \mu_0'(\mathbf{k},\omega) + i\mu_0"(\mathbf{k},\omega)$, which will allow $Re\,\omega \neq 0$; and then the shear mode may be able to propagate as well as just damp out.

<u>Example</u>: Compressional wave – To analyze the compressional wave we take the dot product of (2.f.211) with $\mathbf{k}$

$$\omega\rho_0\mathbf{u} \cdot \mathbf{k} = \mathbf{k} \cdot \delta\mathbf{\Pi} \cdot \mathbf{k} = k^2\delta P + \mathbf{k} \cdot [-\zeta\mathbf{I}i\mathbf{k}\cdot\mathbf{u} - \mu(i\mathbf{k}\mathbf{u} + i\mathbf{u}\mathbf{k} - \tfrac{2}{3}\mathbf{I}i\mathbf{k}\mathbf{u}] \cdot \mathbf{k} =$$

$$k^2\left(\delta P - i\mathbf{k}\cdot\mathbf{u}(\zeta + \tfrac{4}{3}\mu)\right) \tag{2.f.215}$$

which has solution

$$\mathbf{k} \cdot \mathbf{u} = \frac{k^2 \delta P}{\omega \rho_0 + ik^2(\zeta + \frac{4}{3}\mu)} \tag{2.f.216}$$

**Definition**: Let $D_\zeta \equiv \frac{(\zeta + \frac{4}{3}\mu)}{\rho_0}$

Hence,
$$\mathbf{k} \cdot \mathbf{u} \omega \rho_0 = \frac{k^2 \delta P}{1 + i\frac{k^2}{\omega} D_\zeta} \tag{2.f.217}$$

From the linearized continuity equation to lowest order:

$$-\frac{\partial \rho}{\partial t} = \nabla \cdot (\rho \mathbf{u}) = \rho \nabla \cdot \mathbf{u} + \mathbf{u} \cdot \nabla \rho = \rho_0 \nabla \cdot \mathbf{u} \;\rightarrow\; i\omega \delta \rho = \rho_0 i \mathbf{k} \cdot \mathbf{u} \tag{2.f.218}$$

Combining (2.f.217) and (2.f.218)

$$\mathbf{k} \cdot \mathbf{u} = \frac{\omega \delta \rho}{\rho_0} = \frac{\delta P}{\omega \rho_0} \frac{k^2}{1 + i\frac{k^2}{\omega} D_\zeta} \;\rightarrow\; \frac{\delta P}{\delta \rho} = \frac{\omega^2}{k^2}\left(1 + i\frac{k^2}{\omega} D_\zeta\right) \tag{2.f.219}$$

From the entropy equation (2.f.207) $\frac{D}{Dt}\mathcal{S}_m = -\beta \nabla \cdot \mathbf{Q} - \beta \mathbf{P}^{visc} : \nabla \mathbf{u}$ and from (2.f.202) $\mathbf{Q} = -K \nabla T$. We linearize and keep only lowest-order terms:

$$\rho_0(-i\omega)\delta \mathcal{S}_m = -\beta_0 i\mathbf{k} \cdot (-K) i\mathbf{k} \delta T = -k^2 K \frac{\delta T}{T_0} \tag{2.f.220}$$

from which we have

$$\delta \mathcal{S}_m = \frac{k^2 K}{i\omega \rho_0} \frac{\delta T}{T_0} = \left(\frac{k^2 D_T}{i\omega}\right)\frac{C_V \delta T}{T_0}$$

where we introduce the definition $D_T \equiv \frac{K}{C_V \rho_0}$. Next we supplement (2.f.220) with the basic thermodynamic relation

$$\mathcal{S}_m(\rho, T) \;\rightarrow\; \delta \mathcal{S}_m = \left.\frac{\partial \mathcal{S}_m}{\partial \rho}\right|_T \delta \rho + \left.\frac{\partial \mathcal{S}_m}{\partial T}\right|_\rho \delta T \tag{2.f.221}$$

where $\mathcal{S}_m \equiv -\partial F(\rho,T)/\partial T$ and $dF(\rho,T) = -\mathcal{S}_m dT - Pd\rho/\rho^2$, i.e., $P(\rho,T) \equiv -\rho^2 \partial F(\rho,T)/\partial \rho$. In addition, the specific heat at constant volume is defined as $C_V \equiv T(\partial \mathcal{S}_m/\partial T)_\rho$ so that Eq.(2.f.221) yields

$$\delta \mathcal{S}_m(\rho, T) = \left.\frac{\partial \mathcal{S}_m}{\partial \rho}\right|_T \delta \rho + \frac{C_V \delta T}{T_0}$$

and by equating expressions for $\delta S_m$ we obtain

$$\left.\frac{\partial S_m}{\partial \rho}\right|_T \delta\rho = -\left(1 - \frac{k^2 D_T}{i\omega}\right)\frac{C_V \delta T}{T_0} \tag{2.f.222}$$

which implies

$$\delta S_m = -\frac{k^2 D_T}{i\omega}\left(1 - \frac{k^2 D_T}{i\omega}\right)^{-1}\left.\frac{\partial S_m}{\partial \rho}\right|_T \delta\rho$$

From the equation of state for $P(S_m, \rho)$

$$\delta P = \left.\frac{\partial P}{\partial S_m}\right|_\rho \delta S_m + \left.\frac{\partial P}{\partial \rho}\right|_{S_m} \delta\rho \tag{2.f.223}$$

which now yields

$$\frac{\delta P}{\delta \rho} = C_s^2 \left[1 - \left.\frac{\partial \rho}{\partial P}\right|_{S_m}\left.\frac{\partial P}{\partial S_m}\right|_\rho \left.\frac{\partial S_m}{\partial \rho}\right|_T \frac{k^2}{i\omega}D_T\left(1 - \frac{k^2}{i\omega}D_T\right)^{-1}\right] \tag{2.f.224}$$

where we have introduced the square of the sound speed $C_s^2 \equiv \left.\frac{\partial P}{\partial \rho}\right|_{S_m}$. Lastly, using the Maxwell identity (A.4) shown in the Appendix (*Editor's Addendum*), Eq.(2.f.224) becomes

$$\frac{\delta P}{\delta \rho} = C_s^2 \left[1 + \left(1 - \frac{1}{\gamma}\right)\frac{k^2}{i\omega}D_T\left(1 - \frac{k^2}{i\omega}D_T\right)^{-1}\right] \tag{2.f.225}$$

where $\gamma \equiv C_p/C_V$.

Equating the expressions for $\frac{\delta P}{\delta \rho}$ in (2.f.219) and (2.f.225) one obtains the dispersion relation for compressional waves:

$$\frac{\omega^2}{k^2}\left(1 + i\frac{k^2}{\omega}D_\zeta\right) = C^2\left[1 + \left(1 - \frac{1}{\gamma}\right)\frac{k^2}{i\omega}D_T\left(1 - \frac{k^2}{i\omega}D_T\right)^{-1}\right] \tag{2.f.226}$$

We solve (2.f.226) analytically in two simple limits.

Example: High-frequency sound waves in which $\omega \gg k^2 D_{sound}$, i.e., the wave oscillation period is short compared to the characteristic diffusion time for the sound wave:

$$\frac{\omega}{k} = C - \frac{i}{2}kD_{sound} \tag{2.f.227}$$

where $D_{sound} \equiv D_\zeta + \left(1 - \frac{1}{\gamma}\right) D_T$. The $-\frac{i}{2} k D_{sound}$ term in (2.f.227) sets the temporal or spatial damping rate. The expansion of (2.f.226) leading to the solution in (2.f.227) is valid if the wavelength is long compared to the mean free path; the wave attenuation rate must be weak.

Example: Low-frequency thermal mode $\omega \ll kC$. We solve (2.f.226) by successive approximations to obtain

$$\omega = \frac{-ik^2 D_T}{\gamma} \qquad (2.f.228)$$

Exercise: Show that the sound wave is isentropic. Show that the thermal wave is approximately isobaric. Show that the shear mode is isochoric.

## 2.f.ix Generalized Langevin method for transport relations -- sketch

Here we present a sketch of a generalized Langevin method for obtaining the transport relations. References for the formalism are found in (Landau and Lifshitz, 1969) Secs.121-124 and 126-127, and (Landau and Lifshitz, 1987) Chapter XVII.

We begin with $\dot{\mathbf{g}} = -\nabla \cdot \mathbf{\Pi}$, Eq.(2.f.187), which we expand

$$\frac{\partial \mathbf{g}}{\partial t} = -\nabla \cdot \mathbf{\Pi}(\mathbf{x}|\Gamma) = -\nabla \cdot [\langle \mathbf{\Pi}(\mathbf{x}|t)\rangle + \delta \mathbf{\Pi}(\mathbf{x}|\Gamma)] = -\nabla \cdot [\langle \mathbf{\Pi}\rangle_0 + \delta\langle \mathbf{\Pi}\rangle + \delta \mathbf{\Pi}(\mathbf{x}|\Gamma)] \quad (2.f.229)$$

Just to simplify the analysis we replace $\nabla \cdot \mathbf{\Pi}$ with $\nabla P$ and pretend this is valid, i.e.,

$$\frac{\partial \mathbf{g}}{\partial t} = -\nabla(\langle P\rangle + \delta P) \qquad (2.f.230)$$

Suppose for a linear sound wave $\mathbf{g} = \rho_0 \mathbf{u}$ with $\nabla \times \mathbf{u} = 0$ and $\mathbf{u} = -\nabla \tilde{\phi}$, where $\tilde{\phi}$ is the fluid velocity scalar potential function, then

$$\rho_0 \frac{\partial \tilde{\phi}}{\partial t} = \delta\langle P\rangle + \delta P = \left.\frac{dP}{d\rho}\right|_{S_m} \delta\langle \rho\rangle + \delta P \qquad (2.f.231)$$

where $\delta\langle P\rangle$ is that part of $\langle P\rangle$ that varies in space and time. We Fourier analyze and obtain

$$-i\omega \rho_0 \tilde{\phi}_{\mathbf{k}\omega} = C^2 \langle \rho\rangle_{\mathbf{k}\omega} + \delta P_{\mathbf{k}\omega} \qquad (2.f.232)$$

From continuity

$$-\frac{\partial \langle \rho\rangle}{\partial t} = \rho_0 \nabla \cdot \mathbf{u} \;\rightarrow\; i\omega \langle \rho\rangle_{\mathbf{k}\omega} = \rho_0 k^2 \tilde{\phi}_{\mathbf{k}\omega} \qquad (2.f.233)$$

we solve for $\tilde{\phi}_{\mathbf{k}\omega}$ and reduce (2.f.232) to

$$\langle\rho\rangle_{\mathbf{k}\omega} = \frac{k^2 \delta P_{\mathbf{k}\omega}}{\omega^2 - k^2 C^2} \tag{2.f.234}$$

In terms of spectral densities Eq.(2.f.234) leads to

$$S^\rho(\mathbf{k},\omega) = \frac{k^4 S^p(\mathbf{k},\omega)}{(\omega^2 - k^2 C^2)^2} \tag{2.f.235}$$

In (2.f.235) $\pm kC$ are eigenfrequencies $\omega_\mathbf{k}$ for the linear modes. We claim that with dissipation included the right side of (2.f.235) can be generalized to the form

$$\sum_\alpha \frac{const}{|\omega - \omega_\mathbf{k}^\alpha|^2} \tag{2.f.236}$$

for complex $\omega_\mathbf{k}^\alpha$. This analysis and expressions obtained are the analog of the Langevin method used for Brownian motion in Secs. 2.b.i and 2.b.ii  We can argue using Rayleigh-Jeans that the density fluctuations have an energy spectrum *kT* per mode and evaluate the constant in (2.f.236) indirectly.

In the Kubo/Mori approach (Sec. 2.f.v) we used a Hamiltonian

$$H = H_0 + \int d^3\mathbf{x}\, n(\mathbf{x}|\mathbf{\Gamma})\, \tilde{\phi}^{ext}(\mathbf{x},t) = H_0 + \int d^3\mathbf{x}\, \sum_i \delta(\mathbf{x} - \mathbf{r}_i)\, \tilde{\phi}^{ext}(\mathbf{x},t) \tag{2.f.237}$$

and obtained a kinetic equation for the response

$$\delta\langle n\rangle(\mathbf{x},t) = \int d^3\mathbf{x}' \int dt'\, G\tilde{\phi}^{ext}(\mathbf{x}',t') \tag{2.f.238}$$

where $G$ is a Green's function. We then obtained a fluctuation-dissipation theorem relating $G$ to the spectral density $S^{\delta n}$ having derived $G$ from just the linearized hydrodynamic equations (linearized with respect to the external field strength).

## 2.g Non-equilibrium quantum statistical mechanics

References for this discussion of non-equilibrium quantum statistical mechanics are (Tolman, 1938); (Kubo, 1965) Chapter 2; (de Boer, Uhlenbeck, and McLennan, 1962); (Cohen, 1962); and (de Groot and Suttorp, 1972).

We begin by introducing a representation of the state function $|\psi(t)\rangle$ in terms of a decomposition in terms of basis functions that are assumed to be a complete set

$$|\psi(t)\rangle = \sum_n c_n(t)|\psi_n\rangle \qquad \sum_n |c_n|^2 = 1 \tag{2.g.1}$$

Definition: Consider any Hermitian operator **A**

$$\langle A \rangle \equiv \langle \psi|A|\psi \rangle = \sum_{m,n} c_m^* c_n \langle \psi_m|A|\psi_n \rangle \equiv \sum_{m,n} c_m^* c_n A_{mn} \tag{2.g.2}$$

Definition: The ensemble average of $\langle A \rangle$, i.e., the quantum statistical average, is defined as

$$\langle\langle A \rangle\rangle \equiv \overline{\langle \psi|A|\psi \rangle} = \sum_{m,n} \overline{c_m^*(t) c_n(t)} \, A_{mn} \tag{2.g.3}$$

Definition: The density matrix is defined by

$$\rho_{mn}(t) \equiv \overline{c_m^*(t) c_n(t)} \tag{2.g.4}$$

We note $\rho_{mn}(t)$ has the properties that it is Hermitian, positive-definite, and $\mathrm{Tr}(\rho_{mn}) = 1$.

$$\langle\langle A \rangle\rangle = \sum_{m,n} \rho_{mn}(t) A_{mn} \text{ and } \langle A \rangle(t) = \mathrm{Tr}(\rho(t) A) \tag{2.g.5}$$

How $\rho(t)$ varies in time is determined by the Hamiltonian *H(t)*:

$$\frac{\partial \rho}{\partial t} = -\frac{1}{i\hbar}[\rho, H] \tag{2.g.6}$$

(2.g.6) is the quantum mechanical analog of the Liouville equation. Remember that for **A** a time-independent operator $\dot{\mathbf{A}} = -\frac{1}{i\hbar}[\mathbf{A}, H]$; and if **A** is time dependent then we include $\frac{\partial \mathbf{A}}{\partial t}$ additively.

It follows that 
$$\frac{d}{dt}\langle A \rangle(t) = \mathrm{Tr}\left(\frac{\partial \rho}{\partial t} A\right) = \mathrm{Tr}(\rho \dot{A}) = \langle \dot{A} \rangle \tag{2.g.7}$$

Correspondence due to Wigner and Weyl:

$$H(P, Q); \; [P, Q] = \frac{\hbar}{i} \tag{2.g.8}$$

for one degree of freedom. In (2.g.8) *P* and *Q* are operators. The phase-space coordinates are denoted by *p* and *q*. We next define the Weyl transform |A(P,Q)| but first note that

$$\langle q'|A|q''\rangle \equiv \int dq''' \delta(q''' - q') A\left(\frac{\hbar}{i}\frac{\partial}{\partial q'''}, q'''\right) \delta(q''' - q'') \tag{2.g.9}$$

for *A* in the *q* representation.

Definition: The Weyl transform is defined

$$a(p, q) = \int ds \, e^{\frac{i}{\hbar}ps} \langle q - \tfrac{1}{2}s | A(P, Q) | q + \tfrac{1}{2}s \rangle \tag{2.g.10}$$

and its inverse (Wigner transform) is

$$A(P,Q) \equiv \int dp \int dq\ \delta(q-Q)\delta(p-P)e^{\frac{\hbar}{2i}\frac{\partial^2}{\partial p \partial q}}\,a(p,q) \qquad (2.g.11)$$

with the following prescriptions on the correspondence of independent variables:

$$p \leftrightarrow P \qquad (2.g.12a)$$

$$pq \leftrightarrow \frac{1}{2}(PQ + QP) \qquad (2.g.12b)$$

$$p^2q^2 \leftrightarrow \frac{1}{4}(P^2Q^2 + Q^2P^2 + 2PQ^2P) \qquad (2.g.12c)$$

$$\frac{2}{\hbar}\left[\sin\frac{\hbar}{2}\left(\frac{\partial a}{\partial q}\frac{\partial b}{\partial p} - \frac{\partial b}{\partial q}\frac{\partial a}{\partial p}\right)\right] \leftrightarrow \frac{1}{i\hbar}[A,B] \qquad (2.g.12d)$$

where $\frac{\partial a}{\partial q}$ is the partial derivative with respect to $q$ operating on $a$. We note that as $\hbar \to 0$ the left side of (2.g.12d) recovers the Poisson bracket $\{a,b\}$.

<u>Definition</u>: The Wigner function is defined as the Weyl transform of $\rho(t)$, i.e.,

$$\text{Weyl transform of } \rho(t) \text{ density matrix } \to \rho^{Wigner}(p,q;t) \qquad (2.g.13)$$

The Wigner function is like a density in phase space:

$$\rho(q;t) = \int dp\ \rho^{Wigner}(p,q;t) \qquad (2.g.14)$$

$\rho^{Wigner}(p,q;t)$ is real and normed, but can be negative. $\rho(q;t)$ is the correct quantum mechanical probability distribution with respect to $q$. Similarly,

$$\rho(p;t) = \int dq\ \rho^{Wigner}(p,q;t) \qquad (2.g.15)$$

is the correct quantum mechanical probability distribution with respect to $p$. Furthermore, it can be show that $\rho^{Wigner}$ is bounded:

$$\left|\rho^{Wigner}(p,q;t)\right| \leq \frac{2}{h} \qquad (2.g.16)$$

and using the Wigner function

$$\sigma_p \sigma_q \geq \frac{\hbar}{2} \qquad (2.g.17)$$

The equation of evolution for the Wigner function using the Weyl transformation is

$$\tfrac{\partial}{\partial t}\rho^W(p,q;t) = \tfrac{2}{\hbar}\left[\sin\tfrac{\hbar}{2}\left(\tfrac{\partial^{\mathcal{H}}}{\partial q}\tfrac{\partial \rho^W}{\partial p} - \tfrac{\partial \rho^W}{\partial q}\tfrac{\partial^{\mathcal{H}}}{\partial p}\right)\right]\mathcal{H}(p,q;t)\rho^W(p,q;t) \qquad (2.g.18)$$

where the superscripts on the partial derivatives give guidance on what functions the partial derivatives operate in the expression that follows, $\mathcal{H}$ is the Weyl transform of the quantum mechanical Hamiltonian $H$, and only leading terms have been retained, which means only slow variations in $\mathcal{H}(p,q;t)\rho^W(p,q;t)$ are kept. In the limit $\hbar \to 0$ (2.g.18) becomes $\tfrac{\partial}{\partial t}\rho^W = -\{\rho^W, \mathcal{H}\}$. (2.g.18) is a Liouville equation that allows us to do everything on the Wigner function that we did on the classical probability distribution in Sec. 2 of these lecture notes: all of the methods go through.

[*Editors' Note: This was an elegant conclusion to Kaufman's graduate statistical mechanics lectures.*]

[**Editor's Addendum: Appendix -- Thermodynamic Potentials, Maxwell Relations and Identities**

**Thermodynamic Potentials**

*Classical thermodynamics is expressed in terms of four variables: Pressure P and Volume V as one conjugate pair, Temperature T and Entropy S as a second conjugate pair. Much of the thermodynamic analysis is based on which pair of thermodynamic variables are considered independent, and which pair of thermodynamic variables are considered dependent. Each pair of independent thermodynamic variables $(X,Y)$ is associated with a thermodynamic potential $\Psi(X,Y)$, with its defining differential relation $d\Psi(X,Y) \equiv (\partial \Psi/\partial X)_Y \, dX + (\partial \Psi/\partial Y)_X \, dY$, as follows*

Internal Energy $\quad U(S,V) \to dU = T(S,V)dS - P(S,V)dV$,

Helmholtz Free Energy $\quad F(T,V) \equiv U - ST \to dF(T,V) = -S(T,V)dT - P(T,V)dV$,

Enthalpy $\quad H(S,P) \equiv U + PV \to dH(S,P) = T(S,P)dS + V(S,P)dP$,

Gibbs free energy $\quad G(T,P) = U + PV - ST \to dG(T,P) = -S(T,P)dT + V(T,P)dP$,

*where the thermodynamic potentials $(U, F, H, G)$ are related by Legendre transformation associated with the substitution $S \to T$, or $V \to P$, or both. Hence, the thermodynamic variables $(S,T\,;\,P,V)$ can be seen to be either independent variables or dependent functions.*

**Maxwell Relations**

Because of the symmetry of partial derivatives $\partial^2 \Psi(X,Y)/\partial X \partial Y \equiv \partial^2 \Psi(X,Y)/\partial Y \partial X$, we naturally arrive at the Maxwell relations

$$\left(\frac{\partial T}{\partial V}\right)_S = \frac{\partial^2 U}{\partial V \partial S} = -\left(\frac{\partial P}{\partial S}\right)_V,$$

$$\left(\frac{\partial S}{\partial V}\right)_T = -\frac{\partial^2 F}{\partial V \partial T} = \left(\frac{\partial P}{\partial T}\right)_V,$$

$$\left(\frac{\partial T}{\partial P}\right)_S = \frac{\partial^2 H}{\partial P \partial S} = \left(\frac{\partial V}{\partial S}\right)_P,$$

$$\left(\frac{\partial S}{\partial P}\right)_T = -\frac{\partial^2 G}{\partial P \partial T} = -\left(\frac{\partial V}{\partial T}\right)_P,$$

where $(\partial X/\partial Y)_Z$ denotes the partial derivative of $X(Y,Z)$ with respect to $Y$ at constant $Z$.

### Maxwell Identities

The Maxwell relations lead to the following identity involving three thermodynamic variables $(X,Y,Z)$:

$$\left(\frac{\partial X}{\partial Y}\right)_Z \left(\frac{\partial Y}{\partial Z}\right)_X \left(\frac{\partial Z}{\partial X}\right)_Y = -1$$

For example, consider the identity

$$\left(\frac{\partial V}{\partial P}\right)_S \left(\frac{\partial P}{\partial S}\right)_V \left(\frac{\partial S}{\partial V}\right)_P = -1,$$

which can be proved from the Maxwell relations as follows. First we use the Maxwell relation $(\partial P/\partial S)_V = -(\partial T/\partial V)_S$ so that

$$\left(\frac{\partial V}{\partial P}\right)_S \left(\frac{\partial P}{\partial S}\right)_V = -\left(\frac{\partial V}{\partial P}\right)_S \left(\frac{\partial T}{\partial V}\right)_S = -\left(\frac{\partial T}{\partial P}\right)_S$$

which makes use of the identity $(\partial X/\partial Y)_\xi (\partial Y/\partial Z)_\xi \equiv (\partial X/\partial Z)_\xi$. Next, we use the Maxwell relation $(\partial S/\partial V)_P = (\partial P/\partial T)_S$, and we obtain

$$-\left(\frac{\partial T}{\partial P}\right)_S \left(\frac{\partial S}{\partial V}\right)_S = -\left(\frac{\partial T}{\partial P}\right)_S \left(\frac{\partial P}{\partial T}\right)_S \equiv -1$$

which makes use of the identity $(\partial X/\partial Y)_\xi (\partial Y/\partial X)_\xi \equiv 1$.

### Equations (2.f.224) – (2.f.225)

*In Eq.(2.f.224), we find the triple product (returning to $V = \rho^{-1}$)*

$$\left(\frac{\partial \rho}{\partial P}\right)_S \left(\frac{\partial P}{\partial S}\right)_\rho \left(\frac{\partial S}{\partial \rho}\right)_T = \left[\left(\frac{\partial \rho}{\partial P}\right)_S \left(\frac{\partial P}{\partial S}\right)_\rho \left(\frac{\partial S}{\partial \rho}\right)_P\right] \left(\frac{\partial \rho}{\partial S}\right)_P \left(\frac{\partial S}{\partial \rho}\right)_T = -\left(\frac{\partial \rho}{\partial S}\right)_P \left(\frac{\partial S}{\partial \rho}\right)_T \quad (A.1)$$

*where we use a Maxwell identity for $(\rho, P, S)$ with the identity $(\partial S/\partial \rho)_P (\partial \rho/\partial S)_P \equiv 1$. Next we use the partial derivative identity*

$$\left(\frac{\partial S}{\partial \rho}\right)_T = \left(\frac{\partial S[\rho, P(T,\rho)]}{\partial \rho}\right)_T = \left(\frac{\partial S}{\partial \rho}\right)_P + \left(\frac{\partial S}{\partial P}\right)_\rho \left(\frac{\partial P}{\partial \rho}\right)_T,$$

*so that Eq.(A.1) becomes*

$$-\left(\frac{\partial \rho}{\partial S}\right)_P \left(\frac{\partial S}{\partial \rho}\right)_T = -\left(\frac{\partial \rho}{\partial S}\right)_P \left[\left(\frac{\partial S}{\partial \rho}\right)_P + \left(\frac{\partial S}{\partial P}\right)_\rho \left(\frac{\partial P}{\partial \rho}\right)_T\right]$$

$$= -1 - \left(\frac{\partial \rho}{\partial S}\right)_P \left(\frac{\partial S}{\partial P}\right)_\rho \left(\frac{\partial P}{\partial \rho}\right)_T \quad (A.2)$$

*where we have used the identity $(\partial S/\partial \rho)_P (\partial \rho/\partial S)_P \equiv 1$ again. We now introduce the specific heat capacities at constant volume $C_V \equiv T(\partial S/\partial T)_\rho$ and constant pressure $C_P \equiv T(\partial S/\partial T)_P$, so that we obtain*

$$\left(\frac{\partial \rho}{\partial S}\right)_P \left(\frac{\partial S}{\partial P}\right)_\rho = \frac{C_V}{C_P} \left(\frac{\partial \rho}{\partial T}\right)_P \left(\frac{\partial T}{\partial P}\right)_\rho \equiv \frac{1}{\gamma} \left(\frac{\partial \rho}{\partial T}\right)_P \left(\frac{\partial T}{\partial P}\right)_\rho,$$

*where $\gamma \equiv C_P/C_V$ denotes the ratio of specific heat capacities and Eq.(A.2) becomes*

$$-1 - \left(\frac{\partial \rho}{\partial S}\right)_P \left(\frac{\partial S}{\partial P}\right)_\rho \left(\frac{\partial P}{\partial \rho}\right)_T = -1 - \frac{1}{\gamma} \left(\frac{\partial \rho}{\partial T}\right)_P \left(\frac{\partial T}{\partial P}\right)_\rho \left(\frac{\partial P}{\partial \rho}\right)_T \quad (A.3)$$

*Lastly, we use the Maxwell identity for $(\rho, T, P)$, so that Eqs.(1)-(3) are combined to yield*

$$\left(\frac{\partial \rho}{\partial P}\right)_S \left(\frac{\partial P}{\partial S}\right)_\rho \left(\frac{\partial S}{\partial \rho}\right)_T = -\left(1 - \frac{1}{\gamma}\right) \quad (A.4)$$

*which is now inserted in Eq.(2.f.224) to obtain Eq.(2.f.225).*